\documentclass[amssymb, superscriptaddress, nobibnotes, aps, prd]{revtex4-2}

\usepackage{graphicx}
\usepackage{xcolor,hyperref}
\usepackage{float}
\usepackage{amsmath,amssymb,bm}
\usepackage{subcaption}
\usepackage{adjustbox}

\newcommand{\qsc}{q_\mathrm{sc}}
\newcommand{\nusc}{\Delta\nu_\mathrm{inc}}
\newcommand{\nuic}{\Delta\nu_\mathrm{coh}}
\newcommand{\VRF}{V_\mathrm{rf}}
\newcommand{\fint}{f_\mathrm{int}}
\newcommand{\fmax}{f_\mathrm{max}}
\newcommand{\Np}{N_\mathrm{particle}}
\newcommand{\NAM}{N_\mathrm{AM}}
\newcommand{\chiABS}{\chi_\mathrm{ABS}}
\newcommand{\chiABH}{\chi_\mathrm{ABH}}
\newcommand{\amABS}{a_{m,\mathrm{ABS}}}
\newcommand{\amABH}{a_{m,\mathrm{ABH}}}
\newcommand{\tamABS}{\tilde{a}_{m,\mathrm{ABS}}}
\newcommand{\Zdipeff}{Z_{\mathrm{dip,eff}}}
\newcommand{\ZT}{Z_{\mathrm{T1}}}
\newcommand{\WT}{W_{\mathrm{T1}}}

\newcommand{\s}{\mathrm{s^{-1}}}
\newcommand{\m}{\mathrm{m^{-1}}}

\begin{document}

\title{The space charge effects on the intra-bunch motion under large chromaticity \\ at the Main Ring in the Japan Proton Accelerator Research Complex}

\author{Nobuyuki Yoshimura}
\email{yoshimura.nobuyuki.76w@st.kyoto-u.ac.jp}
\affiliation{Department of Physics, Kyoto University, Kyoto 606-8502, Japan}

\author{Takeshi Toyama}
\author{Yoshihiro Shobuda}
\affiliation{J-PARC Center, JAEA and KEK, 2-4 Shirakata, Tokaimura, Nakagun, Ibaraki 319-1195, Japan}

\begin{abstract}

In general, optimizing chromaticity and transverse feedback parameters is important in suppressing beam instabilities for stable operation in high-intensity proton machines.
Meanwhile, space charge effects impact the intrabunch motion of the proton bunch within a large chromaticity region in the machines, such as the main ring (MR) in the Japan Proton Accelerator Research Complex (J-PARC).
To address this issue, the decoherence and recoherence of the transverse motion of the particles comprising the proton bunch are investigated.
The analysis reveals that the space charge effects have a significant influence on the recoherence period through the chromaticity.
Nevertheless, the relationship between the maximum frequency in the bunch and the chromaticity is not affected by the space charge effects.
These findings are demonstrated with particle tracking simulations including the direct and indirect space charge effects, and the impedance source of the J-PARC MR.
Furthermore, we illustrate the influence of the indirect space charge effect on particle motion by examining the excitation patterns of radial and head-tail modes.

\end{abstract}

\maketitle

\begin{table*}[!htbp]
  \caption{List of symbols.}
  \begin{ruledtabular}
  \centering
  \small
  \begin{adjustbox}{width=\textwidth,totalheight=\textheight,keepaspectratio}
  \begin{tabular}{c  c  c}
    Term & Unit & Description \\ \hline
    $m$                   &         & Head-tail mode number index                              \\ 
    $\nusc$               &         & Incoherent tune shift from the direct space charge       \\ 
    $\nuic$               &         & Coherent tune shift from the indirect space charge       \\ 

    $c$                   & m $\s$  & Light velocity                                           \\ 
    $e$                   & C       & Electric charge                                          \\ 
    $E_0$                 & eV      & Total energy                                             \\ 
    $\epsilon_0$          & F $\m$  & Vacuum permittivity                                      \\ 
    $r_p$                 & m       & Classical particle radius of the proton                  \\ 
    $Z_0$                 & $\mathrm{\Omega}$ & Impedance of free space                            \\ 

    $C$                   & m       & Circumference of the ring                                \\ 
    $\xi_x$               &         & Horizontal chromaticity                                  \\ 
    $h$                   &         & Harmonic number                                          \\ 
    $\VRF$                & V       & rf voltage                                               \\ 
    $\beta_x$             & m       & Average value of the beta function along the ring        \\ 
    $\beta$               &         & Lorentz beta                                             \\ 
    $\gamma$              &         & Lorentz gamma                                            \\ 
    $T_0$                 & s       & Revolution period                                        \\ 
    $f_0$                 & $\s$    & Revolution frequency                                     \\ 
    $\eta$                &         & Slippage factor                                          \\ 
    $\nu_{\beta,x}$       &         & Horizontal bare betatron tune                            \\ 
    $\nu_s$               &         & Bare synchrotron tune                                    \\ 
    $N_s$                 & turn    & Synchrotron period                                       \\ 
    $n$                   &         & Simulation step number index                             \\ 
    $s$                   & m       & Path length                                              \\ 

    $\hat{z}$             & m       & Half bunch length of ABS and ABH model                   \\ 
    $\delta_0$            &         & Momentum deviation of ABS model                          \\ 
    $\Delta^{(k)}$        & m C     & Dipole moment in $k$th segment                          \\ 
    $\lambda^{(k)}$       & C       & Charge density in $k$th segment                         \\ 
    $\Delta T$            & s       & Advance time per step                                    \\ 

    $\phi$                & rad     & Betatron phase advance                                   \\ 
    $\sigma_x$            & mm      & rms horizontal beam size                                 \\ 
    $\sigma_z$            & mm      & rms longitudinal beam size                               \\ 
    $\sigma_\delta$       & mrad    & rms momentum deviation                                   \\ 
    $z_b$                 & m       & Full bunch length                                        \\ 
    $X_0$                 & m       & Horizontal offset position at injection                  \\ 
    $N_B$                 &         & Bunch population                                         \\ 
    $B_f$                 &         & Bunching factor                                          \\ 
    $\hat{\lambda}$       & C       & Peak line density of electrical charges                  \\ 
    $\overline{x_n}$      & m       & Transverse centroid of the entire bunch                  \\ 

    $\bar{x}(n,z)$        & m       & Transverse centroid                                      \\ 
    $\chi_\sigma$         & rad     & rms chromatic phase                                      \\ 
    $\chi_b$              & rad     & Chromatic phase                                          \\ 
    $\fint(n)$            & $\s$    & Intrabunch frequency                                    \\ 
    $\fmax$               & $\s$    & Maximum value of intrabunch frequency $f_{\mathrm{int}}(n)$      \\ 
    $\NAM$                & turn    & Recoherence time                                         \\ 
    $N_f$                 & turn    & Decoherence time                                         \\ 
    $X(n)$                & m       & Transverse centroid of the entire bunch                  \\ 
    $\qsc$                &         & Space charge parameter                                   \\ 

    $i$                   &         & Macroparticle number index                               \\ 
    $f$                   & m       & Half distance between the two focal points               \\ 
    $h_C$                 & m       & Radius of the cylindrical conductive chamber             \\ 
    $d_\omega$            & m       & Finite thickness of the resistive wall chamber           \\ 
    $L$                   & m       & Length of the resistive wall chamber                     \\ 
    $\rho$                & m       & Resistivity of the resistive wall chamber                \\ 
    $F$                   &         & Filling factor of the elliptical chambers along the ring \\ 
    $\mathcal{G}$         &         & Geometrical factor                                       \\ 
    $\Np$                 &         & Number of macroparticles in the bunch                    \\ 
    $\sigma_\Delta$       & m       & Segment length                                           \\ 
    $\Delta \nu$          &         & Tune interval between the head-tail modes                \\ 
    $\nu_m$               &         & Head-tail mode frequency                                 \\ 
    $q$                   & C       & Allocated charge for one macroparticle                   \\ 
  \end{tabular}
  \end{adjustbox}
  \end{ruledtabular}
  \label{tab:sec1}
\end{table*}

\section{INTRODUCTION}\label{1}

High-intensity proton beams are essential for a wide range of experiments, from high-energy physics to life and material sciences.
Major facilities include the Super Proton Synchrotron (SPS) \cite{benedikt2004lhc} at Conseil Européen pour la Recherche Nucléaire (CERN),
the Main Injector (MI) synchrotron \cite{none1994fermilab} at Fermi National Accelerator Laboratory (FNAL, Fermilab), 
the Alternating Gradient Synchrotron (AGS) \cite{none1966bnl} at Brookhaven National Laboratory (BNL), and
the SIS-18 synchrotron \cite{kornilov2012transverse,singh2013interpretation,karpov2016early,kornilov2010head,kornilov2010simulation,boine2009transverse,boine2006implementation} at GSI Darmstadt.

The main ring (MR) synchrotron at the Japan Proton Accelerator Research Complex (J-PARC)
\cite{igarashi2021accelerator,koseki2012beam,igarashi2018high,koseki2018upgrade,Yasui:2023bfx,yasui2023first}
is also recognized as a major high-intensity proton facility.
It plays a central role in neutrino physics research, supporting the T2K experiment
\cite{abe2011indication,abe2014observation,t2k2020constraint,abe2023updated}
and the future Hyper-Kamiokande experiment \cite{abe2018hyper}.
The high-intensity proton beam and the stable operation of the J-PARC MR contribute to increasing the statistics and improving the precision of measurements of neutrino oscillation parameters.
The operational average beam power of the J-PARC MR is approximately $800$~kW as of 2024
under the number of protons per pulse (ppp) of $2.3\times10^{14}$ 
with 8 bunches, a cycle of $1.36$~s, and an extraction energy of $30$~GeV.
To further improve the performance of the accelerator toward an intensity frontier, it is important to manipulate the intrabunch motions of the beam adequately.

While enlarging chromaticity is a standard method to suppress beam instabilities, head-tail (azimuthal) and radial modes \cite{Chao:1993zn,Ng:2002gr} 
could be excited by the wakefields in the ring.
An upgrade of the transverse intrabunch feedback (IBFB)
\cite{chin2013analysis,Konstantinova:2013uba,nakamura2014intra,nakamura2014performance,Kurimoto2011THEBB,toyamaanalysis,Toyama:2022gli,Yoshimura:2022idd,Nakamura2024intra,Nakamura:823297,Nakamura:2018orr}
is currently underway to achieve operation at $1.3$~MW.

Needless to say, space charge effects must play a significant role in the intrabunch motion in high-intensity beams.
Theoretically, Blaskiewicz investigated the direct space charge effect on beam instabilities by introducing ``the space charge modes'' \cite{blaskiewicz1998fast}, which is equivalent to the head-tail modes \cite{kornilov2012transverse,singh2013interpretation}.

This paper aims to clarify the behavior of the modes with high-intensity beams caused by the combined effects of the chromaticity, impedance, and the direct and indirect space charge to achieve stable operation.
One pioneering work was done to investigate the effects of the intrabunch motion in the SIS-18 but within limited parameters.
In those studies, the chromatic phase $\chi_b$, defined as
\begin{equation}
  \chi_b=2\pi \frac{\xi_x}{\eta C}z_b,
\end{equation}
was lower than 5, where $\xi_x=\Delta\nu_{\beta,x}/\Delta\delta$ is the horizontal chromaticity, $z_b$ is the full bunch length, $\eta$ is the slippage factor, and $C$ is the circumference of the ring,
$\nu_{\beta,x}$ is the bare betatron tune, and $\delta$ is the longitudinal momentum deviation.
The space charge parameter $\qsc$ \cite{karpov2016early,kornilov2010head,singh2013interpretation,kornilov2010simulation,kornilov2012transverse} is defined as 
\begin{equation}
  \qsc=\frac{|\nusc|}{\nu_s},
\end{equation}
was lower than 10, where $\nusc$ is the incoherent tune shift from the direct space charge and $\nu_s$ is the synchrotron tune.
Among the head-tail modes excited by the transverse bunch displacement during the injection process, the only lower modes $m\leq3$ survived after the injection.

For machines aiming for higher intensities, such as the J-PARC MR, more extensive work is indispensable because the chromatic phase $\chi_b$ could change in wider regions to around 50 to suppress the beam instabilities,
and the space charge parameter $\qsc$ could become higher than 50.
Then, we present a more comprehensive study of the effects of space charge on particle motion within the bunch across the wider parameter range, using theoretical analysis, simulations, and experimental measurements at the J-PARC MR.

One major objective of this study is to understand the fundamental aspects of the intrabunch motion under large chromaticity in high-intensity proton rings.
Specifically, we explore the impact of direct and indirect space charge effects on the head-tail modes through the chromaticity within the wider parameter region when no beam instability is present by choosing suitable machine parameters.

To gain deeper insight into the complex factors of intrabunch dynamics, reconciling simulation and experimental results is essential.
However, full three-dimensional simulations are often impractical as a primary approach due to their time-consuming and overly complex nature.
To address this issue, we developed a simplified particle tracking simulation code to model the time evolution of transverse dipole moments and space charge effects.
This code tracks particles in two dimensions---horizontal and longitudinal---using a linear approximation.
This simplification enables us to better understand the combined effects of the chromaticity and space charge driven head-tail modes.

To validate this code, we provide a robust foundation for handling direct and indirect space charge effects from a fundamental perspective.
During this derivation process, we clarify the limits of the theory's applicability to the direct and indirect space charge effects within the linear approximation.
For a benchmark test, the simulation results are compared with theoretical predictions addressing Gaussian beams in a harmonic potential with weak space charge forces, extended from the airbag square-well (ABS) model originally developed by Blaskiewicz \cite{blaskiewicz1998fast}.

We also estimate the effect of longitudinal and transverse impedances on the intrabunch motion in the J-PARC MR.
In the process, we establish a framework for incorporating the long-range resistive wall impedance into simulation codes by accounting for the chamber thickness.
Finally, we demonstrate that the impedance does not influence intrabunch motion both theoretically and simulationally unless the beam instability occurs.
In other words, our simulation study reveals that the indirect space charge effect stabilizes the beams \cite{shobuda2017theoretical,saha2018simulation} by suppressing the radial and negative head-tail modes, although the direct space charge excites the beam instability \cite{burov2019convective}.

Ultimately, we demonstrate that the simplified simulation tools are valuable for characterizing and understanding experimental measurements dealing with the high-intensity beams.

This paper is structured as follows:
In Sec. \ref{2}, we introduce the basics of the algorithm without space charge effects.
In Sec. \ref{3}, we review the theoretical decoherence model without space charge to understand the characteristics of intrabunch motion.
In Sec. \ref{4}, we establish a robust foundation for incorporating space charge effects into the code.
In Sec. \ref{5}, we address the transverse resistive wall impedance.
In Sec. \ref{6}, we partially generalize the ABS theoretical model to account for Gaussian beams in a harmonic potential, incorporating both weak direct and indirect space charge effects.
In Sec. \ref{7}, we provide a comprehensive analysis of intrabunch motion by comparing the simulation with the theoretical results.
In Sec. \ref{8}, we perform supplementary simulation to survey how the radial and negative head-tail modes are suppressed by the indirect space charge effect.
In Sec. \ref{9}, we examine the effects of space charge on intrabunch motion through chromaticity and rf voltage in the time domain from simulation and analytical points of view in preparation for comparison with the experimental measurements.
In Sec. \ref{10}, the measured data are analyzed and compared with the simulation results.
Finally, the paper is summarized in Sec. \ref{11}.

The symbols used throughout this paper are defined in Table \ref{tab:sec1}.

\section{Basic Algorithm}\label{2}

\begin{table*}[!h]
  \caption{List of symbols in Sec. \ref{2}.}
  \begin{ruledtabular}
  \centering
  \small
  \begin{tabular}{c  c  c}
    Term & Unit & Description \\ \hline
    $x$                   & m       & Horizontal position                                      \\ 
    $x'$                  & rad     & Horizontal position derivative with respect to path length \\ 
    $z$                   & m       & Longitudinal position                                    \\ 
    $\delta$              &         & Longitudinal momentum deviation                          \\ 
    $\hat{T_n}$           & s       & Time at the $n$th step                 \\ 
    $\epsilon_x$          & mm mrad & Unnormalized rms horizontal emittance                    \\ 
    $\sigma_{x'}$         & mrad    & rms horizontal beam divergence                           \\ 
    $x^{(k)}$             & m       & Transverse centroid in $k$th segment      \\ 
    $z^{(k)}$             & m       & Longitudinal position in $k$th segment    \\ 
    $N^{(k)}$             &         & Number of macroparticles in $k$th segment \\ 
    $S_i^{(k)}$           &         & Shape function                                           \\ 
  \end{tabular}
  \end{ruledtabular}
  \label{tab:sec2}
\end{table*}

In this section, let us focus solely on implementing basic kinematics into our code, neglecting the space charge and wakefield effects.
To simplify the situation, the simulation only deals with 1D transverse (horizontal) + 1D longitudinal motion by tracking macroparticle motions in discrete time steps.
In this paper, only the storage mode, not the acceleration mode, is considered to highlight the feature of the space charge effect on the intrabunch motion in the following sections after generalizing this algorithm.
The symbols used in this section are defined in Table \ref{tab:sec2}.

The horizontal coordinates $x$ are defined as the position relative to that of the synchronous particle,
and $x'=dx/ds$ is the derivative of $x$ with respect to $s$, path length along the design trajectory.
The horizontal motion terms are calculated as
\begin{equation}\label{eq:3}
  x_{n+1,i} = \cos (\phi_{n,i}) x_{n,i} + \beta_x \sin (\phi_{n,i}) x'_{n,i},
\end{equation}
\begin{equation}\label{eq:4}
  x'_{n+1,i} = -\frac{1}{\beta_x} \sin (\phi_{n,i}) x_{n,i} + \cos (\phi_{n,i}) x'_{n,i},
\end{equation}
where $i$ is the macroparticle number index,
$n$ is the simulation step number index,
$x_{n,i}$ is the horizontal position,
$x'_{n,i}$ is the horizontal position derivative with respect to $s=\beta c \hat{T}_n$,
$\hat{T}_n$ is the $n$th time,
the betatron phase advance is
\begin{equation}\label{eq:5}
  \phi_{n,i} = 2\pi \left(\nu_{\beta,x} + \xi_x\delta_{n,i}\right)\frac{\Delta T}{T_0},
\end{equation}
$\Delta T=\hat{T}_{n+1}-\hat{T}_n$,
$\delta_{n,i}$ is the longitudinal momentum deviation from the synchronous particle,
the bare betatron tune $\nu_{\beta,x}=21.35$, 
the circumference $C=1567.5$~m,
the slippage factor $\eta=-0.0578$ for a total energy $E_0=3.938$~GeV 
with Lorentz-$\beta=0.971$ and Lorentz-$\gamma=4.198$, 
the revolution period $T_0=C/\beta c=5.384$~$\mu$s, and
$\beta_x=C / 2\pi\nu_{\beta,x}$ is the average value of the beta function along the J-PARC MR.

For the transverse phase space distribution, we adopt a Gaussian distribution with $\sigma_x=7.5$ mm and $\sigma_{x'}=0.64$ mrad as its initial condition.
The condition for this simulation matches the condition of the J-PARC MR because the beams injected into the ring typically have an unnormalized root mean square (rms) emittance of $\epsilon_x=4.8\pi$ mm mrad.
The horizontal offset position at injection is $X_0=0.01$~m.

The longitudinal coordinates $z$ and $\delta$ are defined as the position and the momentum deviation relative to those of the synchronous particle, respectively.
The longitudinal motion terms are calculated as
\begin{equation}\label{eq:6}
  z_{n+1,i} = z_{n,i} - \eta C \delta_{n,i}\frac{\Delta T}{T_0},
\end{equation}
\begin{equation}\label{eq:7}
  \delta_{n+1,i} = \delta_{n,i} - \frac{e\VRF}{\beta^2 E_0} \frac{2\pi h }{C} z_{n+1,i} \frac{\Delta T}{T_0},
\end{equation}
where $z_{n,i}$ is the longitudinal position,
the electric charge $e=1.6\times10^{-19}$~C, and
the harmonic number $h=9$.
Unless otherwise specified, the rf voltage is $\VRF=263$~kV in this paper.

The small-amplitude bare synchrotron tune $\nu_s$ is given by
\begin{equation}
  \nu_{s} = \sqrt{\frac{e\VRF h|\eta|}{2\pi \beta^2 E_0}},
\end{equation}
which is 0.0024 by the given parameters.
The synchrotron period is defined as $N_s=1/\nu_s(=410)$.

For the longitudinal phase space distribution, we adopt a Gaussian distribution with $\sigma_z=10$~m and
\begin{equation}\label{eq:9}
  \sigma_\delta=\frac{2\pi\nu_s}{|\eta| C}\sigma_z,
\end{equation}
as its initial condition.

In this paper, the bunching factor is defined as 
\begin{equation}\label{eq:10}
  B_f=\frac{e N_B}{C \hat{\lambda}},
\end{equation}
where $N_B$ is the bunch population, $\hat{\lambda}$ is the peak line density of electrical charges, typically featuring $B_f=0.015$ in this simulation.

Particularly under the square-well potential in the airbag model (ABS model), which is used for the theoretical analysis in the following sections, 
Eq. \eqref{eq:7} is replaced by 
\begin{equation}\label{eq:11}
  \delta_{n+1,i} =
  \begin{cases}
    \delta_{n,i},  & ~\text{for} ~-\hat{z} \leq z_{n+1,i} \leq \hat{z},\\
    -\delta_{n,i}, & ~\text{for} ~z_{n+1,i} < -\hat{z}, \hat{z} < z_{n+1,i},
  \end{cases}
\end{equation}
to ensure the consistent longitudinal motion with the momentum deviation at a constant velocity in the region of $-\hat{z}<z<\hat{z}$
($\hat{z}$ is the half bunch length).
We set $\hat{z}=\sqrt{\pi/2}\sigma_z=12.5$~m to match the peak density of the uniform distribution and that of the Gaussian distribution for the same bunch intensity $eN_B$,
and the momentum is distributed as $\delta=\pm\delta_0=\pm0.13\%$ to satisfy
\begin{equation}\label{eq:12}
  \delta_0=\frac{4\nu_s}{|\eta|C}\hat{z},
\end{equation}
to guarantee the same synchrotron period as that of the harmonic potential.

The transverse centroid of the entire bunch $\overline{x_n}$ is defined as
\begin{equation}\label{eq:13}
  \overline{x_n} = \frac{1}{\Np} \sum_i x_{n,i},
\end{equation}
where $\Np$ is the number of macroparticles in the bunch.

Subsequently, let us divide the ring with the segment length $\sigma_\Delta$, where $k$th segment is identified with $z^{(k)}=-C / 2h + k\sigma_\Delta$ in the interval $[-C/2h,C/2h]$,
and investigate the intrabunch frequency by calculating the dipole moment $\Delta^{(k)}$ and the charge density $\lambda^{(k)}$ for the $k$th segment when the segment is filled with macroparticles comprising the bunch.
The number of macroparticles $N^{(k)}$ in $k$th segment is calculated as
\begin{equation}
  N^{(k)} = \sum_i S_i^{(k)},
\end{equation}
where $(k)$ and $i$ specify the segment and macroparticle indices, respectively, and the shape function is introduced as follows
\cite{sabbi1995simulation,sabbi1994trisim}
\begin{equation}
  S_i^{(k)}=
  \begin{cases}
    1-\frac{|z_i-z^{(k)}|}{\sigma_\Delta}, & ~\text{for}~z^{(k-1)}\leq z_i\leq z^{(k+1)}, \\
    0,   & ~\text{for}~z_i<z^{(k-1)}, z^{(k+1)}<z_i.
  \end{cases}
\end{equation}

Finally, the dipole moment $\Delta^{(k)}$ and the charge density $\lambda^{(k)}$ of $k$th segment are respectively described as
\begin{equation}\label{eq:16}
  \Delta^{(k)} = \frac{q}{\sigma_\Delta} \sum_i x_i S_i^{(k)},
\end{equation}
\begin{equation}
  \lambda^{(k)} = \frac{q}{\sigma_\Delta} \sum_i S_i^{(k)},
\end{equation}
with $q=eN_B / \Np$ being the allocated charge for one macroparticle.
Accordingly, the transverse centroid $x^{(k)}$ of $k$th segment is given by
\begin{equation}
  x^{(k)} = \frac{\Delta^{(k)}}{\lambda^{(k)}}.
\end{equation}

In the following simulations concerning the transverse oscillation in the bunch, we will conduct tracking simulations with both the ABS model with Eqs. \eqref{eq:6} and \eqref{eq:11}, and the harmonic potential bunch model with Eqs. \eqref{eq:6} and \eqref{eq:7}.
The horizontal chamber boundary is set to $\pm0.065$~m, while the longitudinal bucket size is $\pm C/2h$.
We assume that any macroparticles outside this region should be lost.

\section{Transverse linear decoherence}\label{3}


In this section, to overview a typical behavior of intrabunch motion, 
we review the linear decoherence model \cite{kornilov2012transverse,Meller:1987ug}, neglecting the space charge and wakefield effects, and assuming the tune shift is solely caused by the chromaticity.
The longitudinal and transverse motions of particles within a bunch are correlated through the synchrotron motion.
To explore how the chromaticity affects the motions of the particles comprising the bunch,
we examine a situation where the particles with identical initial transverse offsets are injected into the ring, exciting the intrabunch motions.
In general, the disappearance of the coherent oscillation signal, observed as the beam circulates the ring, reflects the betatron phase spread caused by the chromaticity, (space charge, and wake effects).

The intrabunch oscillation describes the behavior of the transverse centroid at the respective longitudinal position along the bunch for a specific turn in the ring, which is closely related to the decoherence signals of the bunch.
The Fourier transform of this centroid position with respect to the longitudinal direction yields a frequency spectrum in the bunch.
When this frequency spectrum becomes pronounced, it can manifest the instability of the beam.
Proper identification and analysis of intrabunch oscillations are essential for determining the performance of the IBFB because it functions by detecting the frequency signals.

Following this model,
we examine the transverse centroid $\bar{x}(n,z)$ for a group of particles with the same longitudinal position $z$ in a given turn $n$.
For simplicity, all equations in this section and in Sec. \ref{6} are expressed in complex notation.
Observable quantities correspond to the real parts of $\bar{x}(n,z)$ and $X(n)$, i.e., $\mathrm{Re}[\bar{x}(n,z)]$ and $\mathrm{Re}[X(n)]$, respectively.

The initial condition is set as $\bar{x}(n=0,z)=X_0$ regardless of the longitudinal position of the particles, with an initial betatron phase of $\phi(n=0,z)=0$.
Assuming a Gaussian distribution ($\sigma_\delta=2\pi \nu_s / |\eta| C\times \sigma_z$: Eq. \eqref{eq:9}) in the synchrotron phase space, 
the transverse centroid $\bar{x}(n,z)$ for the respective parts of the bunch is calculated as
\cite{karpov2016early,yoshimura2025space}
\begin{equation}\label{eq:19}
  \bar{x}(n,z) = X_0 \exp\left[-\frac{1}{2}\chi_\sigma^2 \sin^2(2\pi\nu_s n)\right] \exp\left[-\textbf{i}(2\pi\nu_{\beta,x} n-2\chi_\sigma \sin^2(\pi\nu_s n)\frac{z}{\sigma_z})\right],
\end{equation}
where $\textbf{i}$ is the imaginary unit, and the rms chromatic phase $\chi_\sigma$ for the bunched beam is defined as
\begin{equation}\label{eq:20}
  \chi_\sigma = 2\pi \frac{\xi_x}{\eta C} \sigma_z = \frac{|\xi_x|}{\nu_s} \sigma_\delta.
\end{equation}

Since the envelope amplitude term is $X_0 \exp(-\chi_\sigma^2 \sin^2(2\pi\nu_s n)/2)$ in Eq. \eqref{eq:19}, 
it is approximated as $X_0 \exp(- 2 (\pi \chi_\sigma \nu_s)^2 \\ (n-m N_s/2)^2)$ up to the second order after expansion around $n=mN_s/2$ ($m$ is an integer), 
regardless of the longitudinal positions $z$ in the bunch.
In other words, the higher the absolute chromaticity, the sharper the recoherence signals.

Since the intrabunch frequency $\fint(n)$ can be interpreted as the betatron phase advance within the bunch per unit time, it can be defined as
\begin{equation}
  \begin{split}
    \fint(n)&= \left|\frac{1}{2\pi}\frac{d}{dt} \left[ 2\pi\nu_{\beta,x} n-4\pi\frac{\xi_x}{\eta C} z \sin^2(\pi\nu_s n) \right]\right| \\
              &= \left|2 \frac{\xi_x}{\eta C} \frac{dz}{dt} \sin^2(\pi\nu_s n)\right| \\
              &= \left|2 f_0 \frac{\xi_x}{\eta} \sin^2(\pi\nu_s n)\right|,
  \end{split}
\end{equation}
by recognizing the exponential factor in Eq. \eqref{eq:19}, where $f_0=1/T_0$ is the revolution frequency.
The maximum value of $\fint(n)$ at the specified longitudinal position $z$ in the bunch is given by
\begin{equation}\label{eq:22}
  \fmax= \fint\left(\frac{N_s}{2}\right)=\left|2 f_0 \frac{\xi_x}{\eta}\right|.
\end{equation}

The intrabunch frequency $\fint(n)$ is proportional to $\xi_x$ and independent of $z$ in the linear decoherence model,
which means any portion of the beam can oscillate at the maximum intrabunch frequency $\fmax$.
Moreover, the maximum intrabunch frequency is remarkably enhanced by increasing the chromaticity.

The transverse centroid of the entire bunch $X(n)$ is calculated by convolving Eq. \eqref{eq:19} with a Gaussian distribution:
\begin{equation}\label{eq:23}
  \rho_z(z) = \frac{1}{\sqrt{2\pi}\sigma_z} \exp\left(-\frac{z^2}{2\sigma_z^2}\right),
\end{equation}
resulting in \cite{Meller:1987ug}
\begin{equation}\label{eq:24}
  X(n) = X_0 \exp(- 2\chi_\sigma^2 \sin^2(\pi\nu_s n)) \exp(- \textbf{i} 2\pi\nu_{\beta,x} n).
\end{equation}
Thus, the amplitude of this betatron oscillation $A(n)$ follows
\begin{equation}\label{eq:25}
  A(n) = X_0 \exp(- 2\chi_\sigma^2 \sin^2(\pi\nu_s n)).
\end{equation}

As the chromaticity increases, the betatron phase spread widens through the chromatic phase, demonstrating that the transverse centroid amplitude decreases.
This represents the decoherence of particles comprising the bunch.
However, after a synchrotron period from injection, $\fint(N_s)=0$, 
the recoherence of particles emerges because of $A(N_s) = X_0$.
It is noticeable that this recoherence phenomenon occurs with the synchrotron period
only when the effect of the chromaticity is taken into consideration.

Similarly to Eq. \eqref{eq:19}, let us expand Eq. \eqref{eq:25} around $n=mN_s$ (where $m$ is an integer), resulting in $A(n)\simeq X_0 \exp(- 2 (\pi \chi_\sigma \nu_s)^2 (n-m N_s)^2)$ up to second order.
The higher the absolute chromaticity, the sharper the recoherence signals.
Notably, the recoherence period for the entire bunch is $N_s$,
although the dipole moments of the particles comprising the bunch align to create sharp signals with a period of $N_s/2$, as implied by Eq. \eqref{eq:19}.

\section{Formulation of the space charge effect}\label{4}

\begin{table*}[!h]
  \caption{List of symbols in Sec. \ref{4}.}
  \begin{ruledtabular}
  \centering
  \small
  \begin{tabular}{c  c  c}
    Term & Unit & Description \\ \hline
    $\bar{x}$             & m       & Horizontal centroid of each segment                      \\ 
    $a_x,a_y$             & m       & Horizontal and vertical beam sizes                       \\ 
    $I(x,y)$              &         & Region occupied by the beam on the transverse plane      \\ 
    $\lambda$             & C       & Charge line density in the laboratory frame              \\ 
    $\bar{\lambda}$       & C       & Charge line density in the rest frame                    \\ 
    $h_{E_x},h_{E_y}$     & m       & Major horizontal and minor vertical axis of the elliptical conductive chamber \\ 
    $g$                   & m       & Half distance between the magnetic pole surfaces of parallel plates placed perpendicularly to the vertical axis \\ 
    $\Phi$                & V       & Scalar potential in the laboratory frame                 \\ 
    $\bar{\Phi}$          & V       & Scalar potential in the rest frame                       \\ 
    $E_x^0$               & V $\m$  & Horizontal component of the electric field excited by the point charge in the laboratory frame \\ 
    $E^0_{x,\mathrm{DS}},E^0_{x,\mathrm{im}}$ & V $\m$   & $E_x^0$ excited by the direct and indirect space charge                   \\ 
    $E_{x,\mathrm{DS}}$   & V $\m$  & Superposition of the horizontal component of the electric field by the direct space charge     \\ 
    $E_{x,\mathrm{im}}$   & V $\m$  & Superposition of the horizontal component of the electric field by the indirect space charge   \\ 
    $F_x^{\mathrm{ele}}$  & N       & Electric contributions to the Lorentz force             \\ 
    $F_x^{\mathrm{mag}}$  & N       & Magnetic contributions to the Lorentz force             \\ 
    $\epsilon_1$          &         & Incoherent electric image coefficient                    \\ 
    $\xi_1$               &         & Coherent electric image coefficient                      \\ 
    $\epsilon_2$          &         & Incoherent magnetic image coefficient                    \\ 
    $\xi_2$               &         & Coherent magnetic image coefficient                      \\ 
    $A_z$                 & V s $\m$ & Vector potential in the laboratory frame                \\ 
    $F_B$                 &         & Filling factor of the bending magnet along the ring      \\ 
  \end{tabular}
  \end{ruledtabular}
  \label{tab:sec4}
\end{table*}

To investigate the space charge effects on the intrabunch motion, 
this section formulates the effects, which are implemented into our tracking simulation, based on the Poisson equation.
In this context, we will directly observe the direct and indirect space charge effects are characterized by the incoherent tune shift $\nusc$ and coherent tune shift $\nuic$, respectively, within the linear approximation,
although most papers assume this characteristic without specifying any particular conditions required.
We will find that this linear approximation helps manage the space charge effects efficiently when the transverse beam size is sufficiently smaller than the chamber radius.
The symbols used in this section are defined in Table \ref{tab:sec4}.

In the case of a proton bunch like the J-PARC MR, since the longitudinal bunch length is much larger than the transverse beam size, 
we simplify the three-dimensional beam to the two-dimensional one by slicing the beam perpendicularly to the longitudinal direction before treating the respective segments independently.
This approach not only streamlines the computational process but also allows us to effectively capture the transverse dynamics of the bunch.
Within each segment, we assume that the beam is represented with a charge line density $\lambda(z)$ in the longitudinal direction and a uniform-density elliptical beam with a horizontal beam size $a_x$ and a vertical beam size $a_y$ on the transverse plane, following the Kapchinski-Vladimirskij (K-V) distribution \cite{kapchinskij1959limitations}.
Since the simulation tracks only horizontal motion, $a_x$ is dynamically updated at each time step with $a_y$ fixed.

In the tracking simulation, since the transverse distribution is not always uniform, we need to evaluate the representative value of $a_x$.
Here, the transverse charge distribution is assumed to follow a Gaussian beam, as shown in 
\begin{equation}
  \rho(x,y)=\frac{\lambda}{2\pi\sigma_x \sigma_y} \exp\left(-\frac{(x-\bar{x})^2}{2\sigma_x^2}-\frac{y^2}{2\sigma_y^2}\right),
\end{equation}
with the horizontal rms beam size $\sigma_x$ and the vertical rms beam size $\sigma_y$, producing the line density $\lambda$ after integrating over the transverse plane.
Therefore, we set $a_x=\sqrt{2}\sigma_x$ with $a_y=\sqrt{2} \sigma_y=0.014$~m to reconcile the following space charge calculation based on the uniform elliptical distribution with that on the Gaussian distribution.
This is because the peak densities of the elliptical beam with $a_x=a_y$ and the Gaussian beam with $\sigma_x=\sigma_y$ are identified as $\lambda/\pi a_x^2$ and $\lambda/2\pi \sigma_x^2$ respectively, for the same line density $\lambda$.
In the case of the J-PARC MR, the typical transverse beam sizes are given by $\sigma_x=0.009$~m and $\sigma_y=0.010$~m, which are almost symmetric in accordance with the assumption.

To formulate the space charge effects in the elliptical beam, we will proceed by separately addressing the elliptical chamber in open space and the one placed inside the bending magnet, simplified by parallel magnetic pole faces.
First, we consider the case of two-dimensional beam passing through the elliptical chamber in open space.
Let us define the region $I$ occupied by the beam on the transverse plane as 
\begin{equation}
  I(x,y) = 
  \begin{cases}
    1, &~\text{for}~D(x,y,\bar{x})\leq1, \\
    0, &~\text{for}~D(x,y,\bar{x})>1,
  \end{cases}
\end{equation}
where
\begin{equation}
  D(x,y,\bar{x}) = \frac{(x-\bar{x})^2}{a_x^2}+\frac{y^2}{a_y^2},
\end{equation}
where $\bar{x}$ is the horizontal centroid of each segment.

The charge line density $\bar{\lambda}$ in the rest frame and the charge line density $\lambda$ in the laboratory frame are related by
\begin{equation}
  \lambda=\gamma\bar{\lambda},
\end{equation}
after the Lorentz transformation.

We assume an elliptical conductive chamber with the major horizontal axis $2h_{E_x}=2f\cosh{u_0}$ and the minor vertical axis $2h_{E_y}=2f\sinh{u_0}$.
Here, the half distance between the two focal points $f$ is calculated as $f=\sqrt{h_{E_x}^2-h_{E_y}^2}$ for the chamber of $h_{E_x}>h_{E_y}$.

When a point particle is located at $(x_s,y_s)$, the scalar potential $\bar{\Phi}$ in the rest frame is given by
\cite{zotter1975tune} 
\begin{equation}
  \bar{\Phi}(x,y,x_s,y_s)=-\frac{\bar{\lambda}}{4\pi\epsilon_0}\log\left|\frac{D_s^x}{D_s}\frac{CD_s-C_sD}{CD_s^x-C_s^xD}\frac{C-C_s}{D+D_s}\frac{C-C_s^x}{D+D_s^x} \right|,
\end{equation}
where $\epsilon_0$ is vacuum permittivity,
\begin{equation}\label{eq:31}
  \begin{split}
    S &= S(\zeta) = \mathrm{sn}(u(\zeta),k),\\
    C &= C(\zeta) = \mathrm{cn}(u(\zeta),k),\\
    D &= D(\zeta) = \mathrm{dn}(u(\zeta),k),
  \end{split}
\end{equation}
sn, cn, and dn are the Jacobian elliptic functions,
$C_s=C(\zeta_s),D_s=D(\zeta_s)$,
$S^x=S(\overline{\zeta})=\overline{S}, C^x=C(\overline{\zeta})=\overline{C}, D^x=D(\overline{\zeta})=\overline{D}$,
$C_s^x=C(\overline{\zeta_s})=\overline{C_s}$, $D_s^x=D(\overline{\zeta_s})=\overline{D_s}$ \cite{abramowitz1968handbook},
$\zeta=x+\textbf{i}y$,
and the complex conjugate $\bar{\zeta}$.

Their arguments are calculated as
\begin{equation}\label{eq:32}
  u(\zeta) = \frac{2K(k)}{\pi}\cos^{-1}{\frac{\zeta}{f}},
\end{equation}
by the complete elliptic integral of the first kind $K(k)$ \cite{abramowitz1968handbook}.
The ``modulus'' $k$ in Eqs. \eqref{eq:31} and \eqref{eq:32} is determined by $K(\sqrt{1-k^2})/K(k)=2/\pi \tanh^{-1} (h_{E_y}/h_{E_x})$.
From now on, we will simplify the description of $K(k)$ as $K$ by omitting $k$ in the argument of the complete elliptic integral of the first kind.

Once the scalar potential in the rest frame is obtained,
the scalar potential $\Phi$ in the laboratory frame is calculated as
\begin{equation}
  \Phi = \gamma \bar{\Phi}.
\end{equation}

Accordingly, the horizontal component of the electric field $E_x^0$ excited by the point charge in the laboratory frame is expressed as 
\begin{equation}\label{eq:34}
  \begin{split}
    &E^0_x=-\frac{\partial \Phi}{\partial x} \\
    &=\frac{\lambda}{4\pi\epsilon_0}\frac{KS}{\pi W} \left[\frac{D}{C-C_1}+\frac{D}{C-C_1^x}-\frac{C k^2}{D+D_1}-\frac{C k^2}{D+D_1^x}+\frac{DD_1-C_1 C k^2}{CD_1-C_1 D}-\frac{D D_1^x-C_1^x C k^2}{CD_1^x-C_1^x D} \right]\\
    &+\frac{\lambda}{4\pi\epsilon_0}\frac{KS^x}{\pi W^x} \left[\frac{D^x}{C^x-C_1^x}+\frac{D^x}{C^x-C_1}-\frac{C^x k^2}{D^x+D_1^x}-\frac{C^x k^2}{D^x+D_1} +\frac{D^x D_1^x-C_1^x C^x k^2}{C^x D_1^x -C_1^x D^x}-\frac{D^x D_1-C_1 C^x k^2}{C^x D_1-C_1 D^x} \right],
  \end{split}
\end{equation}
where
\begin{equation}
  W =W(\zeta) = \sqrt{f^2-\zeta^2},
\end{equation}
and $W^x\equiv W(\bar{\zeta})=\overline{W}$.

Subsequently, the electric fields by the elliptic uniform beam is calculated after superposing Eq. \eqref{eq:34} on the beam cross-section as follows
\begin{equation}
  \begin{split}
    E_x &= \frac{1}{\pi a_x a_y} \int \!\!\! \int dx_sdy_s E^0_x I(x_s,y_s) \\
        &= \frac{1}{\pi a_x a_y} \int \!\!\! \int dx_sdy_s (E^0_{x,\mathrm{DS}}+E^0_{x,\mathrm{im}}) I(x_s,y_s),
  \end{split}
\end{equation}
where $E^0_{x,\mathrm{DS}}$ and $E^0_{x,\mathrm{im}}$ are the direct and indirect space charge contributions from the point particle, respectively.

When the witness particles are on the horizontal axis ($y=0$), $E^0_{x,\mathrm{DS}}$ and $E^0_{x,\mathrm{im}}$ are simplified as \cite{yoshimura2025space}
\begin{equation}
  E^0_{x,\mathrm{DS}} = \frac{\lambda}{2\pi\epsilon_0} \frac{x-x_s}{(x-x_s)^2+y_s^2},
\end{equation}
\begin{equation}
  \begin{split}
    E^0_{x,\mathrm{im}} 
    &= \frac{\lambda}{2\pi\epsilon_0} \left[\frac{KS}{\pi W} \left(\frac{D}{C-C_s}+\frac{D}{C-\bar{C_s}}-\frac{C k^2}{D+D_s}-\frac{C k^2}{D+\bar{D_s}}\right) - \frac{x-x_s}{(x-x_s)^2+y_s^2} \right] \\
    &\simeq \frac{\lambda}{2\pi\epsilon_0} \left(A_1 x_s + A_2 x + \frac{A_3}{3}x_s^3 -A_3 x_sy_s^2 + A_4 x_s^2 x -A_4 y_s^2 x + A_3 x_s x^2 \right),
  \end{split}
\end{equation}
in the first order approximation, where
\begin{equation}
  \begin{split}
    A_1 =& \frac{1}{6f^2}\left(1-\frac{4(1-2k^2)K^2}{\pi^2}\right), \\
    A_2 =& \frac{1}{6f^2}\left(2-\frac{4(2-k^2)K^2}{\pi^2}\right), \\
    A_3 =& \frac{1}{120f^4}\left(17 - \frac{40(1-2k^2)K^2}{\pi^2} - \frac{16(7+8k^2-8k^4)K^4}{\pi^4}\right), \\
    A_4 =& \frac{1}{120f^4}\left(8 - \frac{16(8-8k^2-7k^4)K^4}{\pi^4}\right),
  \end{split}
\end{equation}
which respectively approach
\begin{equation}\label{eq:40}
  \begin{split}
    A_1 &\to\frac{1}{h^2_C}, \\
    A_2 &\to0, \\
    A_3 &\to0, \\
    A_4 &\to\frac{1}{h^4_C},
  \end{split}
\end{equation}
equivalent to those of a cylindrical chamber with radius $h_C$, as $h_{E_x}$ approaches $h_{E_y}=h_C$.

The superposition of the electric field on the elliptical beam distribution provides $E_{x,\mathrm{DS}}$ due to the beam itself and $E_{x,\mathrm{im}}$ due to the chamber boundary as follows,
\begin{equation}
  \begin{split}
    E_{x,\mathrm{DS}} 
    &= \frac{1}{\pi a_x a_y} \int \!\!\! \int dx_s dy_s E^0_{x,\mathrm{DS}} I(x_s,y_s) \\
    &= \frac{\lambda}{2\pi^2\epsilon_0 a_x a_y} \int_{-a_x+\bar{x}}^{a_x+\bar{x}} dx_s \int_{-\frac{a_y}{a_x}\sqrt{a_x^2-(x_s-\bar{x})^2}}^{\frac{a_y}{a_x}\sqrt{a_x^2-(x_s-\bar{x})^2}} dy_s \frac{x-x_s}{(x-x_s)^2+y_s^2} \\
    &= \frac{\lambda}{\pi^2\epsilon_0 a_x a_y} \int_{-a_x+\bar{x}}^{a_x+\bar{x}} dx_s \tan^{-1}\left(\frac{a_y}{a_x}\frac{\sqrt{a_x^2-(x_s-\bar{x})^2}}{x-x_s}\right) \\
    &= \frac{\lambda}{\pi\epsilon_0} \frac{x-\bar{x}}{a_x(a_x+a_y)},
  \end{split}
\end{equation}
\begin{equation}\label{eq:42}
  \begin{split}
    E_{x,\mathrm{im}} 
    &=\frac{1}{\pi a_x a_y} \int \!\!\! \int dx_sdy_s E^0_{x,\mathrm{im}} I(x_s,y_s) \\
    &\simeq \frac{\lambda}{2\pi\epsilon_0} \left( A_1 \bar{x} + A_2 x + \frac{1}{4} (a_x^2-a_y^2)(A_3\bar{x} +A_4 x) + x\bar{x}(A_3 x+A_4\bar{x})+ \frac{1}{12}A_3 \bar{x}^3 \right) \\
    &\simeq \frac{\lambda}{2\pi\epsilon_0} \left( A_2 (x-\bar{x}) + (A_1 + A_2)\bar{x} \right) \\
    &= \frac{\lambda}{\pi\epsilon_0} \left( \frac{\epsilon_1}{h^2_{E_y}} (x-\bar{x}) + \frac{\xi_1}{h^2_{E_y}}\bar{x} \right),
  \end{split}
\end{equation}
in the first order approximation, after the third term on the second line of Eq. \eqref{eq:42} is neglected for a beam whose size is considerably smaller than the chamber size, as indicated by Eq. \eqref{eq:40}.
As a result, the coefficients $A_2$ and $A_1+A_2$ are respectively related to the well-known incoherent and coherent electric image coefficients $\epsilon_1$ and $\xi_1$ for the elliptical chamber \cite{laslett1963intensity}.

Since the vector potential $A_z$ in the laboratory frame is obtained by
\begin{equation}
  A_z = \frac{\beta}{c} \gamma \bar{\Phi},
\end{equation}
the electric $F_x^{\mathrm{ele}}$ and the magnetic $F_x^{\mathrm{mag}}$ contributions to the Lorentz force are respectively expressed with the scalar potential $\Phi$ as
\begin{equation}
  F_{x}^{\mathrm{ele}} = - e \nabla_x \Phi = e E_x,
\end{equation}
\begin{equation}
  F_{x}^{\mathrm{mag}} = e \nabla_x A_z \beta c = - \beta^2 e E_x.
\end{equation}

Finally, when an elliptic beam with a horizontal offset passes through the elliptic chamber placed in open space, the contributions to the direct space charge force from the electric field $F^{\mathrm{ele}}_{x,\mathrm{DS}}$, and the magnetic field $F^{\mathrm{mag}}_{x,\mathrm{DS}}$, 
as well as the contributions to the indirect space charge from the electric field $F^{\mathrm{ele}}_{x,\mathrm{im}}$, and the magnetic field $F^{\mathrm{mag}}_{x,\mathrm{im}}$, are summarized as 
\begin{equation}\label{eq:46}
  F_{x,\mathrm{DS}}^{\mathrm{ele}} = \frac{e\lambda}{\pi \epsilon_0 }\frac{x-\bar{x}}{a_x(a_x+a_y)},
\end{equation}
\begin{equation}\label{eq:47}
  F_{x,\mathrm{DS}}^{\mathrm{mag}} = -\beta^2 \frac{e\lambda}{\pi \epsilon_0 } \frac{x-\bar{x}}{a_x(a_x+a_y)},
\end{equation}
\begin{equation}\label{eq:48}
  F_{x,\mathrm{im}}^{\mathrm{ele}} = \frac{e\lambda}{\pi \epsilon_0} \left(\frac{\epsilon_1}{h^2_{E_y}} (x-\bar{x}) + \frac{\xi_1}{h^2_{E_y}} \bar{x}\right),
\end{equation}
\begin{equation}\label{eq:49}
  F_{x,\mathrm{im}}^{\mathrm{mag},\mathrm{ac}} = -\beta^2 \frac{e}{\pi \epsilon_0} \left(\lambda -\frac{eN_B}{C}\right) \left(\frac{\epsilon_1}{h^2_{E_y}} (x-\bar{x}) + \frac{\xi_1}{h^2_{E_y}} \bar{x}\right).
\end{equation}
Here, the dc component of the beam is subtracted in Eq. \eqref{eq:49}, as illustrated by Laslett \cite{laslett1963intensity}, because the magnetic field with the dc component can penetrate the metal chamber.

More generally, we deal with the ring composed of cylindrical and elliptical chambers, like the J-PARC MR,
by replacing the coherent electric image coefficients $\epsilon_1/h_{E_y}^2$ and $\xi_1/h_{E_y}^2$ in Eqs. \eqref{eq:48} and \eqref{eq:49} as
\begin{equation}
  \frac{\epsilon_{1}}{h_{E_y}^2} \to F \frac{\epsilon_{1,E}}{h^2_{E_y}} + (1-F) \frac{\epsilon_{1,C}}{h_C^2} = F \frac{\epsilon_{1,E}}{h^2_{E_y}},
\end{equation}
\begin{equation}\label{eq:51}
  \frac{\xi_{1}}{h_{E_y}^2} \to F \frac{\xi_{1,E}}{h^2_{E_y}} + (1-F) \frac{\xi_{1,C}}{h_C^2} = \mathcal{F} \frac{\xi_{1,C}}{h_C^2},
\end{equation}
where $F$ is the filling factor of the elliptical chambers along the ring because the Laslett parameters are given by $\epsilon_{1,C} =0$ and $\xi_{1,C} =1/2$ for the cylindrical chamber \cite{laslett1963intensity}.
In the case of the J-PARC MR, $h_C=0.065$~m; $F=0.358$; $\epsilon_{1,E}=-0.10$ and $\xi_{1,E}=0.25$, because of $2h_{E_x}=2\times0.065$~m and $2h_{E_y}=2\times0.050$~m, producing $\mathcal{F} = 0.94$ in Eq. \eqref{eq:51}.
Finally, we modify Eqs. \eqref{eq:48} and \eqref{eq:49} as
\begin{equation}\label{eq:52}
  F_{x,\mathrm{im}}^{\mathrm{ele}} = \frac{e\lambda}{\pi \epsilon_0} \left(F\frac{\epsilon_{1,E}}{h^2_{E_y}} (x-\bar{x}) + \mathcal{F}\frac{\xi_{1,C}}{h^2_C} \bar{x}\right),
\end{equation}
\begin{equation}\label{eq:53}
  F_{x,\mathrm{im}}^{\mathrm{mag},\mathrm{ac}} = -\beta^2 \frac{e}{\pi \epsilon_0} \left(\lambda-\frac{eN_B}{C}\right) \left(F\frac{\epsilon_{1,E}}{h^2_{E_y}} (x-\bar{x}) + \mathcal{F}\frac{\xi_{1,C}}{h^2_C} \bar{x}\right),
\end{equation}
respectively.

Next, let us consider the chamber placed in the bending magnet to derive the dc component in the Lorentz force by presuming a coasting beam with line density $eN_B/C$
with the horizontal centroid of the entire bunch $X$.
In the case, we assume only the magnetic pole surfaces of parallel plates placed perpendicularly to the vertical axis with a distance of $2g$ as the boundary condition, neglecting the chamber boundary.

Because the vector potential $A_z$ by a point charge located at ($x_s,y_s$) in the laboratory frame is given by
\cite{zotter1975tune}
\begin{widetext}
\begin{equation}
  A_z=-\frac{\beta}{c}\frac{1}{4\pi\epsilon_0} \frac{eN_B}{C} \log\left[(\cosh{\kappa(x-x_s)}-\cos{\kappa(y-y_s)})(\cosh{\kappa(x-x_s)}+\cos{\kappa(y+y_s)})\right],
\end{equation}
\end{widetext}
where $\kappa=\pi/2g$, 
the corresponding Lorentz force $F_x^{\mathrm{mag},0}$ is calculated as
\begin{widetext}
\begin{equation}
  \frac{F_x^{\mathrm{mag},0}}{\beta^2e}=\frac{\partial A_z}{\partial x}\frac{c}{\beta} = -\frac{1}{2\pi\epsilon_0} \frac{eN_B}{C} \frac{\kappa(\cosh{\kappa(x-x_s)}-\sin{\kappa y}\sin{\kappa y_s})\sinh{\kappa(x-x_s)}}{(\cosh{\kappa(x-x_s)}-\cos{\kappa(y-y_s)})(\cosh{\kappa(x-x_s)}+\cos{\kappa(y+y_s)})} ,
\end{equation}
\end{widetext}
which is simplified as
\begin{widetext}
\begin{equation}
  \begin{split}
    \frac{F_{x,\mathrm{im}}^{\mathrm{mag},0}}{\beta^2e} &= -\frac{1}{2\pi\epsilon_0} \frac{eN_B}{C} \left(\frac{\kappa\cosh{\kappa(x-x_s)}\sinh{\kappa(x-x_s)}}{(\cosh{\kappa(x-x_s)}-\cos{\kappa y_s})(\cosh{\kappa(x-x_s)}+\cos{\kappa y_s})} - \frac{x-x_s}{(x-x_s)^2+y_s^2}\right) \\
                                      &\simeq -\frac{1}{2\pi\epsilon_0} \frac{eN_B}{C} \left( -\frac{\kappa^2}{3}x_s +\frac{\kappa^2}{3} x +\frac{\kappa^4}{45} x_s^3 -\frac{\kappa^4}{15} x_s y_s^2 -\frac{\kappa^4}{15} x_s^2 x +\frac{\kappa^4}{15} y_s^2 x +\frac{\kappa^4}{15} x_s x^2\right),
  \end{split}
\end{equation}
\end{widetext}
on the horizontal axis ($y=0$).
Similar to deriving Eq. \eqref{eq:42},
the straightforward superposition of the magnetic field on the elliptical beam provides the contribution to the Lorentz force $F^{\mathrm{mag},\mathrm{DC}}_{x,\mathrm{im}}$ only from the magnetic boundary as 
\begin{equation}
  \begin{split}
    \frac{F_{x,\mathrm{im}}^{\mathrm{mag},\mathrm{dc}}}{\beta^2e} 
    &=\frac{1}{\pi a_x a_y} \int \!\!\! \int dx_sdy_s \frac{F_{x,\mathrm{im}}^{\mathrm{mag},0}}{\beta^2e} I(x_s,y_s) \\
    &\simeq -\frac{1}{2\pi\epsilon_0} \frac{eN_B}{C} \left( \frac{\kappa^2}{3}(x-X) -\frac{\kappa^4}{60} (a_x^2-a_y^2)(x-X) +\frac{\kappa^4}{15} x\bar{x}(x-X) +\frac{\kappa^4}{45} X^3 \right) \\
    &\simeq -\frac{1}{\pi\epsilon_0} \frac{eN_B}{C} \frac{\pi^2}{24} \frac{1}{g^2} (x-X) \\
    &= \frac{1}{\pi\epsilon_0} \frac{eN_B}{C} \left( \frac{\epsilon_2}{g^2} (x-X) + \frac{\xi_2}{g^2}X \right),
  \end{split}
\end{equation}
where the higher-order terms are neglected by assuming the large magnetic pole surfaces.
Again, the well-known incoherent and coherent magnetic image coefficients for the parallel plate boundary, $\epsilon_2$ and $\xi_2$, appear in the Lorentz force.
Note that the Laslett parameter $\xi_2=0$ in the horizontal direction when the parallel magnetic planes \cite{laslett1963intensity} are oriented perpendicular to the vertical axis.

Thus, the Lorentz force due to the magnetic image is given by
\begin{equation}\label{eq:58}
  F_{x,\mathrm{im}}^{\mathrm{mag},\mathrm{dc}} = \beta^2 \frac{e}{\pi \epsilon_0 } \frac{eN_B}{C} F_B \left(\frac{\epsilon_2}{g^2} (x-X) + \frac{\xi_2}{g^2} X \right),
\end{equation}
where $F_B$ is the filling factor of the bending magnet along the ring.
In the case of the J-PARC MR, the half-height of the ferromagnetic material $g$ is equal to $0.053$~m, and the filling factor $F_B=F$, because the elliptical chamber only occupies the bending region.

Finally, we can simplify the equation of motion by summing up
Eqs. \eqref{eq:46}, \eqref{eq:47}, \eqref{eq:52}, \eqref{eq:53}, and \eqref{eq:58} as
\begin{widetext}
\begin{equation}\label{eq:59}
  \begin{split}
    x''&= \frac{1}{\gamma m_p c^2 \beta^2} \left(F_{x,\mathrm{DS}}^{\mathrm{ele}}+F_{x,\mathrm{DS}}^{\mathrm{mag}}+F_{x,\mathrm{im}}^{\mathrm{ele}}+F_{x,\mathrm{im}}^{\mathrm{mag},\mathrm{ac}}+F_{x,\mathrm{im}}^{\mathrm{mag},\mathrm{dc}}\right) \\
       &= \frac{e}{\pi \epsilon_0 m_p c^2 \gamma \beta^2} \left[\left(\frac{\lambda}{\gamma^2} \frac{1}{a_x(a_x+a_y)} + \frac{\lambda}{\gamma^2} F\frac{\epsilon_{1,E}}{h^2_{E_y}} + \beta^2 \frac{eN_B}{C} F\frac{\epsilon_{1,E}}{h^2_{E_y}} \right) (x-\bar{x}) + \beta^2 \frac{eN_B}{C} F\frac{\epsilon_2}{g^2} (x-X) \right] \\
       &+ \frac{e}{\pi \epsilon_0 m_p c^2 \gamma \beta^2} \left(\frac{\lambda}{\gamma^2} \mathcal{F} \frac{\xi_{1,C}}{h^2_C} + \beta^2 \frac{eN_B}{C} \mathcal{F}\frac{\xi_{1,C}}{h^2_C} \right) \bar{x}.
  \end{split}
\end{equation}
\end{widetext}

Accordingly, the horizontal kick at the mesh grid point $k$ by the space charge implemented into the code is simplified as
\begin{equation}\label{eq:60}
 \Delta x'_i = x'' c\beta \Delta T = -\frac{4\pi}{\beta_x} \left(\nusc^{(k)} (x_i-x^{(k)}) + \nuic^{(k)} x^{(k)}\right) \frac{\Delta T}{T_0},
\end{equation}
where the incoherent tune shift $\nusc^{(k)}$ in $k$th segment approximated as
\begin{equation}\label{eq:61}
  \begin{split}
    \nusc^{(k)} \simeq -\frac{r_p C^2}{2\pi^2 e \gamma^3 \beta^2 \nu_{\beta,x}}\frac{\lambda^{(k)}}{a_x^{(k)}(a_x^{(k)}+a_y)},
  \end{split}
\end{equation}
with the classical particle radius of the proton $r_p=e^2/4\pi\epsilon_0 m_p c^2=1.5\times10^{-18}$~m
by neglecting the contribution of terms proportional to $\epsilon_{1,E}$ and $\epsilon_2$, 
because the beam size in the J-PARC MR, which is typically $a_{x,y}\leq0.015$~m, is considerably smaller than the typical chamber size: $0.065$~m,
with the coherent tune shift $\nuic^{(k)}$ in $k$th segment summarized as
\begin{equation}\label{eq:62}
  \nuic^{(k)} = - \frac{r_p C^2}{2\pi^2 e \gamma^3\beta^2 \nu_{\beta,x}} \lambda^{(k)} \mathcal{F}\frac{\xi_{1,C}}{h^2_C} - \frac{N_B r_p C}{2\pi^2 \gamma \nu_{\beta,x}} \mathcal{F}\frac{\xi_{1,C}}{h^2_C}.
\end{equation}

Note that the respective macroparticles must follow the symplectic transformation even when the space charge effects are incorporated into the code.
This is summarized as
\begin{equation}
  x_{n+1} =\cos \left(2\pi (\nu_{\beta,x} + \xi_x \delta_n) \frac{\Delta T}{T_0}\right) x_n + \beta_x \sin \left(2\pi (\nu_{\beta,x} + \xi_x \delta_n) \frac{\Delta T}{T_0}\right) x'_n,
\end{equation}
\begin{equation}\label{eq:63}
  \begin{split}
     x'_{n+1} =& -\frac{1}{\beta_x} \sin \left(2\pi (\nu_{\beta,x} + \xi_x \delta_n) \frac{\Delta T}{T_0}\right) x_n + \cos \left(2\pi (\nu_{\beta,x} + \xi_x \delta_n) \frac{\Delta T}{T_0}\right) x'_n \\
               & -\frac{4\pi}{\beta_x}\nusc \frac{\Delta T}{T_0} (x_{n+1}-\overline{x_{n+1}}) -\frac{4\pi}{\beta_x}\nuic \frac{\Delta T}{T_0} \overline{x_{n+1}},
  \end{split}
\end{equation}
after combining Eqs. \eqref{eq:3}, \eqref{eq:4}, \eqref{eq:5}, and \eqref{eq:60}, where $\nusc$ and $\nuic$ must be assigned to those in the segment in which the transferred macroparticles are located, and $\overline{x_{n+1}}$ is the transverse centroid of macroparticles in the segment after $(n+1)$th step.
To ensure the transformation remains symplectic, it is crucial that the third and fourth terms on the right-hand side of Eq. \eqref{eq:63} are proportional to $x_{n+1}$ rather than $x_n$.

\section{Implementation of the horizontal wake potential into the code}\label{5}


If the space charge force affects the recoherence time, the wakefields could potentially have an impact on it as well.
Since the authors have identified that the effect of the longitudinal impedance on the decoherence time is negligible
\cite{Yoshimura:2023idd,yoshimura2025space},
this section describes how to implement the horizontal wakefields due to the resistive wall chambers into the code to simulate the impedance source in the J-PARC MR.

The horizontal resistive wall wake function $\WT(s)$ is given by
\cite{shobuda2002resistive,yokoya1993resistive,gluckstern1993coupling,kobayashiinvestigation}
\begin{equation}\label{eq:65}
  \begin{split}
    \WT(s) &= \frac{L D_{1x}(u_0)}{h_{E_y}^3} \frac{2 c Z_0 \rho}{\pi d_\omega} \sum^\infty_{n=0} \exp\left(-\frac{\pi^2\rho}{d_\omega^2} s \left(n+\frac{1}{2}\right)^2\right)\\
      &\approx
      \begin{cases}
        \frac{L D_{1x}(u_0)}{h_{E_y}^3} \frac{c Z_0 \sqrt{\rho}}{\pi\sqrt{\pi}} \sqrt{\frac{1}{s}}, & \text{for}~0<s<500\mathrm{m}, \\
        \frac{L D_{1x}(u_0)}{h_{E_y}^3} \frac{2 c Z_0 \rho}{\pi d_\omega} \exp\left(-\frac{\pi^2\rho}{d_\omega^2}\frac{s}{4}\right), & \text{for}~s>500\mathrm{m},
      \end{cases}
  \end{split}
\end{equation}
for an elliptic chamber with the major horizontal axis $2h_{E_x} = 2f \cosh{u_0}$, the minor vertical axis $2h_{E_y} = 2f \sinh{u_0}$, ($f=\sqrt{h_{E_x}^2-h_{E_y}^2}$), a finite thickness $d_\omega$, a length $L$,
SUS316L resistivity $\rho=1.96\times10^{-9}$~m, and the impedance of free space $Z_0=120\pi \mathrm{\Omega}$,
where
\begin{equation}
  D_{1x}(u_0)=\frac{\sinh^3{u_0}}{4\pi} \int_0^{2\pi} dv \frac{Q_{1x}^2(v)}{\sqrt{\sinh^2{u_0}+\sin^2{v}}},
\end{equation}
\begin{equation}
  Q_{1x}(v) = 2 \sum_{m=0}^\infty (-1)^m (2m+1) \frac{\cos{(2m+1)v}}{\cosh{(2m+1)u_0}}.
\end{equation}

The factor $D_{1x}(u_0)$ characterizes the asymmetry of the elliptic chamber and approaches one for a cylindrical chamber.
Notice that the wake function $\WT(s)$ is proportional to $L D_{1x}/h_{E_y}^3$.
As the thickness $d_\omega$ goes to infinity, Eq. \eqref{eq:65} becomes proportional to $1/\sqrt{s}$, which is the typical behavior of the resistive wall impedance.
In other words, we can make use of the damping effect due to the chamber thickness in Eq. \eqref{eq:65} to calculate beam behavior affected by the resistive wall impedance.

In the J-PARC MR, where the cylindrical chamber with an inner radius $h_C$(=0.065 m) dominates, the thickness of the chamber is $d_\omega$(=0.002 m).
The bending magnet region, where the elliptical ducts with a major axis $2h_{E_x}$($= 2 \times 0.065$~m) and a minor axis $2h_{E_y}$($= 2 \times 0.050$~m), equivalent to $D_{1x} = 0.615$, are placed, occupies $F$(= 0.358) of the total circumference of the ring.
Hence, the total resistive wall impedance along the ring, including the effect of the elliptic chambers, is enhanced by the factor $\mathcal{G}$:
\begin{equation}
  \mathcal{G} = F D_{1x} \frac{h_C^3}{h_{E_y}^3}+(1-F)=1.1,
\end{equation}
compared to the case where the ring is assumed to be perfectly composed of cylindrical chambers.
This indicates that the assumption that the resistive wall wake for the cylindrical chamber:
\begin{equation}
  \WT(s) = \frac{C}{h_C^3} \frac{2 c Z_0 \rho}{\pi d_\omega} \sum^\infty_{n=0} \exp\left(-\frac{\pi^2\rho}{d_\omega^2} s \left(n+\frac{1}{2}\right)^2\right),
\end{equation}
perfectly occupies the J-PARC MR, estimates the resistive wall impedance of the ring within about 10\% accuracy, which is applied for our formulation.

The resistive wall impedance provides the kick on the witness particle in a similar way as the longitudinal wake case.
Concretely, the horizontal dipole kick $\mathrm{kick}_x^{(k)}$ at the mesh grid point $k$ with a single-turn wake is calculated into the code as
\begin{equation}
  \mathrm{kick}_x^{(k)} = \sigma_\Delta \sum_j \WT(z_j-z^{(k)})\Delta^{(j)},
\end{equation}
where the dipole moment $\Delta^{(j)}$ of $j$th segment is calculated as Eq. \eqref{eq:16}.
Using this kick at each segment, the horizontal dipole kick $\mathrm{kick}_{x,i}$ on $i$th macroparticle between $k$th and $(k+1)$th grid points are estimated as
\begin{equation}
  \mathrm{kick}_{x,i} = \left(1-\frac{z_i-z^{(k)}}{\sigma_\Delta}\right) \mathrm{kick}_x^{(k)} + \frac{z_i-z^{(k)}}{\sigma_\Delta} \mathrm{kick}_x^{(k+1)}.
\end{equation}

Finally, the horizontal dipole kick at each time step is provided as
\begin{equation}\label{eq:72}
  \Delta x'_i = \frac{1}{E_0} \mathrm{kick}_{x,i} \frac{\Delta T}{T_0}.
\end{equation}

Taking into account the thickness of the chamber facilitates dealing with the multiturn long-range effects on the beams with the rapid decay of the resistive wall impedance, according to Eq. \eqref{eq:65}.
To see the convergence of the accumulated multiturn wake, 
let us simply sum the resistive wall impedance up to $N_\lambda$, assuming the positions of beams exciting the wakes in the previous turns are identical, as
\begin{equation}\label{eq:73}
  V(N_\lambda) =\sum_{l=1}^{N_\lambda} \WT(lC).
\end{equation}

\begin{figure}[!h]
  \begin{tabular}{cc}
    \begin{minipage}[t]{0.45\hsize}
      \centering
      \includegraphics[width=3.0in]{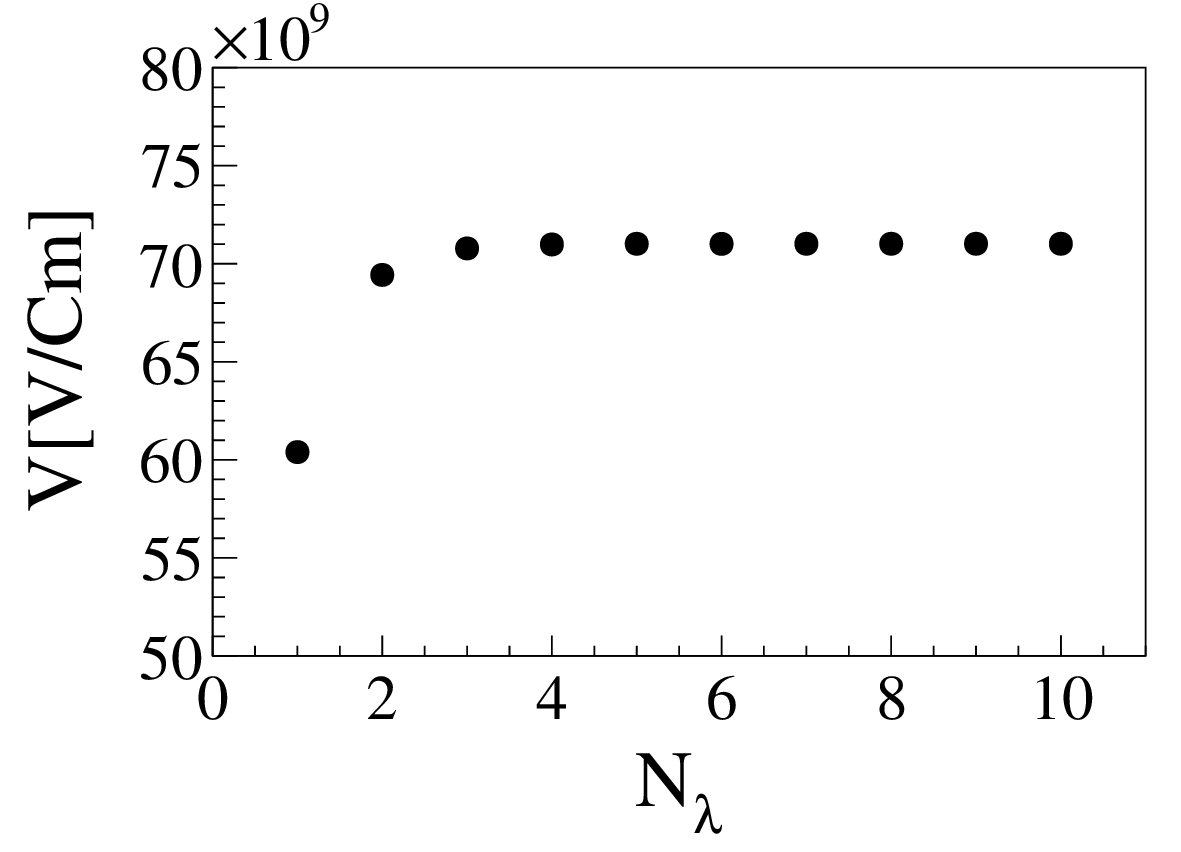}
      \subcaption{}
    \end{minipage} &
    \begin{minipage}[t]{0.45\hsize}
      \centering
      \includegraphics[width=3.0in]{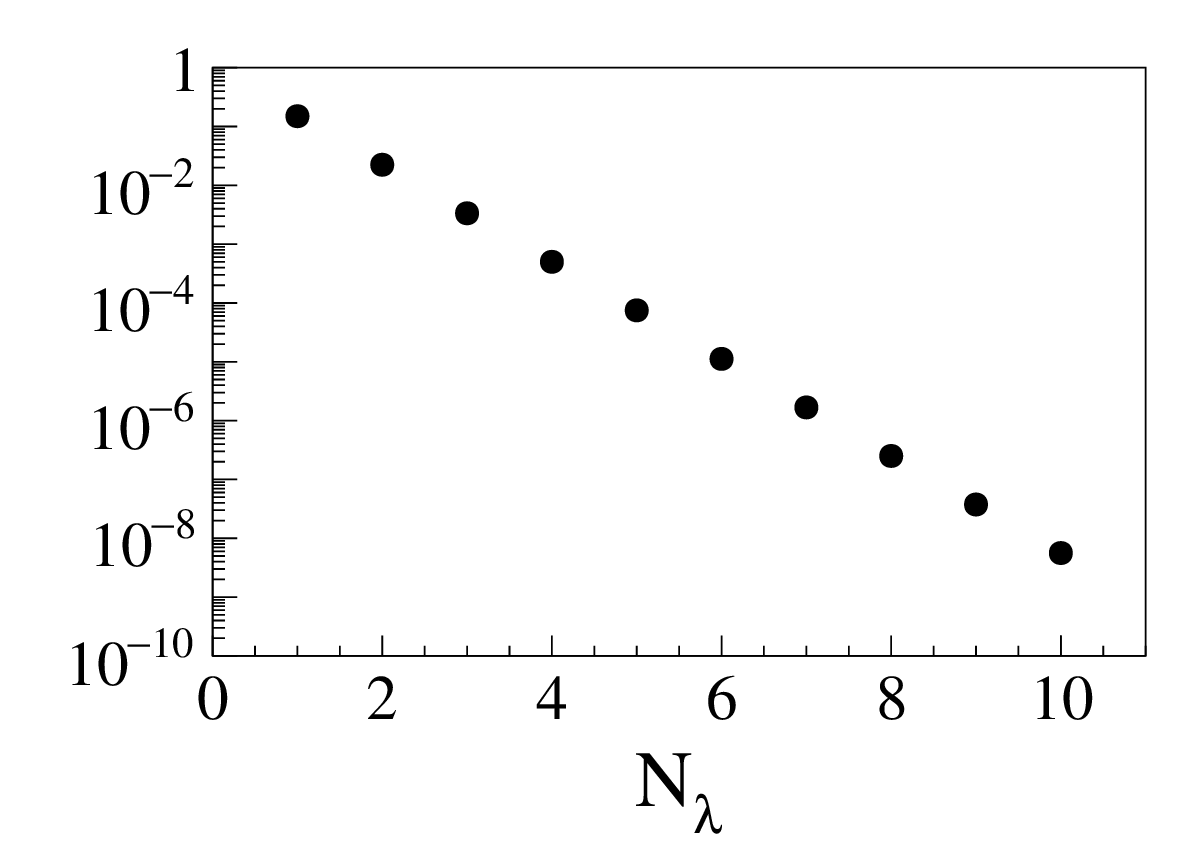}
      \subcaption{}
    \end{minipage}
  \end{tabular}
  \caption{Sum of multiturn wakes for different $N_\lambda$(a) and the deviation from the analytical estimate with infinite $N_\lambda$(b). Here, $V$ represents the accumulated wake function values, assuming they are stored only for the past $N_\lambda$ turns. (a) $V(N_\lambda)$ and (b) $1-V(N_\lambda)/V(\infty)$}
  \label{fig:1}
\end{figure}

Figure \ref{fig:1} shows Eq. \eqref{eq:73} for different $N_\lambda$ and the deviation of $V(N_\lambda)$ from the analytical estimate of Eq. \eqref{eq:73} with infinite $N_\lambda$, respectively.
In the J-PARC MR with circumference $C$(=1567.5 m), the deviation reduces to nearly 0.1\% after $N_\lambda=4$ turns as already revealed in Ref. \cite{kobayashiinvestigation}, because the thickness of the installed chamber is typically $d_\omega$(=0.002 m).
Therefore, we conclude that $N_\lambda=4$ is sufficient to adequately incorporate the accumulation effects of the long-range wakefields in the code.

More explicitly, the accumulated horizontal kicks are calculated as
\begin{equation}\label{eq:74}
  \Delta x' = \frac{1}{E_0} \mathrm{kick}_x \frac{\Delta T}{T_0},
\end{equation}
where 
\begin{equation}
  \mathrm{kick}_x = \sum_{l=1}^{4} \WT(lC) \Delta_{n-l},
\end{equation}
and the dipole moment of the respective bunches $\Delta_n$ is
\begin{equation}\label{eq:76}
  \Delta_n = q \sum_i x_{n,i} = eN_B \overline{x_n}.
\end{equation}
Here, the transverse centroid of the entire bunch $\overline{x_n}$ in Eq. \eqref{eq:76} is calculated by averaging the macroparticle positions in the bunch following Eq. \eqref{eq:13}.

As a result, the short-range [Eq. \eqref{eq:72}] and long-range [Eq. \eqref{eq:74}] wakes are simultaneously implemented into the code by making use of the feature of resistive wall impedance with chamber thickness.

\section{Definition of head-tail modes}\label{6}

\begin{table*}[!h]
  \caption{List of symbols in Sec. \ref{6}.}
  \begin{ruledtabular}
  \centering
  \small
  \begin{tabular}{c  c  c}
    Term & Unit & Description \\ \hline
    $F_{\mathrm{beam}}$   & N       & Space charge force and wakefield                         \\ 
    $x_u,x_d$             & m       & Transverse position of each stream                       \\ 
    $\phi_u,\phi_d$       & rad     & Betatron phase advance of each stream                    \\ 
    $\delta_u,\delta_d$   &         & Momentum deviation of each stream                        \\ 
    $\chiABS$             & rad     & Chromatic phase in the ABS model                         \\ 
    $\chiABH$             & rad     & Chromatic phase in the ABH model                         \\ 
    $k$                   &         & Eigenmode number index                                   \\ 
    $\Delta \nu_x, \Delta \nu_{\pm,k}$ &       & Eigenvalue of the head-tail mode              \\ 
    $A_m$                 &         & Head-tail mode coefficient                               \\ 
    $\amABS$              &         & Relative amplitude in the ABS model                      \\ 
    $\amABH$              &         & Relative amplitude in the ABH model                      \\ 
    $\tamABS(z)$          &         & Relative amplitude in the ABS model in each segment      \\ 
    $\theta$              & rad     & Synchrotron phase                                        \\ 
    $\theta_0$            & rad     & Synchrotron initial phase                                \\ 
    $a_{m,b}$             &         & Relative amplitude for the Gaussian beam                 \\ 
    $\tilde{a}_{m,b}(z)$  &         & Relative amplitude for the Gaussian beam in each segment \\ 
  \end{tabular}
  \end{ruledtabular}
  \label{tab:sec6}
\end{table*}

The space charge intrinsic eigenmodes, head-tail modes, originally introduced by Blaskiewicz \cite{blaskiewicz1998fast}, are quite beneficial to interpret the following simulation results describing the intrabunch motion, though the head-tail modes are assumed to be excited by wakes in most of the cases.
In this section, we delineate the intrabunch transverse motions with respect to the space charge intrinsic modes, based on the ABS model,
and an airbag harmonic model (ABH model) which replaces the square-well potential in the original ABS model with a realistic harmonic potential for the longitudinal motions.
Finally, the airbag distribution will be replaced by a Gaussian distribution under a weak space charge force as an approximation.

The detailed derivation is beneficial for understanding how the space charge intrinsic modes are dynamically excited by the synchrotron motions in the bunch, and for qualitatively understand the measurements.
The derived formula will be used as a tool for conducting benchmark tests of our simulation code in the following sections.
The symbols used in this section are defined in Table \ref{tab:sec6}.

\subsection{Head-tail modes in the ABS model with space charge effects}\label{6.1}
The horizontal motion of a beam particle in the beam obeys the equation of motion as
\cite{Ng:2002gr}
\begin{equation}\label{eq:77}
  \frac{d^2}{ds^2}x + \frac{4\pi^2}{C^2}(\nu_{\beta,x} + \delta\xi_x)^2x = \frac{F_{\mathrm{beam}}(x,\bar{x})}{\gamma m_p c^2 \beta^2},
\end{equation}
where $F_{\mathrm{beam}}$ represents the space charge force and wakefields.
When the effects of the wakefields can be neglected, Eq. \eqref{eq:77} simplifies to
\begin{equation}\label{eq:78}
  \frac{d^2}{dn^2}x + 4\pi^2(\nu_{\beta,x} + \delta\xi_x)^2 x = - 8\pi^2\nu_{\beta,x} (\nusc (x-\bar{x}) + \nuic \bar{x}),
\end{equation}
after transforming the variable $s$ to the turn number $n$ by using the relation $s=Cn$,
where $\nusc$ and $\nuic$ are assumed to be representative of the values at the bunch center, with their $z$-dependence negligible.

This simplified equation of motion
is realized with the validity of the linear approximation by referring to Eqs. \eqref{eq:59}, \eqref{eq:61}, and \eqref{eq:62}, noting that the transverse beam size must be significantly smaller than the chamber size.
Here, we should also note that even an indirect space charge effect can be addressed through the coherent space charge tune shift in Eq. \eqref{eq:78}.

In the ABS model, Eqs. \eqref{eq:6}, \eqref{eq:11}, and \eqref{eq:12} follow
\begin{equation}\label{eq:79}
  \frac{dz}{dn}= -\eta C \delta,
\end{equation}
\begin{equation}\label{eq:80}
  \delta = \pm \delta_0 = \pm \frac{4\hat{z}\nu_s}{|\eta| C}.
\end{equation}

Let us define the horizontal positions of particles with momentum deviation $\delta_0$ and those with $-\delta_0$ as $x_u$ and $x_d$, respectively.
When $\eta<0$, we obtain the total derivative with respect to $n$ for $x_u$ and $x_d$ as follows
\begin{equation}\label{eq:81}
  \frac{d}{dn}x_{u,d} = \frac{\partial}{\partial n} x_{u,d} \pm 4\hat{z} \nu_s \frac{\partial}{\partial z} x_{u,d},
\end{equation}
where the plus and minus signs in front of the second term correspond to the equations for $x_u$ and $x_d$, respectively.

On the other hand, the equations of motion for particles Eq. \eqref{eq:78} with positive momentum $\delta=\delta_0$ and negative momentum $\delta=-\delta_0$ are described as follows
\begin{equation}\label{eq:82}
  \frac{d}{dn}x_{u,d} = - \textbf{i} 2\pi\nu_{\beta,x} x_{u,d} \mp \textbf{i} 2\pi\delta_0 \xi_x x_{u,d} - \textbf{i} 2\pi\nusc (x_{u,d}-\bar{x}) - \textbf{i} 2\pi\nuic \bar{x},
\end{equation}
after neglecting the second-order terms for the coherent and incoherent tune shifts.
Here, $\bar{x}=(x_u+x_d)/2$ denotes the transverse centroid of the beam.
The minus and plus signs in front of the second term in Eq. \eqref{eq:82} correspond to the equations for $x_u$ and $x_d$, respectively.

During synchrotron motion in the square-well potential, the transverse positions do not change abruptly with an instantaneous change in momentum.
Hence, the boundary conditions are imposed as
\begin{equation}
  \begin{cases}
    x_u\left(z=\hat{z}\right) = x_d\left(z=\hat{z}\right), \\
    x_u\left(z=-\hat{z}\right) = x_d\left(z=-\hat{z}\right),
  \end{cases}
\end{equation}
which provides the eigenvalue $\Delta \nu_x$ as \cite{yoshimura2025space}
\begin{equation}\label{eq:84}
  \begin{cases}
    \Delta \nu_{\pm,k} = \frac{\nusc+\nuic}{2} \pm \sqrt{\frac{(\nusc-\nuic)^2}{4} + k^2 \nu_s^2}, \\
    \Delta \nu_0 = \nuic,
  \end{cases}
\end{equation}
which satisfies $\sqrt{(\Delta \nu_x-\nusc)(\Delta \nu_x-\nuic)} = k \nu_s$.
Notice that the eigenvalue $\Delta \nu_{\pm,k}$ and the corresponding amplitude $A_{\pm,k}$ are independently defined for the respective branches $\pm$ in $\Delta \nu_{\pm,k}$ for the specified $k\in \mathbb{N} =\lbrace 1,2,3, \cdots\rbrace$,
and the eigenvalue $\Delta \nu_0$ and the amplitude $A_0$ express the $m=0$ mode, having only one branch.
Equation \eqref{eq:84} reproduces the result found in Ref. \cite{blaskiewicz1998fast} as $\nuic$ approaches zero neglecting the indirect space charge effect.

Finally, the superposition of these solutions provides the horizontal positions of particles with $+\delta_0$ and $-\delta_0$ as \cite{yoshimura2025space}
\begin{equation}\label{eq:85}
  \begin{split}
    x_u(n,z) =& \exp\left(\textbf{i} \chiABS \frac{z}{\hat{z}} - \textbf{i} 2\pi\nu_{\beta,x} n \right) A_0 \exp\left(- \textbf{i} 2\pi\Delta \nu_0 n \right)\\
             +& \exp\left(\textbf{i} \chiABS \frac{z}{\hat{z}} - \textbf{i} 2\pi\nu_{\beta,x} n \right) \sum_{\pm,k\in \mathbb{N}} A_{\pm,k} \exp\left(- \textbf{i} 2\pi\Delta \nu_{\pm,k} n \right) \times \left( \cos(k \theta') - \textbf{i} \frac{\Delta \nu_{\pm,k}-\nuic}{k\nu_s} \sin(k \theta') \right),
  \end{split}
\end{equation}
\begin{equation}\label{eq:86}
  \begin{split}
    x_d(n,z) =& \exp\left(\textbf{i} \chiABS \frac{z}{\hat{z}} - \textbf{i} 2\pi\nu_{\beta,x} n \right) A_0 \exp\left(- \textbf{i} 2\pi\Delta \nu_0 n \right)\\
             +& \exp\left(\textbf{i} \chiABS \frac{z}{\hat{z}} - \textbf{i} 2\pi\nu_{\beta,x} n \right) \sum_{\pm,k\in \mathbb{N}} A_{\pm,k} \exp\left(- \textbf{i} 2\pi\Delta \nu_{\pm,k} n \right) \times \left( \cos(k \theta') + \textbf{i} \frac{\Delta \nu_{\pm,k}-\nuic}{k\nu_s} \sin(k \theta') \right),
  \end{split}
\end{equation}
where $\theta' = -\pi/2\times (z/\hat{z}-1)$ with ($0<\theta'<\pi$),
and the chromatic phase $\chiABS$ in the ABS model is defined by
\begin{equation}\label{eq:87}
  \chiABS=2\pi \frac{\xi_x}{\eta C} \hat{z},
\end{equation}
which differs from the definition for the Gaussian beam [refer to Eq. \eqref{eq:20}].

Equations \eqref{eq:85} and \eqref{eq:86} produce
the transverse centroid of each segment in the bunch as
\begin{widetext}
\begin{equation}\label{eq:88}
  \begin{split}
    \bar{x}(n,\theta') 
                 =&  X_0 \sum_{m \in \mathbb{Z}} \amABS \exp\left( \textbf{i} \chiABS \frac{z}{\hat{z}} - \textbf{i} 2\pi \nu_m n \right) \cos\left( m \frac{\pi}{2} \left(\frac{z}{\hat{z}}-1\right) \right) \\
                 =&  X_0 \sum_{m \in \mathbb{Z}} \tamABS(z) \exp\left(- \textbf{i} 2\pi\nu_m n\right),
  \end{split}
\end{equation}
\end{widetext}
with the space charge intrinsic modes $\nu_m$, the head-tail modes.
Here, the relative amplitude $\tamABS(z)$ in each segment is defined as
\begin{equation}\label{eq:89}
  \tamABS(z) = \amABS \exp\left( \textbf{i} \chiABS \frac{z}{\hat{z}}\right) \cos\left( m \frac{\pi}{2} \left(\frac{z}{\hat{z}}-1\right) \right),
\end{equation}
where
\begin{equation}\label{eq:90}
  \begin{split}
    \nu_m=&
    \begin{cases}
    \nu_0 = \nu_{\beta,x} + \Delta \nu_0, \\
    \nu_{\pm k} = \nu_{\beta,x} + \Delta \nu_{\pm,k}, \\
    \end{cases}\\
    A_m=&
    \begin{cases}
    A_0, \\
    A_{\pm k} = A_{\pm,k}, \\
    \end{cases}\\
    \amABS &= \frac{A_m}{X_0},
  \end{split}
\end{equation}
and $m\in \mathbb{Z}$ is an integer.
The expansion coefficient $A_m$ is normalized to $\amABS$ by dividing by the initial horizontal position of the beam $\bar{x}(n=0,z)=X_0$ (refer to Sec. \ref{3}), 
where all macroparticles are aligned regardless of $z$; and the movable region of mode $m$ in the final expression of Eq. \eqref{eq:88} is extended to the entire integer region by combining the minus branch with the original mode $k$.

\subsection{Derivation of the oscillation amplitude with weak space charge force under the ABS model}\label{6.2}
In Eqs. \eqref{eq:85}-\eqref{eq:88}, the amplitude $A_m$ is still an unknown factors.
In general, determining these parameters analytically becomes quite difficult when space charge effects are significant.
However, when this effect is neglected or within the minor level, the amplitudes can be explicitly obtained by comparing Eqs. \eqref{eq:85}-\eqref{eq:88} with the horizontal positions derived independently based on the linear decoherence scenario in Sec. \ref{3}, after being applied to the ABS model.

Let us examine the transverse centroid $\bar{x}(n,z)$ for a group of particles with the same longitudinal position $z$ in a given turn $n$.
As described in Sec. \ref{3}, the initial condition is set as $\bar{x}(n=0,z)=X_0$ regardless of the longitudinal position of the particles, with an initial betatron phase of $\phi(n=0,z)=0$.

The equation of motion for a single particle corresponding to Eqs. \eqref{eq:6} and \eqref{eq:11} is given by Eqs. \eqref{eq:79} and \eqref{eq:80}.
Since the synchrotron phase is followed as $\theta=2\pi\nu_s n+\theta_0$ with initial phase $\theta_0$,
the equation of motion for a single particle is described as
\begin{equation}
  \frac{dz}{d\theta} = \frac{dz}{dn}\frac{dn}{d\theta} = \mp\frac{2\hat{z}}{\pi}.
\end{equation}

Therefore, each particle motion is followed as
\begin{equation}
  z(n,\theta_0) = -\frac{2\hat{z}}{\pi} \cos^{-1}(\cos(\theta(n,\theta_0))) + \hat{z},
\end{equation}
\begin{equation}
  \delta(n,\theta_0) =\delta_0\mathrm{sgn}(\sin(\theta(n,\theta_0))).
\end{equation}

Given $z=z_0$ at $n=n_0$, there are two solutions $\theta_{0,u}(n_0,z_0)$ and $\theta_{0,d}(n_0,z_0)$ for the initial synchrotron phase $\theta_0$, as
\begin{equation}
  \theta_{0,(u,d)}(n_0,z_0) = \mp \frac{\pi}{2\hat{z}}\left(z_0 - \hat{z}\right) - 2\pi\nu_s n_0.
\end{equation}

Hence, the betatron phase advance $\phi_u(n_0,z_0)$ and $\phi_d(n_0,z_0)$ can be calculated for the respective $\theta_{0,(u,d)}(n_0,z_0)$ as 
\begin{widetext}
\begin{equation}
  \begin{split} 
    \phi_{u,d}(n_0,z_0) = 2\pi\nu_{\beta,x} n_0 - \chiABS \left(\frac{z_0}{\hat{z}} - 1 + \frac{2}{\pi} \cos^{-1}(\cos\left(\mp \frac{\pi}{2}\left(\frac{z_0}{\hat{z}}-1\right) - 2\pi\nu_s n_0\right))\right).
  \end{split}
\end{equation}
\end{widetext}

Because of the transverse oscillation with initial condition $\bar{x}(n=0,z)=X_0$,
the transverse positions of particles with $\pm\delta_0$ and the transverse centroid of each segment in the bunch are expressed as
\begin{equation}
  \begin{split}
    x_{u,d}(n,z) = X_0 \exp(- \textbf{i} 2\pi\nu_{\beta,x} n) \exp\left( \textbf{i} \chiABS \left(\frac{z}{\hat{z}} - 1\right)\right) \exp\left[ \textbf{i} \frac{2}{\pi} \chiABS \cos^{-1}(\cos\left(\mp \frac{\pi}{2}\left(\frac{z}{\hat{z}}-1\right) - 2\pi\nu_s n\right)) \right],
  \end{split}
\end{equation}
\begin{equation}\label{eq:97}
  \begin{split}
    \bar{x}(n,z) &= \frac{1}{2} X_0 \exp(- \textbf{i} 2\pi\nu_{\beta,x} n) \exp\left( \textbf{i} \chiABS \left(\frac{z}{\hat{z}} - 1\right)\right) \\
            &\times \left( \exp\left[ \textbf{i} \frac{2}{\pi} \chiABS \cos^{-1}(\cos\left(- \frac{\pi}{2}\left(\frac{z}{\hat{z}}-1\right) - 2\pi\nu_s n\right)) \right] \right.\\
                &+ ~~ \left. \exp\left[ \textbf{i} \frac{2}{\pi} \chiABS \cos^{-1}(\cos\left(+ \frac{\pi}{2}\left(\frac{z}{\hat{z}}-1\right) - 2\pi\nu_s n\right)) \right] \right),
  \end{split}
\end{equation}
respectively.

Now, let us equate Eq. \eqref{eq:88}, the result of the mode expansion method, to Eq. \eqref{eq:97}, the result of the betatron phase tracking method, producing
\begin{equation}\label{eq:98}
  \begin{split}
     & 2 \sum_{m\in \mathbb{Z}} \amABS \exp\left( \textbf{i} \chiABS \right) \exp(- \textbf{i} m \varphi) \cos(m\theta') \\
    =& \exp\left[ \textbf{i} \frac{2}{\pi} \chiABS \cos^{-1}(\cos(+\theta' - \varphi)) \right] + \exp\left[ \textbf{i} \frac{2}{\pi} \chiABS \cos^{-1}(\cos(-\theta' - \varphi)) \right] ,
  \end{split}
\end{equation}
where
$\theta' = -\pi/2\times(z/\hat{z}-1)$ 
with ($0<\theta'<\pi$),
and $\varphi = 2\pi\nu_s n$.

After multiplying Eq. \eqref{eq:98} by $\exp(im\varphi)$ and then integrating it over $\varphi$ from $0$ to $2\pi$, 
the coefficient $\amABS$ is expressed as
\begin{equation}\label{eq:99}
  \amABS = (-\textbf{i})^m \frac{\chiABS}{\chiABS^2-(m\frac{\pi}{2})^2} \sin\left(\chiABS - m\frac{\pi}{2}\right).
\end{equation}

The transverse centroid of the entire bunch $X(n)$, the average value of Eq. \eqref{eq:88} over the longitudinal positions of the bunch, is calculated as
\begin{equation}
  X(n) = \int_{-\hat{z}}^{\hat{z}} \frac{dz}{2\hat{z}} \bar{x}(n,z) = X_0 \sum_{m\in \mathbb{Z}} |\amABS|^2 \exp\left(- \textbf{i} 2\pi \nu_m n\right),
\end{equation}
where the amplitude $\amABS$ is determined by Eq. \eqref{eq:99}.

\subsection{Head-tail modes in the ABH model with space charge effects}\label{6.3}
Next, let us consider the ABH model under any strength of space charge by replacing the square-well potential with a harmonic potential to more realistically describe the motion of particles in the bunch.
Equation \eqref{eq:7} instead of Eqs. \eqref{eq:11} and \eqref{eq:12} follows
\begin{equation}\label{eq:101}
  \frac{d\delta}{dn}=\frac{(2\pi\nu_s)^2}{\eta C}z.
\end{equation}

When we discuss only the case of $\eta<0$,
the momentum deviations $\delta_u$ and $\delta_d$ are uniquely determined for $z$ and are denoted by 
\begin{equation}
  \delta_{u,d}=\mp\frac{2\pi\nu_s}{\eta C} \sqrt{\hat{z}^2-z^2},
\end{equation}
where the minus and plus signs correspond to $\delta_u$ and $\delta_d$, respectively and $\hat{z}$ is the half bunch length at the ABH model.

Rigorously, $\nusc$ and $\nuic$ are distributed along the longitudinal direction within the bunch.
However, in this model, we simplify the analysis by neglecting the $z$-dependence of $\nusc$ and $\nuic$,
assuming that the space charge effects near the center of the bunch are well described by the waterbag model and the effects are nearly uniform in the entire bunch.

Consequently, the equations corresponding to Eqs. \eqref{eq:81} and \eqref{eq:82} are given by
\begin{equation}
  \frac{d}{dn}x_{u,d} = \frac{\partial}{\partial n} x_{u,d} - \eta C \delta_{u,d} \frac{\partial}{\partial z} x_{u,d},
\end{equation}
\begin{equation}
  \frac{d}{dn}x_{u,d} = - \textbf{i} 2\pi\nu_{\beta,x} x_{u,d} - \textbf{i} 2\pi\delta_{u,d} \xi_x x_{u,d} - \textbf{i} 2\pi\nusc (x_{u,d}-\bar{x}) - \textbf{i} 2\pi\nuic \bar{x},
\end{equation}
where $x_u$ and $x_d$ represent the horizontal positions of particles with momentum deviations $\delta_u > 0$ and $\delta_d < 0$, respectively.

Finally, the head-tail mode expansion formula for the transverse centroid of each segment in the bunch is expressed as
\begin{equation}\label{eq:105}
  \bar{x}(n,z) = X_0 \sum_{m \in \mathbb{Z}} \amABH \exp\left( \textbf{i} \chiABH \frac{z}{\hat{z}} - \textbf{i} 2\pi\nu_m n \right) \cos\left(m \cos^{-1} \left(\frac{z}{\hat{z}}\right)\right),
\end{equation}
in the ABH model.
Notice that $\nu_m$ follows the same form as Eq. \eqref{eq:90} with Eq. \eqref{eq:84} in the ABS model.
When $\nu_m$ is replaced by $\nu_{\beta,x}+m\nu_s$ by neglecting the space charge effect, the outcome reproduces the Formula (6.186) in the Chao's textbook \cite{Chao:1993zn}, which was derived under the assumption that the space charge effects are negligible from the beginning.
In this sense, Eq. \eqref{eq:105} successfully generalizes the previous formula to incorporate the space charge effects.

Here, the chromatic phase $\chiABH$ is defined by
\begin{equation}
  \chiABH=2\pi\frac{\xi_x}{\eta C} \hat{z},
\end{equation}
which is equivalent to Eq. \eqref{eq:87}.

\subsection{Derivation of the oscillation amplitude with weak space charge force under the ABH model}\label{6.4}
In a similar manner explained in Sec. \ref{6.2}, we need to derive $\bar{x}(n,z)$ to determine $\amABH$ in the weak space charge limit, following the linear decoherence scenario under the harmonic potential.
Since the synchrotron phase is followed as $\theta=2\pi \nu_s n +\theta_0$ with initial phase $\theta_0$,
the equations of motion given by Eqs. \eqref{eq:79} and \eqref{eq:101} are rewritten as
\begin{equation}
  \frac{dz}{d\theta} = \frac{dz}{dn}\frac{dn}{d\theta} = -\frac{\eta C}{2\pi\nu_s} \delta,
\end{equation}
\begin{equation}
  \frac{d\delta}{d\theta} = \frac{d\delta}{dn}\frac{dn}{d\theta} = \frac{2\pi\nu_s}{\eta C} z.
\end{equation}

Therefore, each particle motion is followed as 
\begin{equation}
  z(n,\theta_0) = \hat{z} \cos\theta(n,\theta_0),
\end{equation}
\begin{equation}
  \delta(n,\theta_0) = \frac{2\pi\nu_s}{\eta C} \hat{z} \sin\theta(n,\theta_0).
\end{equation}

Given $z=z_0$ at $n=n_0$, there are two solutions $\theta_{0,u}(n_0,z_0)$ and $\theta_{0,d}(n_0,z_0)$ for the initial synchrotron phase $\theta_0$, as
\begin{equation}
  \theta_{0,(u,d)} (n_0,z_0) = \pm \cos^{-1} \left(\frac{z_0}{\hat{z}}\right) - 2\pi\nu_s n_0.
\end{equation}

Hence, the betatron phase advance $\phi_u(n_0,z_0)$ and $\phi_d(n_0,z_0)$ can be calculated for the respective $\theta_{0,(u,d)}(n_0,z_0)$ as 
\begin{equation}
  \phi_{u,d}(n,z) = 2\pi\nu_{\beta,x} n - \chiABH \left(\frac{z}{\hat{z}} - \cos(\pm \cos^{-1}\left(\frac{z}{\hat{z}}\right) - 2\pi\nu_s n)\right),
\end{equation}
where the plus and minus signs on the right-hand side correspond to $\phi_u$ and $\phi_d$, respectively.

Because of the transverse oscillation with initial condition $\bar{x}(n=0,z)=X_0$,
the transverse positions of particles with $\delta_{u,d}$ and the transverse centroid of each segment in the bunch are expressed as

\begin{equation}
  x_{u,d}(n,z) = X_0 \exp(- \textbf{i} 2\pi\nu_{\beta,x} n) \exp\left( \textbf{i} \chiABH \frac{z}{\hat{z}} \right) \exp\left[- \textbf{i} \chiABH \cos(\pm \cos^{-1}\left(\frac{z}{\hat{z}}\right) - 2\pi\nu_s n)\right],
\end{equation}
\begin{equation}\label{eq:114}
  \begin{split}
    \bar{x}(n,z) =&\frac{1}{2} X_0 \exp(- \textbf{i} 2\pi\nu_{\beta,x} n) \exp\left( \textbf{i} \chiABH \frac{z}{\hat{z}} \right) \\
            &\times \left(\exp\left[- \textbf{i} \chiABH \cos(+ \cos^{-1}\left(\frac{z}{\hat{z}}\right) - 2\pi\nu_s n)\right] \right.\\
            &+ ~~ \left. \exp\left[- \textbf{i} \chiABH \cos(- \cos^{-1}\left(\frac{z}{\hat{z}}\right) - 2\pi\nu_s n)\right] \right),
  \end{split}
\end{equation}
respectively.

After equating Eq. \eqref{eq:114} to Eq. \eqref{eq:105}, we obtain
\begin{equation}
  \begin{split}
     & 2\sum_{m\in \mathbb{Z}} \amABH \exp(- \textbf{i} m \varphi) \cos\left( m \theta' \right) \\
    =& \exp\left[- \textbf{i} \chiABH \cos(+ \theta' - \varphi)\right] + \exp\left[- \textbf{i} \chiABH \cos(- \theta' - \varphi)\right] ,
  \end{split}
\end{equation}
corresponding to Eq. \eqref{eq:98} in the ABS model,
where $\theta'=\cos^{-1}(z/\hat{z})$ with ($0<\theta'<\pi$), and $\varphi = 2\pi\nu_s n$.
In the end, the coefficient $\amABH$ is expressed as
\begin{equation}\label{eq:116}
  \amABH = (-\textbf{i})^m J_m(\chiABH),
\end{equation}
where $J_m(x)$ is the Bessel function of the first kind \cite{abramowitz1968handbook}.
If we compare Eq. \eqref{eq:116} of the ABH model with Eq. \eqref{eq:99} of the ABS model, 
we observe similar responses of the dominant head-tail modes to a given chromatic phase, $\chiABH$ or $\chiABS$.

The transverse centroid of the entire bunch $X(n)$, the charge-density-weighted average value of Eq. \eqref{eq:105} with respect to the airbag distribution over the longitudinal positions of the bunch produces, is calculated as
\begin{equation}\label{eq:117}
  \begin{split}
     X(n) = \int_{-\hat{z}}^{\hat{z}} \frac{dz}{\pi\sqrt{\hat{z}^2-z^2}} \bar{x}(n,z) 
    = X_0 \sum_{m\in \mathbb{Z}} |\amABH|^2 \exp(- \textbf{i} 2\pi \nu_m n),
  \end{split}
\end{equation}
where the amplitude $\amABH$ is determined by Eq. \eqref{eq:116}.

\subsection{Derivation of the oscillation amplitude for a Gaussian beam with weak space charge force}\label{6.5}
Now, we consider the transverse centroid of each segment $\bar{x}(n,z)$ in an entire Gaussian bunch by generalizing the ABH model.
If the space charge effect is small and there is no interaction between the respective airbags,
the calculation can proceed by convolving $\bar{x}(n,z,\delta)$, obtained by replacing $\hat{z}$ in Eq. \eqref{eq:105} with $\sqrt{z^2+(\eta C/2\pi\nu_s)^2\delta^2}$,
with the Gaussian distribution in the momentum direction characterized by the momentum deviation $\sigma_\delta$, as follows:
\begin{equation}\label{eq:118}
  \bar{x}(n,z) 
  = \int_{-\infty}^{\infty} d\delta \frac{1}{\sqrt{2\pi}\sigma_\delta} \exp\left(-\frac{\delta^2}{2\sigma_\delta^2}\right) \bar{x}(n,z,\delta)
  = X_0 \sum_{m \in \mathbb{Z}} \tilde{a}_{m,b}(z) \exp\left(- \textbf{i} 2\pi\nu_m n\right).
\end{equation}

Since this result must be the same as Eq. \eqref{eq:19} for the weak space charge limit, the relative amplitude $\tilde{a}_{m,b}(z)$ in each segment is calculated in two different ways as:
\begin{equation}\label{eq:119}
  \begin{split}
            \tilde{a}_{m,b}(z) 
         =& \exp\left( \textbf{i} \chi_\sigma \frac{z}{\sigma_z}\right) (-\textbf{i})^m \sum_{l=-\infty}^{\infty} J_{m+2l}\left[\chi_\sigma \frac{z}{\sigma_z}\right] (-1)^{l} \exp\left[-\frac{\chi_\sigma^2}{4}\right] I_l\left[\frac{\chi_\sigma^2}{4}\right] \\
         =& \exp\left( \textbf{i} \chi_\sigma \frac{z}{\sigma_z}\right) (-\textbf{i})^m \frac{z}{\sqrt{2\pi}\sigma_z} \int_{-\frac{\pi}{2}}^{\frac{\pi}{2}} \frac{d\theta}{\cos^2\theta} J_m\left[\chi_\sigma \frac{z}{\sigma_z} \frac{1}{\cos\theta}\right] \exp\left[-\frac{z^2}{2\sigma_z^2} \tan^2\theta \right] \cos(m \theta).
  \end{split}
\end{equation}

Formally, the transverse centroid of the entire bunch $X(n)$ should be calculated by averaging Eq. \eqref{eq:118} with a charge density $\rho_z(z)$, given by Eq.\eqref{eq:23}, weight over the longitudinal positions of the bunch, 
resulting in
\begin{equation}\label{eq:120}
  X(n)= \int_{-\infty}^{\infty} dz \rho_z(z) \bar{x}(n,z) 
  = X_0 \sum_{m\in \mathbb{Z}} |a_{m,b}|^2 \exp(- \textbf{i} 2\pi \nu_m n).
\end{equation}

Since the straightforward calculation to derive $X(n)$ and $a_{m,b}$ is quite difficult, 
we bypass it by generalizing Eq. \eqref{eq:117}, treating $\hat{z}$ as a variable, 
and integrating it with a Rayleigh distribution over $\hat{z}$, resulting in the expression for $X(n)$ as
\begin{equation}\label{eq:121}
  \begin{split}
     X(n) =& \int_0^{\infty} d\hat{z} \frac{\hat{z}}{\sigma_z^2} \exp\left(-\frac{\hat{z}^2}{2\sigma_z^2}\right) X(n,\hat{z}) \\
          =& X_0 \sum_{m\in \mathbb{Z}} \int_0^{\infty} d\hat{z} \frac{\hat{z}}{\sigma_z^2} \exp\left(-\frac{\hat{z}^2}{2\sigma_z^2}\right) J_m^2\left(\chi_\sigma \frac{\hat{z}}{\sigma_z}\right) \exp(- \textbf{i} 2\pi \nu_m n) \\
          =& X_0 \sum_{m\in \mathbb{Z}} \exp(-\chi_\sigma^2) I_m(\chi_\sigma^2) \exp(- \textbf{i} 2\pi \nu_m n),
  \end{split}
\end{equation}
where $I_m(x)$ is the modified Bessel function of the first kind \cite{abramowitz1968handbook},
and the rms chromatic phase $\chi_\sigma$ relates to the
chromatic phase of ABH model as $\chiABH=\chi_\sigma \hat{z}/\sigma_z$.
This final result is identical to Eq. \eqref{eq:24} after $m$ is summed over.
Here, we use $\nu_m=\nu_{\beta,x}+m\nu_s$, because the space charge is negligible.

The comparison between Eqs. \eqref{eq:120} and \eqref{eq:121} leads to
\begin{equation}\label{eq:122}
  |a_{m,b}| = \sqrt{\exp\left(-\chi_\sigma^2\right) I_m(\chi_\sigma^2)}.
\end{equation}
Though Eq. \eqref{eq:116} features the dominant mode for the chromaticity $\xi_x$ in the airbag model, 
they disappear in Eq. \eqref{eq:122} and there is no longer the dominant mode for the specified chromaticity $\xi_x$ in the low-intensity beam
because the Gaussian bunches fill in the airbag distribution in the $\hat{z}$ direction.

The analytical results of Eqs. \eqref{eq:119} and \eqref{eq:122} support the validity of our code by the tracking simulation in Ref. \cite{yoshimura2025space}.

\section{Response to chromaticity and beam intensity in tracking simulation}\label{7}


In this section, we will scan the beam intensity and chromaticity in the ABS model, and the Gaussian bunch model under the harmonic potential for the longitudinal distribution 
to realize the limit of the ABS model and understand the space charge effects on the Gaussian bunch.

The developed tracking simulation code in Secs. \ref{2}, \ref{4}, and \ref{5} enables us to investigate how the frequency ($\nu_m$), and the amplitude
in the entire bunch ($a_m$) of the head-tail modes are affected by the transverse wake and space charge effects on the intrabunch motion unless instabilities are excited after choosing suitable machine parameters.

The benchmark tests of the basic algorithm ignoring the space charge effect and the transverse wake are provided in Ref. \cite{yoshimura2025space}.
Furthermore, the results of checking the reliability of the parameter settings in the simulation with the space charge effects are surveyed there \cite{yoshimura2025space}.

Roughly, the beam sizes assumed to be $\sigma_x=0.009$~m with fixed $\sigma_y=0.010$~m, (refer to Sec. \ref{4})
and the peak charge density $\hat{\lambda}$ determined by Eq. \eqref{eq:10} through the bunching factor (typically $B_f=0.015$) produce
the minimum incoherent tune shift in the bunch:
\begin{equation}
  \nusc= -157\times10^{-4} \times N_B / 10^{12},
\end{equation}
and the minimum coherent tune shift in the bunch:
\begin{equation}\label{eq:124}
  \nuic= -7.4\times10^{-4} \times N_B / 10^{12},
\end{equation}
for both potential models,
after replacing $a_x^{(k)}$, $a_y$ and $\lambda^{(k)}$ by $\sqrt{2}\sigma_x$, $\sqrt{2}\sigma_y$, and $\hat{\lambda}$ in Eqs. \eqref{eq:61} and \eqref{eq:62}, respectively, 
though the simulation dynamically changes the horizontal and longitudinal beam sizes.

In both simulations in the ABS model and the Gaussian bunch model under the harmonic potential,
the tune spectrum, $\nu_m$, and the absolute relative amplitude, $|a_m|$, of the centroid $\overline{x_n}$ (Eq. \eqref{eq:13}) for the entire bunch oscillation
are obtained by applying a fast Fourier transform (FFT) to the turn-by-turn bunch oscillations from the start of the simulation up to 10,000 turns.
The results of these simulations will be compared with the theoretical predictions.

\begin{figure}[!h]
  \begin{tabular}{ccc}
    \begin{minipage}[t]{0.3\hsize}
      \centering
      \includegraphics[width=2.0in]{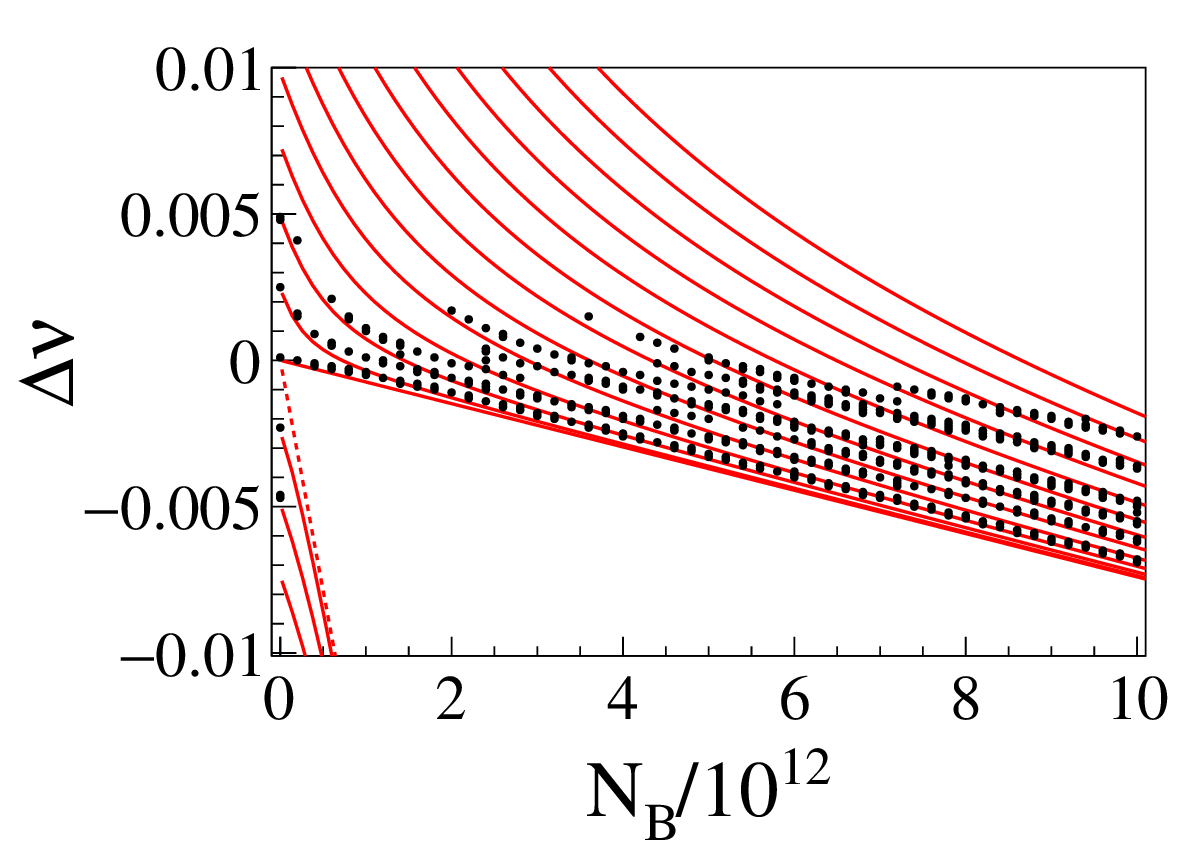}
      \subcaption{}
    \end{minipage} &
    \begin{minipage}[t]{0.3\hsize}
      \centering
      \includegraphics[width=2.0in]{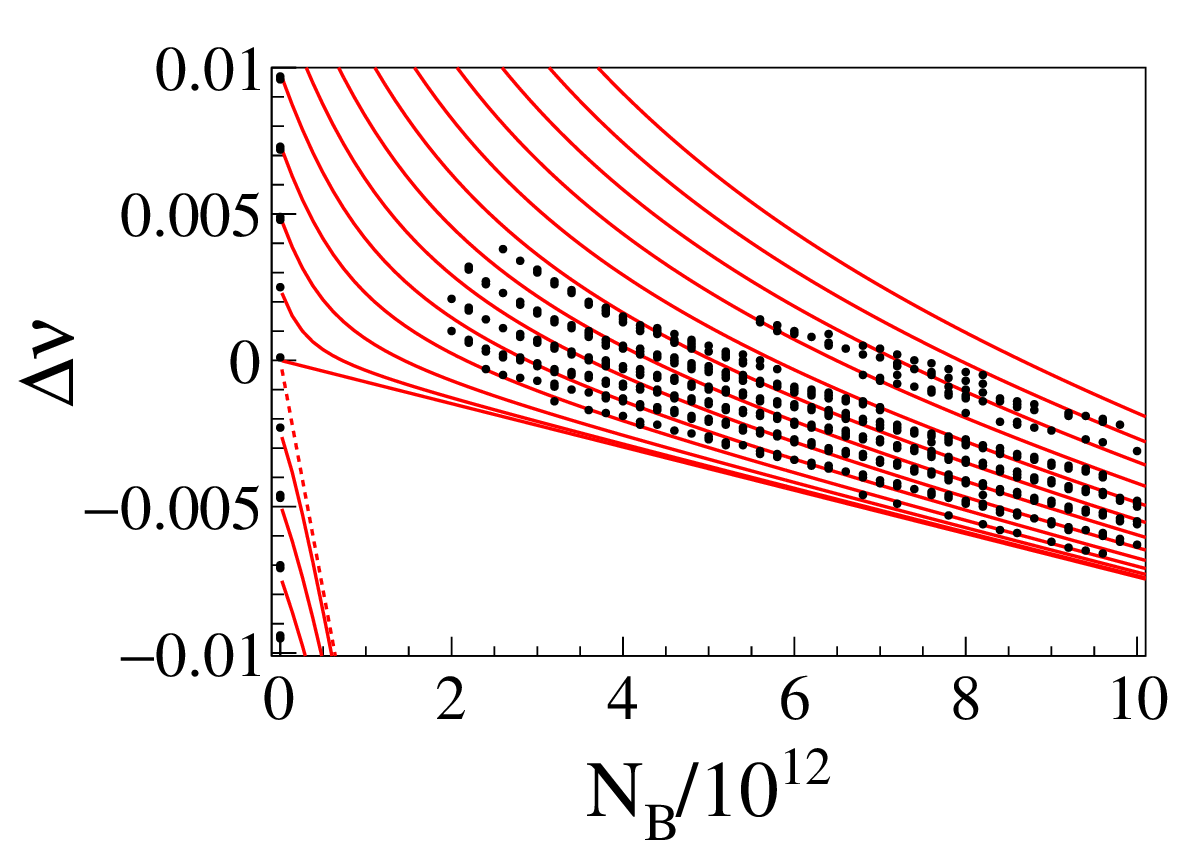}
      \subcaption{}
    \end{minipage} &
    \begin{minipage}[t]{0.3\hsize}
      \centering
      \includegraphics[width=2.0in]{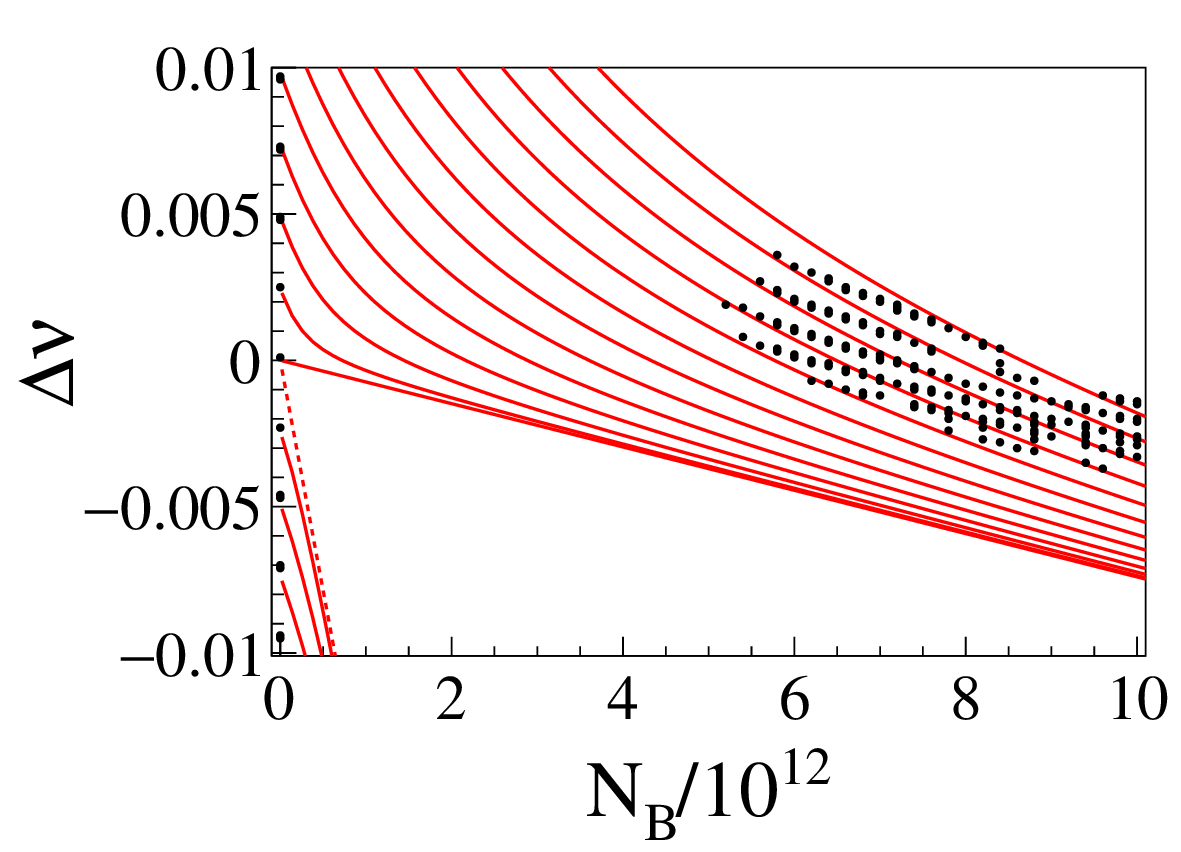}
      \subcaption{}
    \end{minipage} \\
  \end{tabular}
  \caption{Head-tail tune spectra in the harmonic potential model with indirect space charge effect for various values of chromaticity: (a) $\xi_x=-1.5$, (b) $\xi_x=-6.5$, and (c) $\xi_x=-12.5$. The dots indicate only the results for $|a_m|>0.1$. The blue, green and red dashed lines and black, red, green, blue, orange, magenta, cyan, dark green, purple, gray, dark red, dark cyan and yellow lines represent the frequency shifts calculated by Eq. \eqref{eq:84} for each head-tail mode in the range of $m = -3, -2, \cdots 12$, respectively.}
  \label{fig:2}
\end{figure}

Figure \ref{fig:2} shows the intensity dependence of the excited frequency shifts in the Gaussian bunch under the harmonic potential
for the chromaticity values ($\xi_x$)
, where the analytical results of Eq. \eqref{eq:84} based on the ABS model, which is identical to the frequency shift in the ABH model, are superimposed.
The overall behavior of the intensity dependence of the tune shifts can be understood using Eq. \eqref{eq:84}, which incorporates all space charge effects.
Although only positive modes appear in high-intensity regimes, the observed shift aligns with prior simulations considering the direct space charge alone, due to finite coherent tune shift ($\nuic \ne 0$) arising from the indirect space charge.
However, achieving a perfect fit with the simulation for the Gaussian bunch model is impossible because the analytical Eq. \eqref{eq:84} is based on the ABS and ABH models, and we need to rely on simulation studies to rigorously identify the excitation pattern.

\begin{figure}[!h]
  \centering
  \includegraphics[width=3.0in]{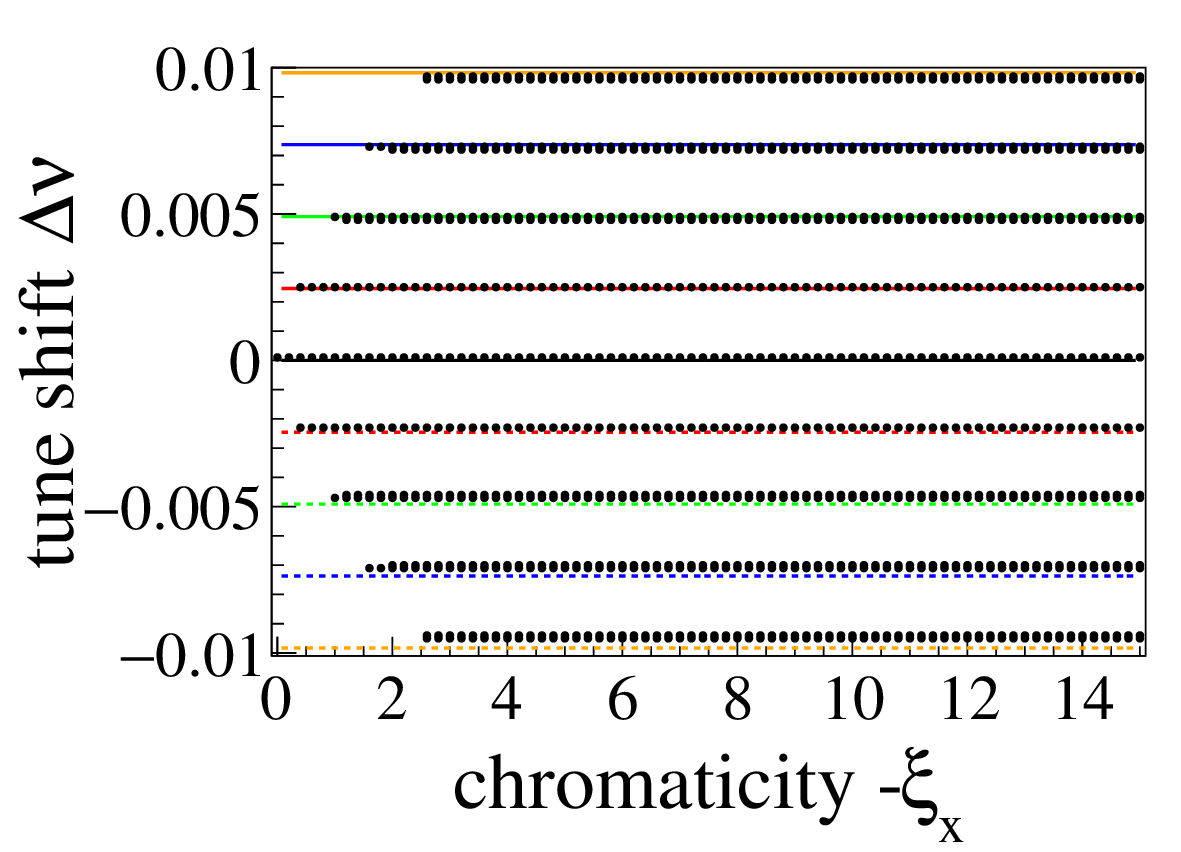}
  \caption{Head-tail tune spectra in the harmonic potential model with weak space charge force. The dots indicate only the results for $|a_m|>0.1$. The orange, blue, green and red dashed lines and black, red, green, blue and orange lines represent the frequency shifts calculated by Eq. \eqref{eq:84} for each head-tail mode in the range of $m = -4, -3, \cdots 4$, respectively.}
  \label{fig:3}
\end{figure}

Figure \ref{fig:3} illustrates the chromaticity dependence of the mode-dependent tune shifts for the low-intensity Gaussian beams under the harmonic potential.
The analytical results from Eq. \eqref{eq:84} are overlaid on the lines in the figures.
The head-tail tune spectra match the theoretical predictions and exhibit no chromaticity dependence.
Simulations also confirm that all modes are excited regardless of chromaticity, consistent with the prediction of Eq. \eqref{eq:122}.

\begin{figure}[!h]
  \begin{tabular}{cc}
    \begin{minipage}[t]{0.45\hsize}
      \centering
      \includegraphics[width=3.0in]{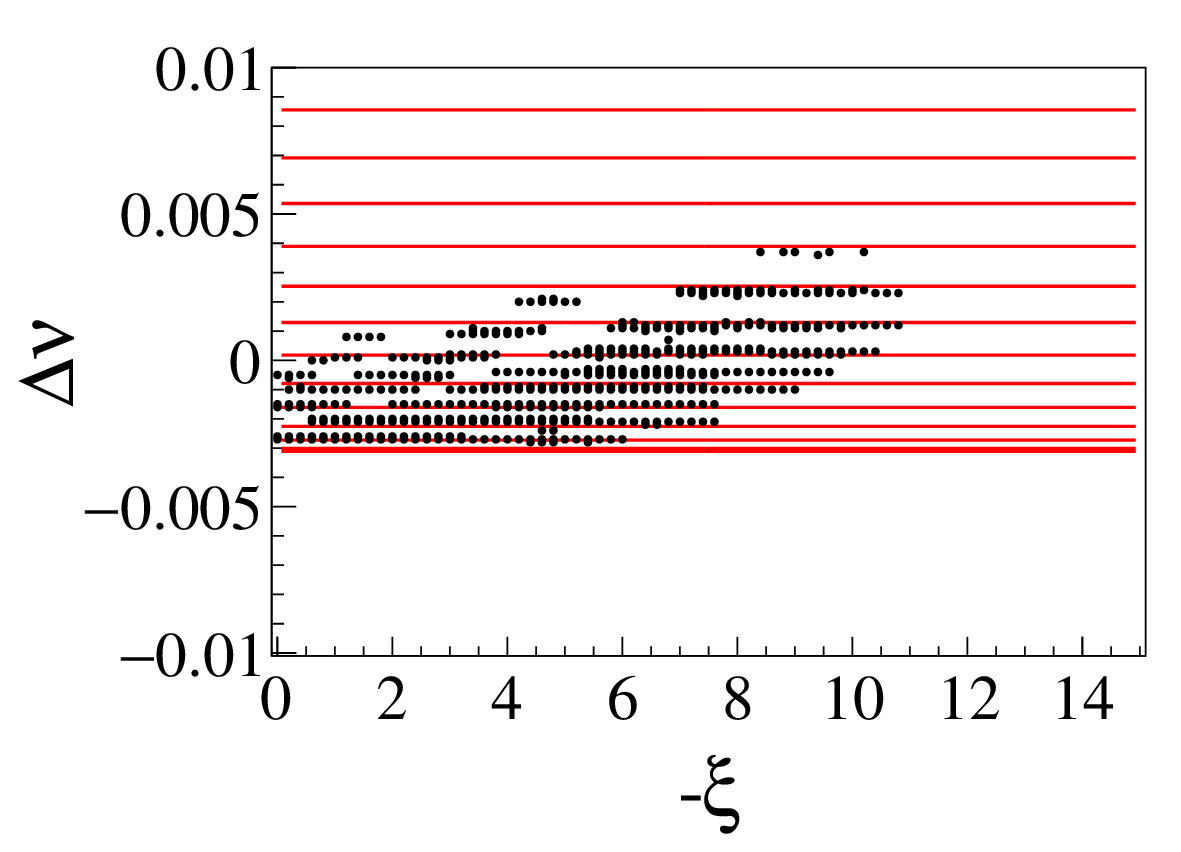}
      \subcaption{}
    \end{minipage} &
    \begin{minipage}[t]{0.45\hsize}
      \centering
      \includegraphics[width=3.0in]{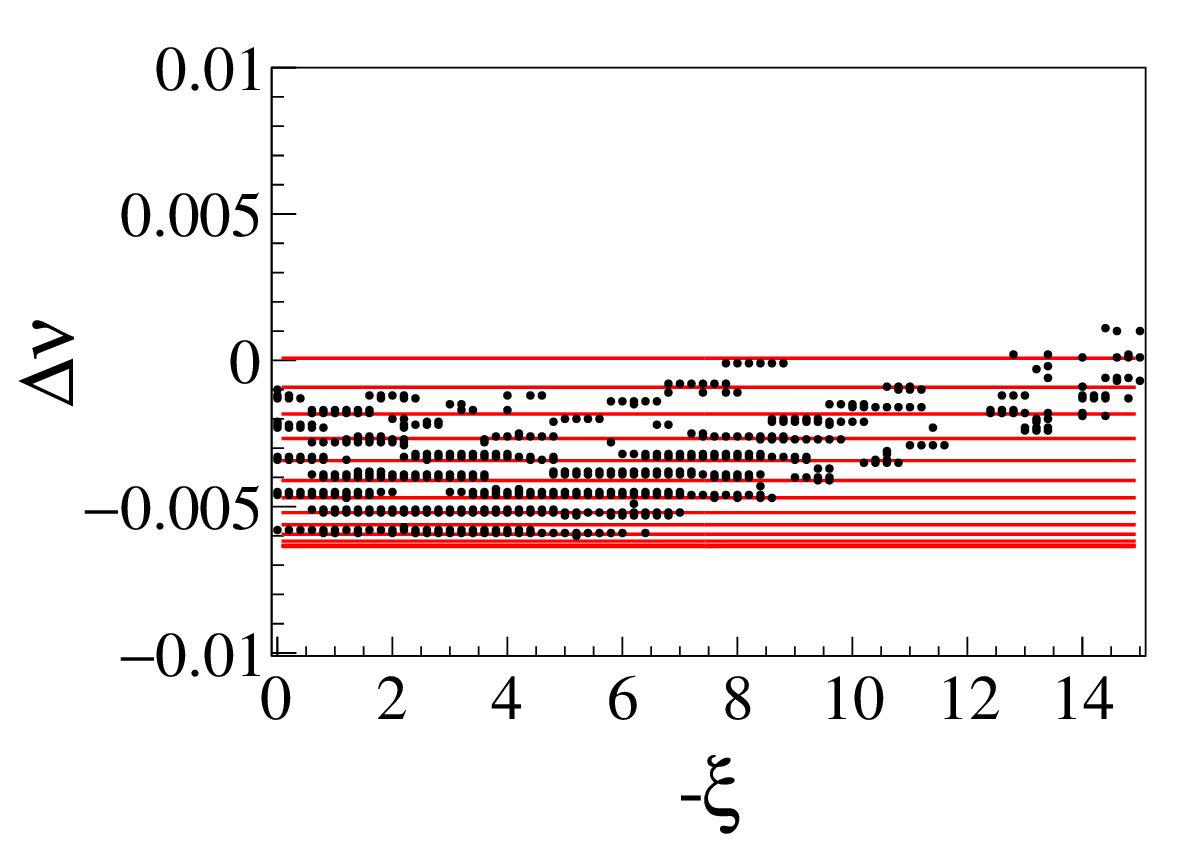}
      \subcaption{}
    \end{minipage} \\
  \end{tabular}
  \caption{Head-tail tune spectra in the harmonic potential model with indirect space charge effect for various values of intensity: (a) $N_B=4.2\times10^{12}$~ppb and (b) $N_B=8.6\times10^{12}$~ppb. The dots indicate only the results for $|a_m|>0.1$. The black, red, green, blue, orange, magenta, cyan, dark green, purple, gray, dark red, dark cyan and yellow lines represent the frequency shifts calculated by Eq. \eqref{eq:84} for each head-tail mode in the range of $m = 0, 1, 2, \cdots 12$, respectively, but due to the resolution they are drawn as if the low modes were degenerate.}
  \label{fig:4}
\end{figure}

For high-intensity Gaussian beams under the harmonic potential, Fig. \ref{fig:4} illustrates the chromaticity dependence of the mode-dependent tune shifts for (a) $N_B=4.2 \times 10^{12}$~ppb and (b) $N_B=8.6\times10^{12}$~ppb.
In contrast to the low-intensity case there is a discrepancy between the theoretical results and simulations: the lower modes remain distinguishable, although the theoretical model predicts degenerate lines.
Still, no significant chromaticity dependence is observed in either case.

\begin{figure}[!h]
  \begin{tabular}{cc}
    \begin{minipage}[t]{0.45\hsize}
      \centering
      \includegraphics[width=3.0in]{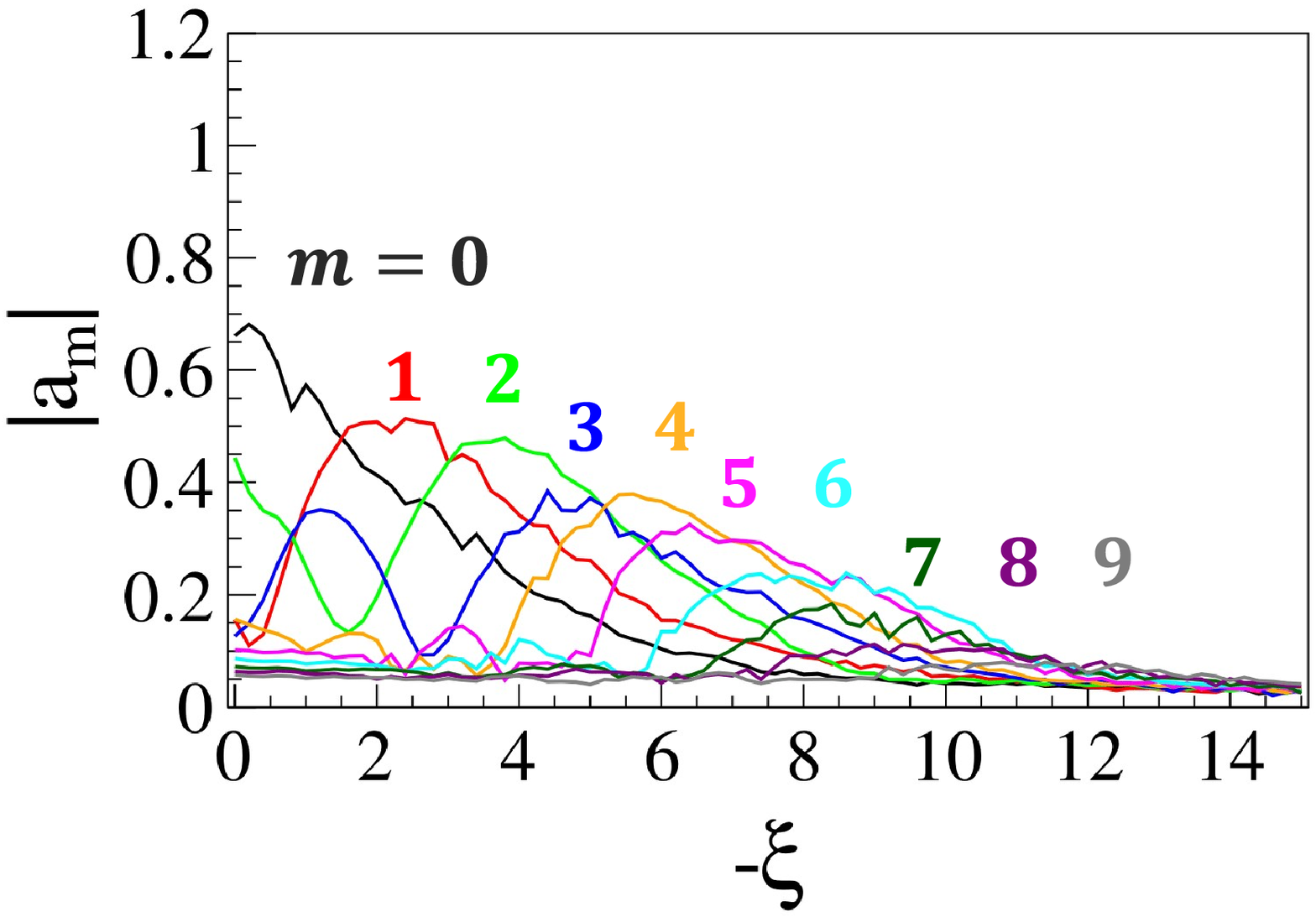}
      \subcaption{}
    \end{minipage} &
    \begin{minipage}[t]{0.45\hsize}
      \centering
      \includegraphics[width=3.0in]{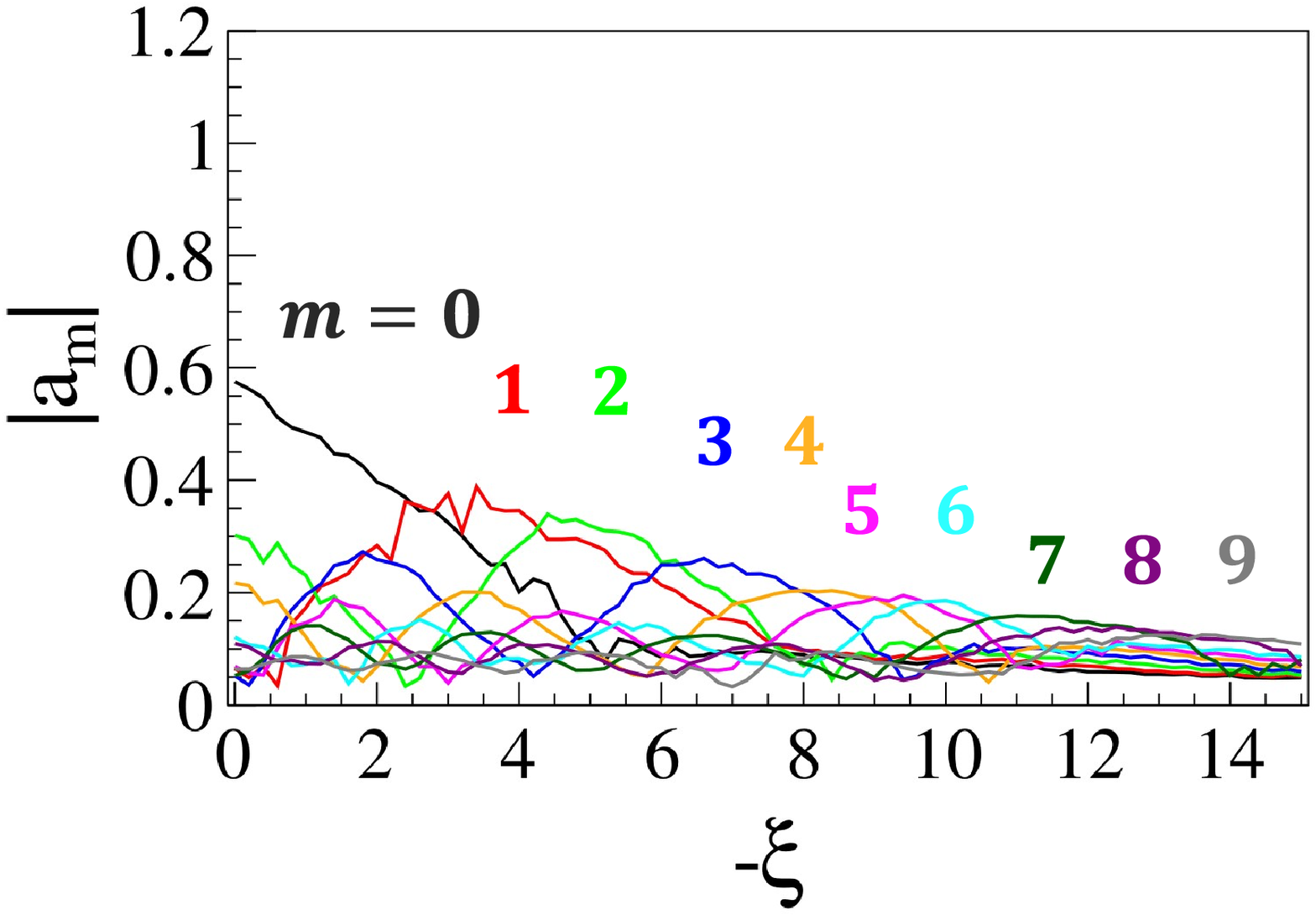}
      \subcaption{}
    \end{minipage} \\
  \end{tabular}
  \caption{Relative amplitudes $|a_m|$ tracking simulation of the harmonic potential model with indirect space charge effect. The black, red, green, blue, orange, magenta, cyan, dark green, purple, and gray curves denote the cases of $m = 0, 1, 2, \cdots 9$, respectively. (a) $N_B=4.2\times10^{12}$~ppb and (b) $N_B=8.6\times10^{12}$~ppb.}
  \label{fig:5}
\end{figure}

More specifically, Fig. \ref{fig:5} depicts the simulations of the relative amplitudes for the respective head-tail modes for the cases of (a) $N_B=4.2\times10^{12}$~ppb and (b) $N_B=8.6\times10^{12}$~ppb, respectively.
In these figures, the black, red, green, blue, orange, magenta, cyan, dark green, purple, and gray curves represent modes $m=0,1,2,\cdots 9$, respectively.
In contrast to weak space charge cases, head-tail mode excitations depend on the chromaticity: e.g., modes $m=0,1,2,3,5$, and $7$ dominate at $\xi_x=-1.5$ while $m=3,4$, and $5$ dominate at $\xi_x=-8.0$ for $N_B=8.6\times 10^{12}$~ppb.
These excitations, characterized by peaks along the chromaticity axis, arise due to the space charge effects.
Such chromaticity-dependent behavior does not appear in weak beams as described by Eq. \eqref{eq:122}, but resembles the predictions of Eqs. \eqref{eq:99} and \eqref{eq:116} under the strong space charge conditions.

\begin{figure}[!h]
  \begin{tabular}{cc}
    \begin{minipage}[t]{0.45\hsize}
      \centering
      \includegraphics[width=3.0in]{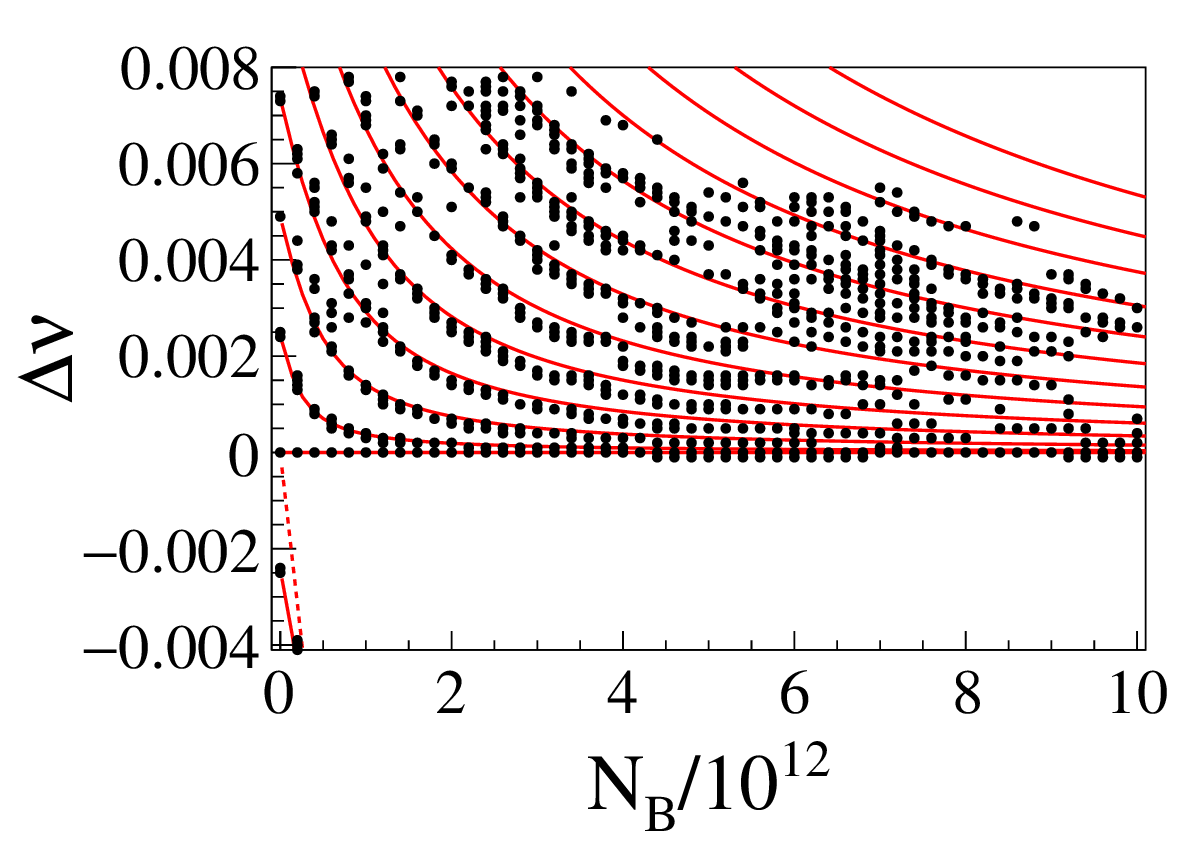}
      \subcaption{}
    \end{minipage} &
    \begin{minipage}[t]{0.45\hsize}
      \centering
      \includegraphics[width=3.0in]{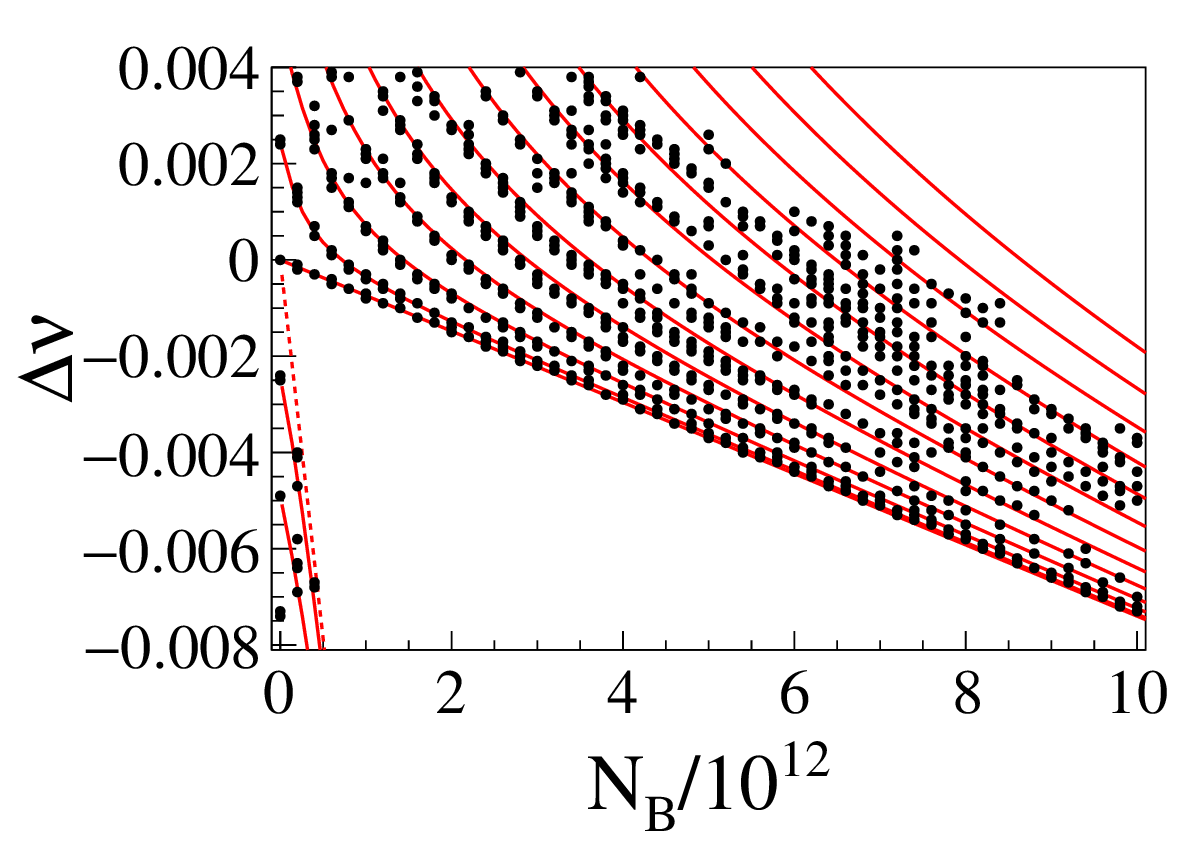}
      \subcaption{}
    \end{minipage} \\
    \begin{minipage}[t]{0.45\hsize}
      \centering
      \includegraphics[width=3.0in]{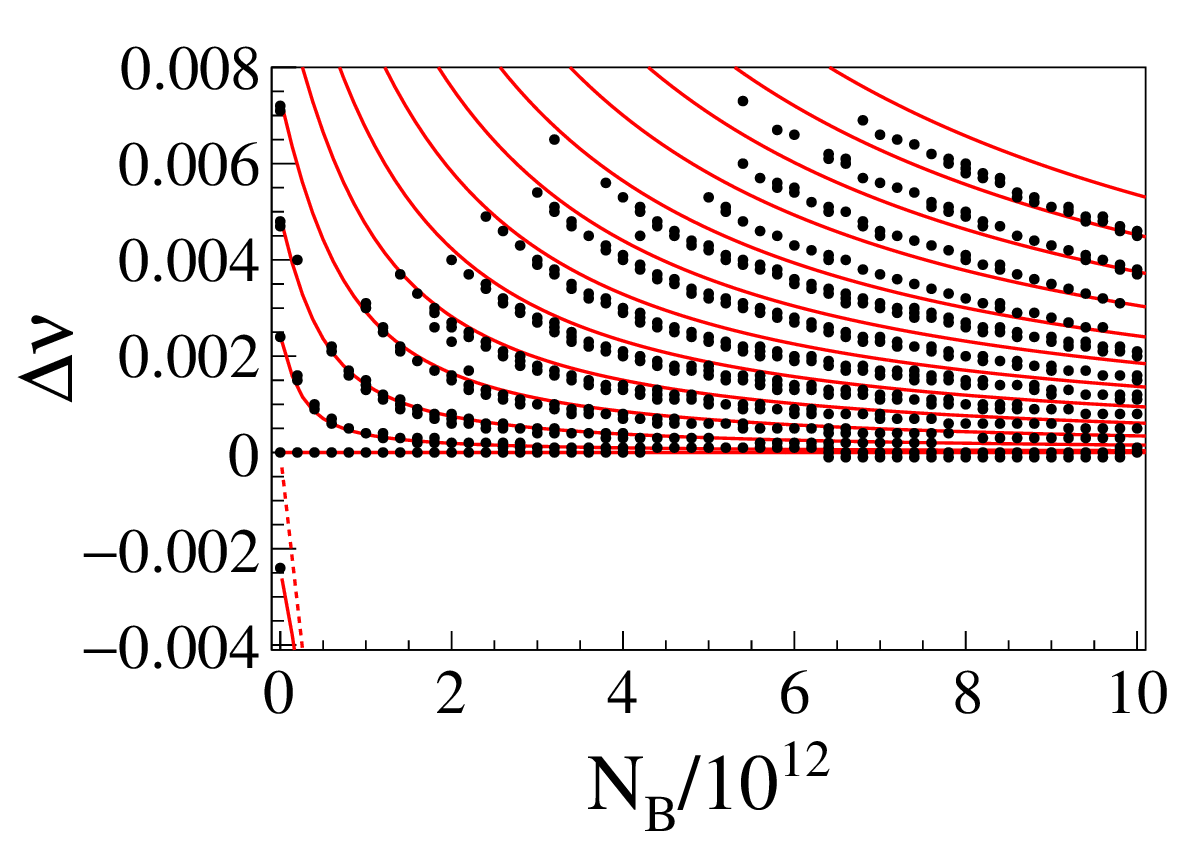}
      \subcaption{}
    \end{minipage} &
    \begin{minipage}[t]{0.45\hsize}
      \centering
      \includegraphics[width=3.0in]{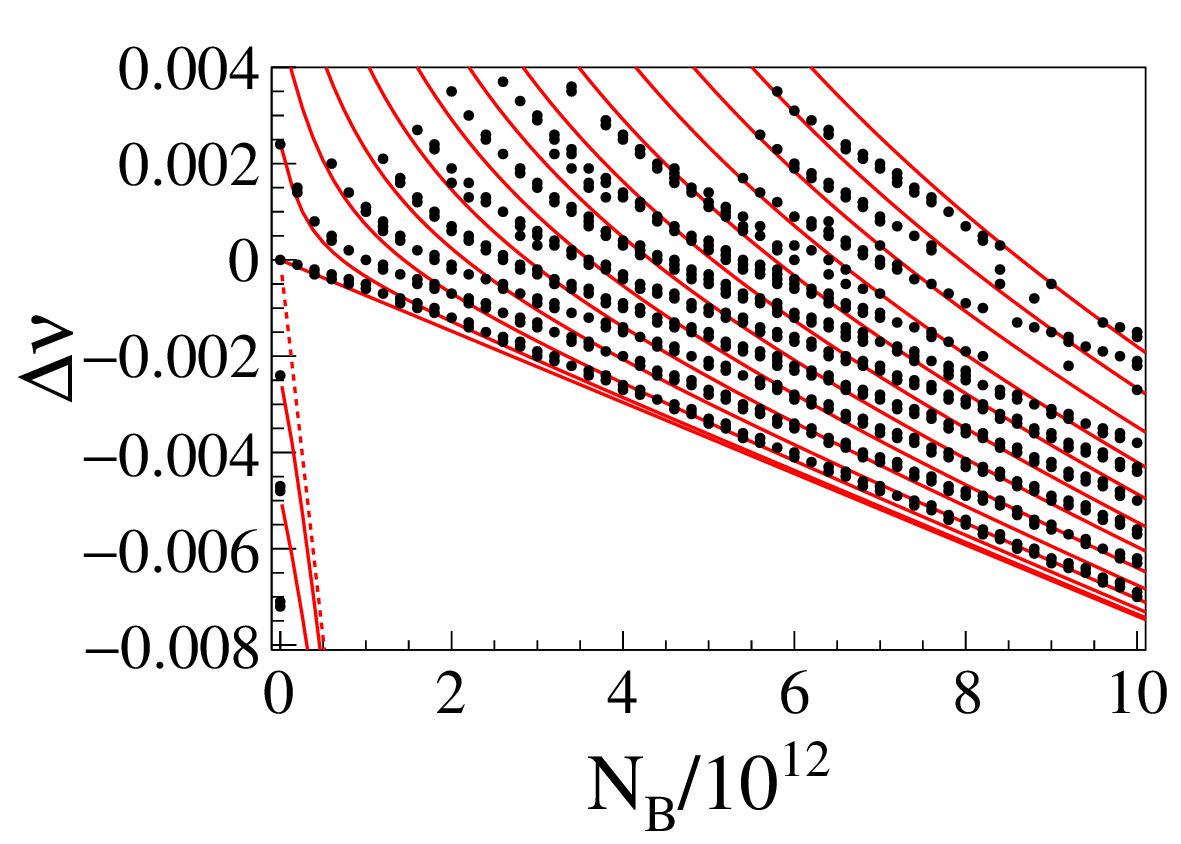}
      \subcaption{}
      \label{fig:6d}
    \end{minipage} \\
  \end{tabular}
  \caption{Tune shift at which the head-tail tune spectrum reaches its maximum peak in the simulation of the beam intensity parameters. The dots indicate only the results for $|a_m|>0.1$. The green and red dashed lines and black, red, green, blue, orange, magenta, cyan, dark green, purple, gray, dark red, dark cyan and yellow lines represent the frequency shifts calculated by Eq. \eqref{eq:84} for each head-tail mode in the range of $m = -2, -1, \cdots 12$, respectively. (a) The ABS model without indirect space charge effect, (b) the ABS model with indirect space charge effect, (c) the harmonic potential model without indirect space charge effect, and (d) the harmonic potential model with indirect space charge effect.}
  \label{fig:6}
\end{figure}

Finally, Fig. \ref{fig:6} summarizes the head-tail frequency shifts by the simulations for four cases:
(a) the ABS model without indirect space charge effect,
(b) the ABS model including all space charge effects, 
(c) a Gaussian beam under the harmonic potential model without indirect space charge effect,
and
(d) a Gaussian beam under the harmonic potential model including all space charge effects.

These figures superimpose the peak positions of each spectrum
with dots for the chromaticity values $\xi_x=-1.5,-3.5,-5.5,-6.5,-8.0$, and $-12.5$, for the respective beam intensities
because the frequency shifts due to the head-tail mode are independent of chromaticity, as previously demonstrated.
Analytical results from Eq. \eqref{eq:84} are overlaid in all panels by the solid lines.

The simulations and analytical results by the ABS model (a) and (b) are excellent in agreement, demonstrating the credibility of our code \cite{yoshimura2025space}.
In both the ABS model and the harmonic potential model, the head-tail modes become degenerate as the beam intensity increases.
Mode 0 exhibits a negative tune shift in (b) and (d) that increases with beam intensity, caused by the indirect space charge effect.
The frequency shifts for the harmonic potential model including all space charge effects (d) show the deviations from the analytical results.
Although the indirect space charge partially resolves the degeneracy of head-tail modes in the ABS model, the effect is more enhanced in the harmonic potential model in the simulation.
This discrepancy arises due to the limitations of the current theoretical model, as the formula is derived under the assumption of a square-well potential with an airbag beam, rather than a harmonic potential with the Gaussian bunch.

The present simulations include both space charge effects and the transverse resistive wall impedance, with algorithm described in Sec. \ref{5}.
The simulations suggest that the resistive wall impedance does not significantly affect the intrabunch motion when no beam instability occurs
because some agreement can be observed between the analytical results, which consider only space charge effects, and the simulations that include the wakefields, as well \cite{yoshimura2025space}.

A theoretical estimate of the tune shift due to wakefields offers insight to justify this negligence.
The coherent tune shift due to the wakefields is given by
\begin{equation}\label{eq:125}
  \nuic = \Delta\nu_\mathrm{single} + \Delta\nu_\mathrm{multi} = -0.24\times10^{-4}\times N_B / 10^{12},
\end{equation}
where the coherent tune shift for a Gaussian bunch caused by the single-turn wake $\Delta\nu_\mathrm{single}$ is calculated as
\begin{equation}
  \begin{split}
    \Delta\nu_\mathrm{single} &= - \textbf{i} \mathcal{G} \frac{C e c}{16 \pi^2 \sqrt{\pi} \sigma_z \nu_{\beta} \beta E_0} N_B \Zdipeff(\omega_\xi) \\
                              &= (-0.24 -0.24\textbf{i}) \times 10^{-4} \times N_B / 10^{12},
  \end{split}
\end{equation}
and the coherent tune shift of the multiturn wake $\Delta\nu_\mathrm{multi}$ is given by 
\begin{equation}
  \begin{split}
    \Delta\nu_\mathrm{multi} &= - \mathcal{G} \frac{C^2 e c \rho}{4 \pi^3 \nu_\beta \beta E_0 h_C^3 d_\omega} N_B \sum_{k=1}^\infty \sum_{n=0}^{\infty} e^{-\frac{\pi^2\rho k C}{d_\omega^2 Z_0}(n+\frac{1}{2})^2- \textbf{i} 2\pi k \nu_\beta} \\
                      &= (0.016+0.018\textbf{i}) \times10^{-4} \times N_B / 10^{12},
  \end{split}
\end{equation}
with the transverse effective impedance
\begin{equation}
  \Zdipeff(\omega) = \frac{\sum_{p=-\infty}^\infty \ZT^D ((p + \nu_\beta) \omega_0) h_0((p + \nu_\beta) \omega_0-\omega)}{\sum_{p=-\infty}^\infty h_0((p + \nu_\beta) \omega_0-\omega)},
\end{equation}
the transverse impedance
\begin{equation}\label{eq:129}
  \ZT^D (\omega) = Z_0 (\mathrm{sgn}(\omega)-\textbf{i}) \tanh\left(\sqrt{2i}\frac{d_\omega}{\delta_{\mathrm{skin}}}\right) \frac{L}{2\pi h_C^3} \delta_{\mathrm{skin}},
\end{equation}
the chromatic frequency
$\omega_\xi=\omega_0 \xi/\eta$
(the definition of chromaticity in Ref. \cite{Chao:1993zn} differs from the one used in this paper),
the power spectrum of the Gaussian zero azimuthal bunch mode $h_0(\omega)=\exp(-\omega^2 \sigma_z^2/c^2)$,
and the skin depth $\delta_{\mathrm{skin}}=\sqrt{2 c \rho/Z_0 |\omega|}$ \cite{shobuda2002resistive}.

When the resistive wall wake from Eq. \eqref{eq:65} or \eqref{eq:129} is used as the input parameter,
Eq. \eqref{eq:125} yields two orders of magnitude smaller than Eq. \eqref{eq:124}.
Thus, we conclude that the theory supports the simulations: the resistive wall impedance of the J-PARC MR does not affect intrabunch motion, provided no beam instability is triggered and the wakefield-induced tune shift is negligibly smaller than that caused by the space charge forces within the bunch.

In conclusion, while Eq. \eqref{eq:84} provides valuable qualitative insights into the head-tail mode-dependent tune shifts of a high-intensity Gaussian beam, simulation studies are crucial for quantitatively capturing these shifts.
This is particularly important for analyzing the impact of the indirect space charge effect on the intensity-dependent tune shifts in the Gaussian beam and the selection of mode excitation patterns, influenced by the direct and indirect space charge effects, under the specified chromaticity.

\section{Relation to the Radial Modes}\label{8}


Up to the previous chapter, we demonstrated through simulations that the head-tail tune shifts are well described by the analytical model in a regime with strong space charge unless beam instabilities are excited.
However, for the realistic Gaussian beam the radial modes may play a crucial role in the beam instabilities such as the transverse mode-coupling instability (TMCI).
A question may arise as to whether our simulation code properly handles these modes.
To address this, we perform supplemental simulations in the beam instability regime.

First, to identify the radial modes by comparing the simulations with the theoretical results, we simulate a case where the space charge effect is turned off, leaving only the transverse wakefield,
because the theoretical approach is already established in this simplest case.
The theoretical results are obtained by using DELPHI \cite{mounet2014delphi,mounet2020direct}, a well-established transverse Vlasov solver, with the assumption of the resistive wall impedance given by Eq. \eqref{eq:129} as its input parameter.

\begin{figure}[!h]
  \begin{tabular}{cc}
    \begin{minipage}[t]{0.45\hsize}
      \centering
      \includegraphics[width=3.0in]{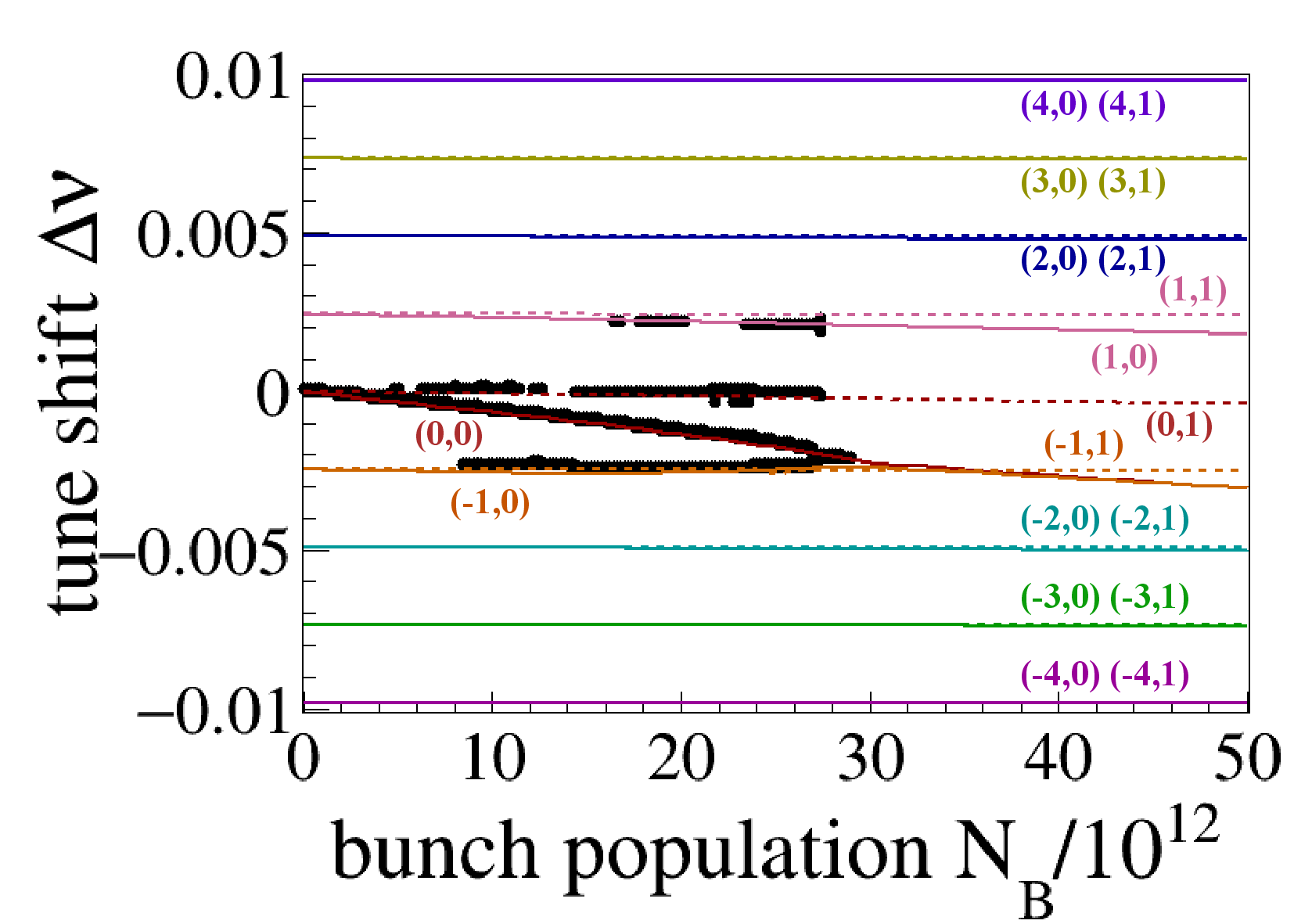}
      \subcaption{}
      \label{fig:7a}
    \end{minipage} &
    \begin{minipage}[t]{0.45\hsize}
      \centering
      \includegraphics[width=3.0in]{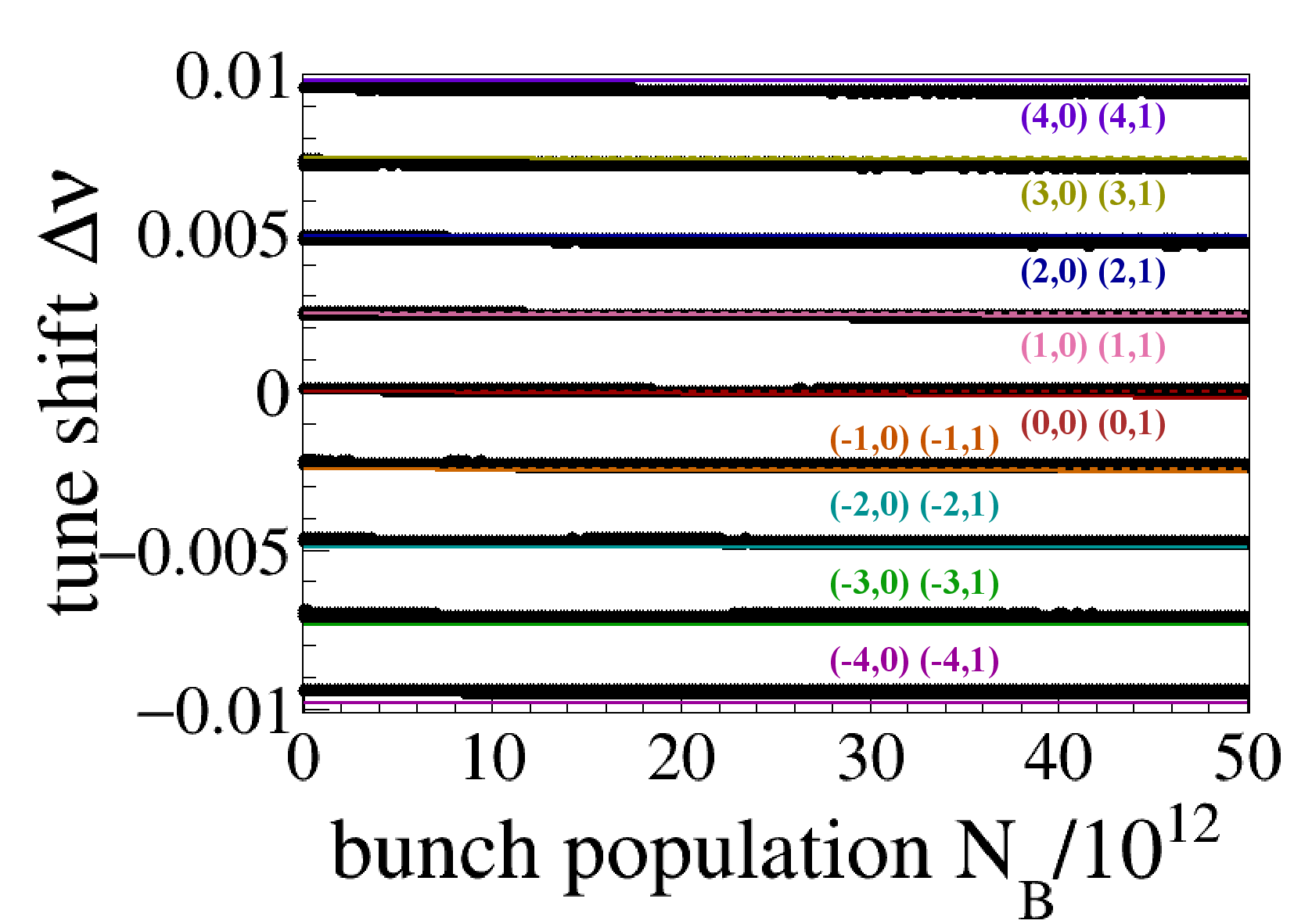}
      \subcaption{}
      \label{fig:7b}
    \end{minipage}
  \end{tabular}
  \caption{Tune shift at which the tune spectrum reaches its maximum peak in the simulation with only the transverse wakefield (space charge off), compared with the results from the DELPHI Vlasov solver. The pair of numbers on the lines describes the excited mode indices ($m,n_r$). (a) Chromaticity $\xi_x=0$ and (b) chromaticity $\xi_x=-8.0$.}
\end{figure}

The simulations are performed up to the high-intensity beam ($N_B\leq50\times10^{12}$~ppb) to clearly excite the radial modes $n_r$.
Figure \ref{fig:7a} shows the tune spectra obtained from our simulation and the theoretical outcome for the wakefield-only case with zero chromaticity ($\xi_x=0$).
Black bullet dots denote the simulation results, while the solid and dashed lines represent the theoretical results with the Vlasov solver.
The solid lines represent modes with $n_r=0$, and the dashed lines represent modes with $n_r=1$.
The magenta, green, cyan, orange, red, pink, blue, yellow, and purple denote the outcomes with $m=-4,-3,-2,-1,0,1,2,3$, and $4$, respectively.
The simulations reveal that only the head-tail modes $-1 \leq m \leq 1$ are excited in this case.
However, as the beam intensity increases, ($m=0,n_r=1$) mode separates from ($m=0,n_r=0$) mode and becomes identifiable.
Furthermore, at a bunch population of $N_B=28\times10^{12}$~ppb, the coupling of ($m=0,n_r=0$) and ($m=-1,n_r=0$) modes leads to TMCI.
The instability threshold observed in our simulation perfectly matches that predicted by the DELPHI.
This provides definitive evidence that our simulation code fully reproduces the dynamics of the radial modes.

Figure \ref{fig:7b} illustrates the case when a large chromaticity ($\xi_x=-8.0$) was introduced to stabilize the beam, where the labeling is the same as in Fig. \ref{fig:7a}.
The simulations perfectly explain the results by DELPHI, all radial modes are degenerate and merge with the primary head-tail modes, which is consistent with our previous simulations.
This indicates that the radial modes no longer play a crucial role in the large chromaticity region where no beam instability is excited.

\begin{figure}[!h]
  \begin{tabular}{cc}
    \begin{minipage}[t]{0.45\hsize}
      \centering
      \includegraphics[width=3.0in]{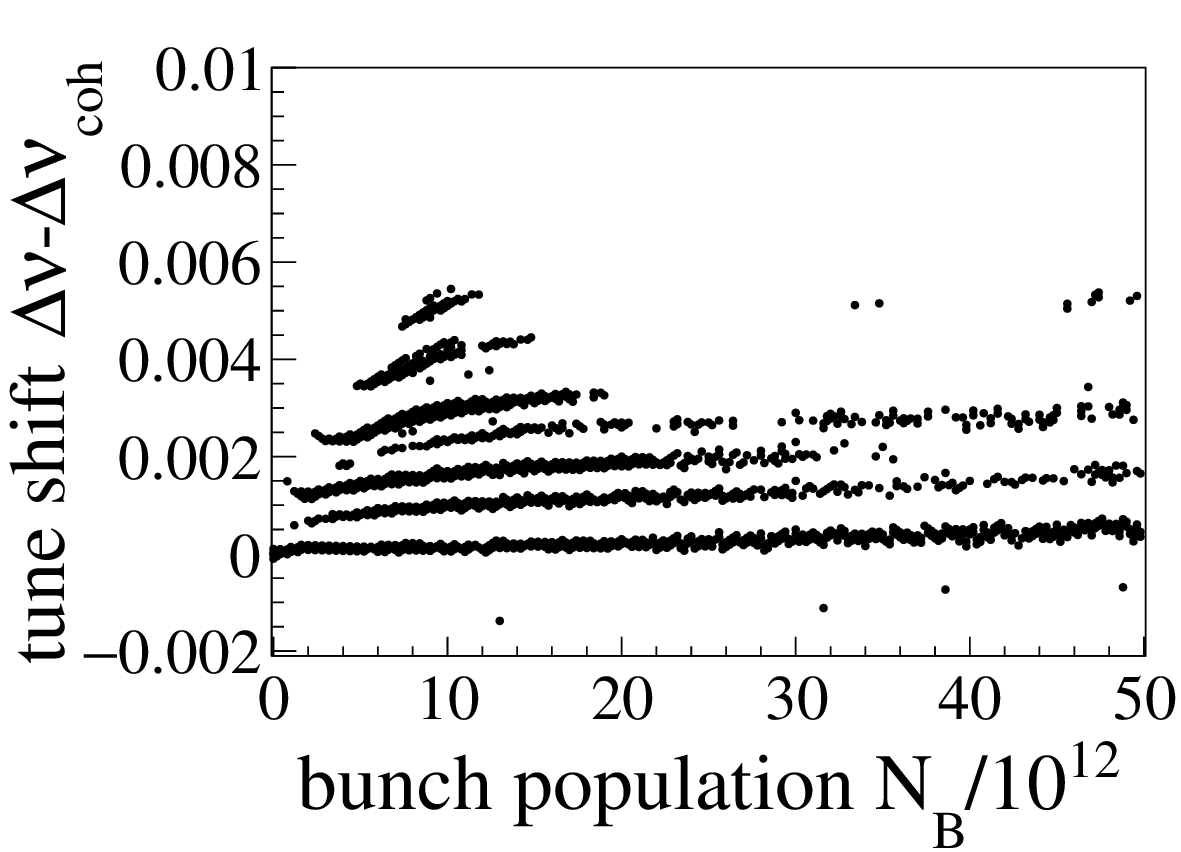}
      \subcaption{}
      \label{fig:8a}
    \end{minipage} &
    \begin{minipage}[t]{0.45\hsize}
      \centering
      \includegraphics[width=3.0in]{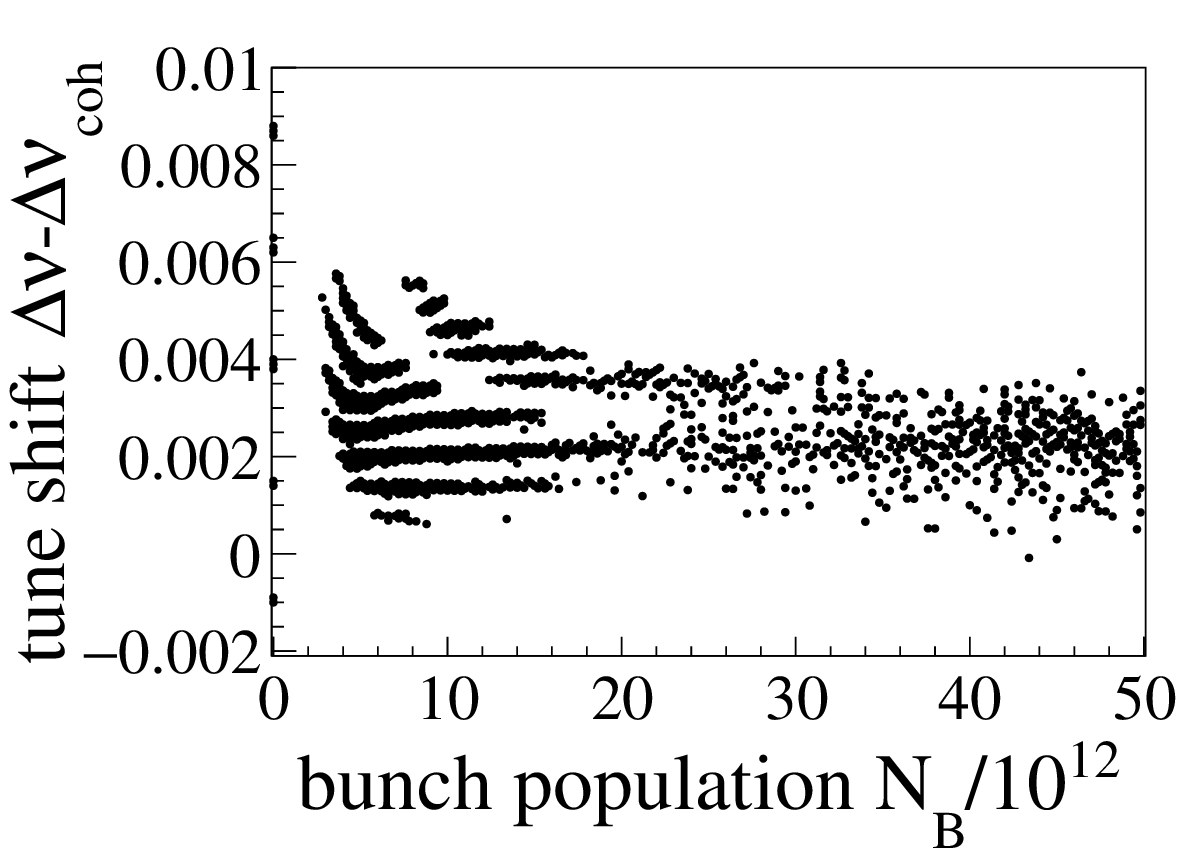}
      \subcaption{}
      \label{fig:8b}
    \end{minipage}
  \end{tabular}
  \caption{Tune shift at which the tune spectrum reaches its maximum peak in the simulation, including the transverse wakefield and the direct and indirect space charge effects. (a) Chromaticity $\xi_x=0$ and (b) chromaticity $\xi_x=-8.0$.}
  \label{fig:8}
\end{figure}

Having confirmed the code's credibility concerning the wakefields,
we then transition to the more physical regime by reintroducing both the direct and indirect space charge effects with the same transverse wakefield.
The tune spectrum is shown in Fig. \ref{fig:8}, with the contribution from Eq. \eqref{eq:124} eliminated to clearly show the excitation modes.
The higher head-tail modes with $m\geq0$ are more excited compared to the results in Figs. \ref{fig:7a} and \ref{fig:7b} regardless of the choice of chromaticity, with the degeneracy of the radial modes maintained in the case of $\xi_x=-8.0$ (Fig. \ref{fig:8b}).
The spectrum in Fig. \ref{fig:8a} does not show a clear bifurcation signature of the radial and head-tail modes, nor the TMCI as identified in Fig. \ref{fig:7a}, indicating the beam instability is suppressed even in the case of $\xi_x=0$.

\begin{figure}[!h]
  \begin{tabular}{cc}
    \begin{minipage}[t]{0.45\hsize}
      \centering
      \includegraphics[width=3.0in]{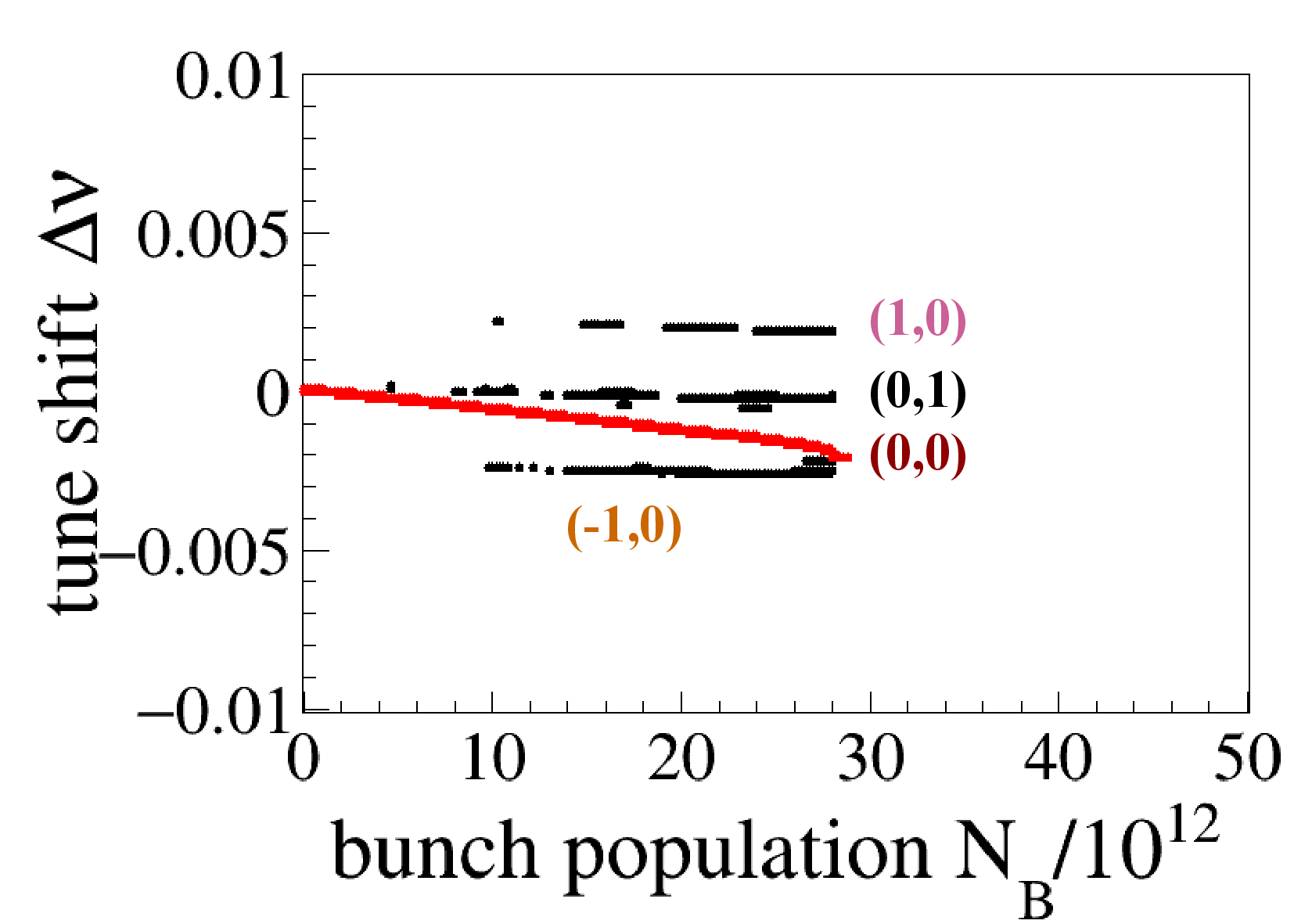}
      \subcaption{}
    \end{minipage} &
    \begin{minipage}[t]{0.45\hsize}
      \centering
      \includegraphics[width=3.0in]{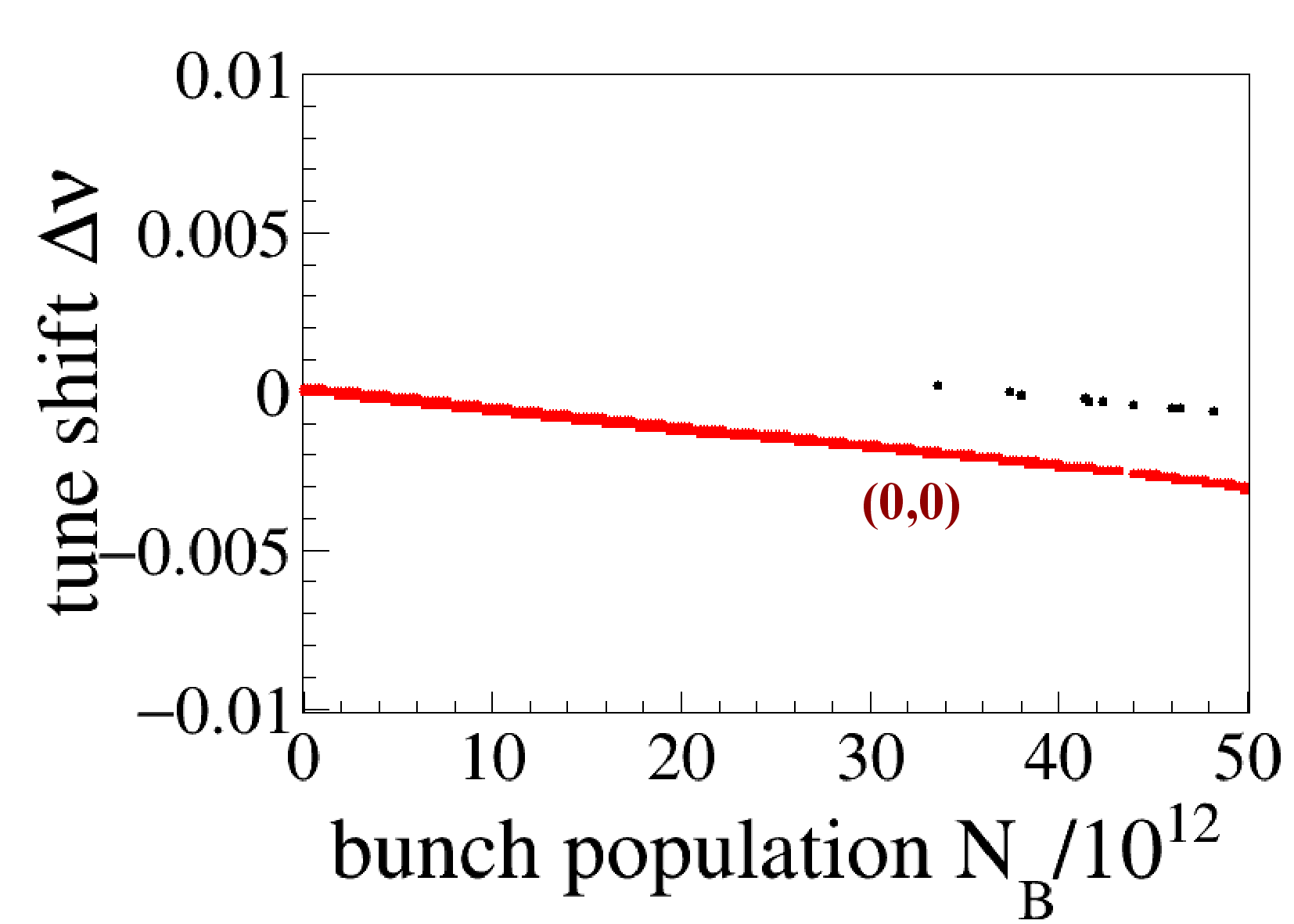}
      \subcaption{}
    \end{minipage} \\
    \begin{minipage}[t]{0.45\hsize}
      \centering
      \includegraphics[width=3.0in]{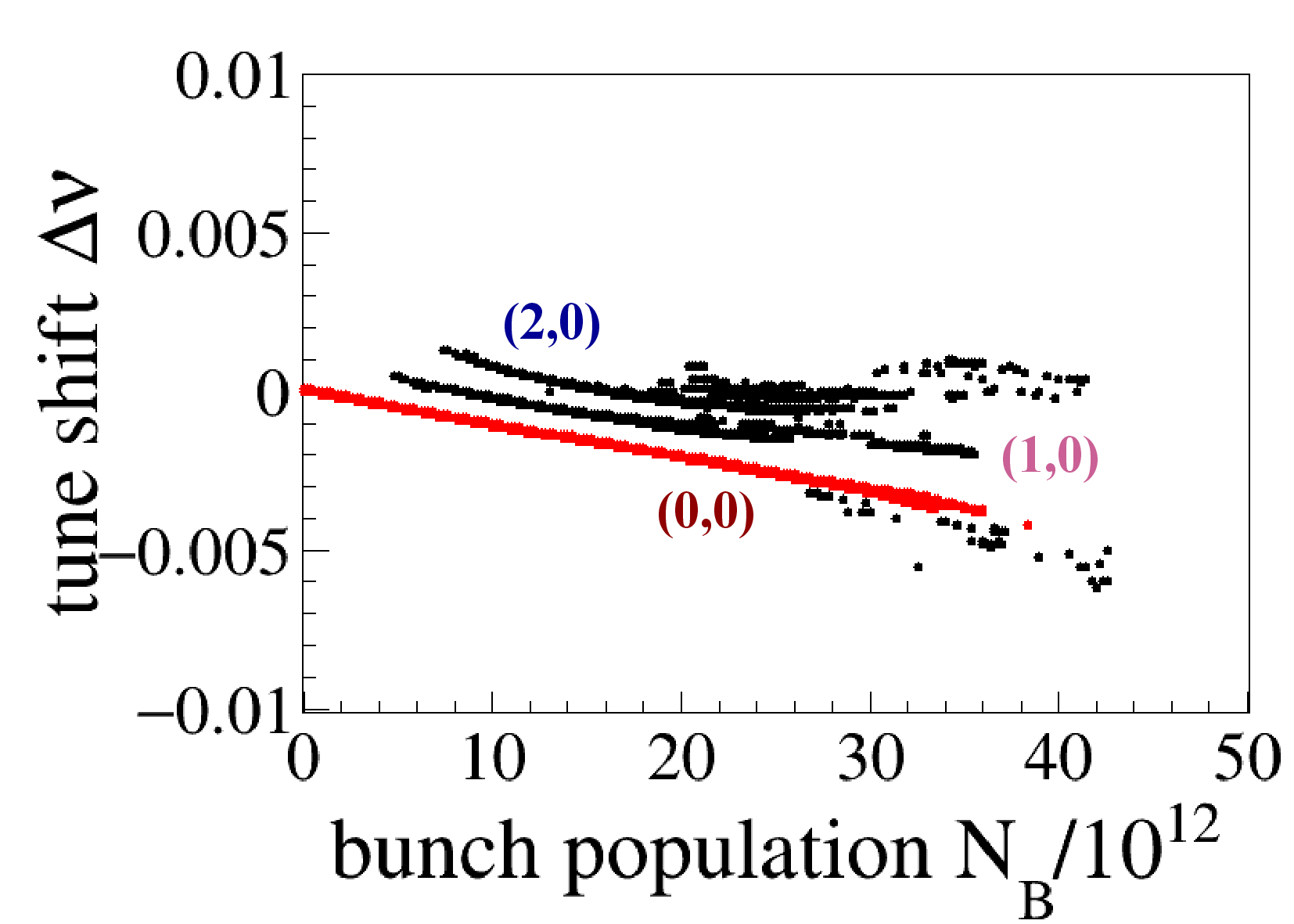}
      \subcaption{}
    \end{minipage} &
    \begin{minipage}[t]{0.45\hsize}
      \centering
      \includegraphics[width=3.0in]{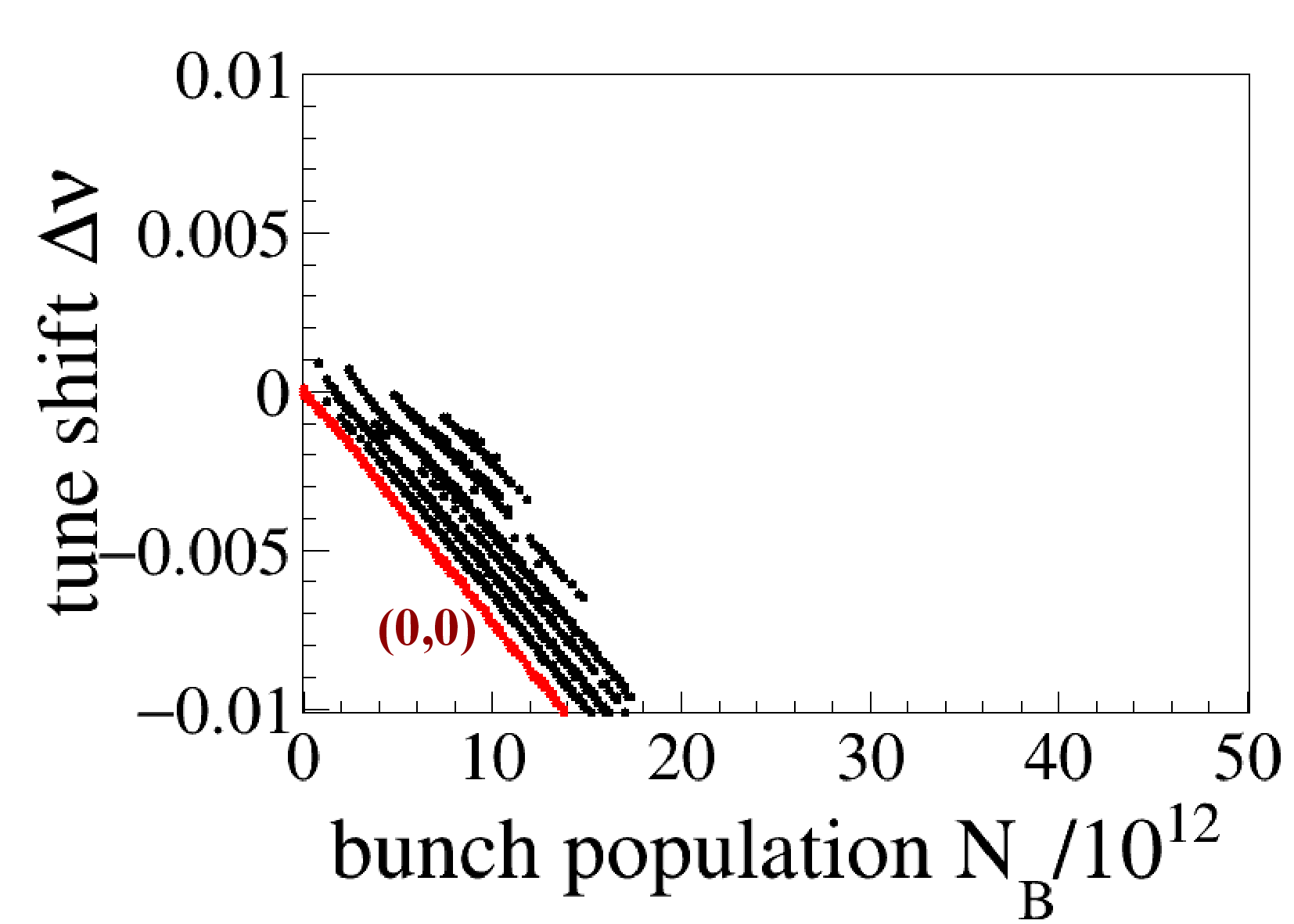}
      \subcaption{}
    \end{minipage}
  \end{tabular}
  \caption{Tune shift at which the tune spectrum reaches its maximum peak in the simulation with $\xi_x=0$. The pair of numbers describes the excited mode indices ($m, n_r$). (a) $r=0.001$, (b) $r=0.01$, (c) $r=0.1$, and (d) $r=1$.}
  \label{fig:9}
\end{figure}

Let us examine how the suppression of radial and negative head-tail modes occurs by intentionally changing the ratio, $r$, of the space charge strength to the wake strength, with it multiplied with the horizontal kick originated from the space charge effect (Eq. \eqref{eq:60}), considering the case of $\xi_x=0$.
The results are shown in Fig. \ref{fig:9}, where the ($m=0,n_r=0$) mode is represented by the red bullet.
We observe that the higher radial modes and negative head-tail modes gradually disappear as the ratio $r$ (space charge strength) increases, and the higher head-tail modes become more enhanced because they are space charge intrinsic modes explained by the extended ABS theory.
This demonstrates that the space charge dismisses TMCI between ($m=0,n_r=0$) and ($m=-1,n_r=0$) at $\xi_x=0$ in Fig. \ref{fig:7a} by suppressing the negative head-tail modes.

To directly confirm that the beam is more stable when the space charge effect is considered, let us investigate the dependence of the beam intensity on the turn-by-turn horizontal amplitude (Courant-Snyder invariant) $\overline{A_n}$:
\begin{equation}
  \overline{A_n}=\sqrt{\overline{x_n}^2+\beta_x^2 \overline{x'_n}^2},
\end{equation}
for the cases of zero chromaticity,
where the Twiss parameter $\alpha$ is assumed to be zero for simplicity, and $\overline{x'_n} = \sum_i x'_{n,i}/\Np$.

\begin{figure}[!h]
  \begin{tabular}{ccc}
    \begin{minipage}[t]{0.3\hsize}
      \centering
      \includegraphics[width=2.0in]{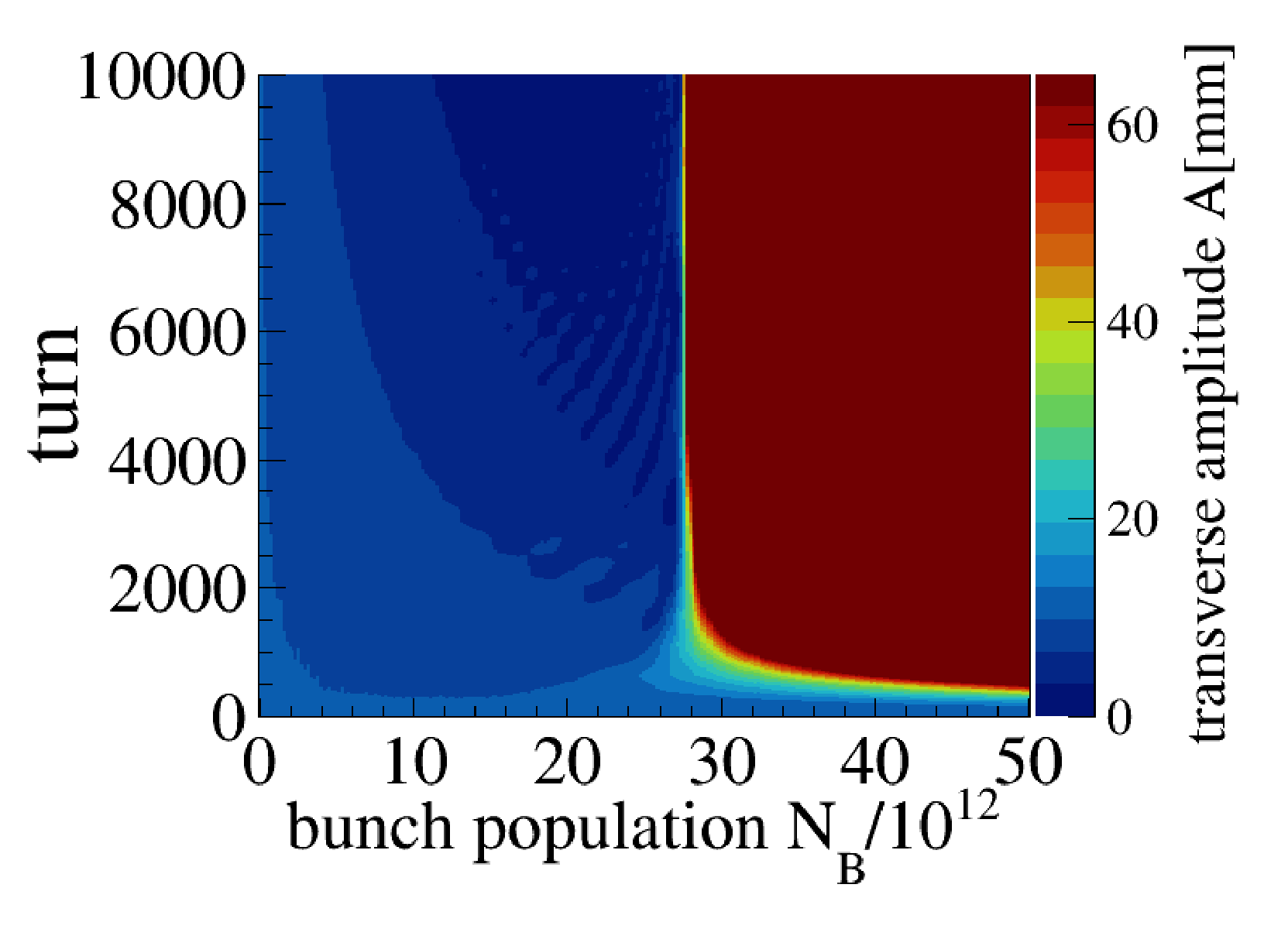}
      \subcaption{}
      \label{fig:10a}
    \end{minipage} &
    \begin{minipage}[t]{0.3\hsize}
      \centering
      \includegraphics[width=2.0in]{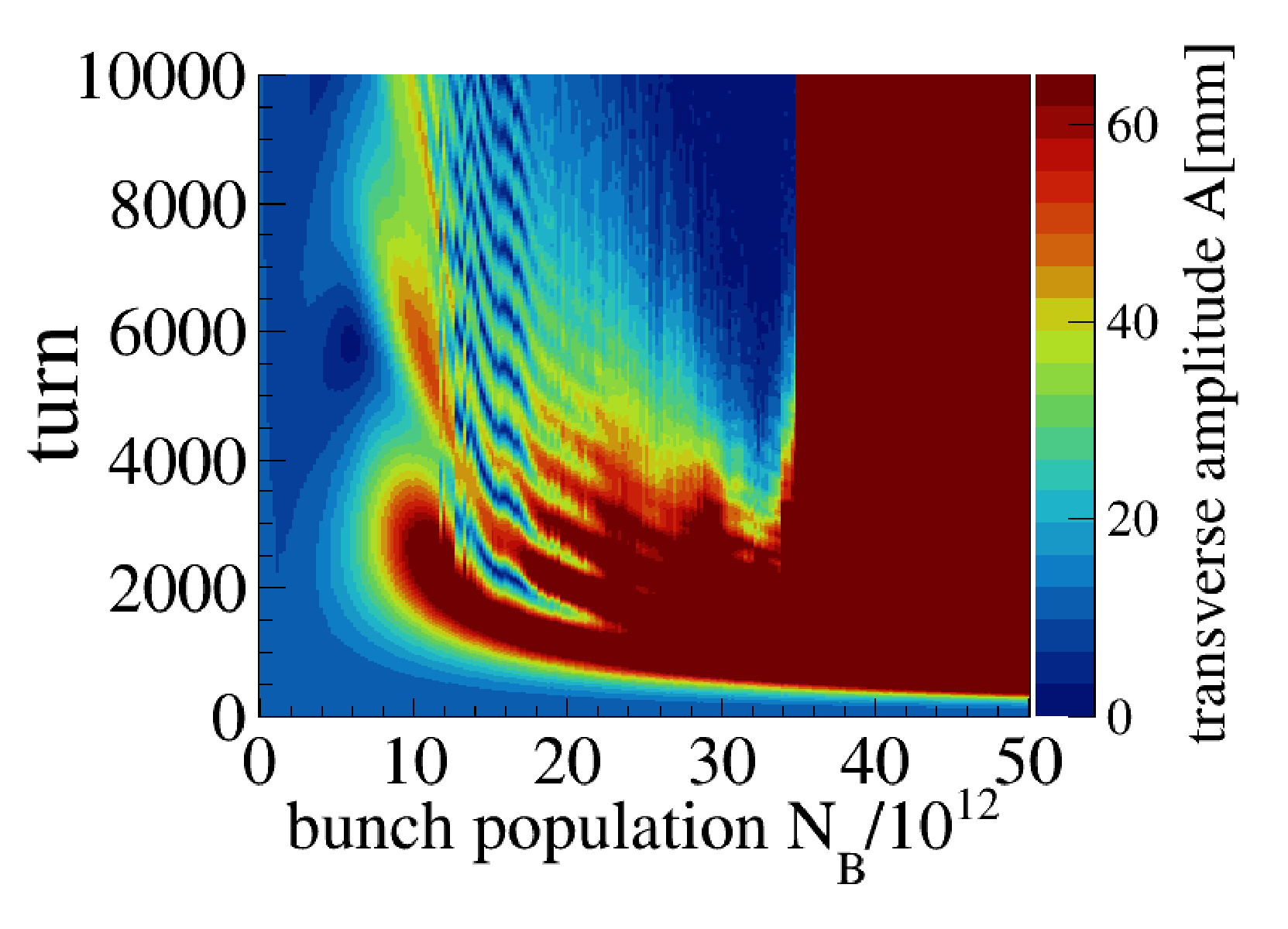}
      \subcaption{}
      \label{fig:10b}
    \end{minipage} &
    \begin{minipage}[t]{0.3\hsize}
      \centering
      \includegraphics[width=2.0in]{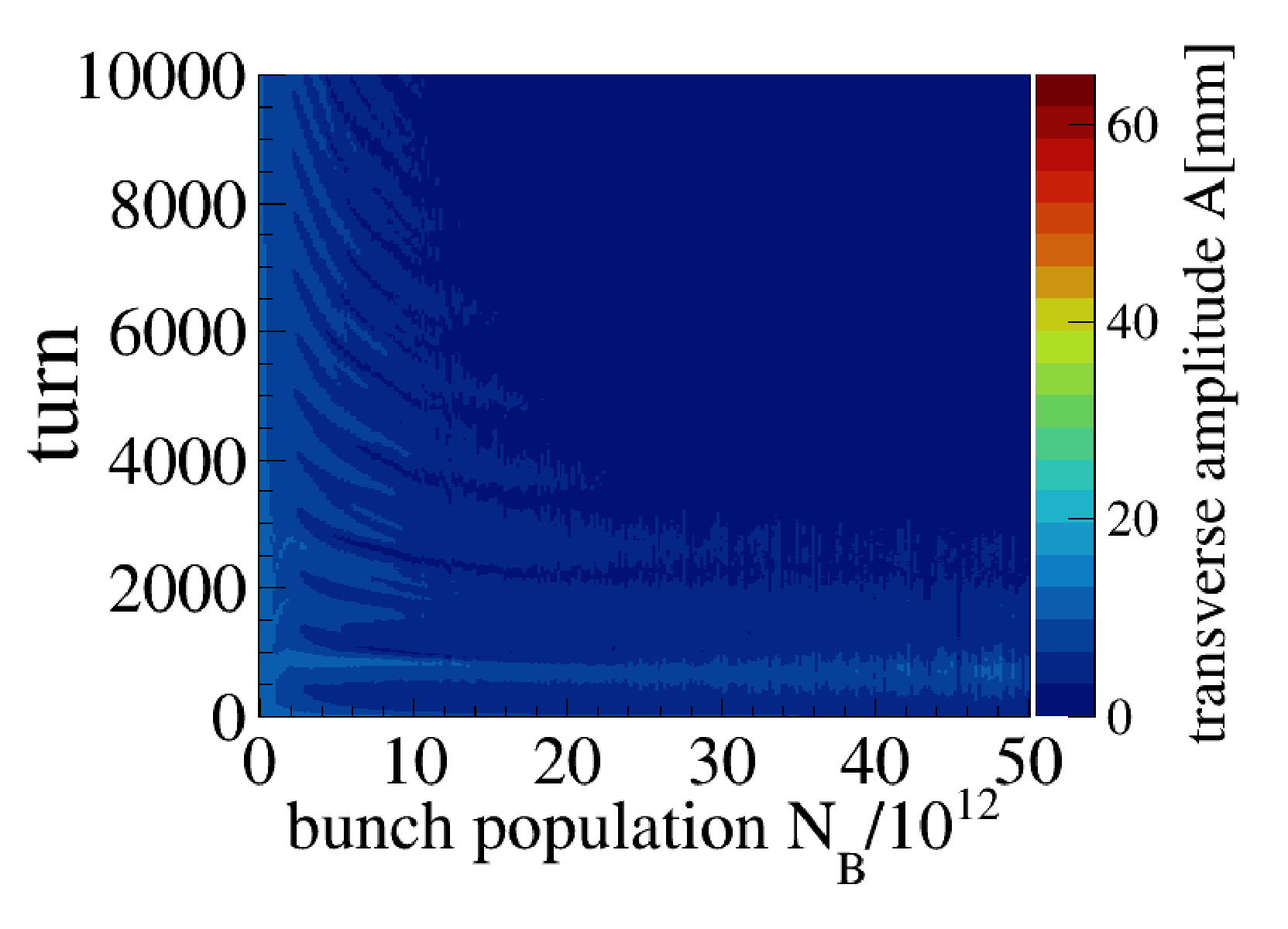}
      \subcaption{}
      \label{fig:10c}
    \end{minipage}
  \end{tabular}
  \caption{Transverse amplitude $\overline{A_n}$ scanned by varying beam intensity $N_B$ for the case of chromaticity $\xi_x=0$. (a) Only the transverse wakefield, (b) the transverse wakefield and the direct space charge, and (c) the transverse wakefield and the direct and indirect space charge.}
  \label{fig:10}
\end{figure}

Figures \ref{fig:10} represents the results with only wakefield effects, wakefield and direct space charge effects, and wakefield with direct and indirect space charge effects, respectively.
The simulation reveals that the growth rate with both direct and indirect space charge and wakefield effects (Fig. \ref{fig:10c}) is milder than that in the case including only wakefield effects (Fig. \ref{fig:10a}), demonstrating the damping effects due to the indirect space charge force \cite{shobuda2017theoretical,saha2018simulation}.
On the other hand, neglecting the indirect space charge force (Fig. \ref{fig:10b}) would significantly enhance the beam growth rate, compared to the case including only wakefield effects (Fig. \ref{fig:10a}), which is consistent with another theoretical prediction \cite{burov2019convective}.
In conclusion, our code successfully resolves two different theoretical predictions and highlights the importance of including indirect space charge effect in the simulation.

\section{Transverse bunch motion in the time domain in the tracking simulation}\label{9}


Since we have analyzed the frequency spectrum of a Gaussian beam, we will observe how the bunch motions in the Gaussian bunch are affected in the time domain by the space charge effects
through the chromaticity, the beam intensity, and the synchrotron period (rf voltage) by the tracking simulation under the storage mode of the J-PARC MR.

\subsection{Typical results of the tracking simulation}\label{9.1}
\begin{figure}[!h]
  \begin{tabular}{ccc}
    \begin{minipage}[t]{0.3\hsize}
      \centering
      \includegraphics[width=2.0in]{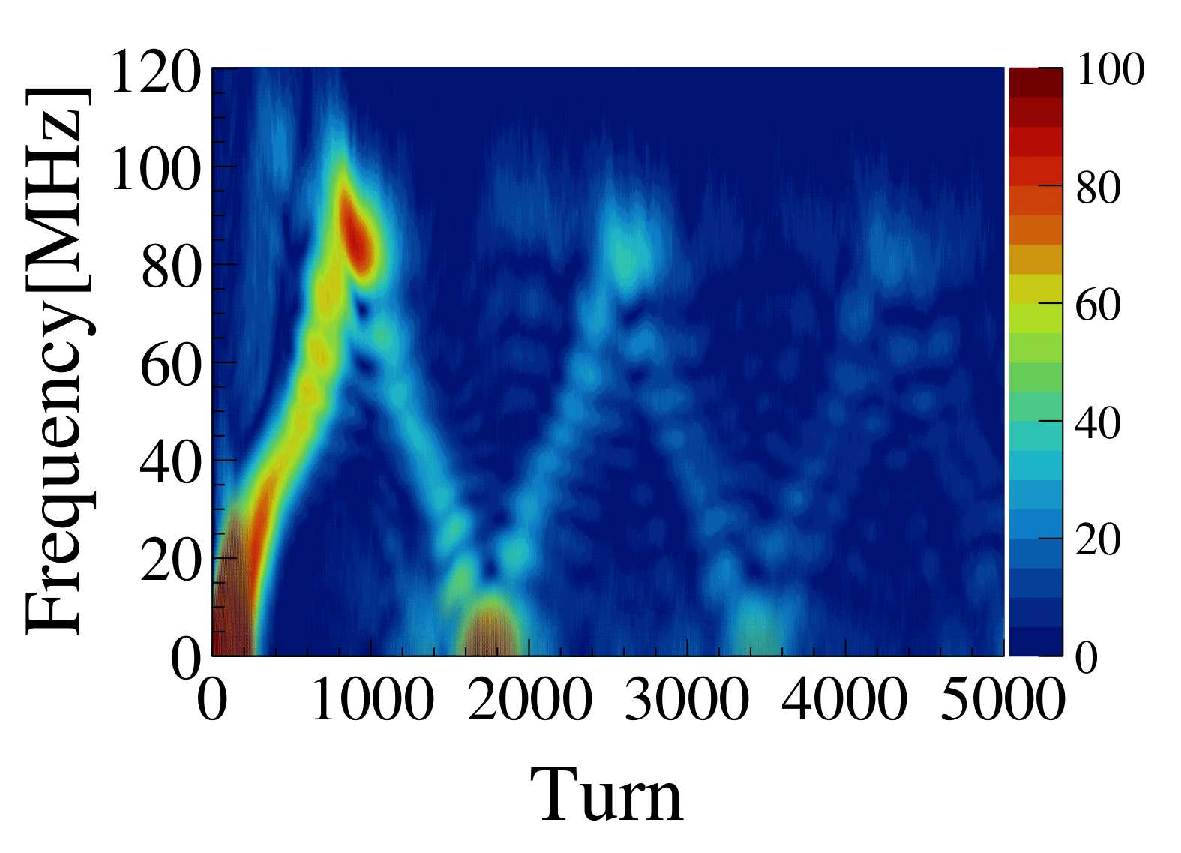}
      \subcaption{}
      \label{fig:11a}
    \end{minipage} &
    \begin{minipage}[t]{0.3\hsize}
      \centering
      \includegraphics[width=2.0in]{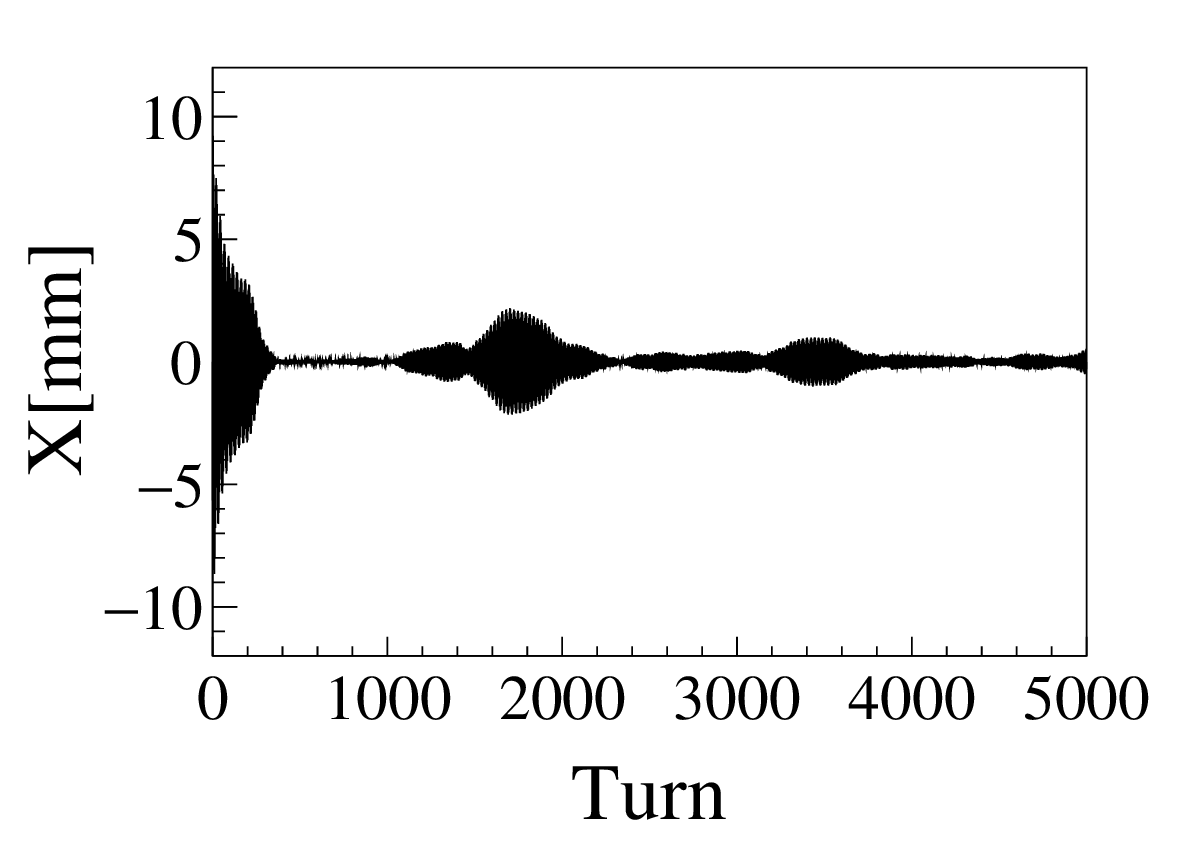}
      \subcaption{}
      \label{fig:11b}
    \end{minipage} &
    \begin{minipage}[t]{0.3\hsize}
      \centering
      \includegraphics[width=2.0in]{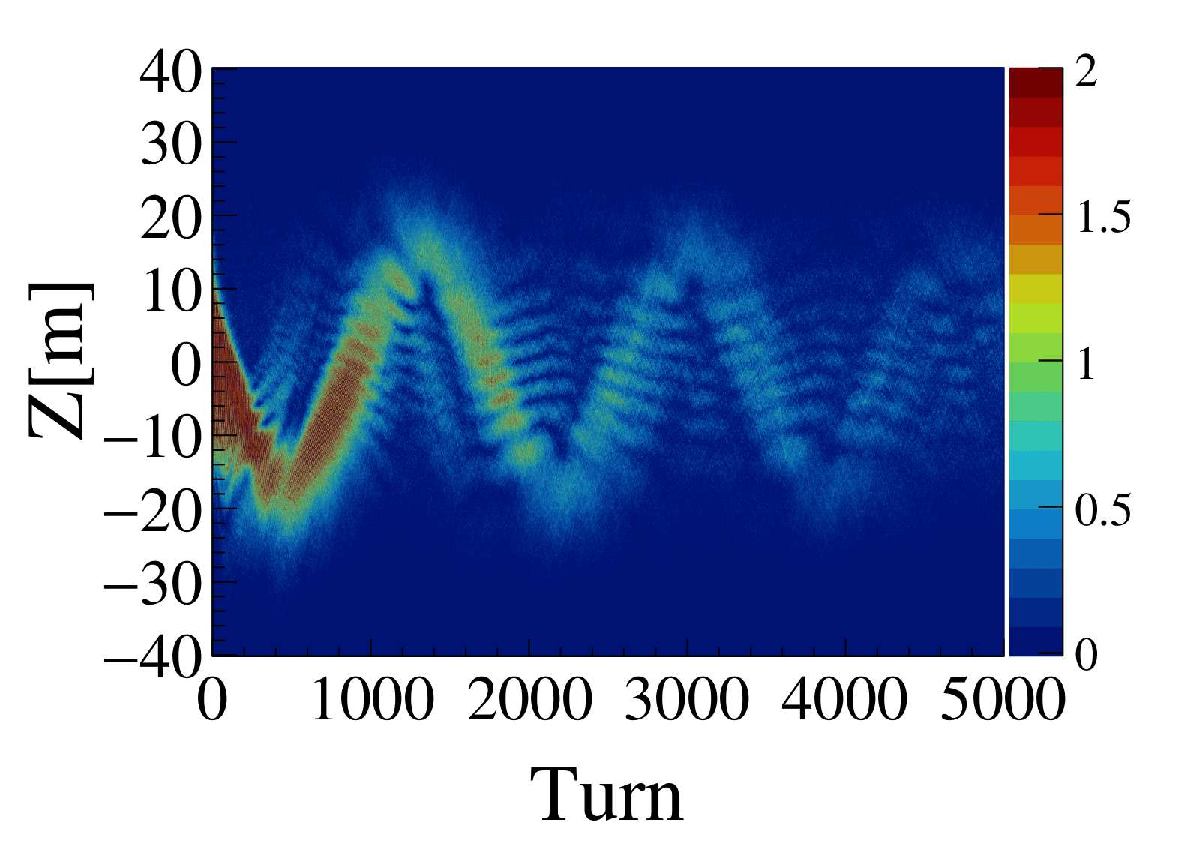}
      \subcaption{}
      \label{fig:11c}
    \end{minipage}
  \end{tabular}
  \caption{Transverse bunch motion in the time domain for the chromaticity $\xi_x=-12.5$ and the beam intensity $N_B=8.6\times10^{12}$~ppb. (a) Frequency component of the dipole moment in the bunch, (b) average of the beam position $\overline{x_n}$, and (c) dipole moment $|\Delta^{(k)}\sigma_\Delta/q|$ in the bunch.}
  \label{fig:11}
\end{figure}

Figure \ref{fig:11} shows typical results in the time domain of the tracking simulation for the case of strong space charge ($N_B=8.6\times10^{12}$~ppb) with the chromaticity $\xi_x=-12.5$ and horizontal injection offset $X_0=0.01$~m.
Figure \ref{fig:11a} illustrates that the frequency components of the dipole moment $\Delta^{(k)}$, alternately exhibiting high and low-frequency components along the turn.
Meanwhile, Fig. \ref{fig:11b} depicts the oscillatory behavior of the horizontal centroid of the entire bunch, representing the decoherent and recoherent motions of the particles comprising the beam.
The comparison between Figs. \ref{fig:11a} and \ref{fig:11b} reveals that when the average transverse position indicates decoherence at the $N_f$ turn (or recoherence at the $\NAM$ turn), 
the intrabunch frequency in Fig. \ref{fig:11a} reaches its maximum ($\fmax$) or minimum, respectively.

To proceed with the following analysis, it is necessary to rigorously define $\fmax$, $N_f$, and $\NAM$ from an observation point of view.
This is because the frequency components in Fig. \ref{fig:11a} distribute with intensities, and some oscillation peaks can be observed in Fig. \ref{fig:11b}, making it challenging to determine these parameters precisely.
Since the high-frequency components of the dipole moment are more clearly identifiable in Fig. \ref{fig:11a}, $N_f$ is defined as the first turn number at which the frequency component reaches its highest value with strongest signal ($\fmax$), sweeping from the injection point.
Finally, the recoherence period is defined as twice the intrabunch oscillation period: $\NAM=2N_f$.
In addition, to compare the recoherence period for different synchrotron periods, the normalized recoherence period $\NAM /N_s$ is introduced.
In Fig. \ref{fig:11} case, we can determine the maximum intrabunch frequency $\fmax=84.1$~MHz, the intrabunch oscillation period $N_f=890$~turns, the recoherence period $\NAM=1780$~turns, and the normalized recoherence period $\NAM/N_s=4.3$, respectively.

Figure \ref{fig:11c} reveals that the dipole moment $|\Delta^{(k)}\sigma_\Delta/q|$ characteristically meanders back and forth in the longitudinal direction, with this cycle being equal to $\NAM$.

\subsection{Modeling of head-tail mode reflecting simulation}\label{9.2}
To understand the simulation results in Fig. \ref{fig:11}, let us simplify the transverse centroid of each segment in the bunch as
\begin{equation}\label{eq:131}
  \bar{x}(n,z) = X_0 \sum_{m=m_\mathrm{min}}^{m_\mathrm{max}} \tilde{a}_{m,\mathrm{sc}}(z) \exp\left(-\textbf{i}2\pi\nu_m n\right),
\end{equation}
where 
\begin{equation}\label{eq:132}
  \tilde{a}_{m,\mathrm{sc}}(z) = (-\textbf{i})^m \exp\left(\textbf{i} 2\pi \frac{\xi_x}{\eta C} z\right) \cos\left(m\left(\tan^{-1}\left(\frac{\pi}{2} \frac{z}{\sigma_z}\right)-\frac{\pi}{2}\right)\right),
\end{equation}
referring to Eq. \eqref{eq:89}.
Here, the relative amplitude in the head-tail mode modeling in each segment is $\tilde{a}_{m,\mathrm{sc}}(z)$.
The factor $\pi/2 \times z/\hat{z}$ is replaced with $\tan^{-1}(\pi/2\times z/\sigma_z)$ in Eq. \eqref{eq:132}.
The excitation modes specified by $m_\mathrm{min}$ and $m_\mathrm{max}$ in Eq. \eqref{eq:131} are determined by the simulations.

Regarding the tune of the head-tail mode $m\geq0$, it is assumed to be expressed as 
\begin{equation}
  \nu_m = \nu_0 + m\Delta\nu,
\end{equation}
with bare betatron tune $\nu_0$ and equal intervals of the head-tail modes $\Delta\nu$, based on the simulation results of Fig. \ref{fig:6d}.

\begin{figure}[!h]
  \begin{tabular}{ccc}
    \begin{minipage}[t]{0.3\hsize}
      \centering
      \includegraphics[width=2.0in]{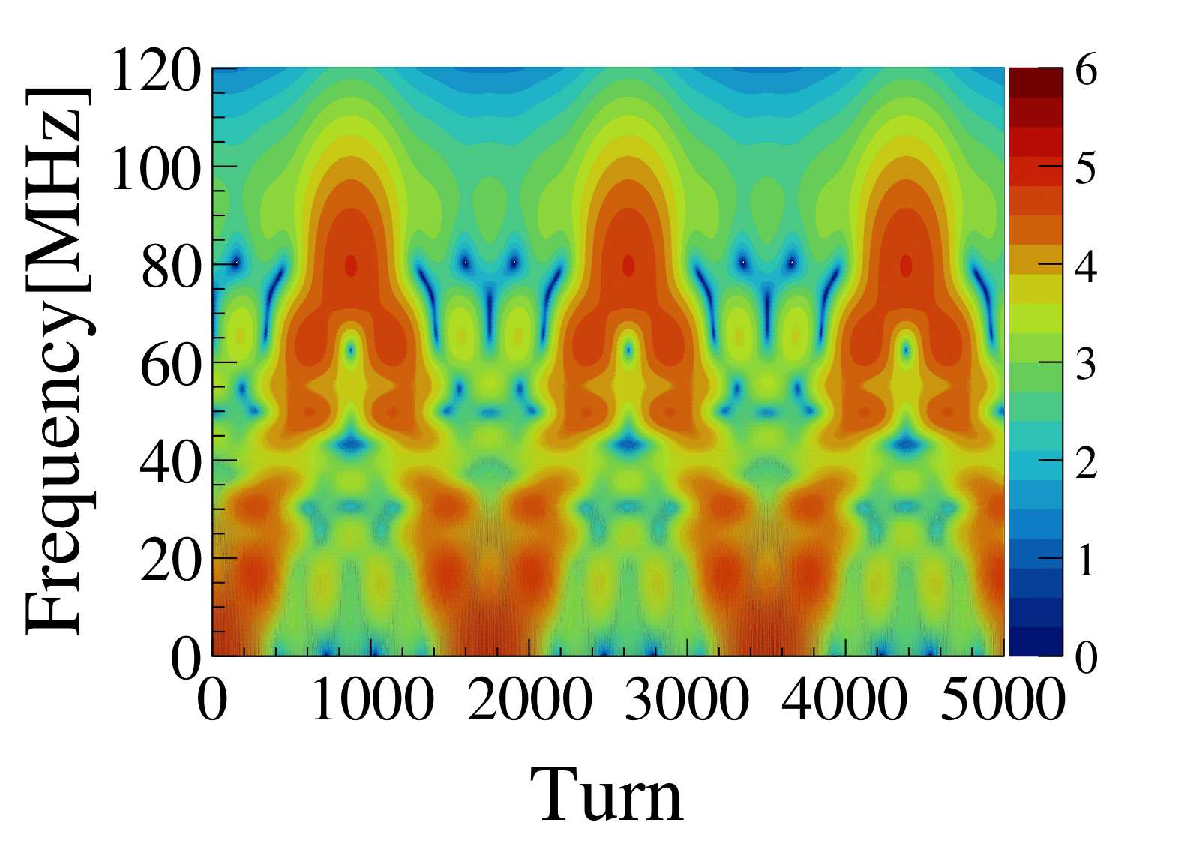}
      \subcaption{}
    \end{minipage} &
    \begin{minipage}[t]{0.3\hsize}
      \centering
      \includegraphics[width=2.0in]{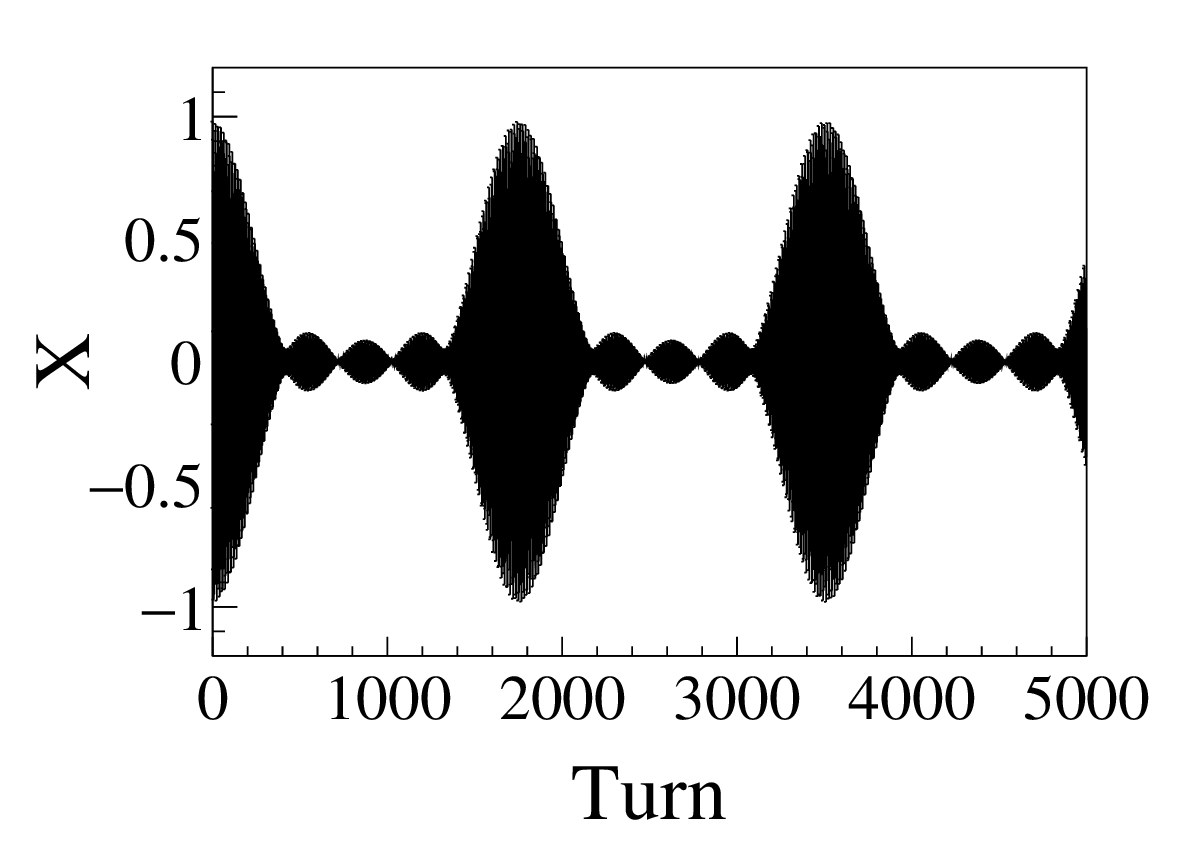}
      \subcaption{}
    \end{minipage} &
    \begin{minipage}[t]{0.3\hsize}
      \centering
      \includegraphics[width=2.0in]{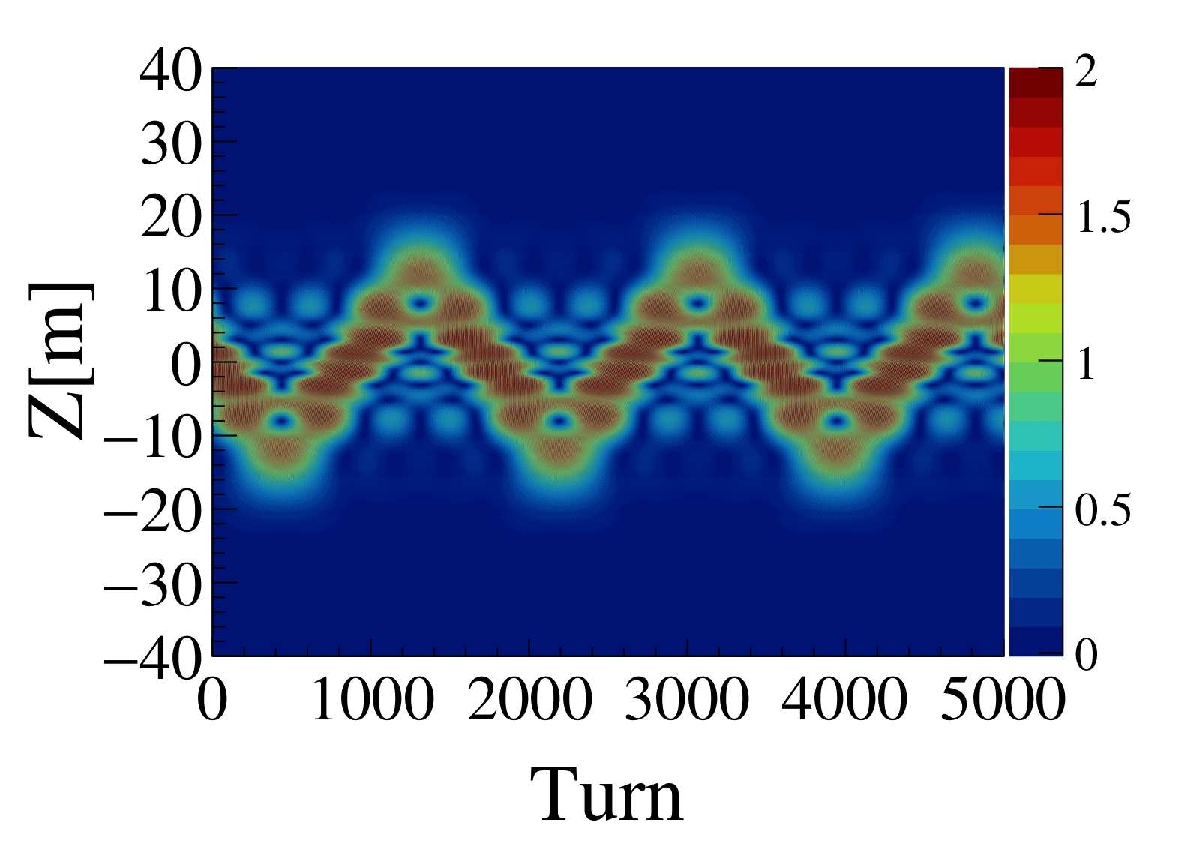}
      \subcaption{}
    \end{minipage}
  \end{tabular}
  \caption{Transverse bunch motion in the time domain under the condition of $\xi_x=-12.5$, $m_\mathrm{min}=5$, $m_\mathrm{max}=9$. (a) Frequency component of the dipole moment in the bunch, (b) average of the beam position $X(n)$, and (c) dipole moment $|\Delta(n,z)|$ in the bunch.}
  \label{fig:12}
\end{figure}

Figure \ref{fig:12} shows the dipole moment:
\begin{equation}
  \Delta(n,z) = \mathrm{Re}[\bar{x}(n,z)] \rho_z(z),
\end{equation}
its frequency component, and the average of the beam position $X(n)= \mathrm{Re}\left[\int_{-\infty}^{\infty} dz \rho_z(z) \bar{x}(n,z)\right]$ for the parameter set of $\xi_x=-12.5$, $m_\mathrm{min}=5$, $m_\mathrm{max}=9$, $\sigma_z=10$~m, $\Delta\nu=0.00057$, $\nu_0=21.35$, and $X_0=1$~mm.
Comparing the results to the tracking simulation (Fig. \ref{fig:11}), 
we can observe that the intrabunch frequency transitions with an interval of $N_f$ between high and low frequencies,
the periodic motion occurs at a cycle of $\NAM =1/\Delta\nu$, and the dipole moment meanders.
The analytical model for the simulations reveals that the space charge effects characterize the motions of particles comprising the beam.

To further analytically understand this phenomenon, we will extend the upper limit of head-tail mode in Eq. \eqref{eq:131} to infinity with $m_\mathrm{min}=0$.
The transverse centroid of each segment is simplified as
\begin{widetext}
\begin{equation}\label{eq:135}
  \begin{split}
      & \bar{x}(n,z)/\exp(-i2\pi\nu_0 n) \\
    = & X_0 \exp\left( \textbf{i} 2\pi \frac{\xi_x}{\eta C} z\right) \sum_{m=0}^{\infty} \exp\left(- \textbf{i} m (2\pi \Delta\nu n+\frac{\pi}{2})\right) \cos\left(m\left(\tan^{-1}\left(\frac{\pi}{2} \frac{z}{\sigma_z}\right)-\frac{\pi}{2}\right)\right) \\
    = & X_0 \exp\left( \textbf{i} 2\pi \frac{\xi_x}{\eta C} z\right) \frac{1}{2}\left(1+\frac{1}{1+\exp(\textbf{i}(\tan^{-1}\left(\frac{\pi}{2} \frac{z}{\sigma_z}\right)-2\pi \Delta\nu n))}-\frac{1}{1-\exp(\textbf{i}(\tan^{-1}\left(\frac{\pi}{2} \frac{z}{\sigma_z}\right)+2\pi \Delta\nu n))}\right).
  \end{split}
\end{equation}
\end{widetext}

The last term characterized by $\Delta\nu$ diverges around
\begin{equation}
  z=
  \begin{cases}
    -\frac{2}{\pi} \sigma_z \tan(2\pi \Delta\nu n), & ~\text{for} ~0<n<\frac{1}{4}\NAM, \frac{3}{4}\NAM< n<\NAM, \\
    +\frac{2}{\pi} \sigma_z \tan(2\pi \Delta\nu n), & ~\text{for} ~\frac{1}{4}\NAM<n<\frac{3}{4}\NAM,
  \end{cases}
\end{equation}
in the range of $0<n<\NAM$, explaining the meandering of the dipole moments along the turn numbers, owing to the space charge effects.

Referring to the derivation of Eq. \eqref{eq:22}, we focus on the phase component of Eq. \eqref{eq:135} to analyze the dependence of the maximum intrabunch frequency on chromaticity, including the effects of space charge.
First, we calculate $\mathrm{Arg}[\bar{x}(n,z)/\exp(-\textbf{i}2\pi\nu_0 n)]$ after substituting
$n=\NAM /2$ into Eq. \eqref{eq:135} for small $z$, resulting in 
\begin{equation}\label{eq:137}
  \begin{split}
      & \mathrm{Arg}\left[ \bar{x}\left(\frac{\NAM}{2},z\right)/\exp\left(-\textbf{i}2\pi\nu_0 \frac{\NAM}{2}\right) \right] - 2\pi \frac{\xi_x}{\eta C} z \\
    = & \mathrm{Arg}\left[ \sum_{m=m_\mathrm{min}}^{m_\mathrm{max}} \textbf{i}^m \cos\left(m\left(\tan^{-1}\left(\frac{\pi}{2} \frac{z}{\sigma_z}\right)-\frac{\pi}{2}\right)\right) \right] \\
    \simeq & \mathrm{Arg}\left[ \sum_{\substack{m_\mathrm{min}\leq m\leq m_\mathrm{max} \\ m:\text{even} }} (+1) + \textbf{i} \sum_{\substack{m_\mathrm{min}\leq m\leq m_\mathrm{max} \\ m:\text{odd} }} m \frac{\pi}{2} \frac{z}{\sigma_z} \right].
  \end{split}
\end{equation}

Here, the second term can be estimated by referring to the condition on
$\mathrm{Arg}[\bar{x}(n,z)/\exp(-\textbf{i}2\pi\nu_0 n)]$ at $n=0$ for small $z$, which must be zero due to the very initial condition.
As a result, this condition is explicitly expressed by
\begin{equation}\label{eq:138}
  \begin{split}
      & \mathrm{Arg}[ \bar{x}(0,z)/\exp(-\textbf{i}2\pi\nu_0 0)] - 2\pi \frac{\xi_x}{\eta C} z \\
    = & \mathrm{Arg}\left[ \sum_{m=m_\mathrm{min}}^{m_\mathrm{max}} (-\textbf{i})^m \cos\left(m\left(\tan^{-1}\left(\frac{\pi}{2} \frac{z}{\sigma_z}\right)-\frac{\pi}{2}\right)\right) \right] \\
    \simeq & \mathrm{Arg}\left[ \sum_{\substack{m_\mathrm{min}\leq m\leq m_\mathrm{max} \\ m:\text{even} }} (+1) - \textbf{i} \sum_{\substack{m_\mathrm{min}\leq m\leq m_\mathrm{max} \\ m:\text{odd} }} m \frac{\pi}{2} \frac{z}{\sigma_z} \right].
  \end{split}
\end{equation}

By combining the complex conjugate of Eq. \eqref{eq:138} with Eq. \eqref{eq:137}, we derive
\begin{equation}
  \mathrm{Arg}\left[\bar{x}\left(\frac{\NAM}{2},z\right)/\exp\left(-\textbf{i}2\pi\nu_0 \frac{\NAM}{2}\right)\right] \simeq 4\pi \frac{\xi_x}{\eta C} z.
\end{equation}

Finally, the maximum value of
$\fint$ at the center of the bunch is given by
\begin{equation}\label{eq:140}
  \fmax=\fint\left(\frac{\NAM}{2}\right) \simeq \left|\frac{1}{2\pi}\frac{d}{dt}\left[2\pi\nu_0 \frac{\NAM}{2}-4\pi \frac{\xi_x}{\eta C} z\right]\right| = \left|2f_0\frac{\xi_x}{\eta}\right|,
\end{equation}
which matches Eq. \eqref{eq:22} in the absence of space charge effects.
The validity of this approximate formula including the space charge effect will be discussed in Sec. \ref{10.5}.

In the end, the simulations featuring the meandering of the dipole moment, the maximum intrabunch frequency, and the recoherence of the entire bunch frequency can be roughly understood by the analytical model for the given excited head-tail modes $m_\mathrm{min}$ and $m_\mathrm{max}$ under the chromaticity condition.

\section{DEMONSTRATION BY BEAM MEASUREMENTS}\label{10}


In this section, we examine the intrabunch motion predicted by the tracking simulation in the previous section by changing the chromaticity, the synchrotron period, and the beam intensity in the storage mode of the J-PARC MR based on beam measurements.
We analyze the horizontal centroid of the entire bunch, the dipole moment in the bunch, and its intrabunch frequency for each turn,
when beams are injected into the MR with the horizontal offset of about $X_0=5$~mm.

The measurements utilized data from a stripline beam position monitor with a 500 MHz sampling rate.
This monitor, comprising two pairs of triangle-shaped electrodes \cite{shobuda2016triangle}, is installed in the J-PARC MR to ensure a wide-band frequency response for the IBFB system.
The horizontal dipole moment, $\Delta(z)$, along the bunch is obtained by integrating the difference signals from the horizontal electrodes.
The horizontal centroid of the entire bunch $\overline{x_n}$ is then calculated by summing all dipole moments in the bunch and multiplying the result by the sensitivity coefficient, 0.046 m, divided by the measured bunch current.
In the 3 GeV storage mode with a single bunch, the beam power corresponds to $0.35\times N_B/10^{12}$~kW for a repetition period of 1.36 s.

\subsection{Conditions of measurements}\label{10.1}

Three sets of measurements were conducted under different machine conditions.

First, we observe the synchrotron tune dependence by scanning the rf voltages $\VRF=80,110,140,200,263$~kV, and $400$~kV and the beam intensities $N_B=2.4\times10^{12},4.2\times10^{12}$, and $8.6\times10^{12}$~ppb
with the fixed chromaticity $(\xi_x,\xi_y)=(-6.5,-7.4)$ and the betatron tune $(\nu_{\beta,x},\nu_{\beta,y})=(21.35,21.41)$.
Here, $\xi_y$ is defined as the vertical chromaticity, and $\nu_{\beta,y}$ is defined as the vertical betatron tune.
Table \ref{tab:1} summarizes the number of measured shots for the analysis.

\begin{table*}[!h]
  \caption{Number of shots for the analysis of the synchrotron period dependence with the fixed chromaticity $(\xi_x,\xi_y)=(-6.5,-7.4)$ and the betatron tune $(\nu_{\beta,x},\nu_{\beta,y})=(21.35,21.41)$.}
  \begin{ruledtabular}
  \centering
  \small
  \begin{tabular}{c  c  c  c }
    $N_B$(ppb)$/10^{12}$ & $~~2.4~~$ & $~~4.2~~$ & $~~8.6~~$ \\ \hline
    $N_s=340$            & 3     & 3     & 3     \\ 
    $N_s=410$            & 3     & 3     & 3     \\ 
    $N_s=480$            & $\cdots$ & 3     & 3     \\ 
    $N_s=570$            & $\cdots$ & 3     & 1     \\ 
    $N_s=650$            & $\cdots$ & 4     & 1     \\ 
    $N_s=760$            & 3     & 2     & 1     \\ 
  \end{tabular}
  \end{ruledtabular}
  \label{tab:1}
\end{table*}

Next, we investigate the beam intensity and chromaticity dependences by scanning the protons per bunch $N_B=1.0\times10^{12}, 2.4\times10^{12}, 4.2\times10^{12}$, and $8.6\times10^{12}$~ppb; and the horizontal chromaticities $\xi_x=-1.5,-7.9$, and $-12.5$;
with the fixed vertical chromaticity $\xi_y=-15$, the betatron tune $(\nu_{\beta,x},\nu_{\beta,y})=(21.35,21.41)$, and the rf voltage $\VRF=263$~kV equivalent to the synchrotron period $N_s=410$~turns.
For the above two conditions, the measured bunching factor at the injection time was $B_f=0.017$ for the beam intensity $N_B\leq4.2\times10^{12}$~ppb, and $B_f=0.022$ for the beam intensity $N_B=8.6\times10^{12}$~ppb.

Alternatively, we scan the beam intensities $N_B=1.3\times10^{11}, 2.5\times10^{11}, 5.0\times10^{11}, 1.0\times10^{12}, 2.4\times10^{12}$, and $4.2\times10^{12}$~ppb; and the chromaticities $(\xi_x,\xi_y)=(-5.7,-6.2)$ and $(-3.3,-4.4)$;
with the fixed betatron tune $(\nu_{\beta,x},\nu_{\beta,y})=(22.30,20.81)$ and the rf voltage $\VRF=155$~kV, corresponding to the synchrotron period $N_s=540$~turns.
For this condition, the bunching factor at the injection timing was $B_f=0.022$ for all beam intensities.
Tables \ref{tab:2} and \ref{tab:3} summarize the number of shots for the analysis, respectively.

\begin{table*}[!h]
  \caption{Number of shots for the analysis of the beam intensity and the chromaticity dependences with the fixed betatron tune $(\nu_{\beta,x},\nu_{\beta,y})=(21.35,21.41)$, and the rf voltage $\VRF=263$~kV.}
  \begin{ruledtabular}
  \centering
  \small
  \begin{tabular}{c  c  c  c  c}
    $N_B$(ppb)$/10^{12}$ & $~~1.0~~$ & $~~2.4~~$ & $~~4.2~~$ & $~~8.6~~$ \\ \hline
    $\xi_x=-1.5$         & 30    & 30    & 30    & $\cdots$ \\ 
    $\xi_x=-7.9$         & 30    & 30    & 30    & 10    \\ 
    $\xi_x=-12.5$        & 30    & 30    & 30    & $\cdots$ \\ 
  \end{tabular}
  \end{ruledtabular}
  \label{tab:2}
\end{table*}

\begin{table*}[!h]
  \caption{Number of shots for the analysis of the beam intensity and the chromaticity dependences with the fixed betatron tune $(\nu_{\beta,x},\nu_{\beta,y})=(22.30,20.81)$, and the rf voltage $\VRF=155$~kV.}
  \begin{ruledtabular}
  \centering
  \small
  \begin{tabular}{c  c  c  c  c  c  c }    
    $N_B$(ppb)$/10^{12}$ & $~~0.13~~$ & $~~0.26~~$ & $~~0.52~~$ & $~~1.0~~$ & $~~2.4~~$ & $~~4.2~~$ \\ \hline
    $\xi_x=-3.3$         & 3      & 3      & 3      & 3     & 3     & $\cdots$ \\ 
    $\xi_x=-5.7$         & 3      & 3      & 3      & 3     & 3     & 3     \\  
  \end{tabular}
  \end{ruledtabular}
  \label{tab:3}
\end{table*}

\subsection{Typical results of the measured intrabunch motion}\label{10.2}
\begin{figure}[!h]
  \begin{tabular}{ccc}
    \begin{minipage}[t]{0.3\hsize}
      \centering
      \includegraphics[width=2.0in]{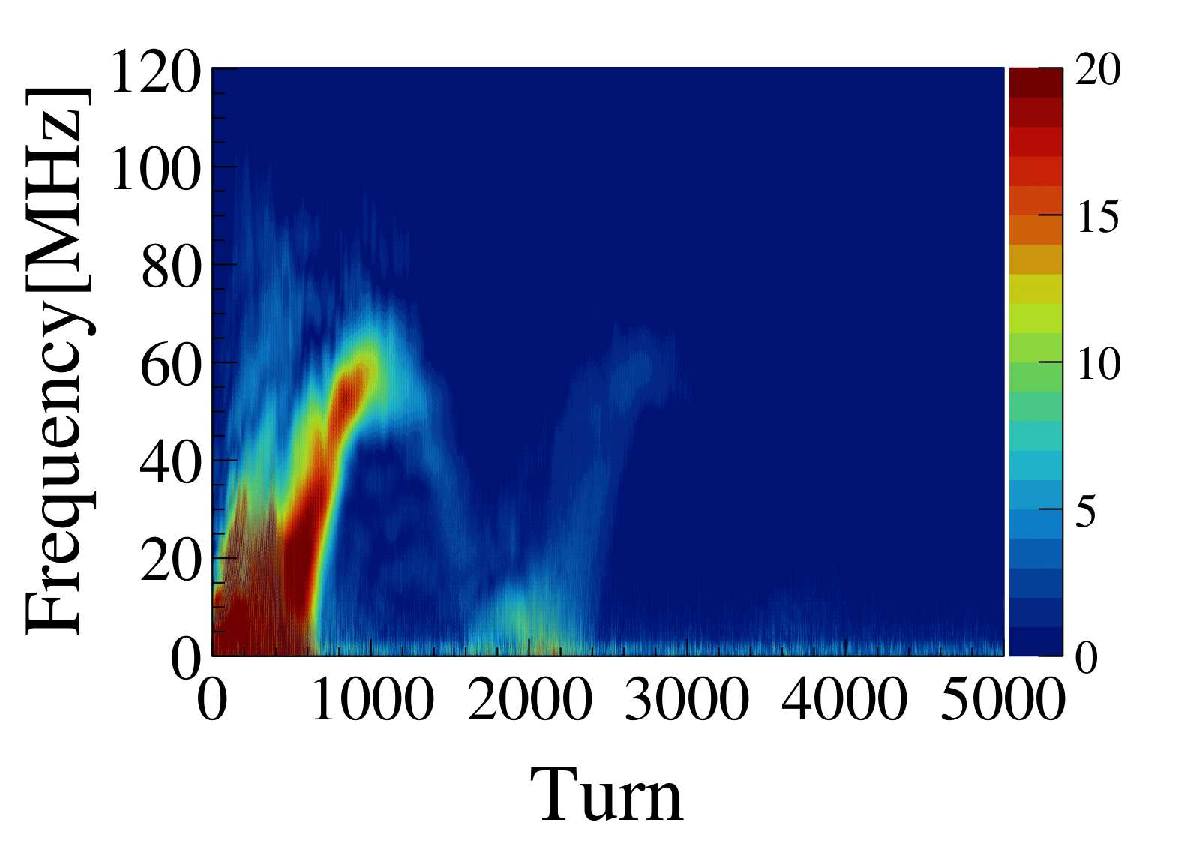}
      \subcaption{}
      \label{fig:13a}
    \end{minipage} &
    \begin{minipage}[t]{0.3\hsize}
      \centering
      \includegraphics[width=2.0in]{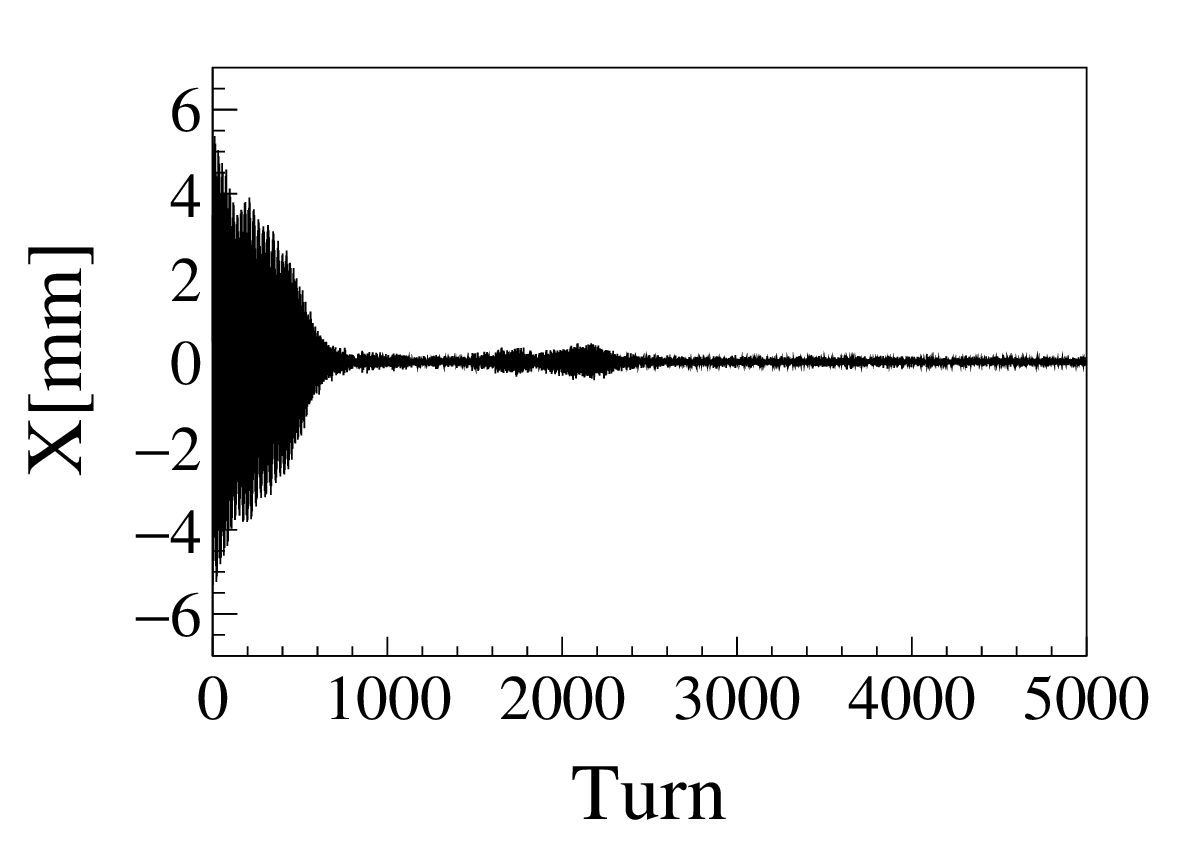}
      \subcaption{}
      \label{fig:13b}
      \end{minipage} &
    \begin{minipage}[t]{0.3\hsize}
      \centering
      \includegraphics[width=2.0in]{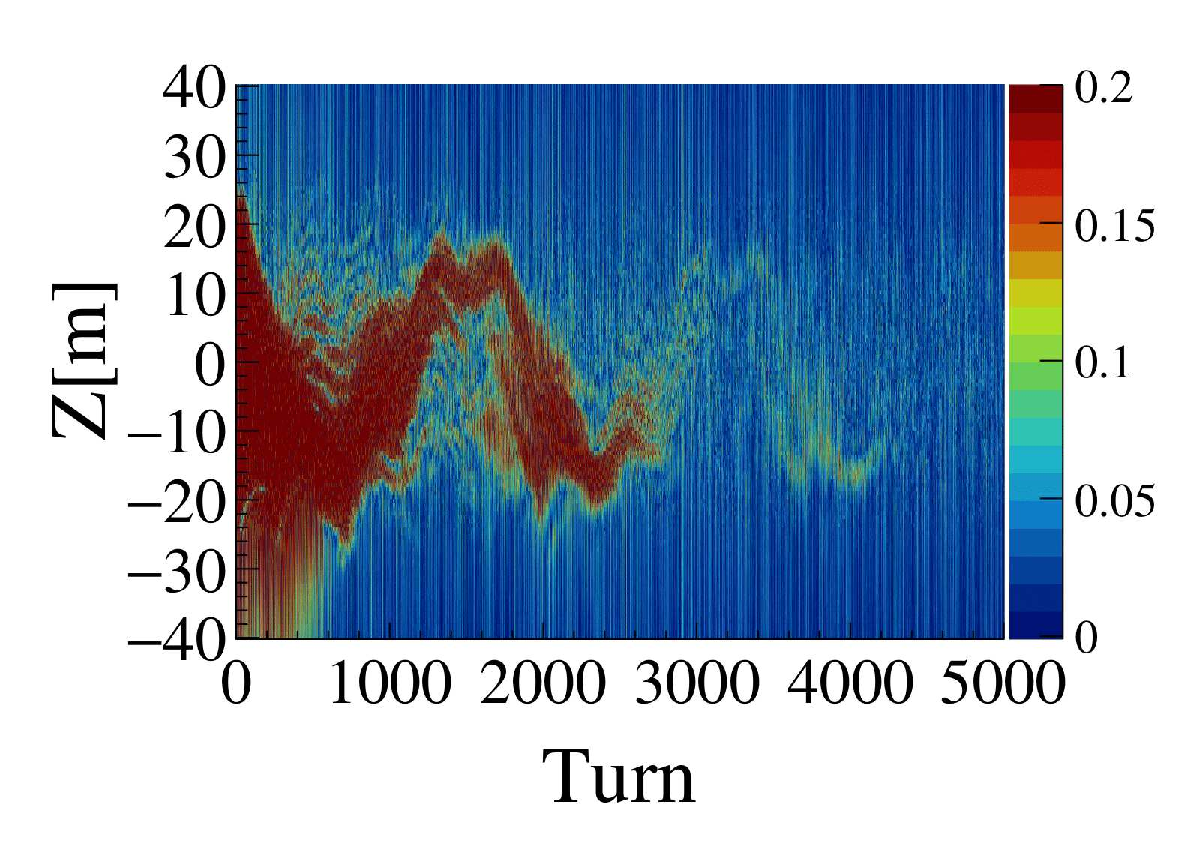}
      \subcaption{}
      \label{fig:13c}
    \end{minipage}
  \end{tabular}
  \caption{Measured transverse bunch motion in the time domain for the chromaticity $\xi_x=-7.9$ and the beam intensity $N_B=8.6\times10^{12}$~ppb and the rf voltage $\VRF=263$~kV. (a) Frequency component of the dipole moment in the bunch, (b) average of the beam position $\overline{x_n}$, and (c) dipole moment $|\Delta(z)|$ in the bunch.}
  \label{fig:13}
\end{figure}

Figure \ref{fig:13} shows a typical result of the measured intrabunch motion in the time domain for a beam intensity of ($N_B=8.6\times10^{12}$~ppb) with a chromaticity of $\xi_x=-7.9$.
As in the case of the tracking simulation, the frequency components of the dipole moment $|\Delta(z)|$ in Figure \ref{fig:13a} reveals the alternating with high-frequency and low-frequency components along the turn.
Figure \ref{fig:13b} shows the oscillatory behavior of the horizontal centroid of the entire bunch, and the level of amplitude represents the decoherence and recoherence motion of the particles that make up the beam.
Here, the definitions of $\fmax$, $N_f$, and $\NAM$ are the same as those used in the tracking simulation.
From the data in Fig.\ref{fig:13}, we determine the maximum bunch frequency to be $\fmax=53.4$~MHz, the bunch oscillation period to be $N_f=850$~turns, the recoherence period to be $\NAM=1700$~turns, and the normalized recoherence period to be $\NAM/N_s=4.1$.
From Fig.\ref{fig:13c}, it can be seen that the dipole moment $|\Delta(z)|$ meanders in the longitudinal direction, and that the period is equal to $\NAM$.

\begin{figure}[!h]
  \begin{tabular}{cccc}
    \begin{minipage}[t]{0.25\hsize}
      \centering
      \includegraphics[width=1.7in]{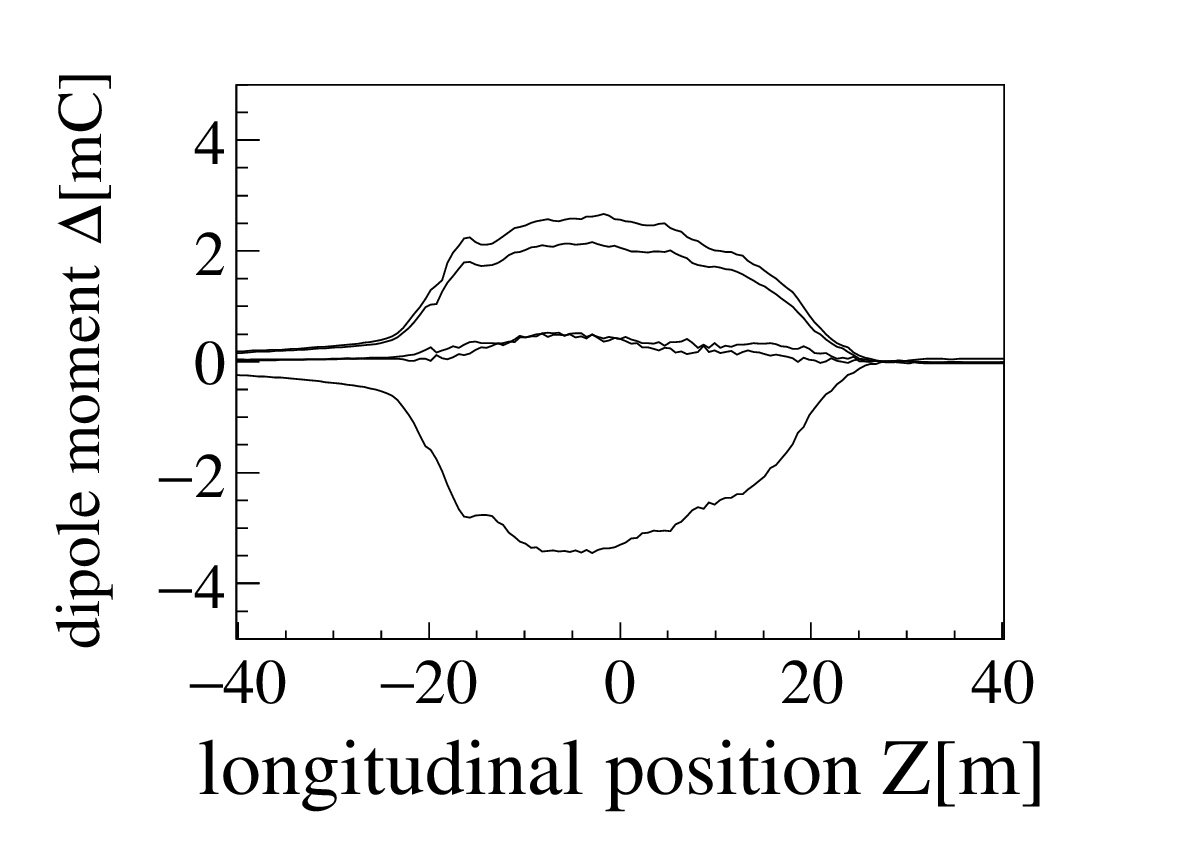}
      \subcaption{}
    \end{minipage} &
    \begin{minipage}[t]{0.25\hsize}
      \centering
      \includegraphics[width=1.7in]{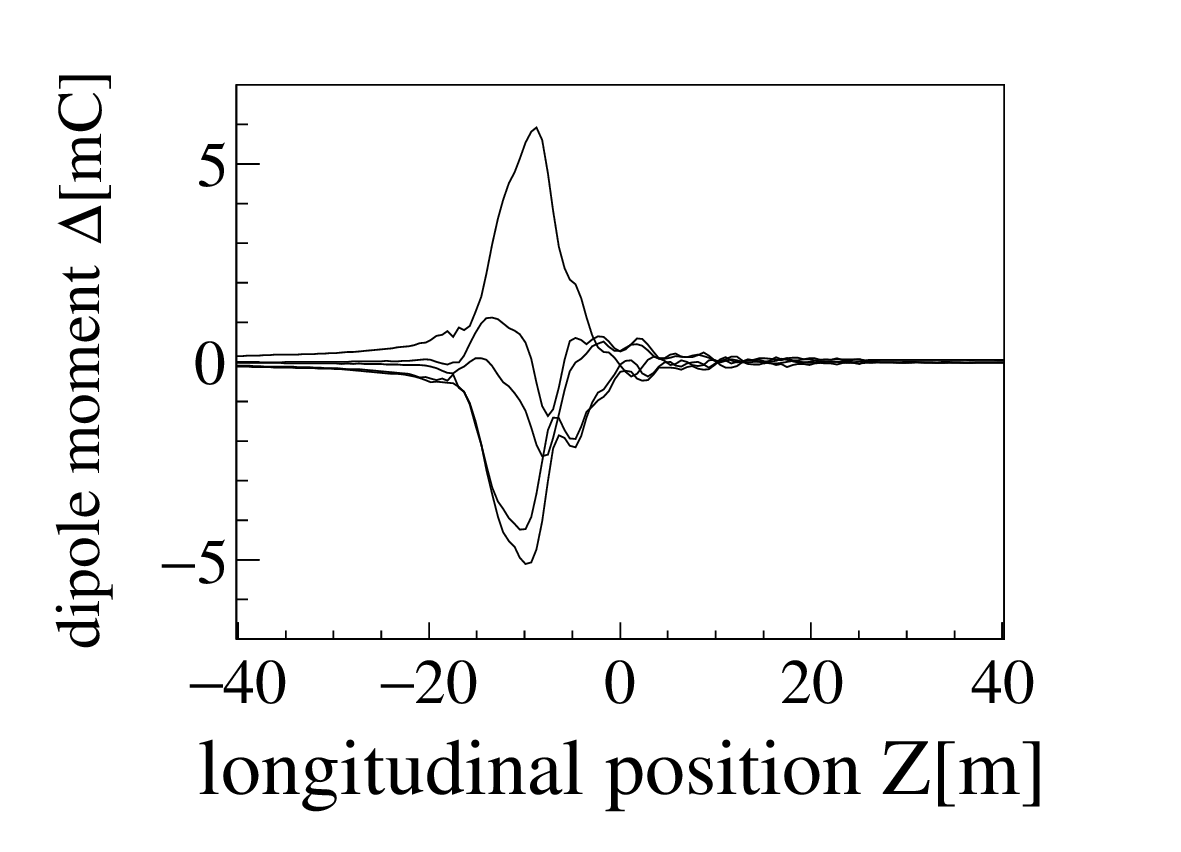}
      \subcaption{}
      \end{minipage} &
    \begin{minipage}[t]{0.25\hsize}
      \centering
      \includegraphics[width=1.7in]{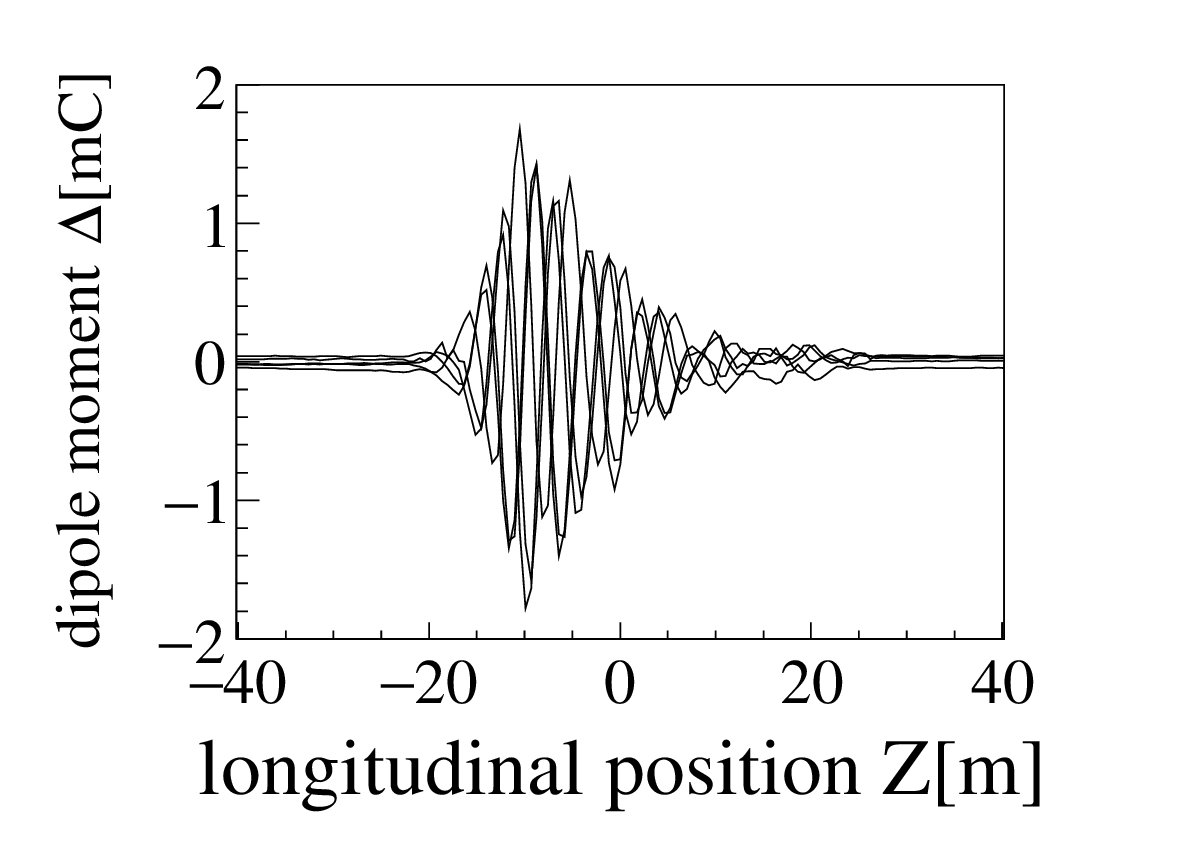}
      \subcaption{}
    \end{minipage} &
    \begin{minipage}[t]{0.25\hsize}
      \centering
      \includegraphics[width=1.7in]{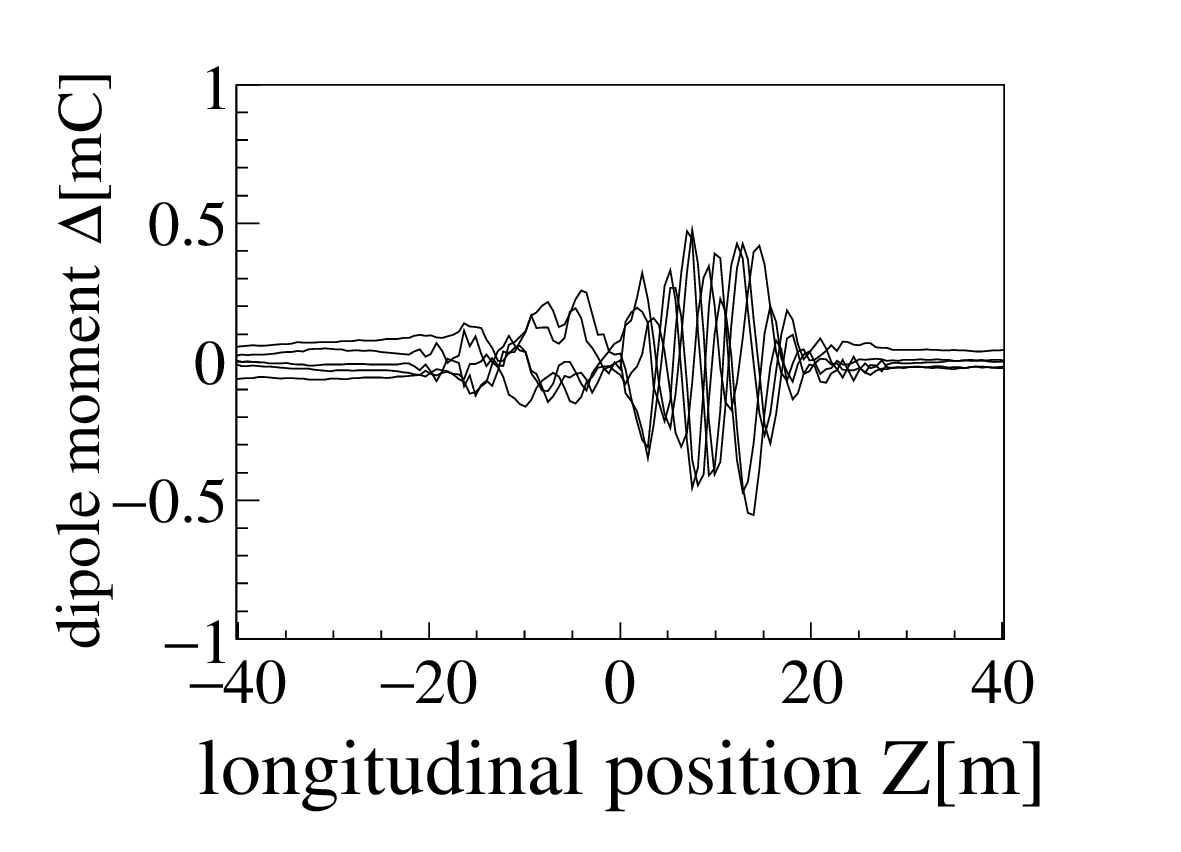}
      \subcaption{}
    \end{minipage} \\

    \begin{minipage}[t]{0.25\hsize}
      \centering
      \includegraphics[width=1.7in]{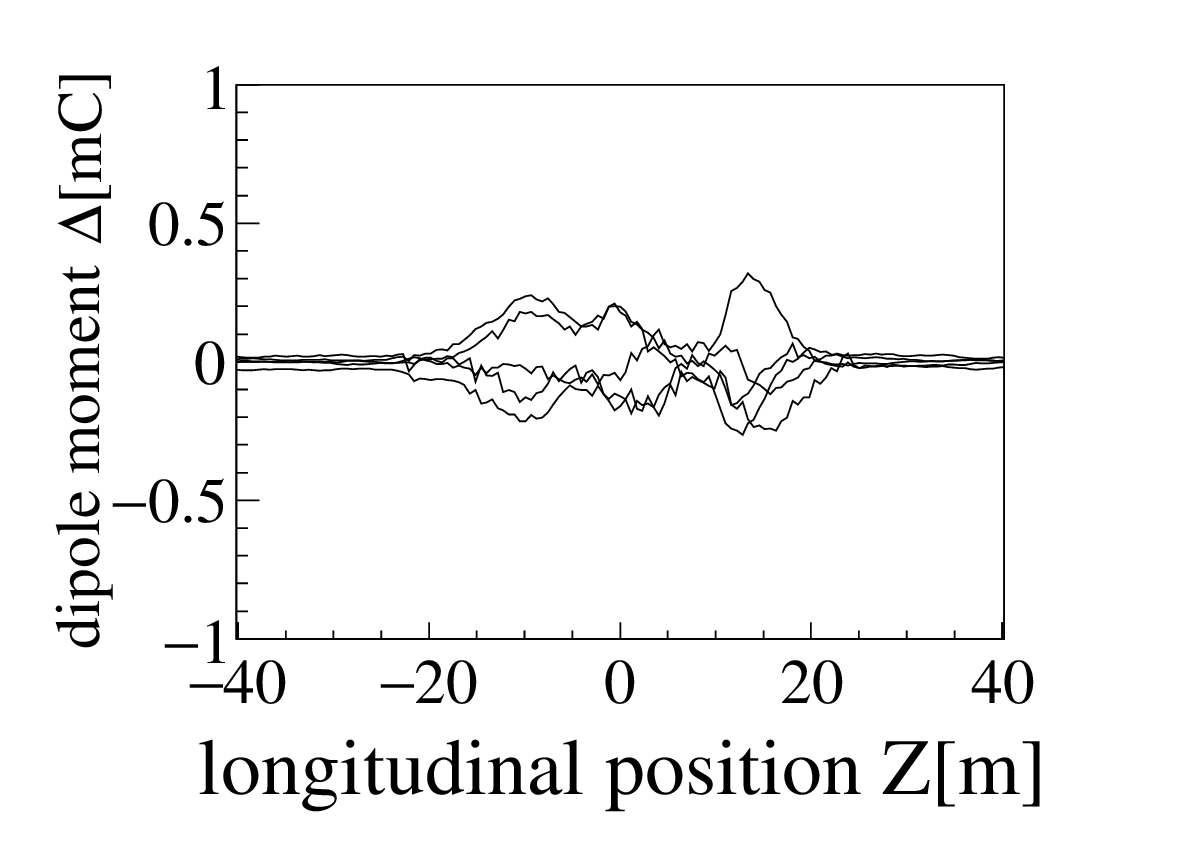}
      \subcaption{}
    \end{minipage} &
    \begin{minipage}[t]{0.25\hsize}
      \centering
      \includegraphics[width=1.7in]{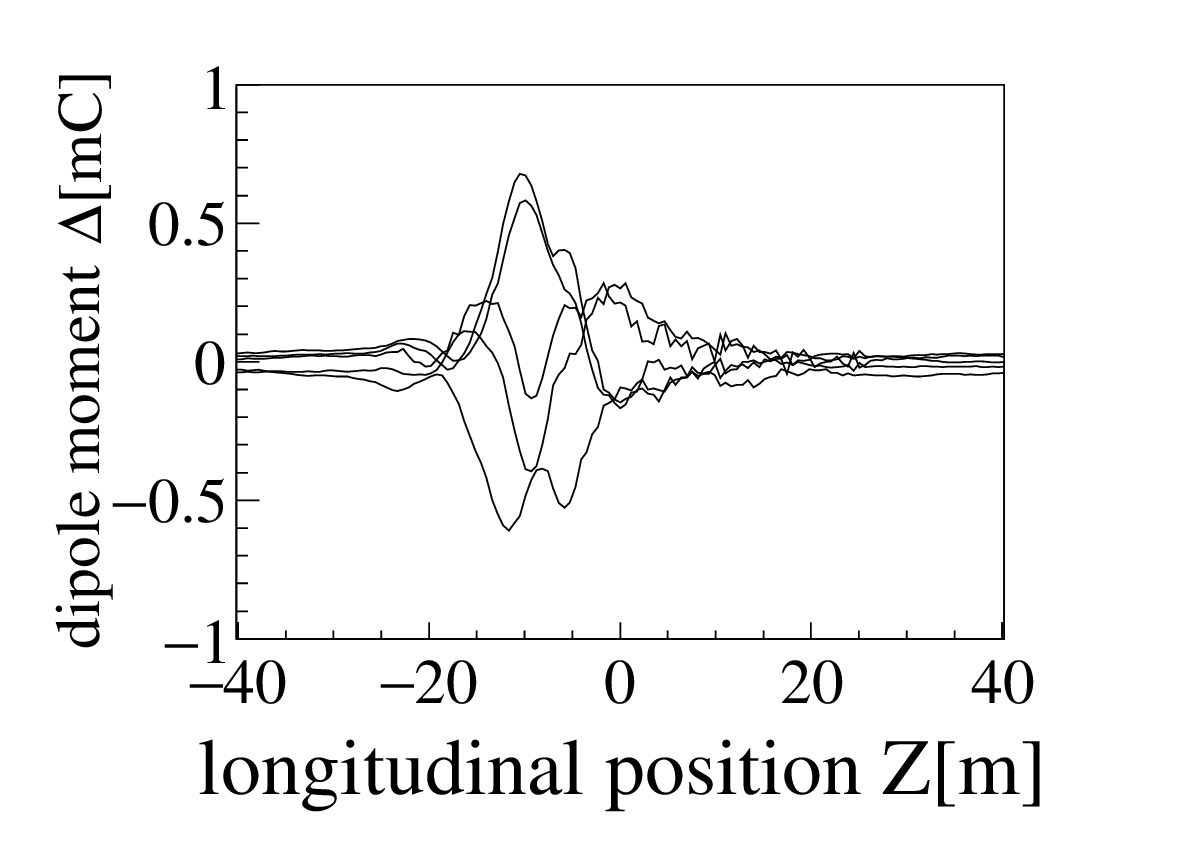}
      \subcaption{}
    \end{minipage} &
    \begin{minipage}[t]{0.25\hsize}
      \centering
      \includegraphics[width=1.7in]{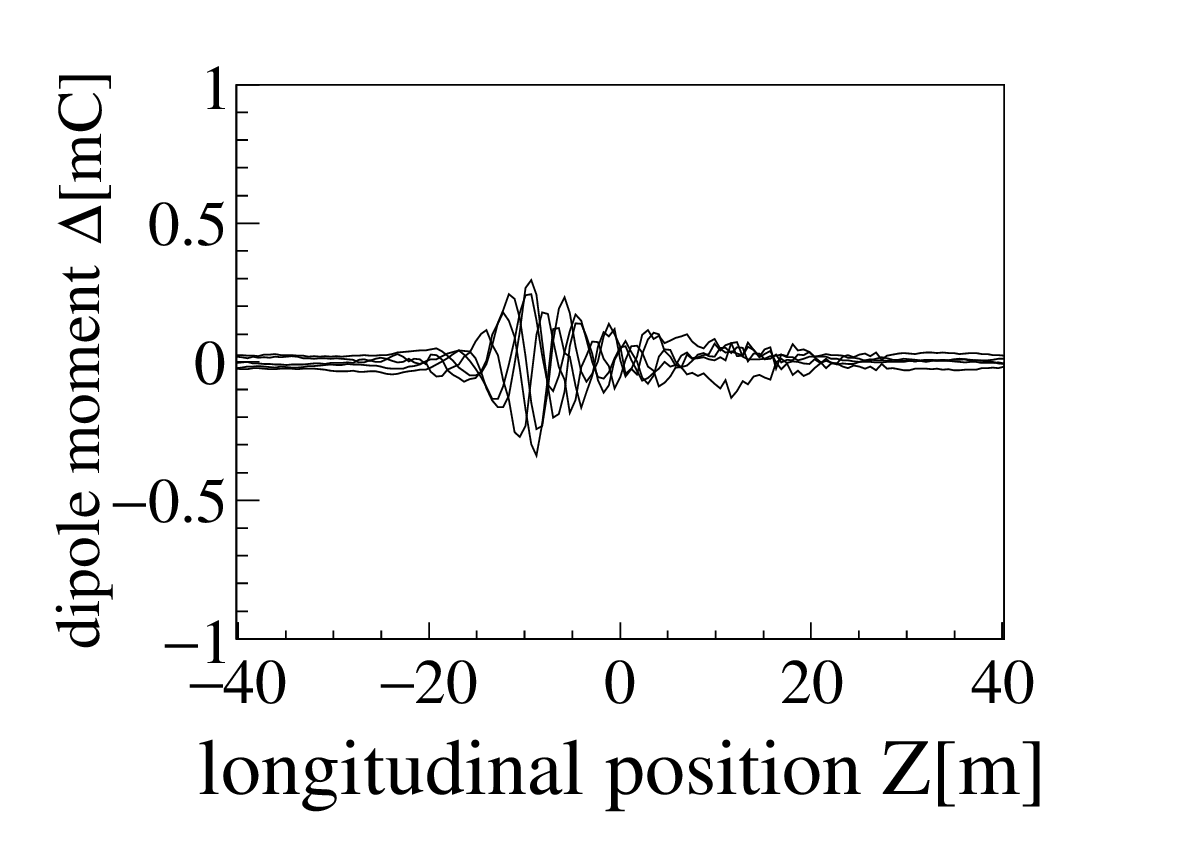}
      \subcaption{}
    \end{minipage} &
    \begin{minipage}[t]{0.25\hsize}
      \centering
      \includegraphics[width=1.7in]{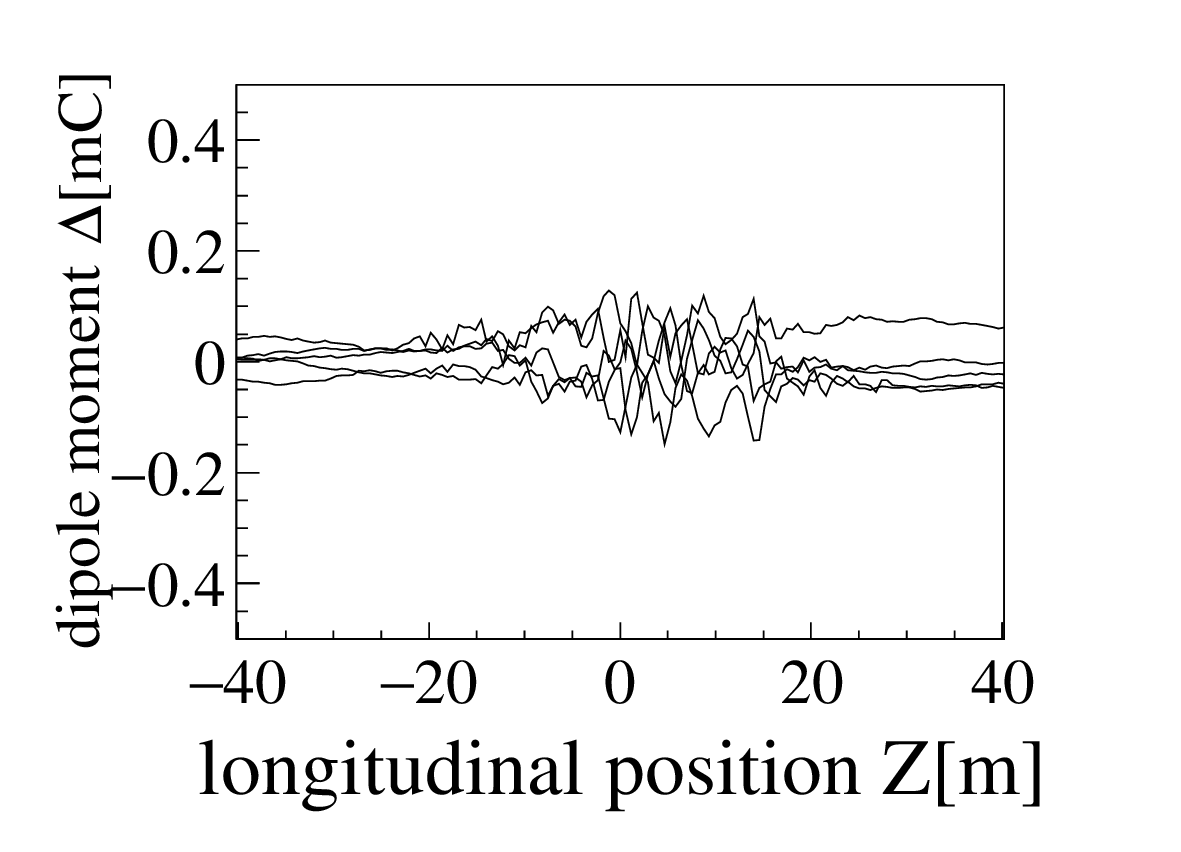}
      \subcaption{}
    \end{minipage}
  \end{tabular}
  \caption{Measured dipole moment $\Delta(z)$ in the bunch at beam measurement for the chromaticity $\xi_x=-7.9$ and the beam intensity $N_B=8.6\times10^{12}$~ppb. (a) 0 turn, (b) $\NAM/4=425$~turns, (c) $\NAM/2=850$~turns, (d) $3\NAM/4=1275$~turns, (e) $\NAM=1700$~turns, (f) $5\NAM/4=2125$~turns, (g) $3\NAM/2=2550$~turns, and (h) $7\NAM/4=2975$~turns.}
  \label{fig:14}
\end{figure}

Figure \ref{fig:14} shows the intrabunch motion (the dipole moment) observed at the beam measurement by superimposing the trajectories of five consecutive turns at intervals of $\NAM/4$.
In the periodic $\NAM$, the oscillation frequency varies between high and low frequencies as mentioned in Fig. \ref{fig:13a}, and the dipole component oscillates with $\NAM$ period between the head and tail as mentioned in Fig. \ref{fig:13c}.

\begin{figure}[!h]
  \begin{tabular}{ccc}
    \begin{minipage}[t]{0.3\hsize}
      \centering
      \includegraphics[width=2.0in]{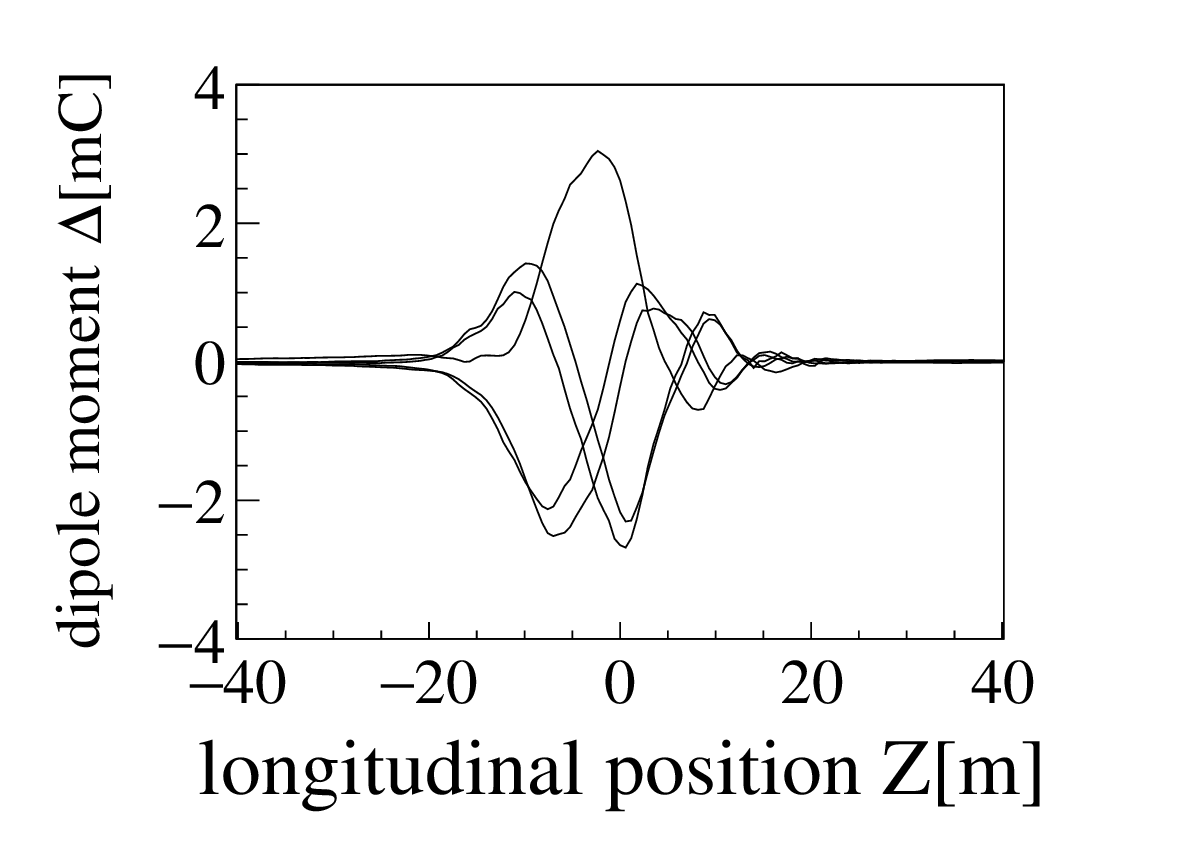}
      \subcaption{}
    \end{minipage} &
    \begin{minipage}[t]{0.3\hsize}
      \centering
      \includegraphics[width=2.0in]{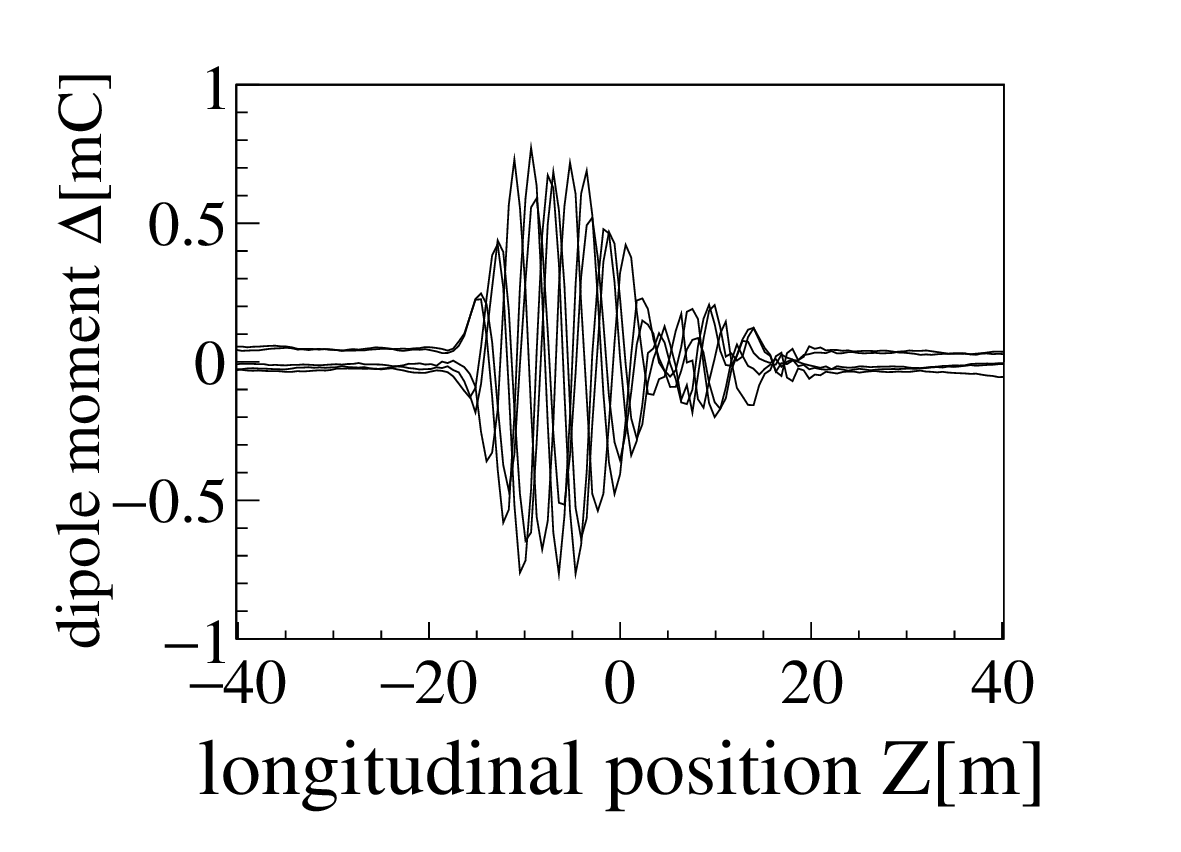}
      \subcaption{}
    \end{minipage} &
    \begin{minipage}[t]{0.3\hsize}
      \centering
      \includegraphics[width=2.0in]{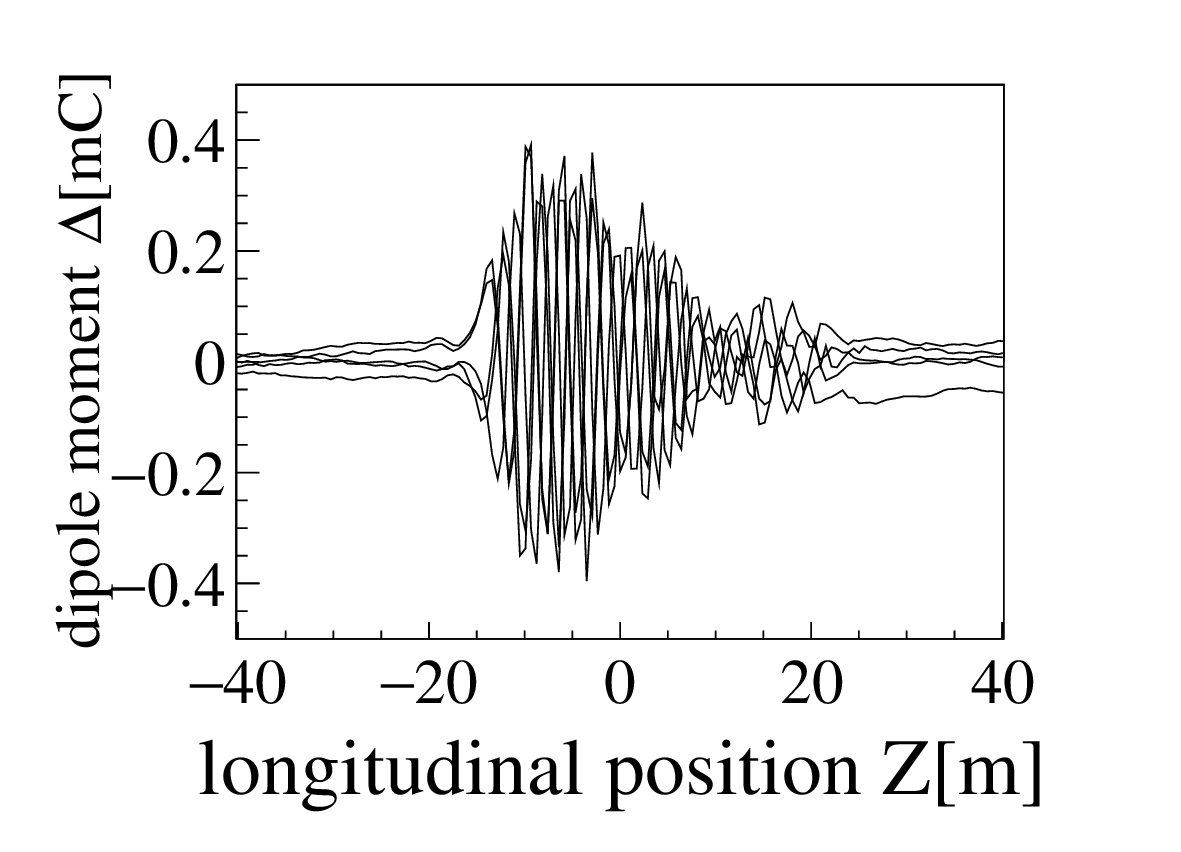}
      \subcaption{}
    \end{minipage} 
  \end{tabular}
  \caption{Measured dipole moment $\Delta(z)$ in the bunch for the beam intensity $N_B=4.2\times10^{12}$~ppb and $\NAM/2=N_f$~turn at beam measurement. (a) $\xi_x=-1.5$, (b) $\xi_x=-7.9$, and (c) $\xi_x=-12.5$.}
  \label{fig:15}
\end{figure}

Figure \ref{fig:15} also shows the intra-beam motion (the dipole moment) at $\NAM/2=N_f$ with various chromaticities.
We can directly observe the higher chromaticity excites the higher head-tail modes, as mentioned in Fig. \ref{fig:4}.

\subsection{Synchrotron period dependence}\label{10.3}
First, the dependence of intrabunch motion on the synchrotron period $N_s$ is investigated by varying the rf voltage.

\subsubsection{Simulation predictions and measurement results}\label{10.3.1}
\begin{figure}[!h]
  \begin{tabular}{ccc}
    \begin{minipage}[t]{0.3\hsize}
      \centering
      \includegraphics[width=2.0in]{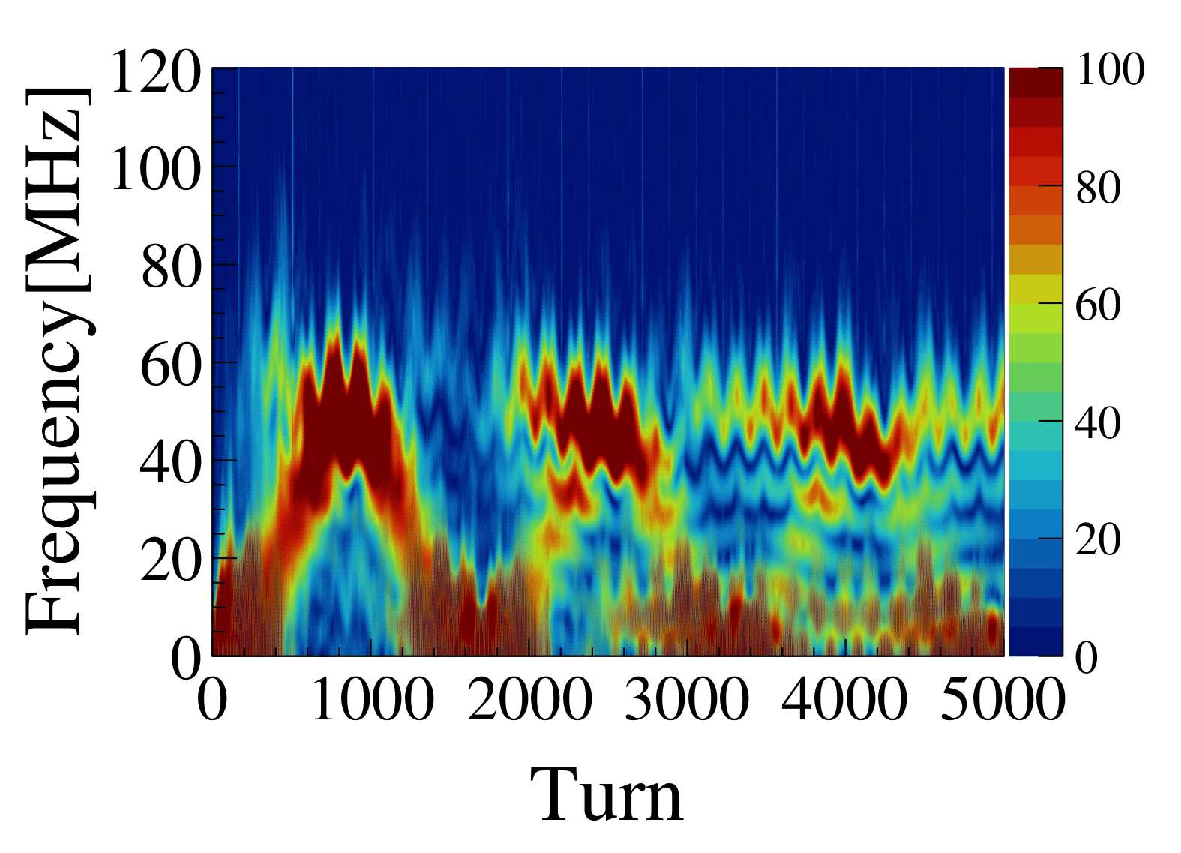}
      \subcaption{}
    \end{minipage} &
    \begin{minipage}[t]{0.3\hsize}
      \centering
      \includegraphics[width=2.0in]{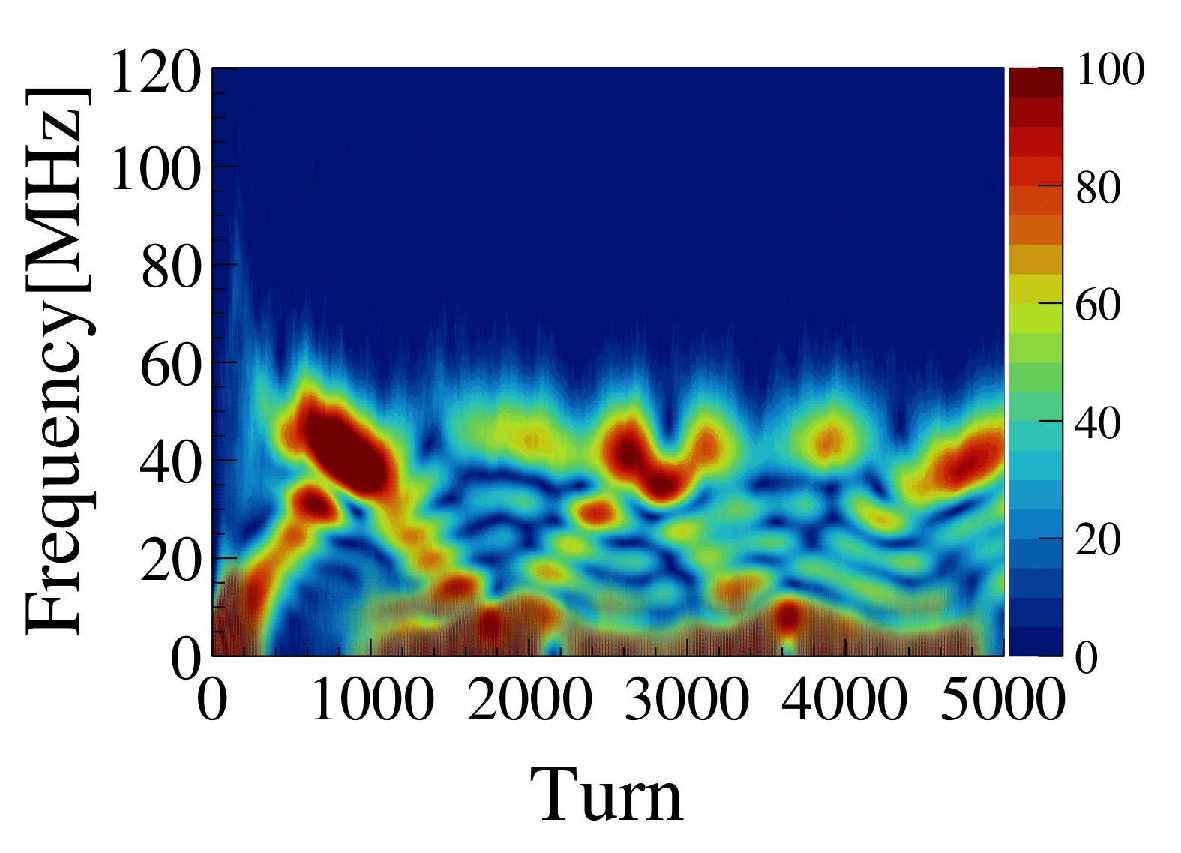}
      \subcaption{F.}
    \end{minipage} &
    \begin{minipage}[t]{0.3\hsize}
      \centering
      \includegraphics[width=2.0in]{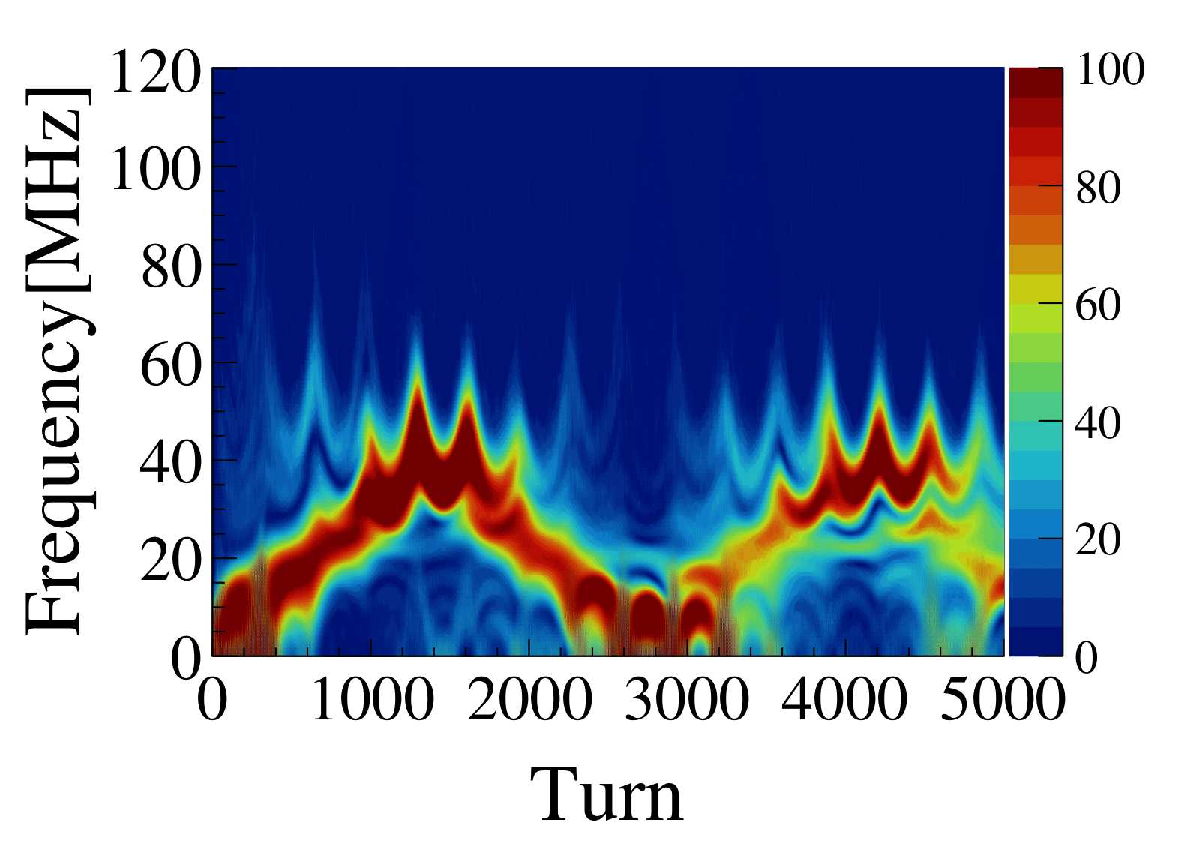}
      \subcaption{}
    \end{minipage} \\

    \begin{minipage}[t]{0.3\hsize}
      \centering
      \includegraphics[width=2.0in]{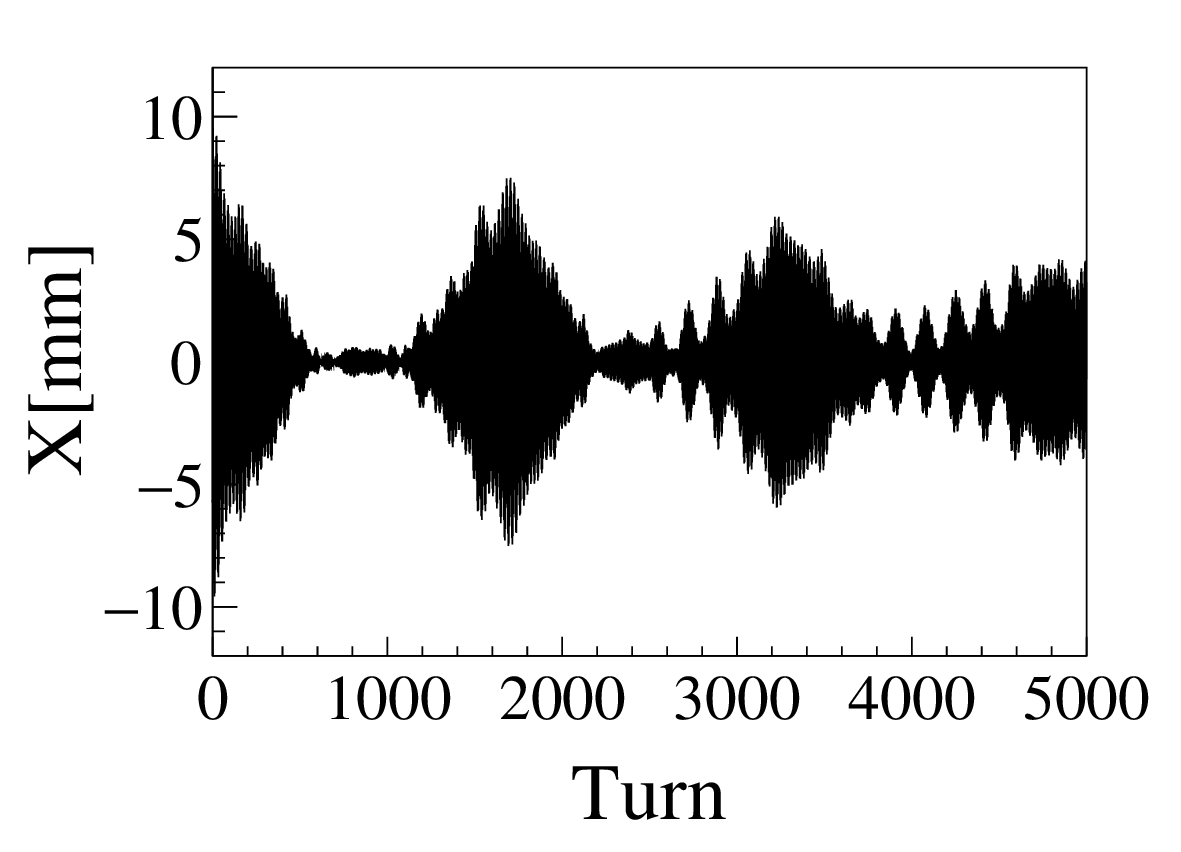}
      \subcaption{}
    \end{minipage} &
    \begin{minipage}[t]{0.3\hsize}
      \centering
      \includegraphics[width=2.0in]{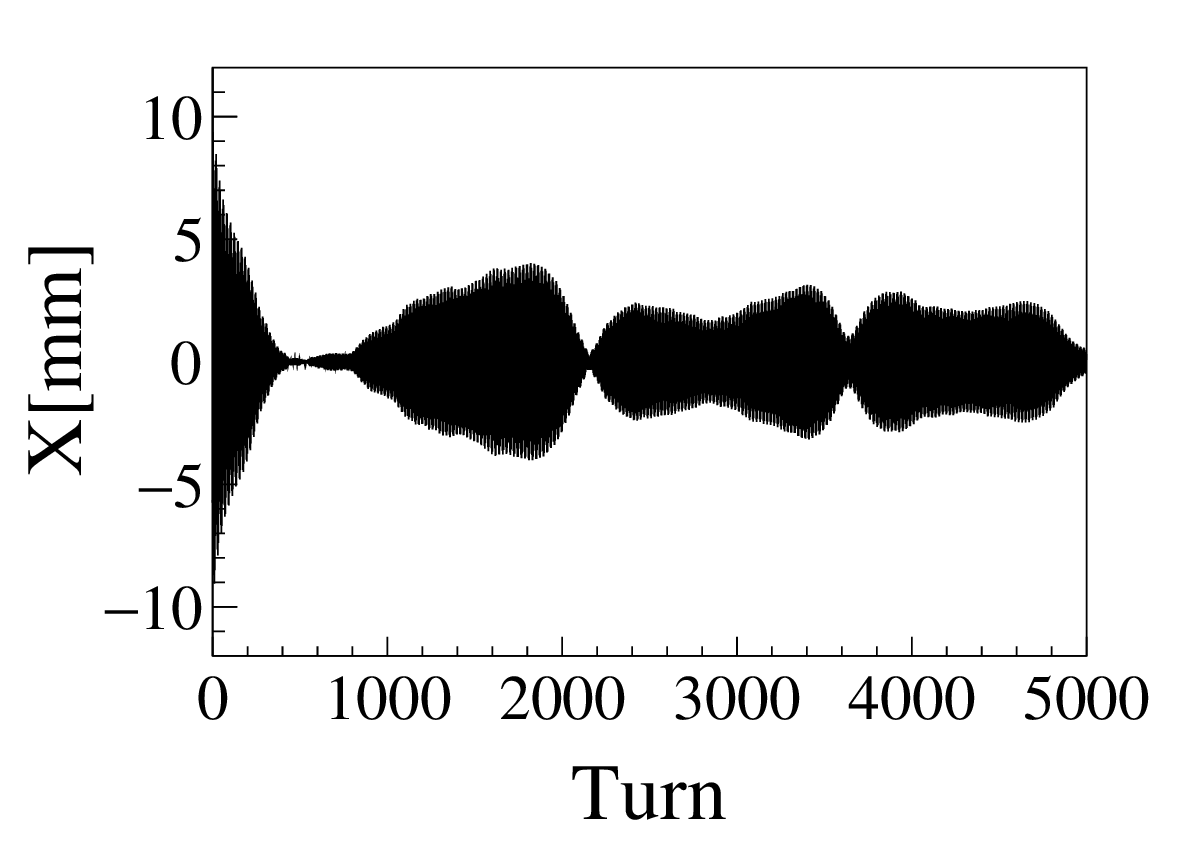}
      \subcaption{}
    \end{minipage} &
    \begin{minipage}[t]{0.3\hsize}
      \centering
      \includegraphics[width=2.0in]{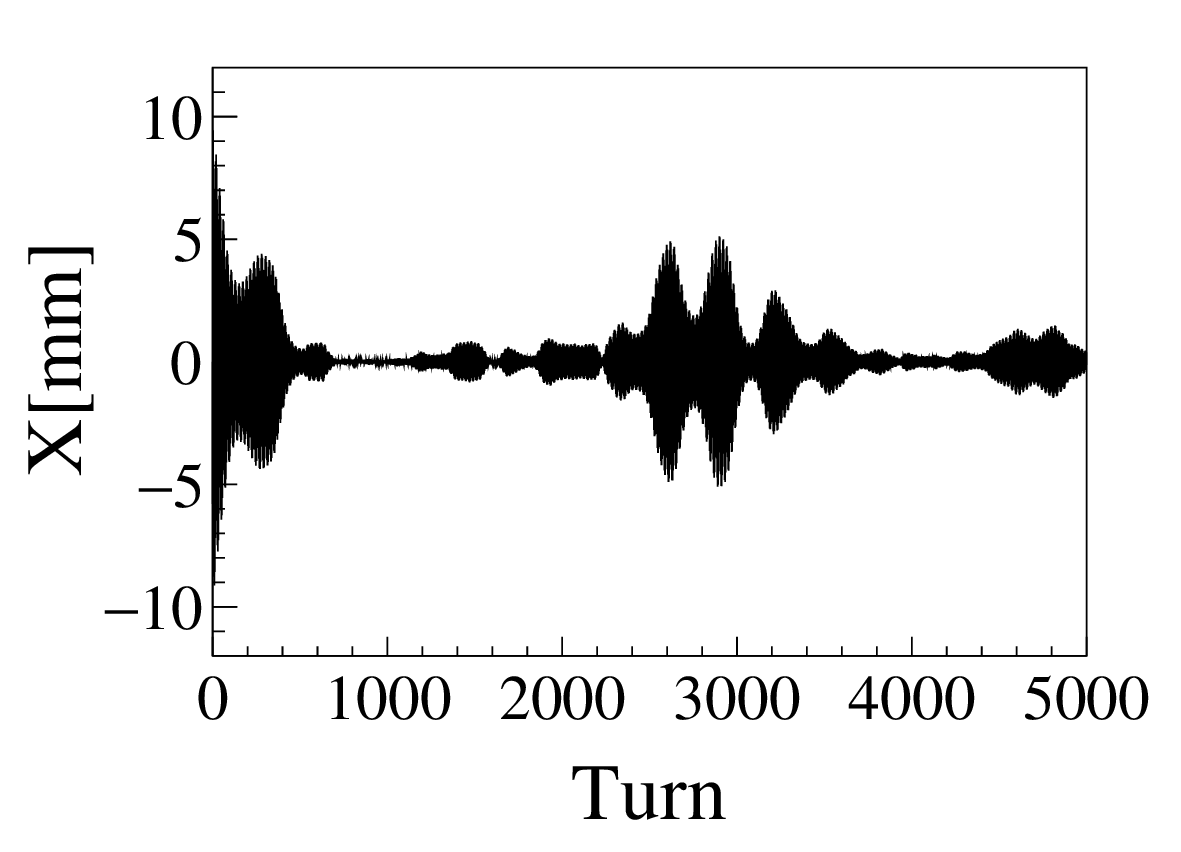}
      \subcaption{}
    \end{minipage} \\

    \begin{minipage}[t]{0.3\hsize}
      \centering
      \includegraphics[width=2.0in]{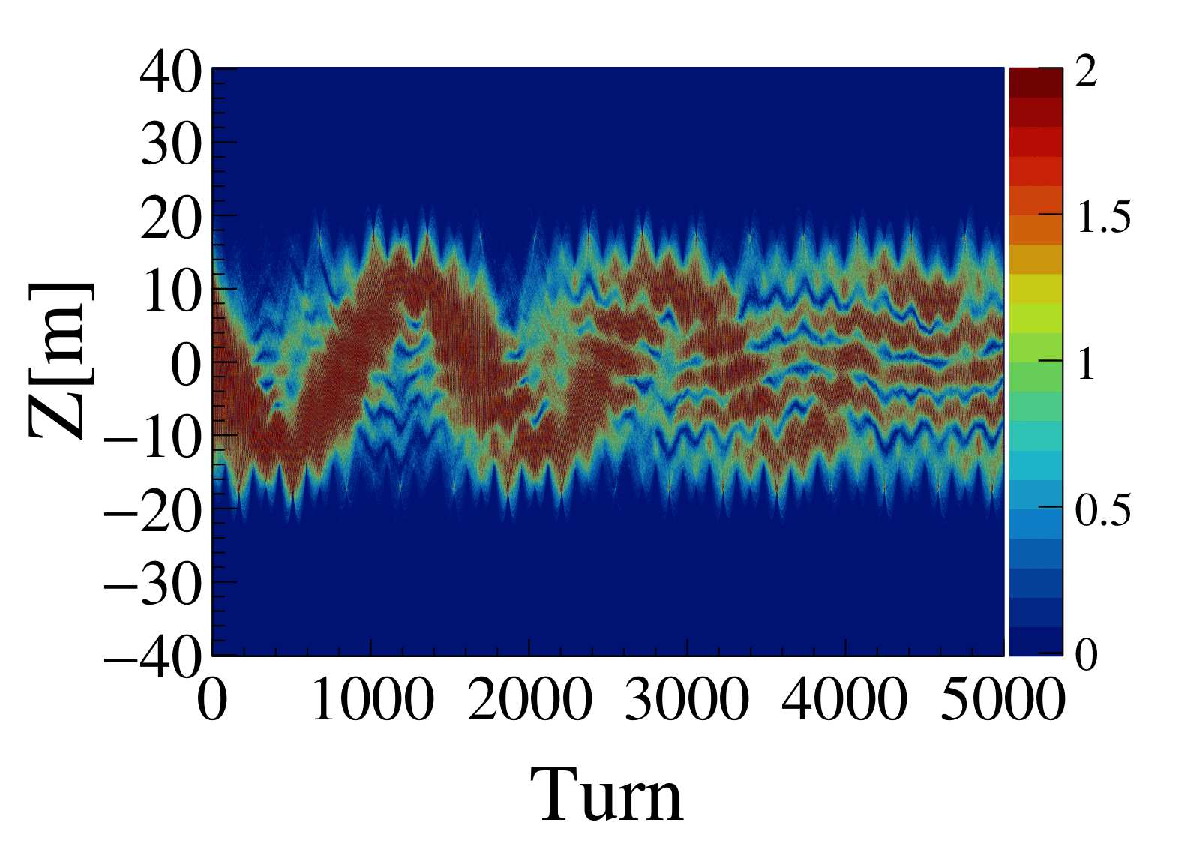}
      \subcaption{}
    \end{minipage} &
    \begin{minipage}[t]{0.3\hsize}
      \centering
      \includegraphics[width=2.0in]{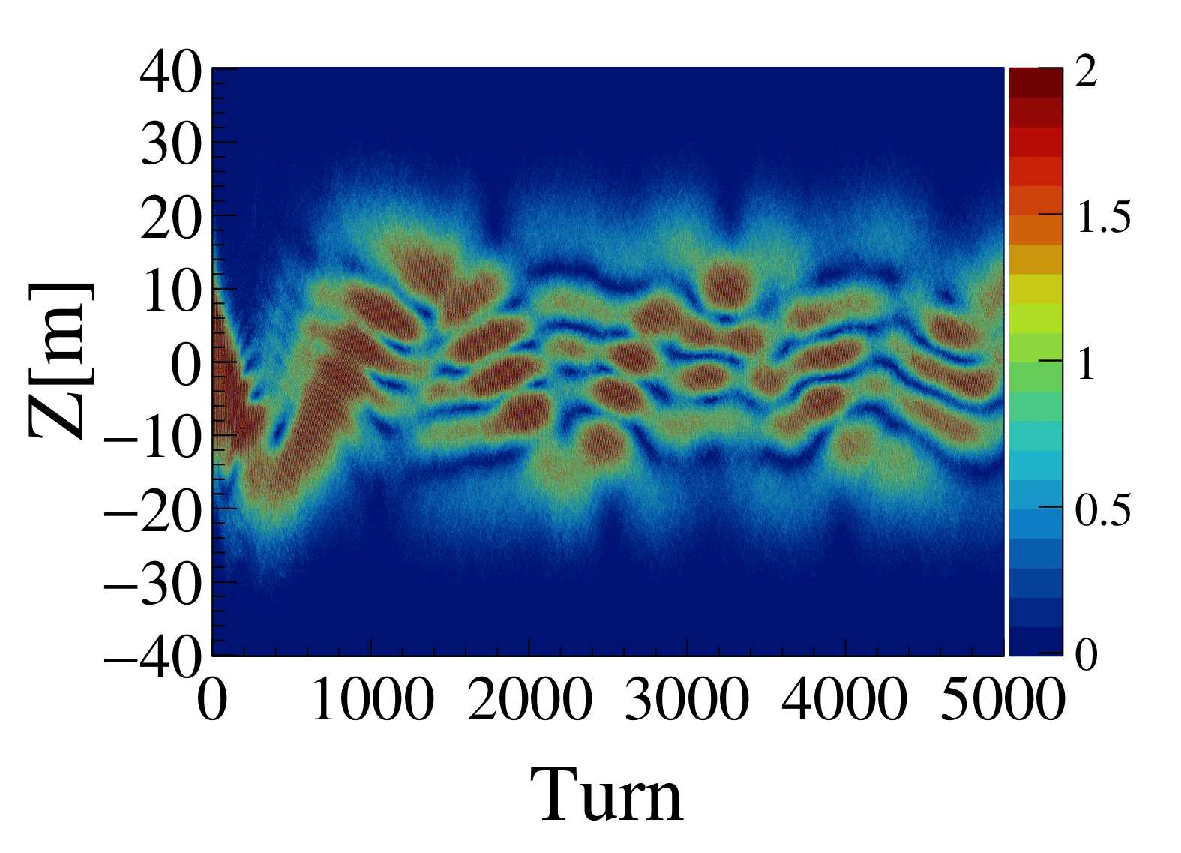}
      \subcaption{}
    \end{minipage} &
    \begin{minipage}[t]{0.3\hsize}
      \centering
      \includegraphics[width=2.0in]{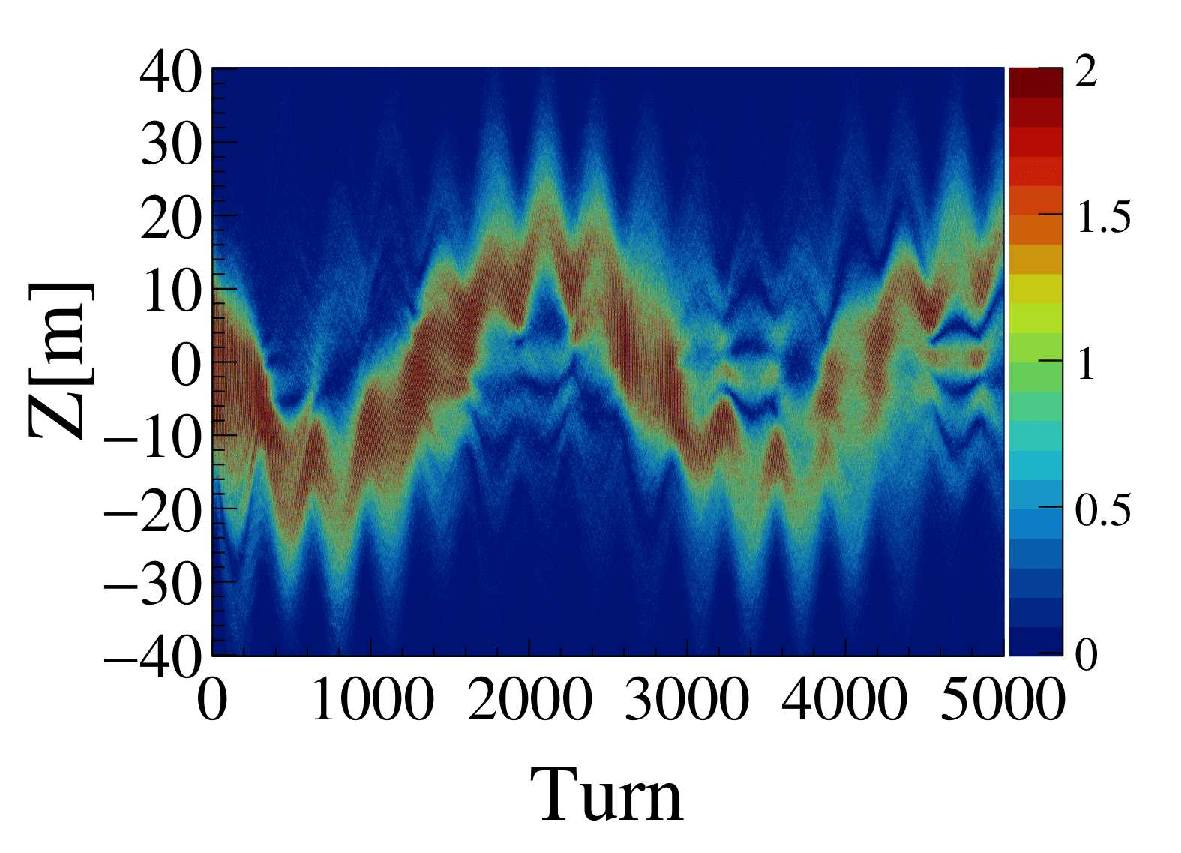}
      \subcaption{}
    \end{minipage}
  \end{tabular}
  \caption{Simulations of the transverse bunch motion in the time domain for the rf voltages: $\VRF=110,263$, and $400$~kV with the chromaticity $\xi_x=-6.5$ and the beam intensity $N_B=4.2\times10^{12}$~ppb. (a) Frequency component of the dipole moment for $\VRF=400$~kV, $\fmax=43.5$~MHz, $\NAM=1700$, $\NAM /N_s=5.0$, (b) frequency component of the dipole moment for $\VRF=263$~kV, $\fmax=39.6$~MHz, $\NAM=1720$, $\NAM /N_s=4.2$, (c) frequency component of the dipole moment for $\VRF=110$~kV, $\fmax=34.2$~MHz, $\NAM=2940$, $\NAM /N_s=4.5$, (d) average of the beam position $\overline{x_n}$ for $\VRF=400$~kV, (e) average of the beam position $\overline{x_n}$ for $\VRF=263$~kV, (f) average of the beam position $\overline{x_n}$ for $\VRF=110$~kV, (g) dipole moment $|\Delta^{(k)}\sigma_\Delta/q|$ in the bunch for $\VRF=400$~kV, (h) dipole moment $|\Delta^{(k)}\sigma_\Delta/q|$ in the bunch for $\VRF=263$~kV, and (i) dipole moment $|\Delta^{(k)}\sigma_\Delta/q|$ in the bunch for $\VRF=110$~kV.}
  \label{fig:16}
\end{figure}

Simulations were performed for various rf voltages corresponding to synchrotron periods from $N_s=340$ to 760~turns.
Specific examples for $\xi_x=-6.5$ and $N_B=4.2\times10^{12}$~ppb are shown in Fig. \ref{fig:16}.
The extracted values for $(\fmax, \NAM)$ were (43.5~MHz, 1700~turns) for $\VRF=400$~kV, (39.6~MHz, 1720~turns) for $\VRF=263$~kV, and (34.2~MHz, 2940~turns) for $\VRF=110$~kV.

\begin{figure}[!h]
  \begin{tabular}{ccc}
    \begin{minipage}[t]{0.3\hsize}
      \centering
      \includegraphics[width=2.0in]{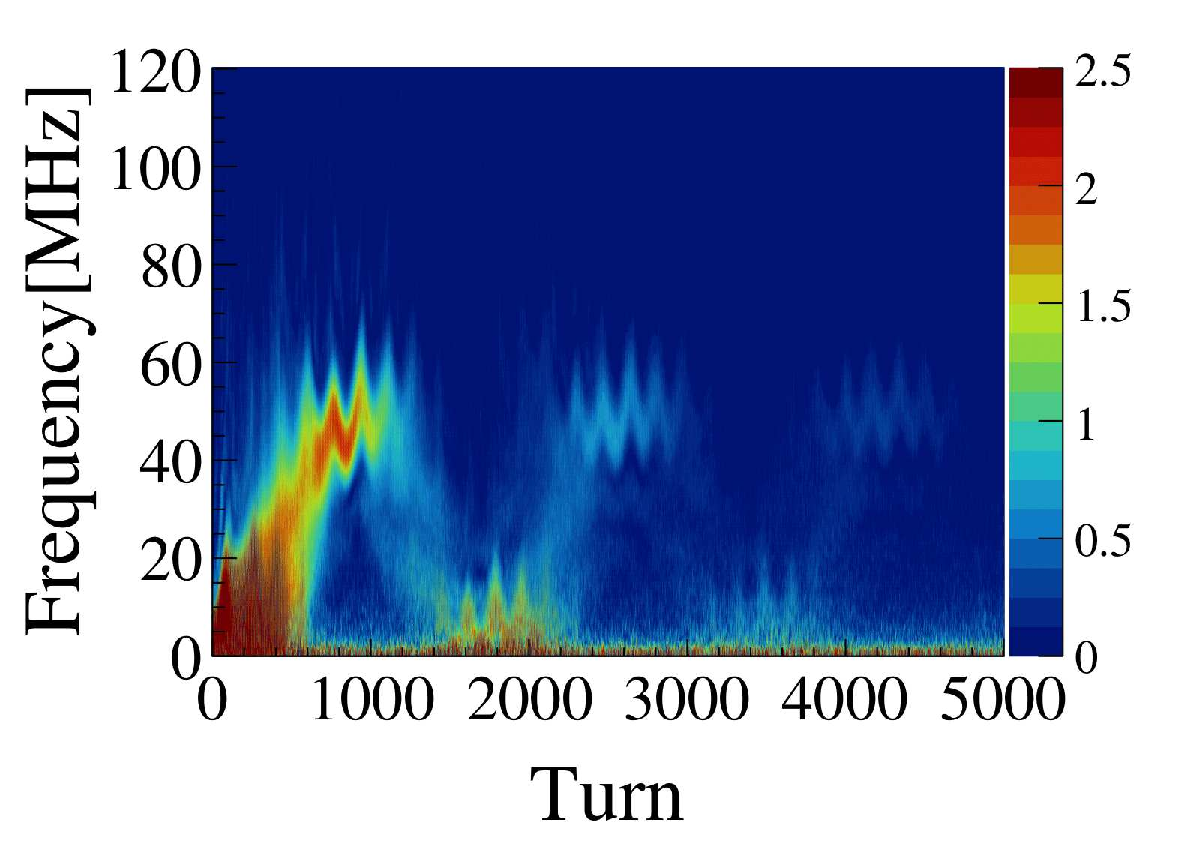}
      \subcaption{}
    \end{minipage} &
    \begin{minipage}[t]{0.3\hsize}
      \centering
      \includegraphics[width=2.0in]{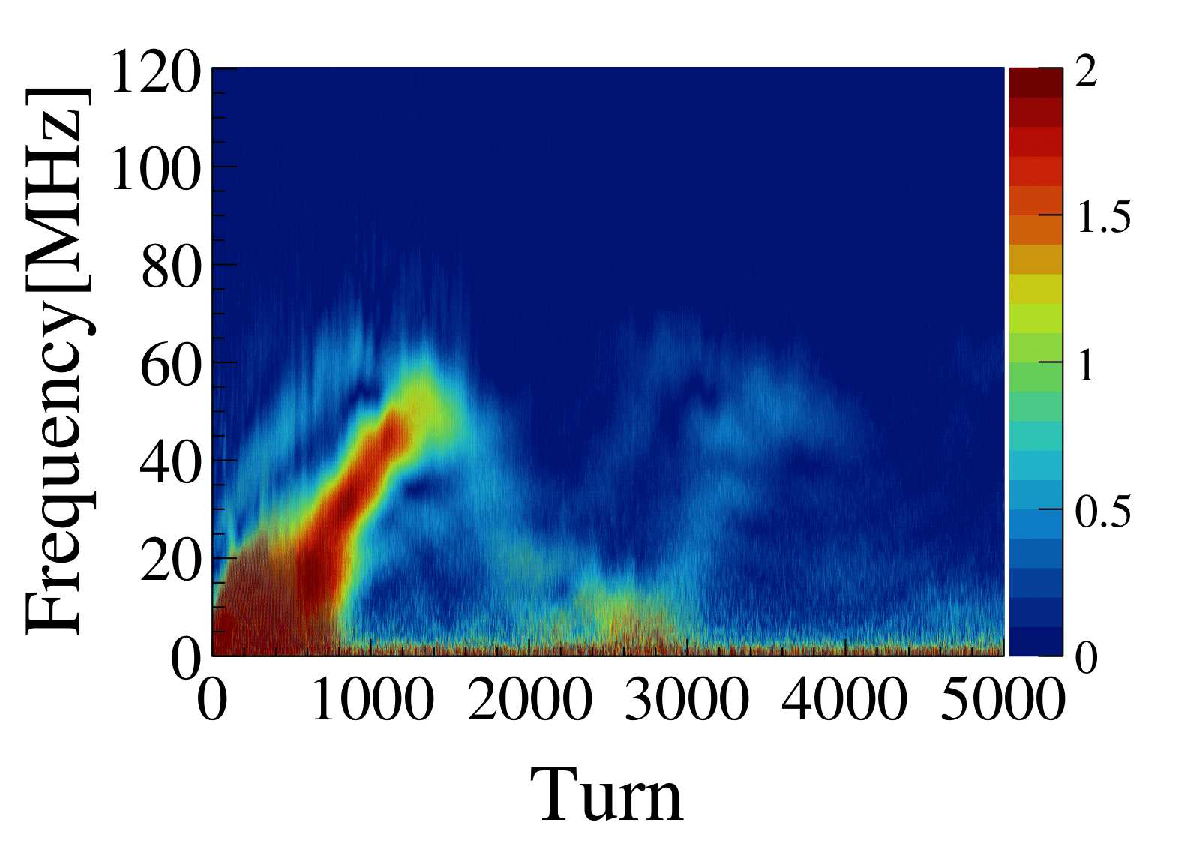}
      \subcaption{}
    \end{minipage} &
    \begin{minipage}[t]{0.3\hsize}
      \centering
      \includegraphics[width=2.0in]{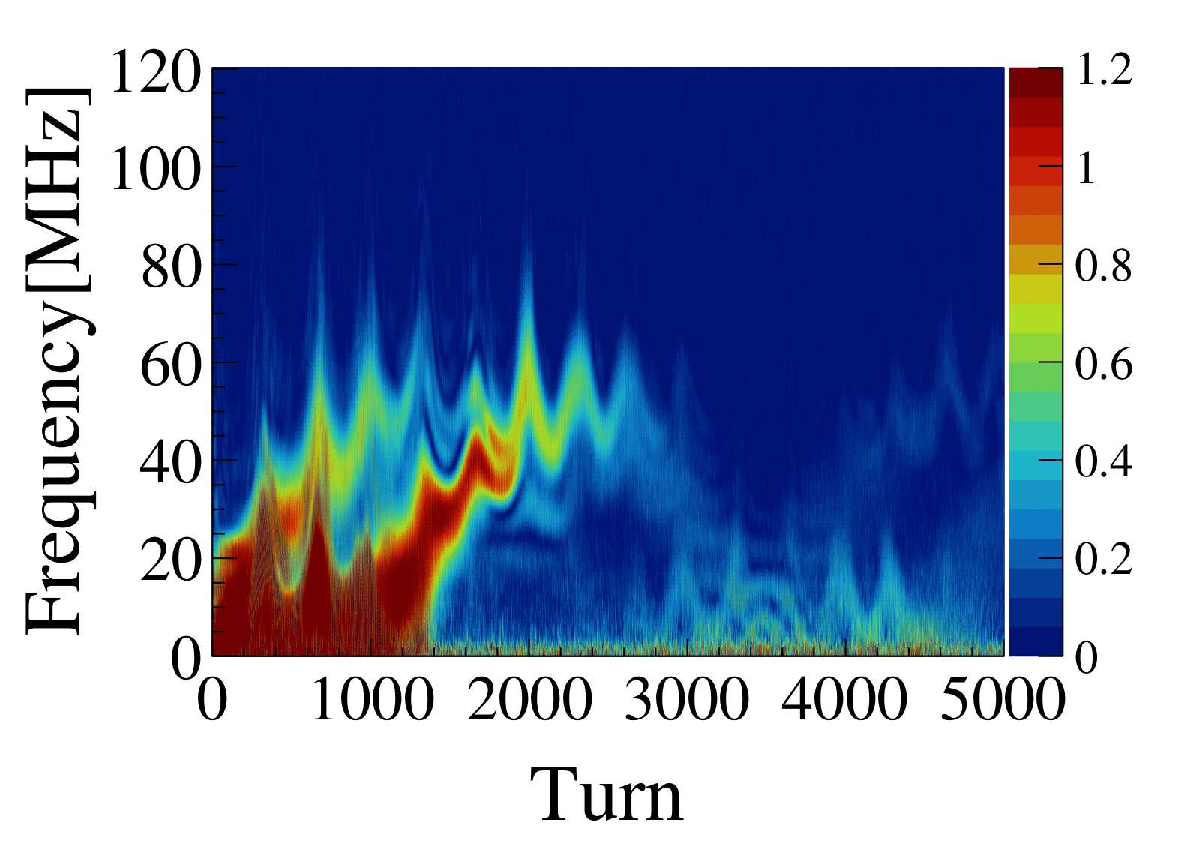}
      \subcaption{}
    \end{minipage} \\

    \begin{minipage}[t]{0.3\hsize}
      \centering
      \includegraphics[width=2.0in]{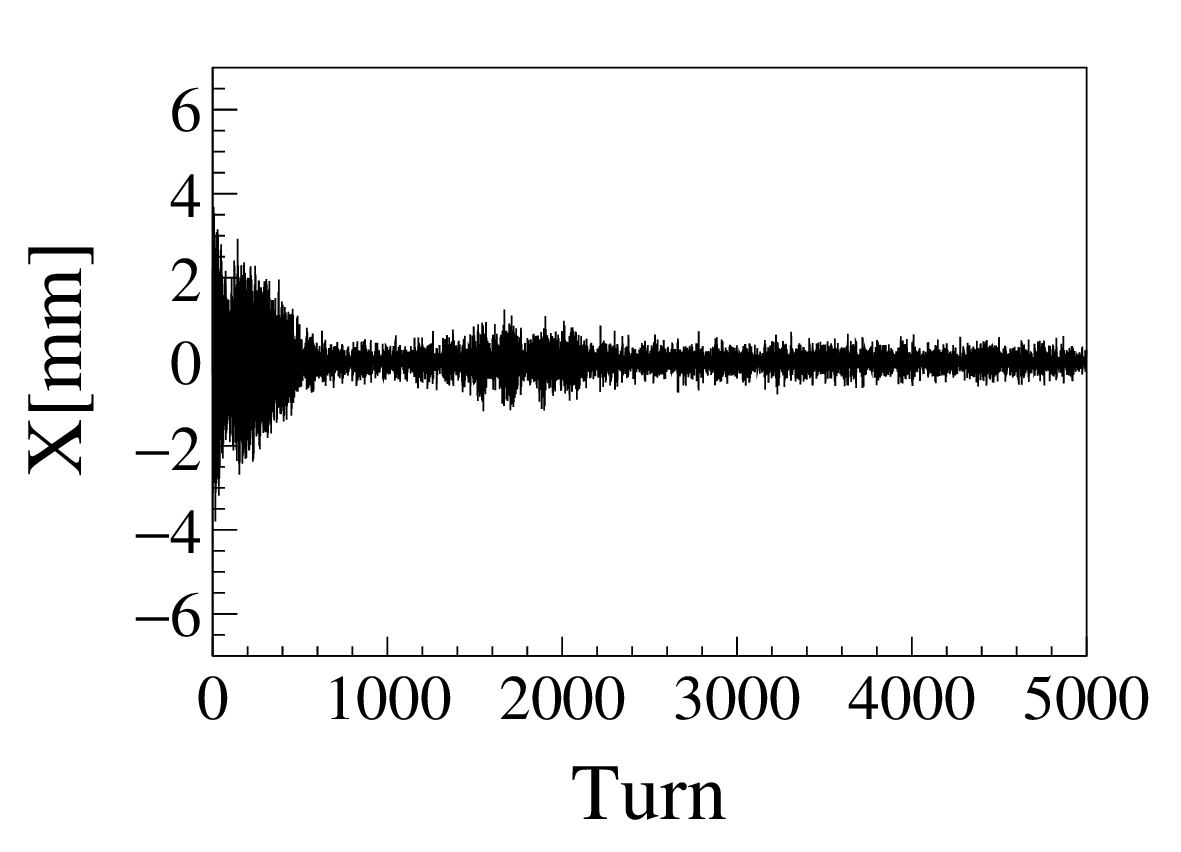}
      \subcaption{}
    \end{minipage} &
    \begin{minipage}[t]{0.3\hsize}
      \centering
      \includegraphics[width=2.0in]{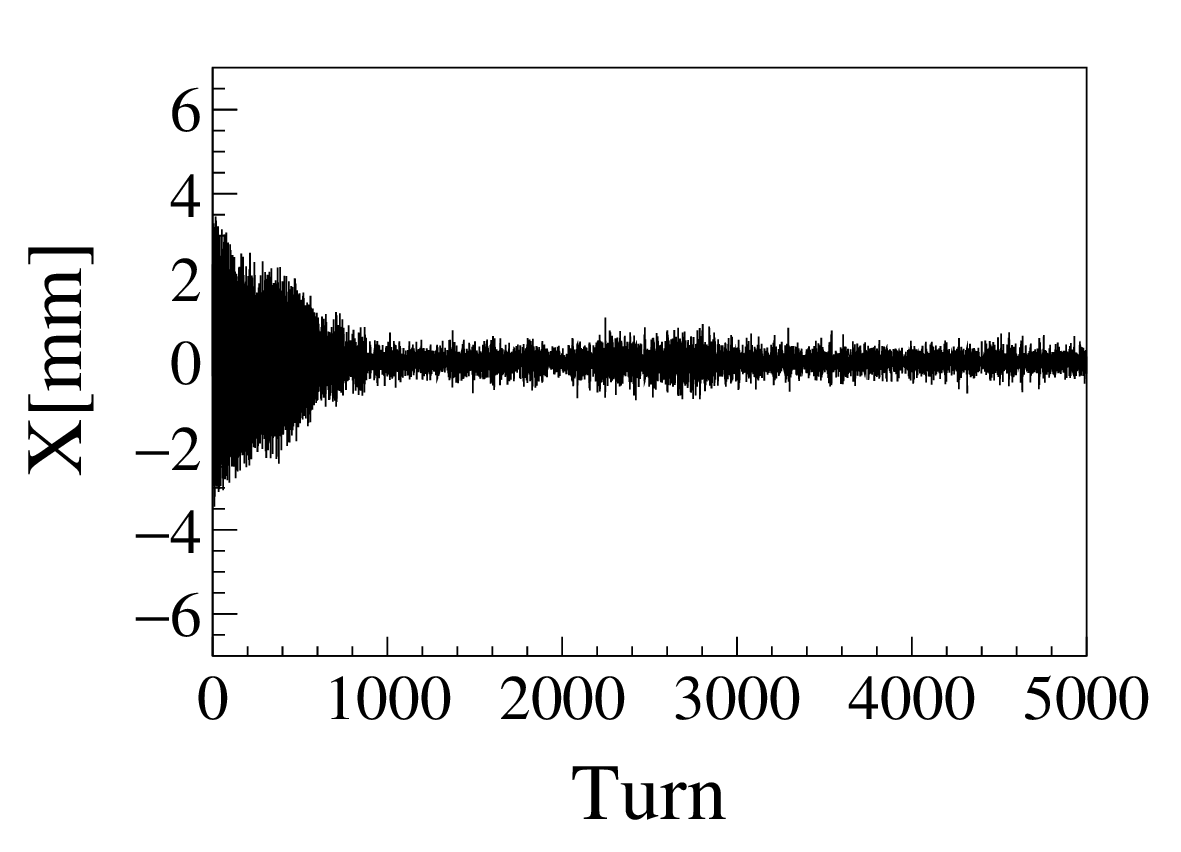}
      \subcaption{}
    \end{minipage} &
    \begin{minipage}[t]{0.3\hsize}
      \centering
      \includegraphics[width=2.0in]{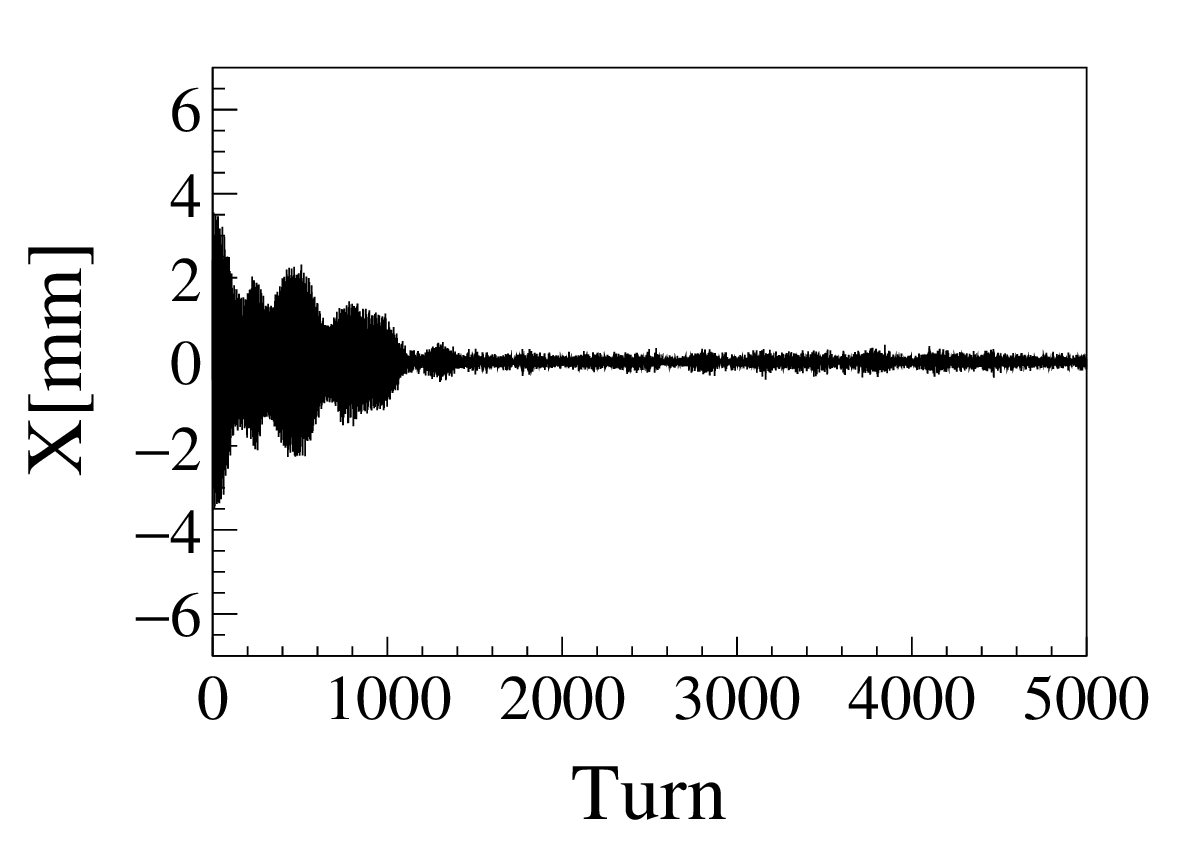}
      \subcaption{}
    \end{minipage} \\

    \begin{minipage}[t]{0.3\hsize}
      \centering
      \includegraphics[width=2.0in]{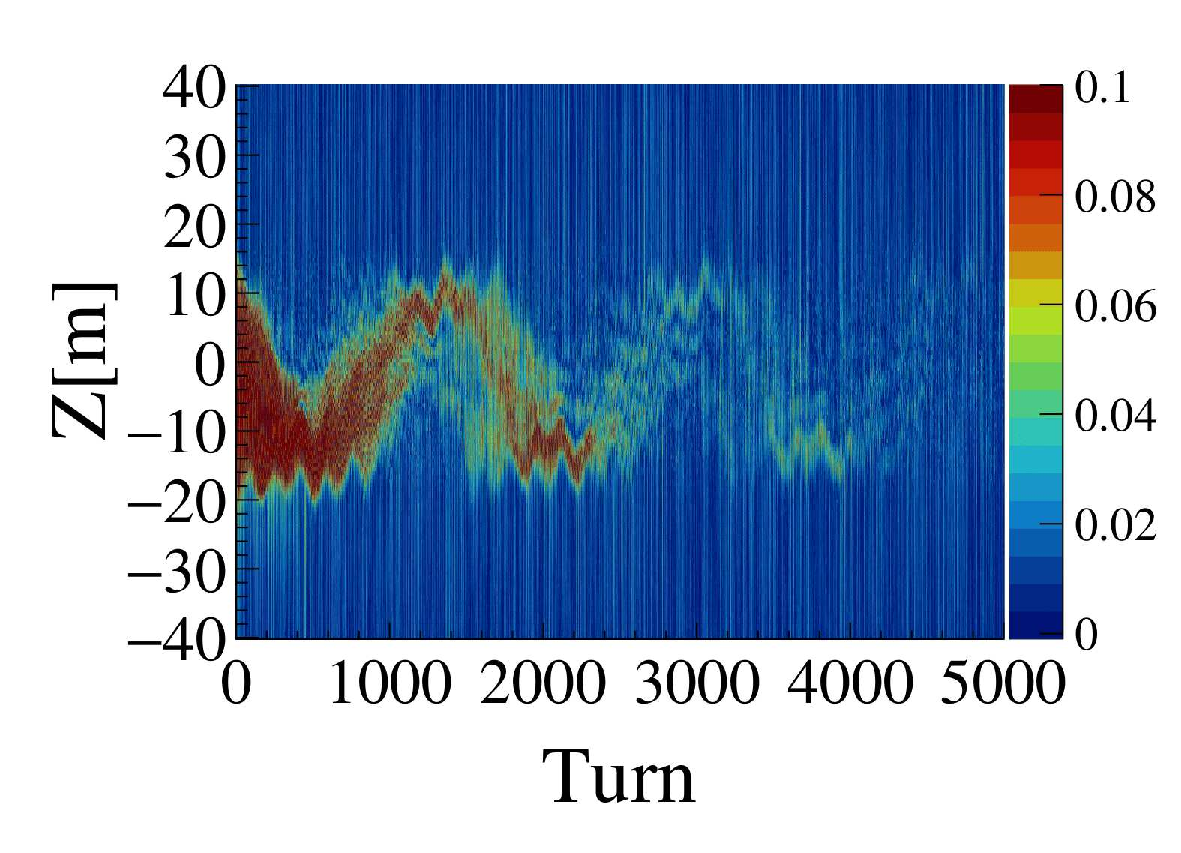}
      \subcaption{}
    \end{minipage} &
    \begin{minipage}[t]{0.3\hsize}
      \centering
      \includegraphics[width=2.0in]{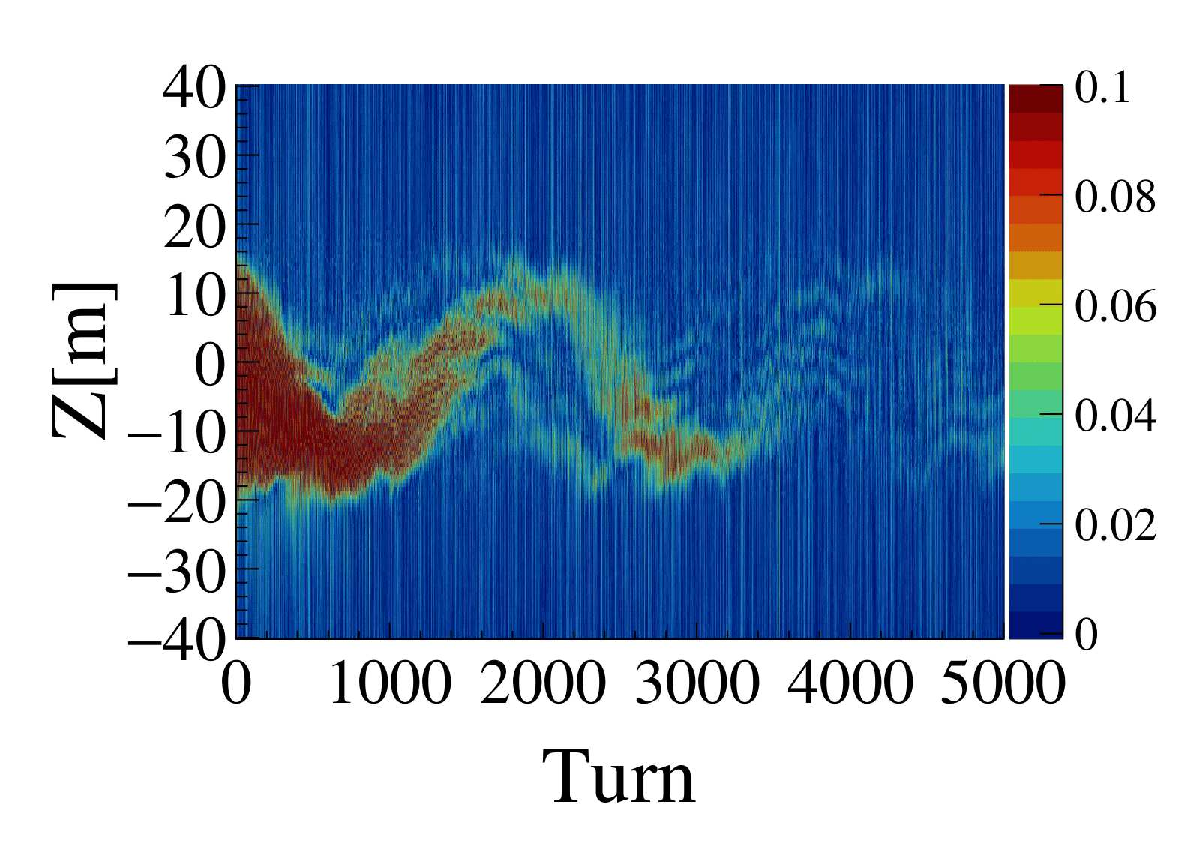}
      \subcaption{}
    \end{minipage} &
    \begin{minipage}[t]{0.3\hsize}
      \centering
      \includegraphics[width=2.0in]{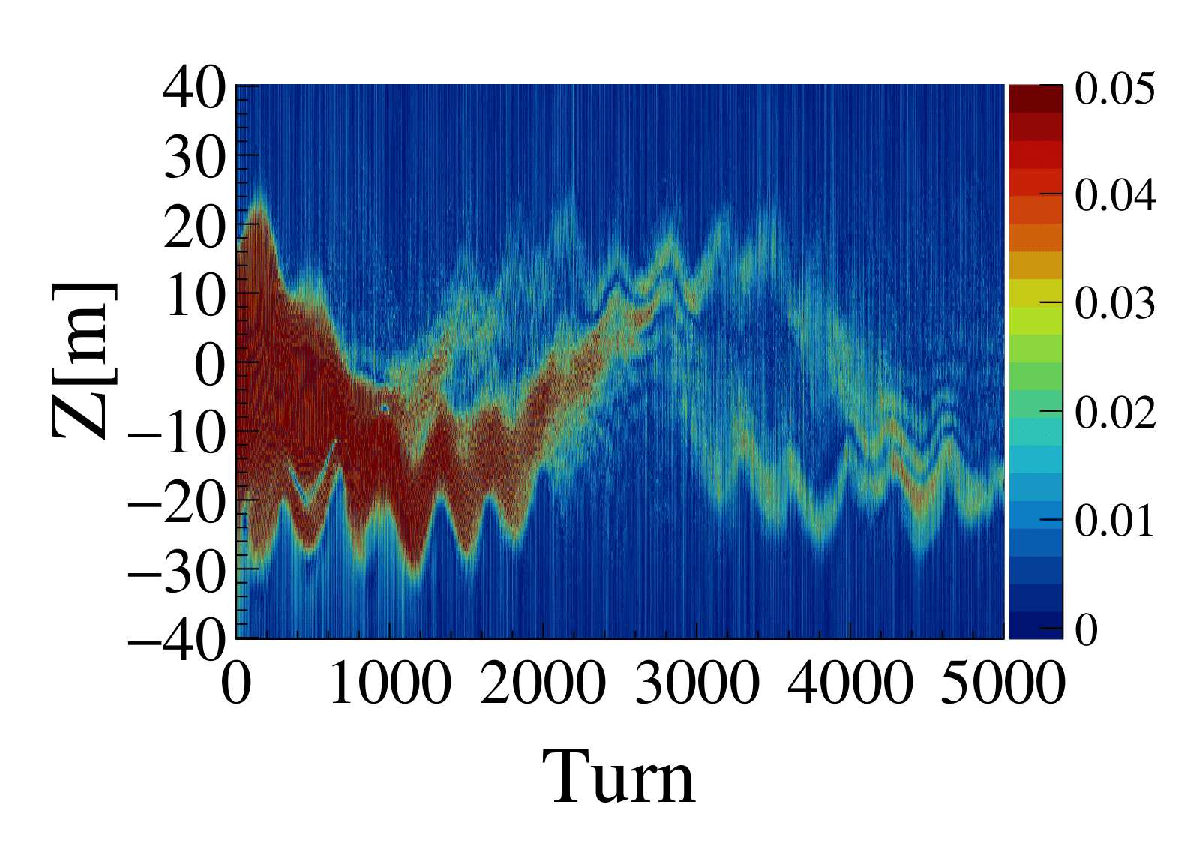}
      \subcaption{}
    \end{minipage}
  \end{tabular}
  \caption{Measurements of the transverse bunch motion in the time domain for the rf voltages: $\VRF=110,263$, and $400$~kV with the chromaticity $\xi_x=-6.5$ and the beam intensity $N_B=4.2\times10^{12}$~ppb. (a) Frequency component of the dipole moment for $\VRF=400$~kV, $\fmax=43.3$~MHz, $\NAM=1600$, $\NAM /N_s=5.4$, (b) frequency component of the dipole moment for $\VRF=263$~kV, $\fmax=45.3$~MHz, $\NAM=2300$, $\NAM /N_s=5.6$, (c) frequency component of the dipole moment for $\VRF=110$~kV, $\fmax=39.8$~MHz, $\NAM=3400$, $\NAM /N_s=5.2$, (d) average of the beam position $\overline{x_n}$ for $\VRF=400$~kV, (e) average of the beam position $\overline{x_n}$ for $\VRF=263$~kV, (f) average of the beam position $\overline{x_n}$ for $\VRF=110$~kV, (g) dipole moment $|\Delta(z)|$ in the bunch for $\VRF=400$~kV, (h) dipole moment $|\Delta(z)|$ in the bunch for $\VRF=263$~kV, and (i) dipole moment $|\Delta(z)|$ in the bunch for $\VRF=110$~kV.}
  \label{fig:17}
\end{figure}

Corresponding measurements were performed under the same parameter set, with the results shown in Fig. \ref{fig:17}.
The measured values for $(\fmax, \NAM)$ were (43.3~MHz, 1600~turns) for $\VRF=400$~kV, (45.3~MHz, 2300~turns) for $\VRF=263$~kV, and (39.8~MHz, 3400~turns) for $\VRF=110$~kV.
The measurements qualitatively reproduce the key features of the simulation, including the periodic frequency structure and the long recoherence period.

The measured $\fmax$ ($39-46$~MHz) and $\NAM/N_s$ ($5.2-5.6$) are comparable to the simulated values of $\fmax$ ($34-44$~MHz) and $\NAM/N_s$ ($4.2-5.0$).

\subsubsection{Comparison and discussion}\label{10.3.2}
\begin{figure}[!h]
  \begin{tabular}{cc}
    \begin{minipage}[t]{0.45\hsize}
      \centering
      \includegraphics[width=3.0in]{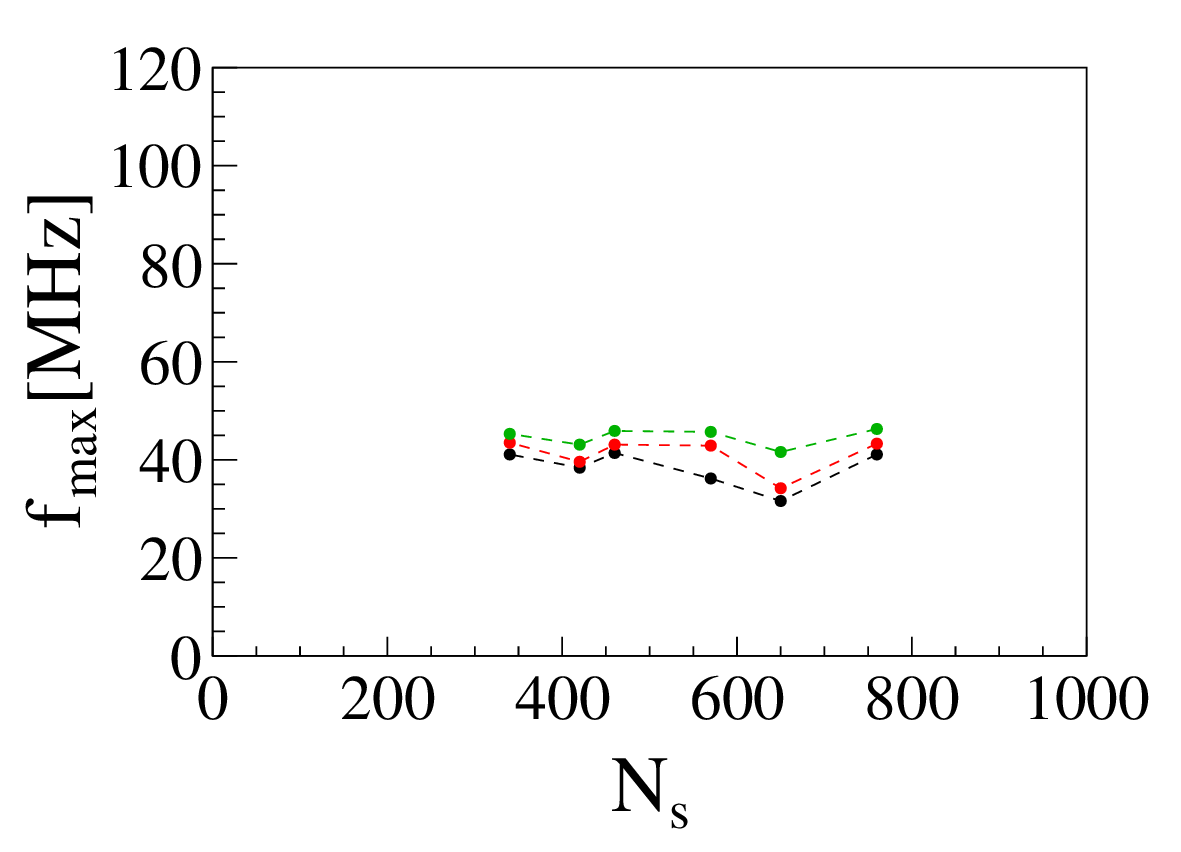}
      \subcaption{}
    \end{minipage} &
    \begin{minipage}[t]{0.45\hsize}
      \centering
      \includegraphics[width=3.0in]{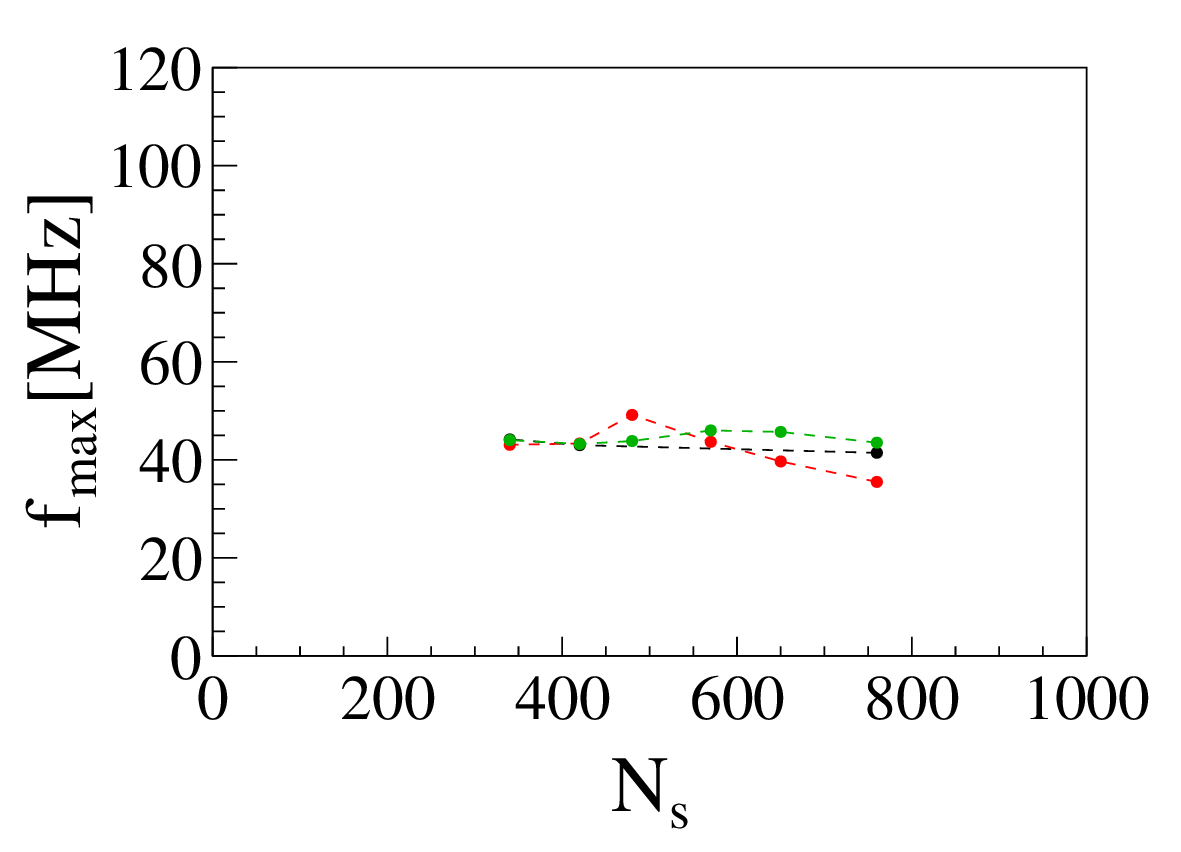}
      \subcaption{}
    \end{minipage}
  \end{tabular}
  \caption{Maximum intrabunch frequency $\fmax$ for the synchrotron period. Here, the black, red, and green denote the cases of $N_B=2.4\times10^{12}$, $4.2\times10^{12}$, and $8.6\times10^{12}$~ppb, respectively. (a) Tracking simulation and (b) measurement.}
  \label{fig:18}
\end{figure}

Figure \ref{fig:18} compares simulation and measurement results to assess the dependence of the maximum intrabunch frequency $\fmax$ on the synchrotron period.
Both datasets show minimal variation, indicating that the maximum intrabunch frequency $\fmax$ is essentially independent of synchrotron tune and beam intensity, as predicted by Eq. \eqref{eq:140}.

\begin{figure}[!h]
  \begin{tabular}{cc}
    \begin{minipage}[t]{0.45\hsize}
      \centering
      \includegraphics[width=3.0in]{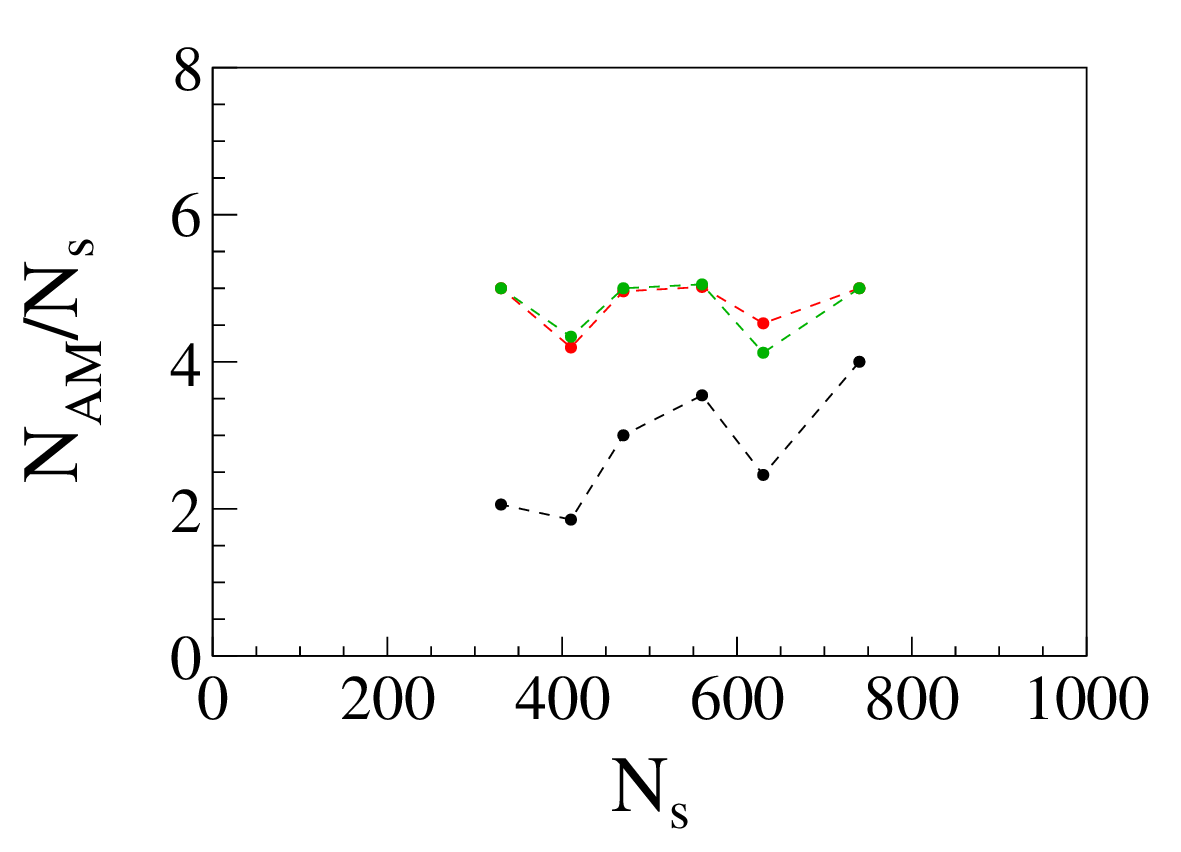}
      \subcaption{}
      \label{fig:19a}
    \end{minipage} &
    \begin{minipage}[t]{0.45\hsize}
      \centering
      \includegraphics[width=3.0in]{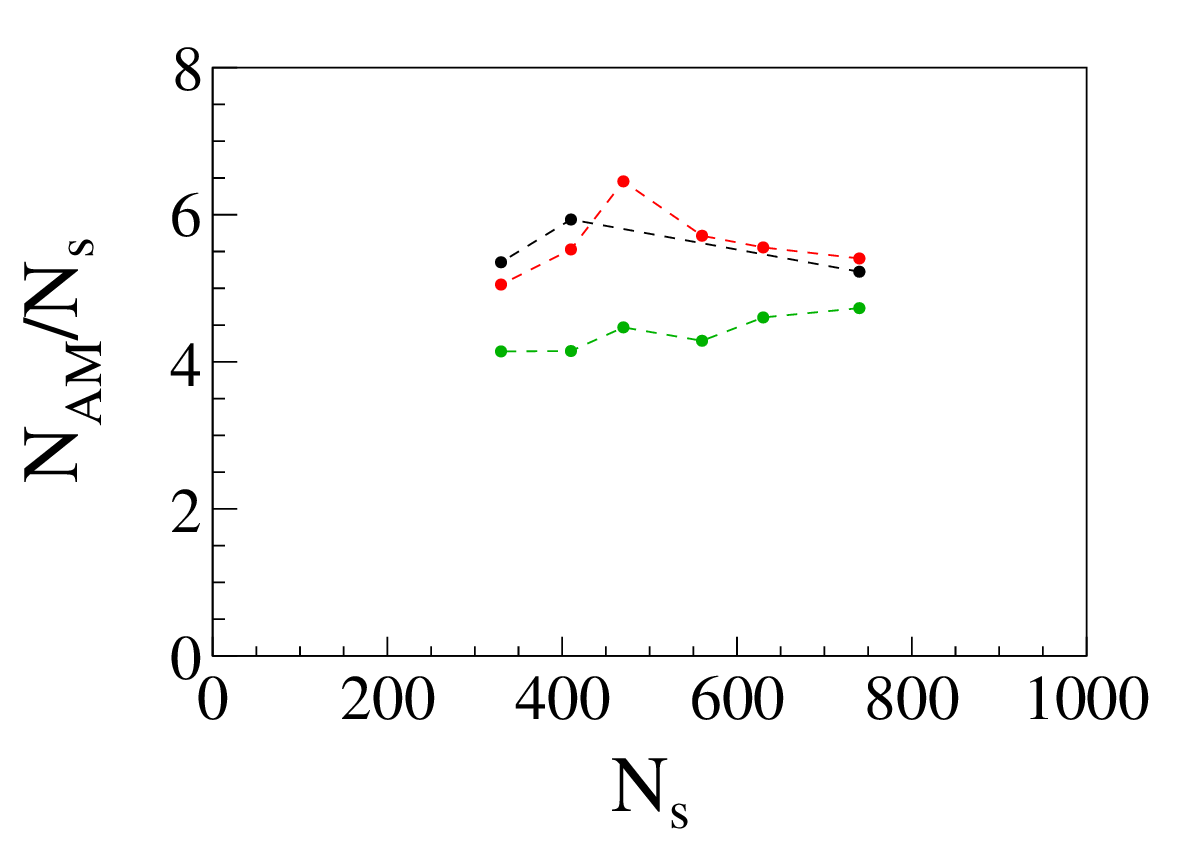}
      \subcaption{}
      \label{fig:19b}
    \end{minipage}
  \end{tabular}
  \caption{Normalized recoherence period $\NAM/N_s$ for the synchrotron period. Here, the black, red, and green denote the cases of $N_B=2.4\times10^{12}, 4.2\times10^{12}$, and $8.6\times10^{12}$~ppb, respectively. (a) Tracking simulation and (b) measurement.}
\end{figure}

Figure \ref{fig:19a} represents the simulations to explain the synchrotron period dependence of the normalized recoherence period $\NAM/N_s$.
When the space charge effect is stronger ($N_B=4.2\times10^{12}$ and $8.6\times10^{12}$~ppb), the normalized recoherence period stays within the region of $4N_s<\NAM<5N_s$ even when the synchrotron period is changed.
Meanwhile, when the space charge effect is moderate ($N_B=2.4\times10^{12}$~ppb), the normalized recoherence period tends to increase as the synchrotron period increases.

\begin{figure}[!h]
  \begin{tabular}{ccc}
    \begin{minipage}[t]{0.3\hsize}
      \centering
      \includegraphics[width=2.0in]{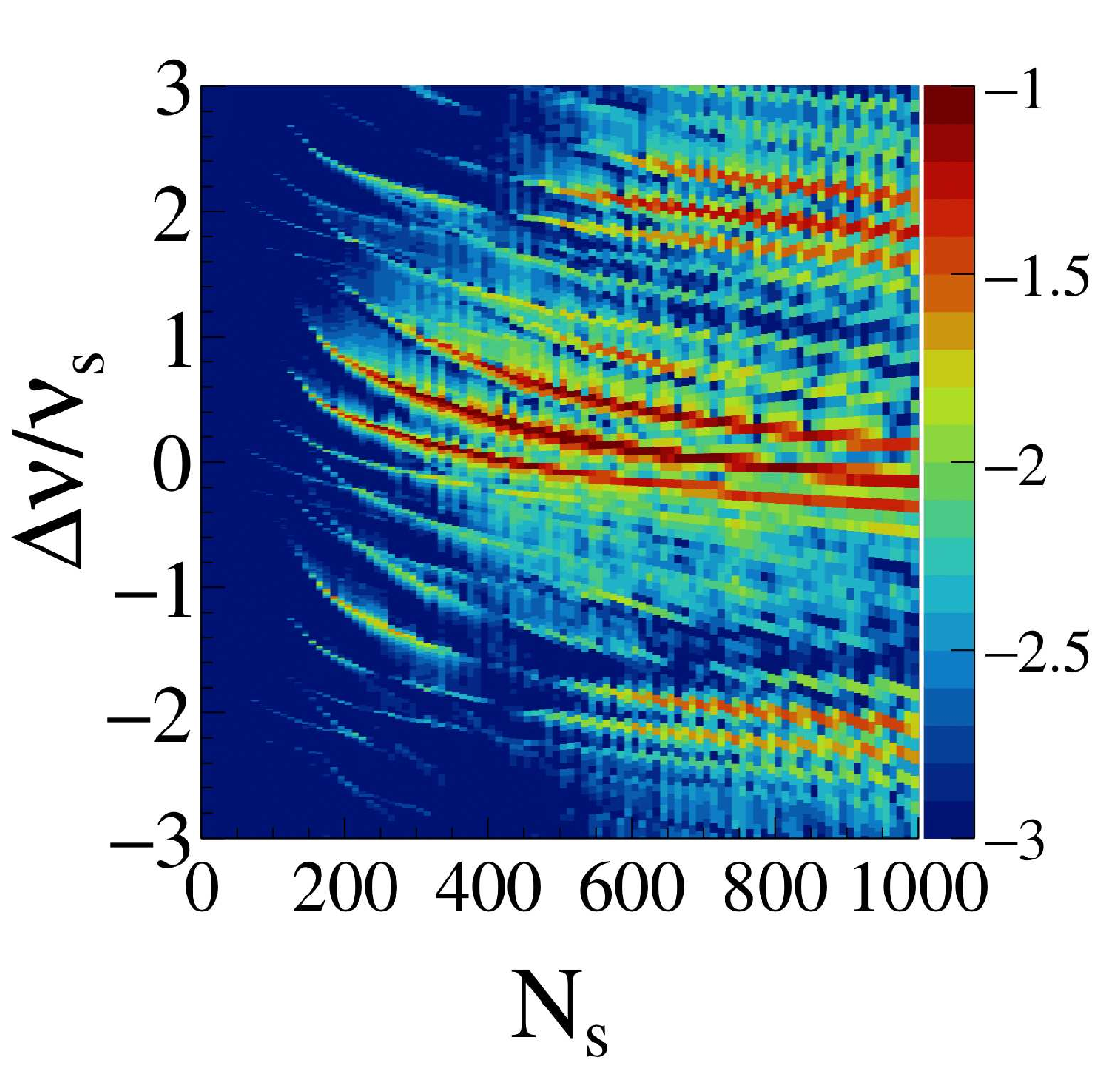}
      \subcaption{}
      \label{fig:20a}
    \end{minipage} &
    \begin{minipage}[t]{0.3\hsize}
      \centering
      \includegraphics[width=2.0in]{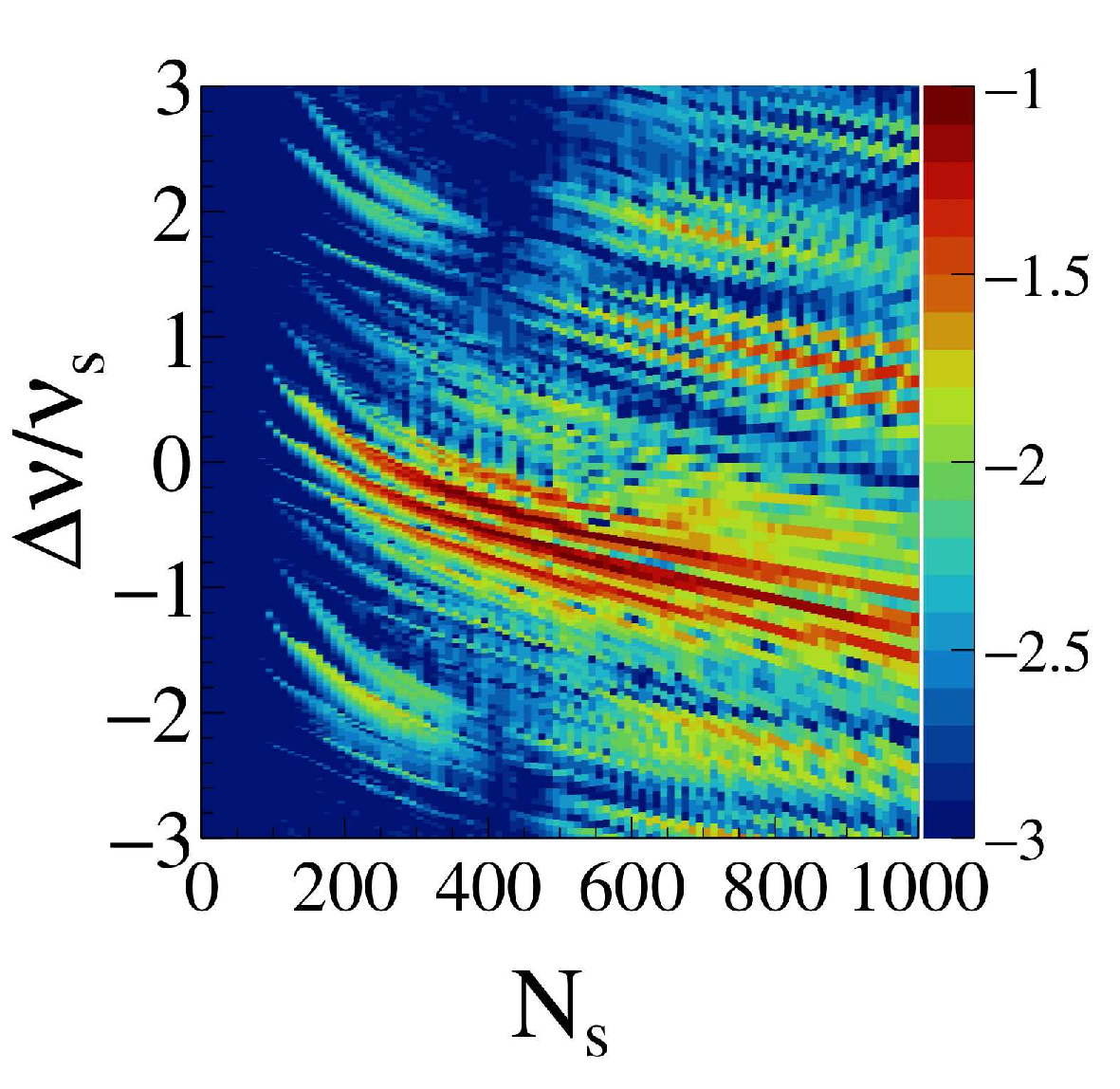}
      \subcaption{}
      \label{fig:20b}
    \end{minipage} &
    \begin{minipage}[t]{0.3\hsize}
      \centering
      \includegraphics[width=2.0in]{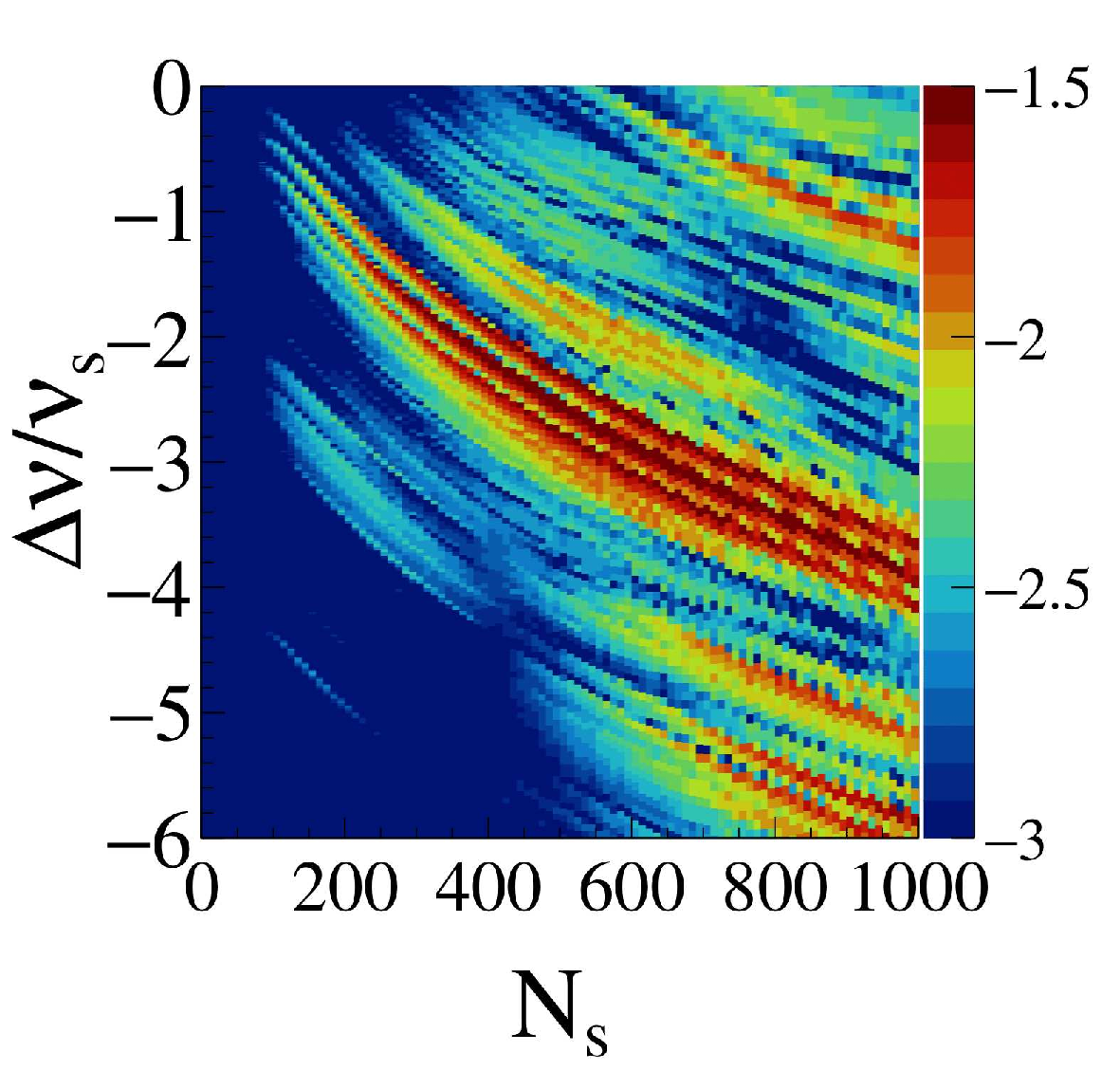}
      \subcaption{}
      \label{fig:20c}
    \end{minipage}
  \end{tabular}
  \caption{Simulated spectra of head-tail tune normalized with synchrotron tune $\Delta \nu/\nu_s$ for the synchrotron period obtained by analyzing $\log|a_m|$ of the harmonic model with the indirect space charge effect. (a) $N_B=2.4\times10^{12}$ ppb, (b) $N_B=4.2\times10^{12}$ ppb, and (c) $N_B=8.6\times10^{12}$ ppb.}
  \label{fig:20}
\end{figure}

This behavior of the recoherence period can be understood in terms of the head-tail modes analysis.
Figure \ref{fig:20} shows the head-tail tune spectrum of the relative amplitude $a_m$, where the synchrotron period is scanned by changing the rf voltage in the tracking simulation.
The vertical axis is the head-tail modes tune normalized by the synchrotron tune: $\Delta\nu/\nu_s$.
In Figs. \ref{fig:20b} and \ref{fig:20c} (in the cases of the strong space charge effect with the respective beam intensities $N_B=4.2\times10^{12}$ and $8.6\times10^{12}$~ppb), we can see that the normalized frequency interval between adjacent head-tail modes is close to a constant of around $0.2<\Delta\nu/\nu_s<0.25$, even when the synchrotron period is changed.
On the other hand, in Fig. \ref{fig:20a} (in the case of the moderate space charge effect with the beam intensity $N_B=2.4\times10^{12}$~ppb), the interval dynamically decreases from $0.5$ to $0.25$ as the synchrotron period increases.
As a result, we can conclude that the normalized recoherence period saturates for the higher synchrotron tune due to the space charge effects.

The measured data for the normalized recoherence period $\NAM/N_s$ is shown in Fig. \ref{fig:19b}.
The period remains within the range of $4 < \NAM/N_s < 7$ and shows no strong dependence on $N_s$, consistent with the saturation behavior predicted for high-intensity beams.
Interestingly, the data for $N_B=8.6\times10^{12}$~ppb deviates slightly from the lower intensity cases.
These measured results indicate that the recoherence period already saturates in the midstrength region of the space charge, suggesting that the current modeling of the indirect space charge, based on the simplified geometry of the J-PARC MR, overestimates its effect on the intrabunch motions.
Meanwhile, the reason the normalized recoherence period falls below the saturated value for beams with midstrength space charge is that the injected beam is broadened from the outset due to the large number of $N_B$.
Note, $B_f=0.017$ for $N_B\leq4.2\times10^{12}$~ppb, but $B_f=0.022$ for $N_B=8.6\times10^{12}$~ppb in the measurements.
Consequently, the effective space charge effect is mitigated in the bunch with $N_B=8.6\times10^{12}$~ppb, leading to a shorter normalized recoherence period compared to the other results.

\subsection{Beam intensity dependence}\label{10.4}
Next, the dependence of intrabunch motion on beam intensity is analyzed using combined data from all three measurement sets (Tables \ref{tab:1}, \ref{tab:2} and \ref{tab:3}), based on the observation in Sec. \ref{10.3.2} that the synchrotron period has little influence on $\fmax$ and $\NAM/N_s$.

\subsubsection{Simulation predictions and measurement results}\label{10.4.1}
\begin{figure}[!h]
  \begin{tabular}{cccc}
    \begin{minipage}[t]{0.20\hsize}
      \centering
      \includegraphics[width=1.5in]{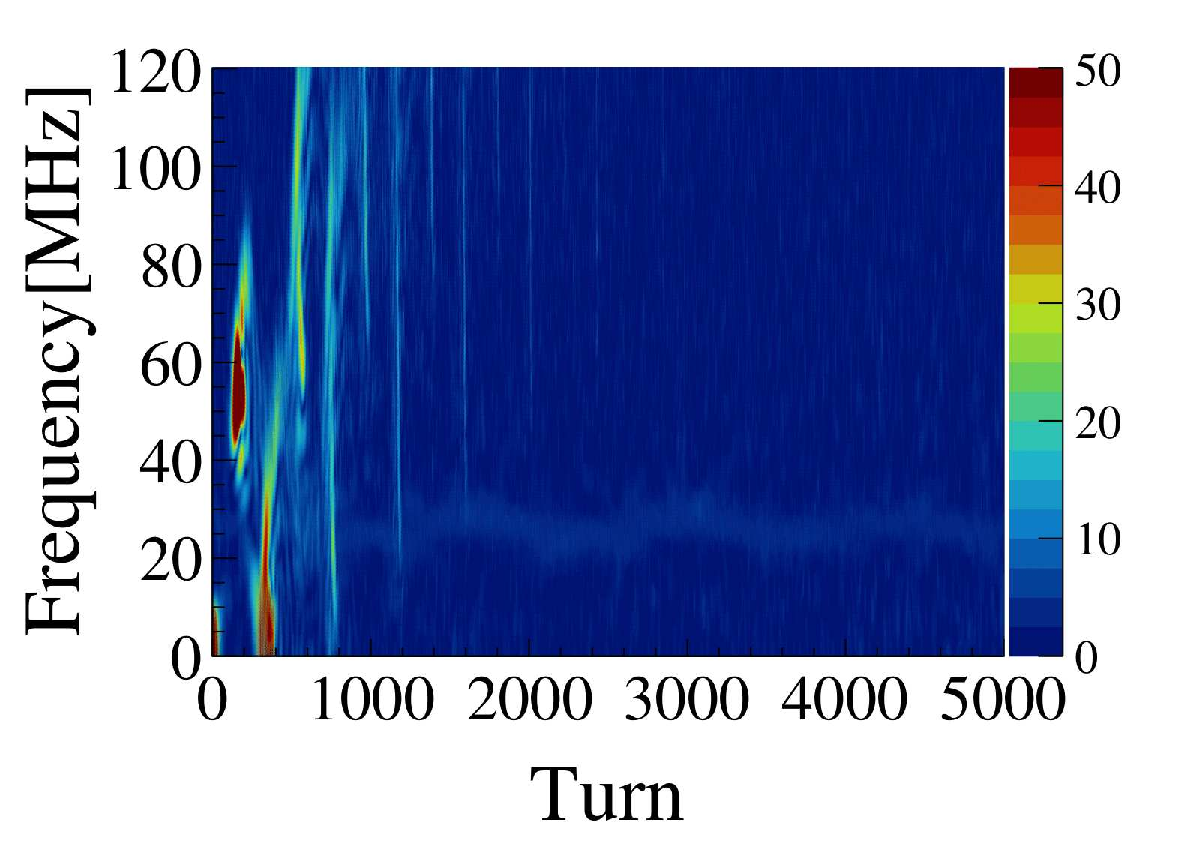}
      \subcaption{}
    \end{minipage} &
    \begin{minipage}[t]{0.20\hsize}
      \centering
      \includegraphics[width=1.5in]{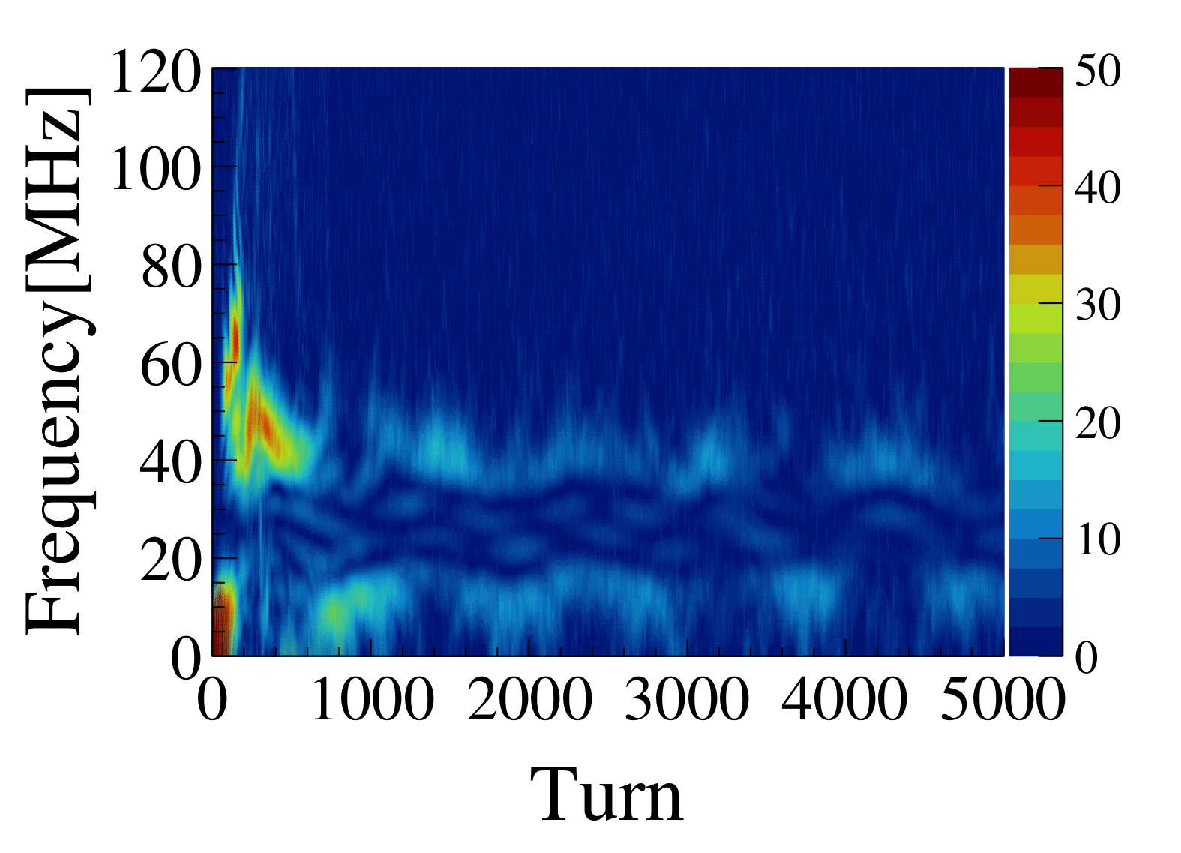}
      \subcaption{}
    \end{minipage} &
    \begin{minipage}[t]{0.20\hsize}
      \centering
      \includegraphics[width=1.5in]{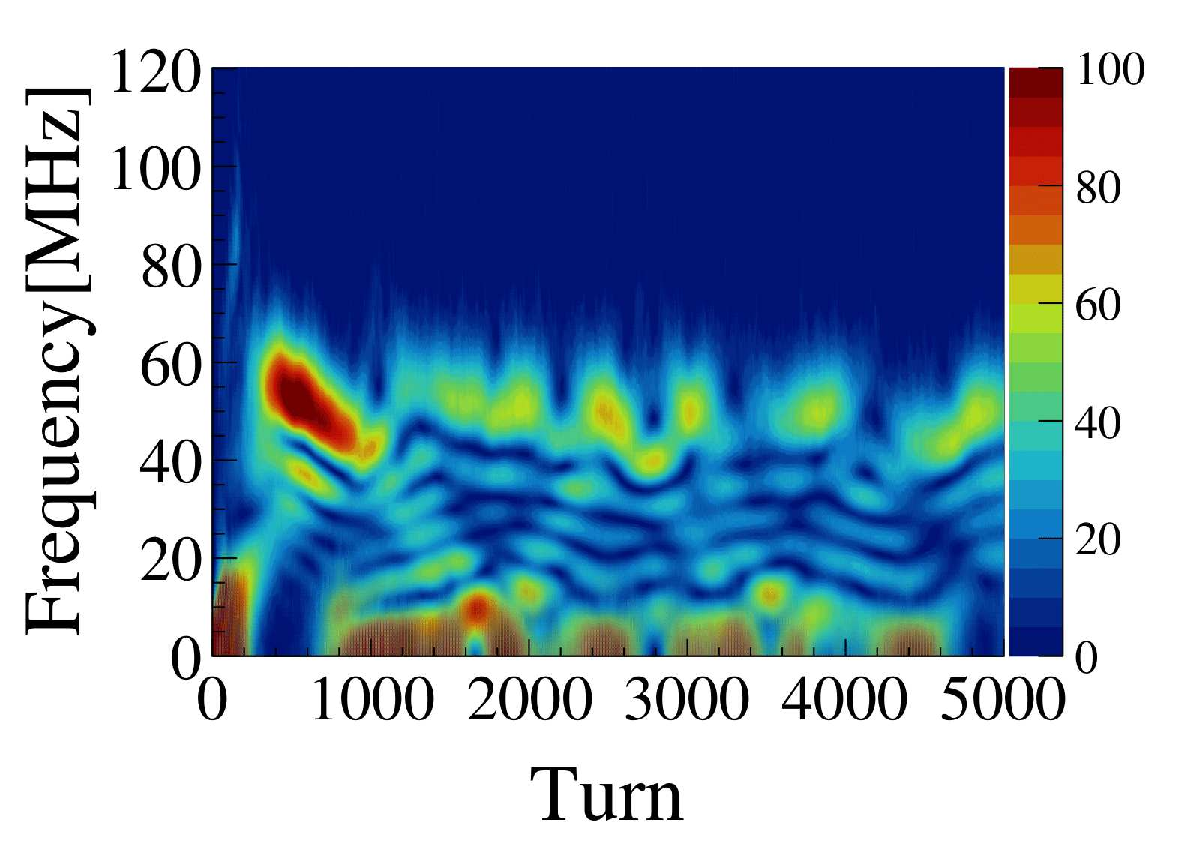}
      \subcaption{}
    \end{minipage} &
    \begin{minipage}[t]{0.20\hsize}
      \centering
      \includegraphics[width=1.5in]{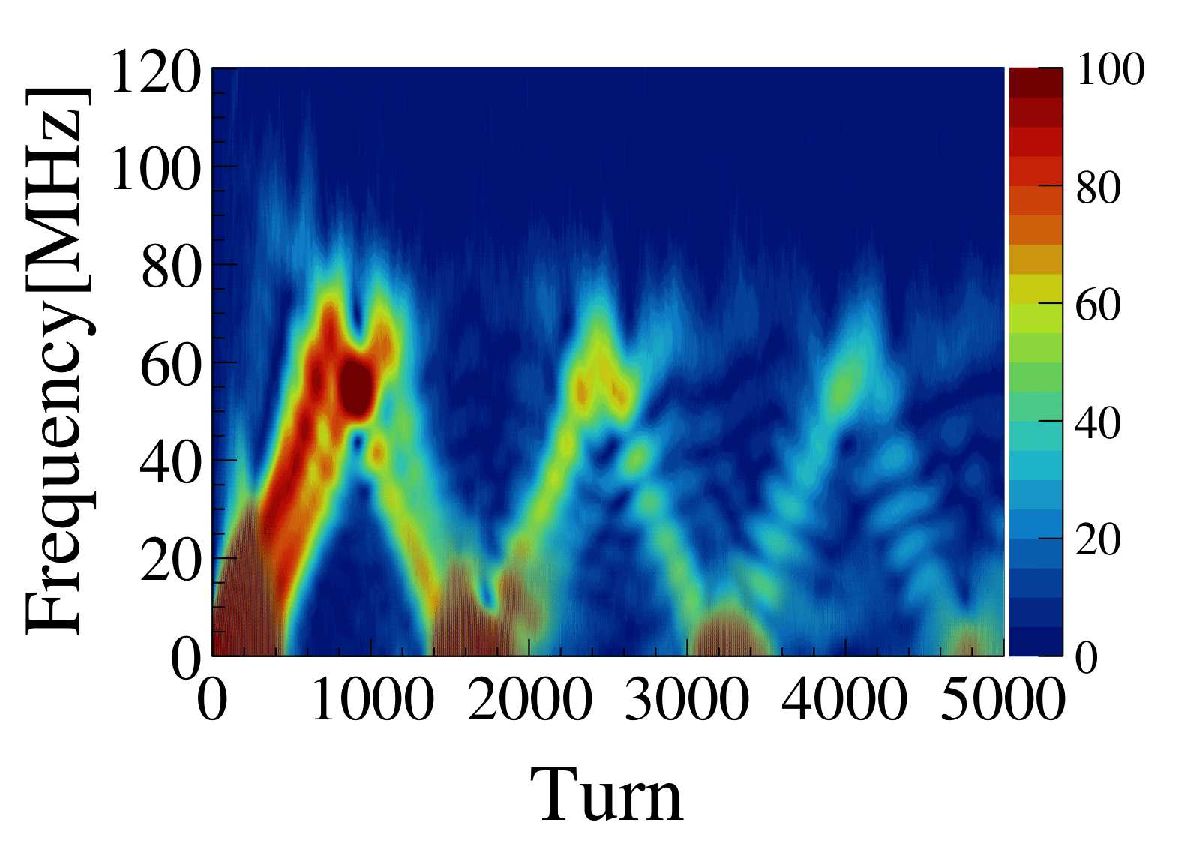}
      \subcaption{}
    \end{minipage} \\

    \begin{minipage}[t]{0.20\hsize}
      \centering
      \includegraphics[width=1.5in]{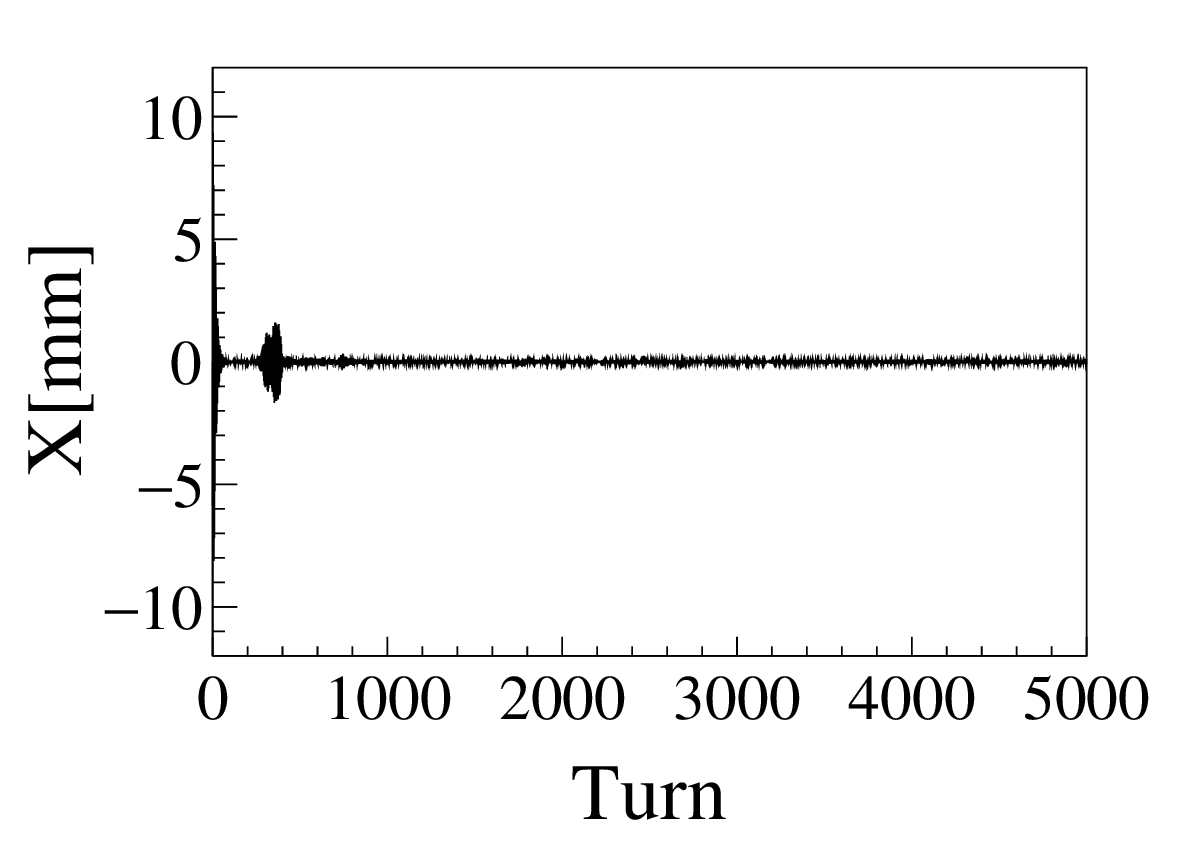}
      \subcaption{}
      \label{fig:21e}
    \end{minipage} &
    \begin{minipage}[t]{0.20\hsize}
      \centering
      \includegraphics[width=1.5in]{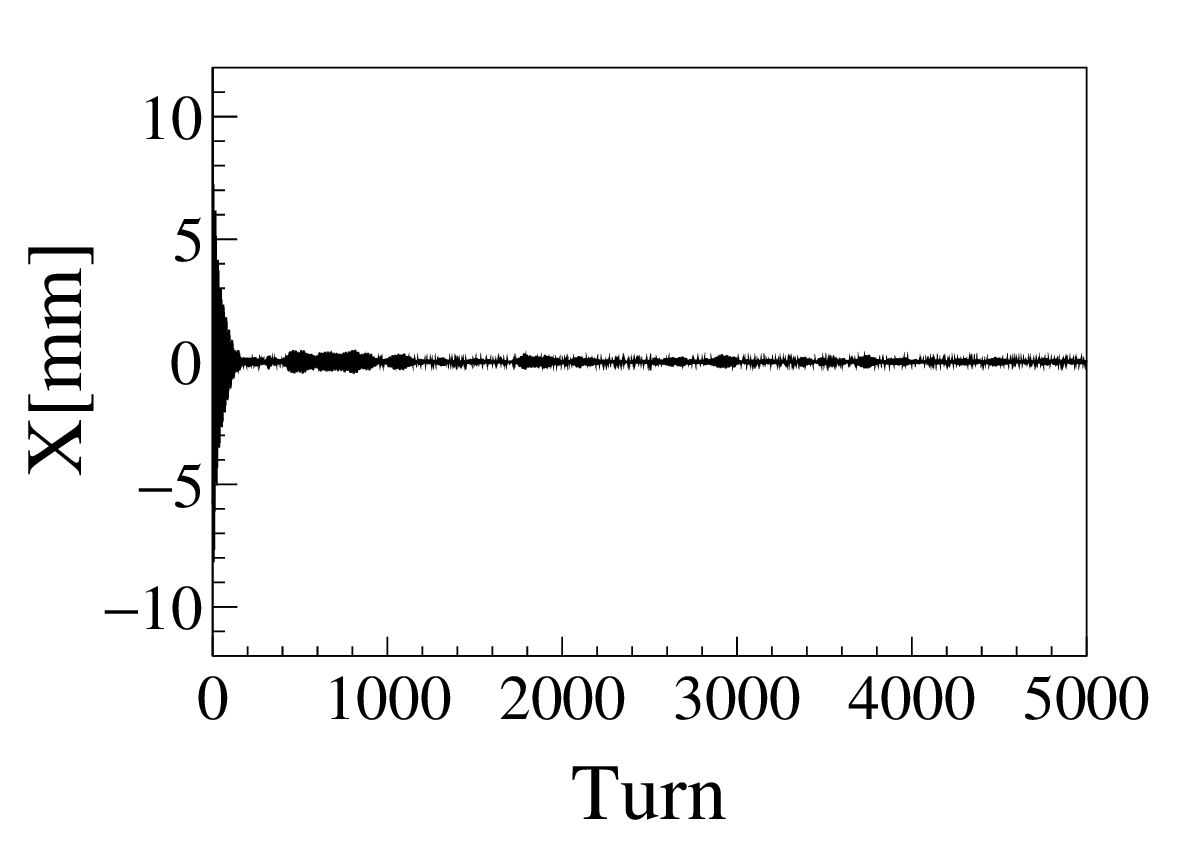}
      \subcaption{}
    \end{minipage} &
    \begin{minipage}[t]{0.20\hsize}
      \centering
      \includegraphics[width=1.5in]{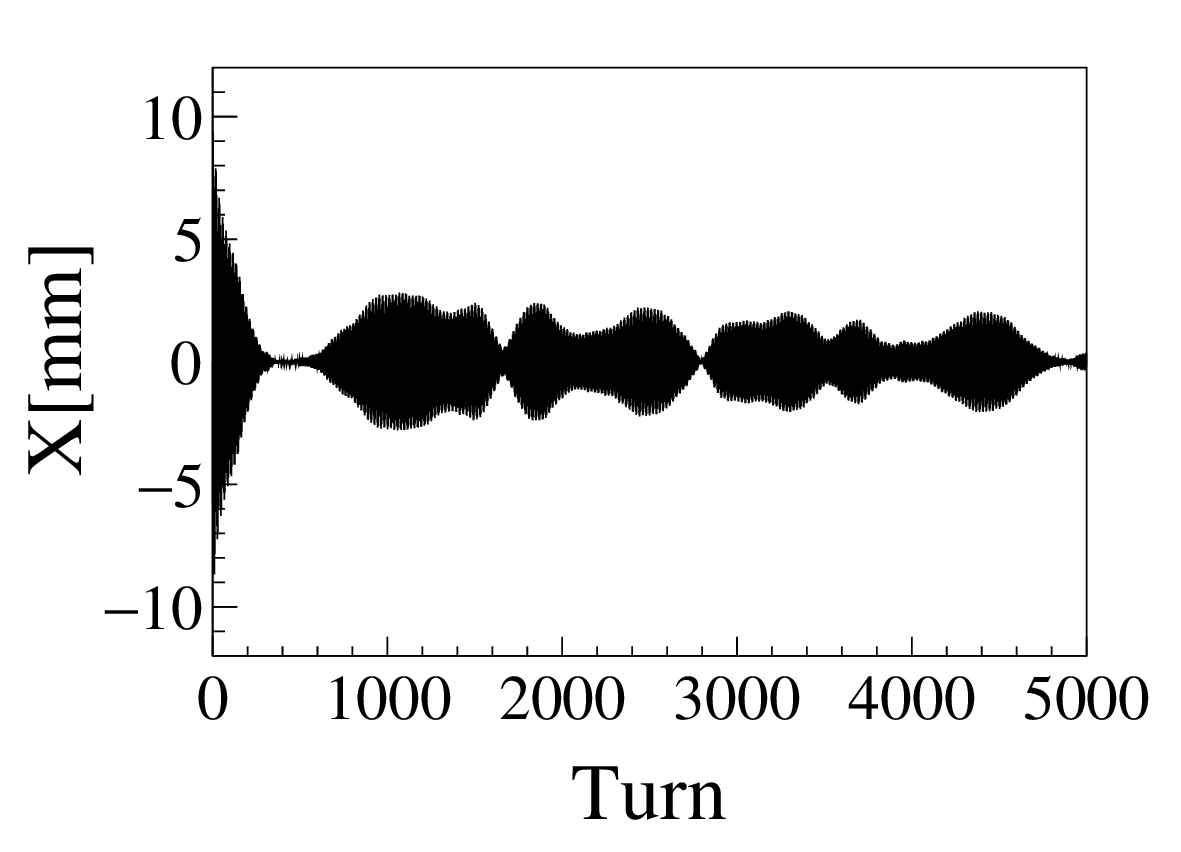}
      \subcaption{}
      \label{fig:21g}
    \end{minipage} &
    \begin{minipage}[t]{0.20\hsize}
      \centering
      \includegraphics[width=1.5in]{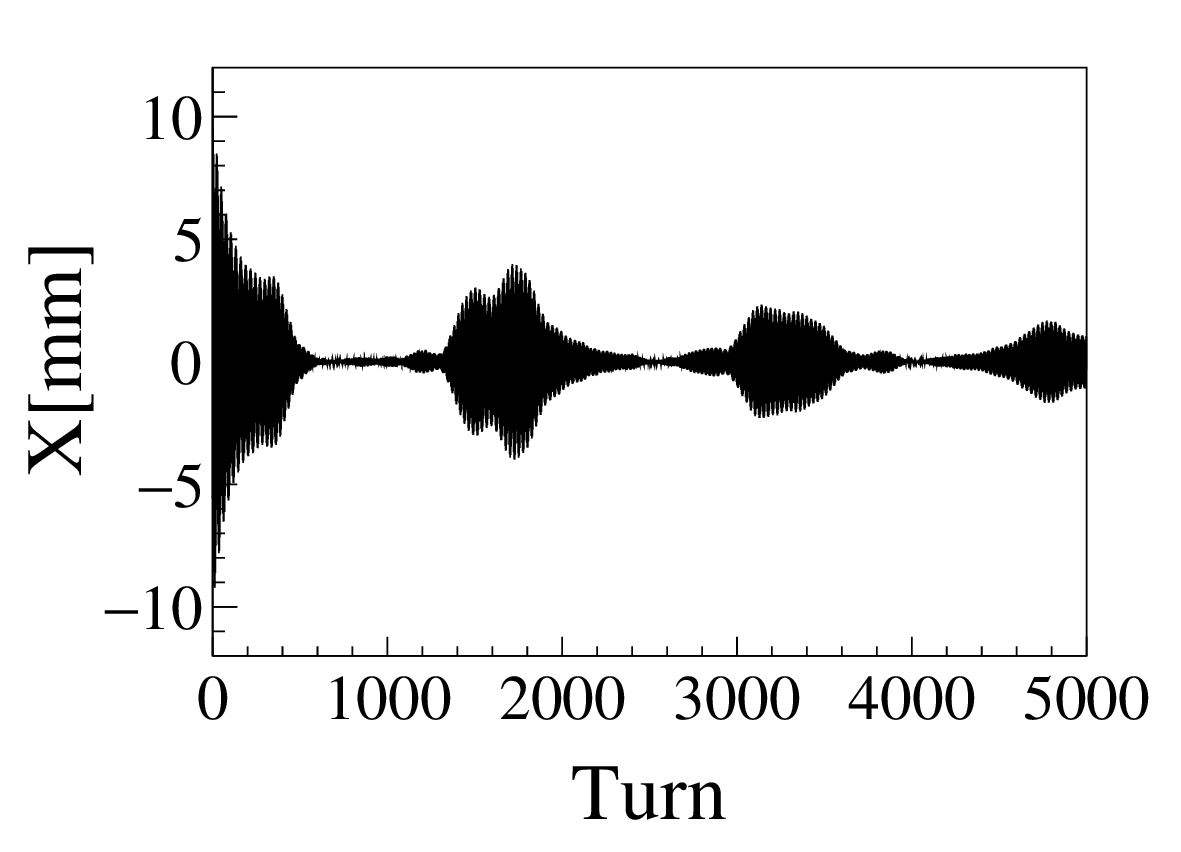}
      \subcaption{}
      \label{fig:21h}
    \end{minipage} \\

    \begin{minipage}[t]{0.20\hsize}
      \centering
      \includegraphics[width=1.5in]{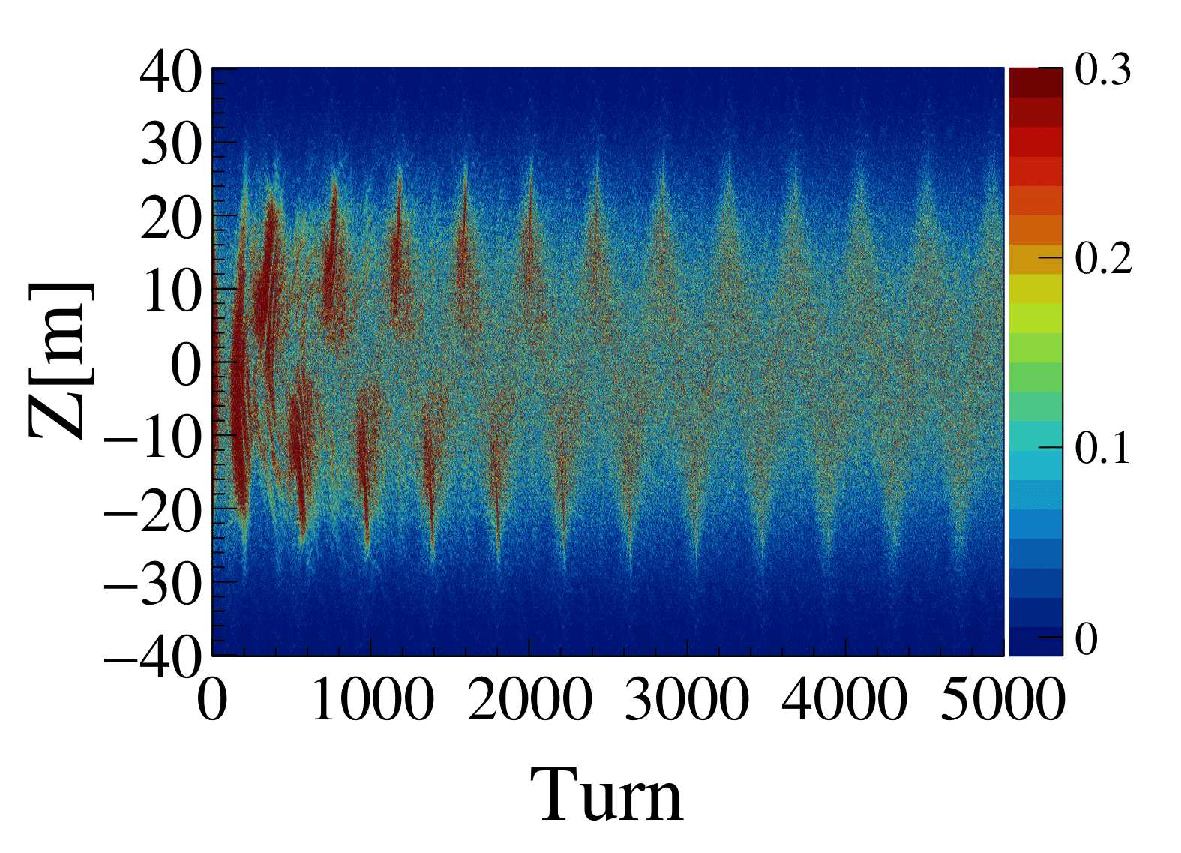}
      \subcaption{}
    \end{minipage} &
    \begin{minipage}[t]{0.20\hsize}
      \centering
      \includegraphics[width=1.5in]{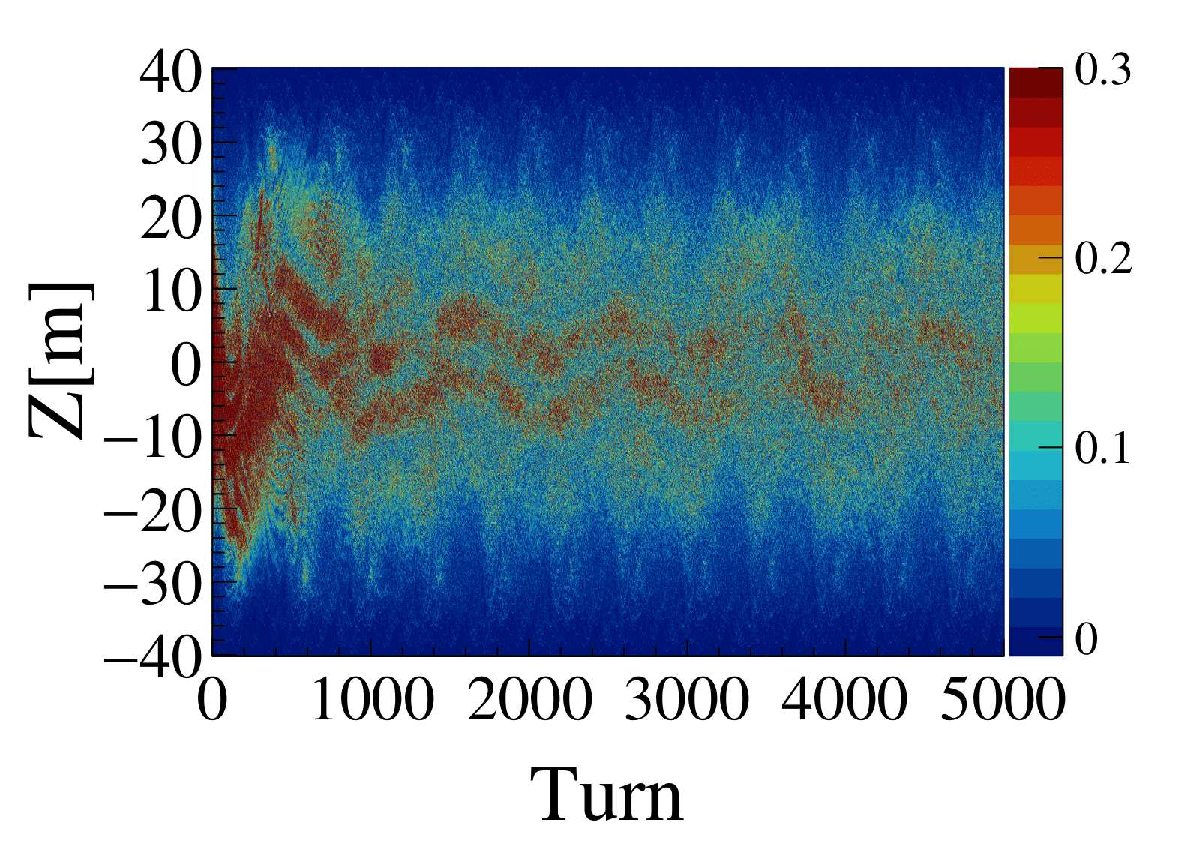}
      \subcaption{}
    \end{minipage} &
    \begin{minipage}[t]{0.20\hsize}
      \centering
      \includegraphics[width=1.5in]{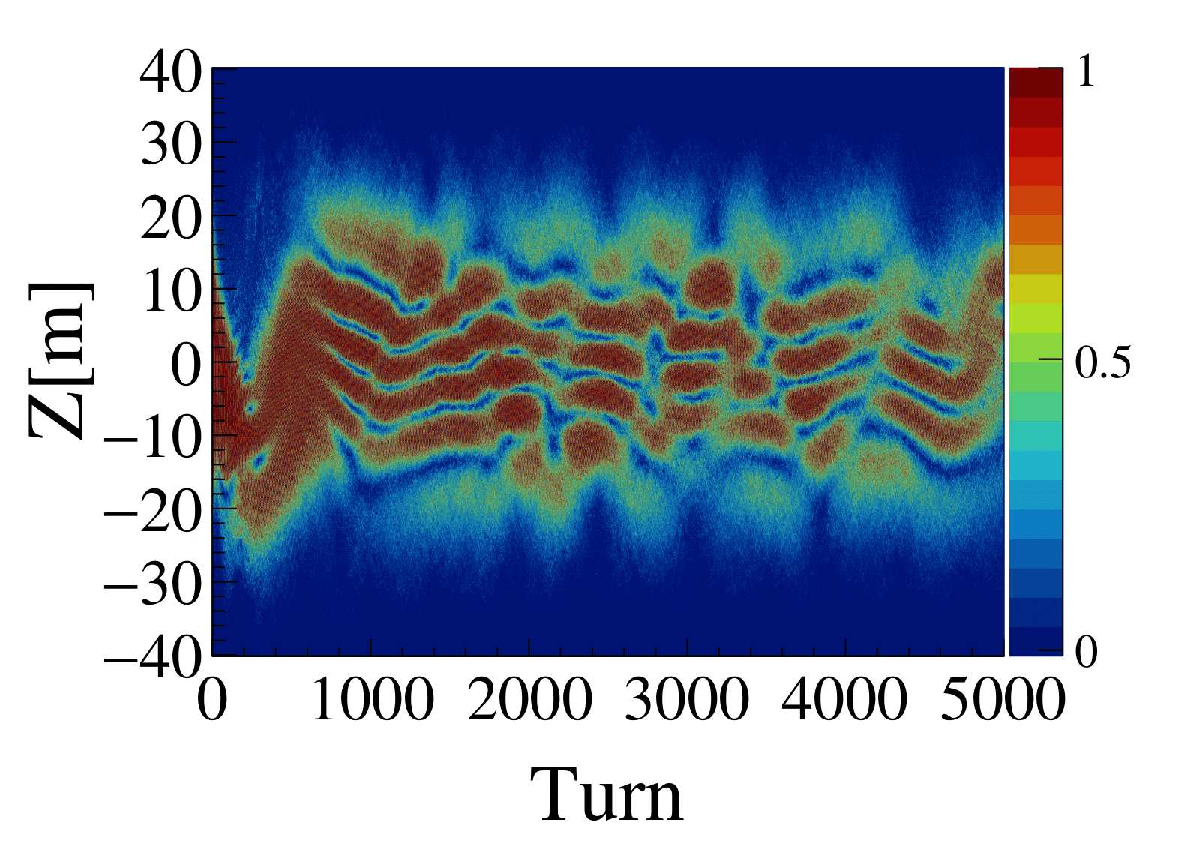}
      \subcaption{}
    \end{minipage} &
    \begin{minipage}[t]{0.20\hsize}
      \centering
      \includegraphics[width=1.5in]{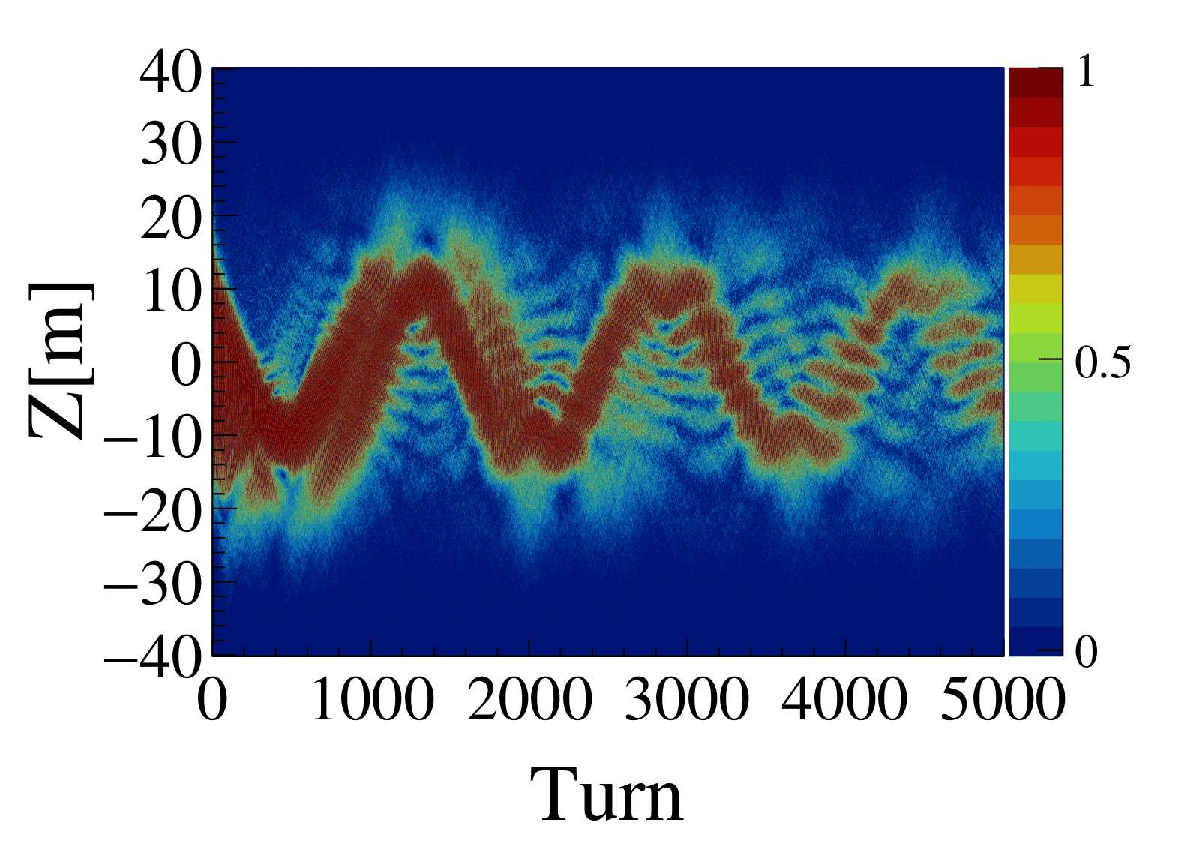}
      \subcaption{}
    \end{minipage}
  \end{tabular}
  \caption{Simulation of the transverse bunch motion in the time domain for the beam intensity: $N_B=1.0\times10^{12},2.4\times10^{12},4.2\times10^{12}$ and $8.6\times10^{12}$~ppb with the chromaticity $\xi_x=-8.0$, and the rf voltage $\VRF=263$~kV. (a) Frequency component of the dipole moment for $N_B=1.0\times10^{12}$~ppb, $\fmax=53.3$~MHz, $\NAM=340$, $\NAM /N_s=0.8$, (b) frequency component of the dipole moment for $N_B=2.4\times10^{12}$~ppb, $\fmax=49.0$~MHz, $\NAM=600$, $\NAM /N_s=1.5$, (c) frequency component of the dipole moment for $N_B=4.2\times10^{12}$~ppb, $\fmax=51.3$~MHz, $\NAM=1040$, $\NAM /N_s=2.5$, (d) frequency component of the dipole moment for $N_B=8.6\times10^{12}$~ppb, $\fmax=53.7$~MHz, $\NAM=1860$, $\NAM /N_s=4.5$, (e) average of the beam position $\overline{x_n}$ for $N_B=1.0\times10^{12}$~ppb, (f) average of the beam position $\overline{x_n}$ for $N_B=2.4\times10^{12}$~ppb, (g) average of the beam position $\overline{x_n}$ for $N_B=4.2\times10^{12}$~ppb, (h) average of the beam position $\overline{x_n}$ for $N_B=8.6\times10^{12}$~ppb, (i) dipole moment $|\Delta^{(k)}\sigma_\Delta/q|$ in the bunch for $N_B=1.0\times10^{12}$~ppb, (j) dipole moment $|\Delta^{(k)}\sigma_\Delta/q|$ in the bunch for $N_B=2.4\times10^{12}$~ppb, (k) dipole moment $|\Delta^{(k)}\sigma_\Delta/q|$ in the bunch for $N_B=4.2\times10^{12}$~ppb, and (l) dipole moment $|\Delta^{(k)}\sigma_\Delta/q|$ in the bunch for $N_B=8.6\times10^{12}$~ppb.}
  \label{fig:21}
\end{figure}

We simulate the beam intensity dependence of intrabunch motions for a fixed rf voltage $\VRF=263$~kV and chromaticity.
Figure \ref{fig:21} shows the simulations of the transverse bunch motion in the time domain for the beam intensity: $N_B=1.0\times10^{12},2.4\times10^{12},4.2\times10^{12}$, and $8.6\times10^{12}$~ppb with fixed chromaticity $\xi_x=-8.0$.
The extracted values for $(\fmax, \NAM)$ were (53.3~MHz, 340~turns) for $N_B=1.0\times10^{12}$~ppb, (49.0~MHz, 600~turns) for $N_B=2.4\times10^{12}$~ppb, (51.3~MHz, 1040~turns) for $N_B=4.2\times10^{12}$~ppb, and (53.7~MHz, 1860~turns) for $N_B=8.6\times10^{12}$~ppb.
When the space charge effect is negligible, the recoherence occurs at $\NAM=N_s$ \cite{yoshimura2025space}.
However, for the moderate beam intensity in Fig. \ref{fig:21e}, the recoherence signals weaken due to the betatron tune spread caused by the space charge effects.
In contrast, as the beam intensity increases significantly, the strong space charge effects bias the particle distributions within the beam, leading to the reemergence of recoherence, as illustrated in Figs. \ref{fig:21g} and \ref{fig:21h}.

\begin{figure}[!h]
  \begin{tabular}{cccc}
    \begin{minipage}[t]{0.20\hsize}
      \centering
      \includegraphics[width=1.5in]{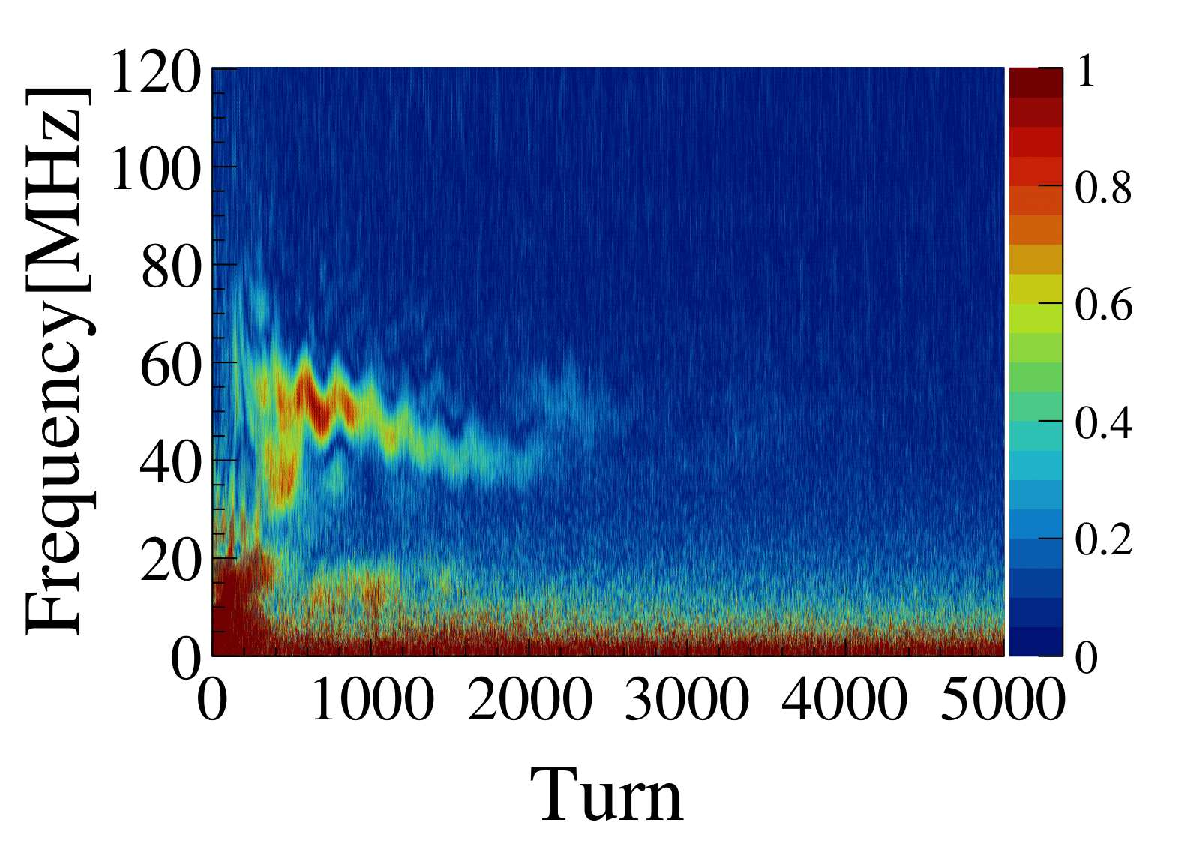}
      \subcaption{}
    \end{minipage} &
    \begin{minipage}[t]{0.20\hsize}
      \centering
      \includegraphics[width=1.5in]{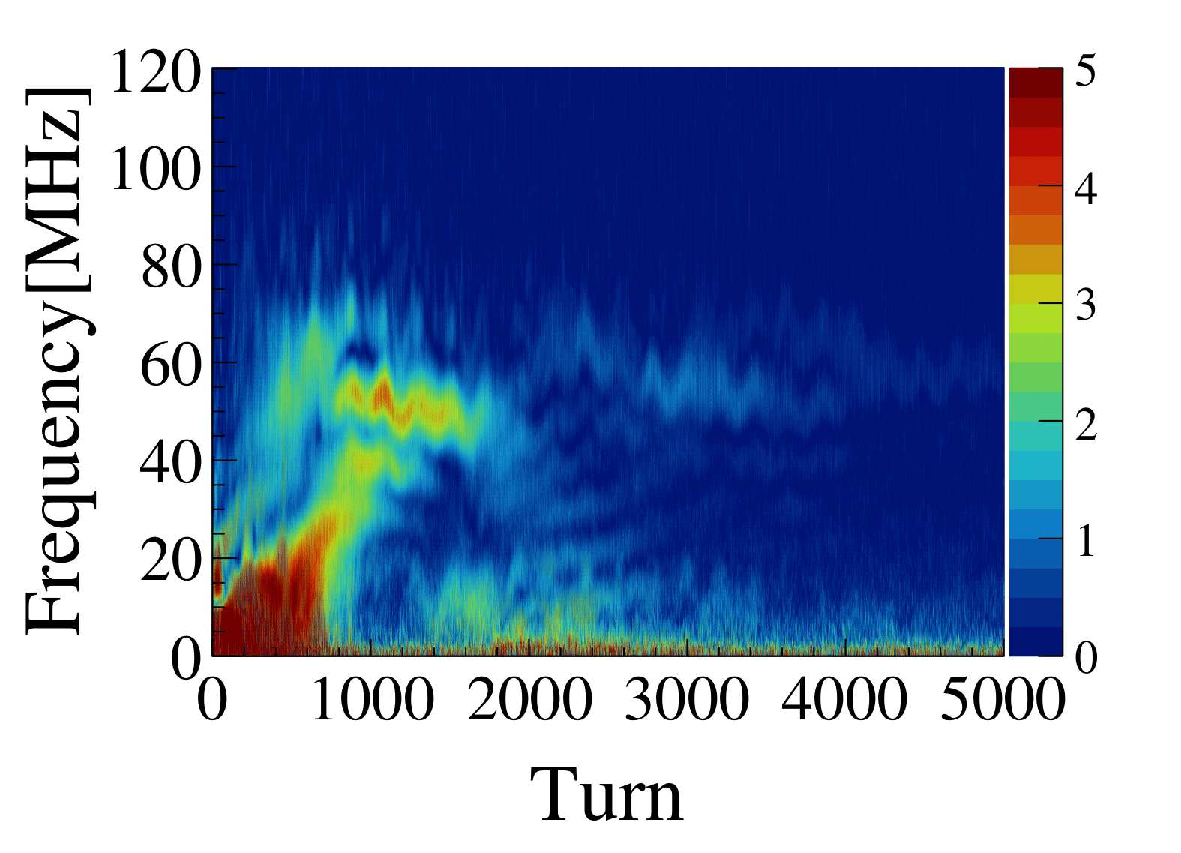}
      \subcaption{}
    \end{minipage} &
    \begin{minipage}[t]{0.20\hsize}
      \centering
      \includegraphics[width=1.5in]{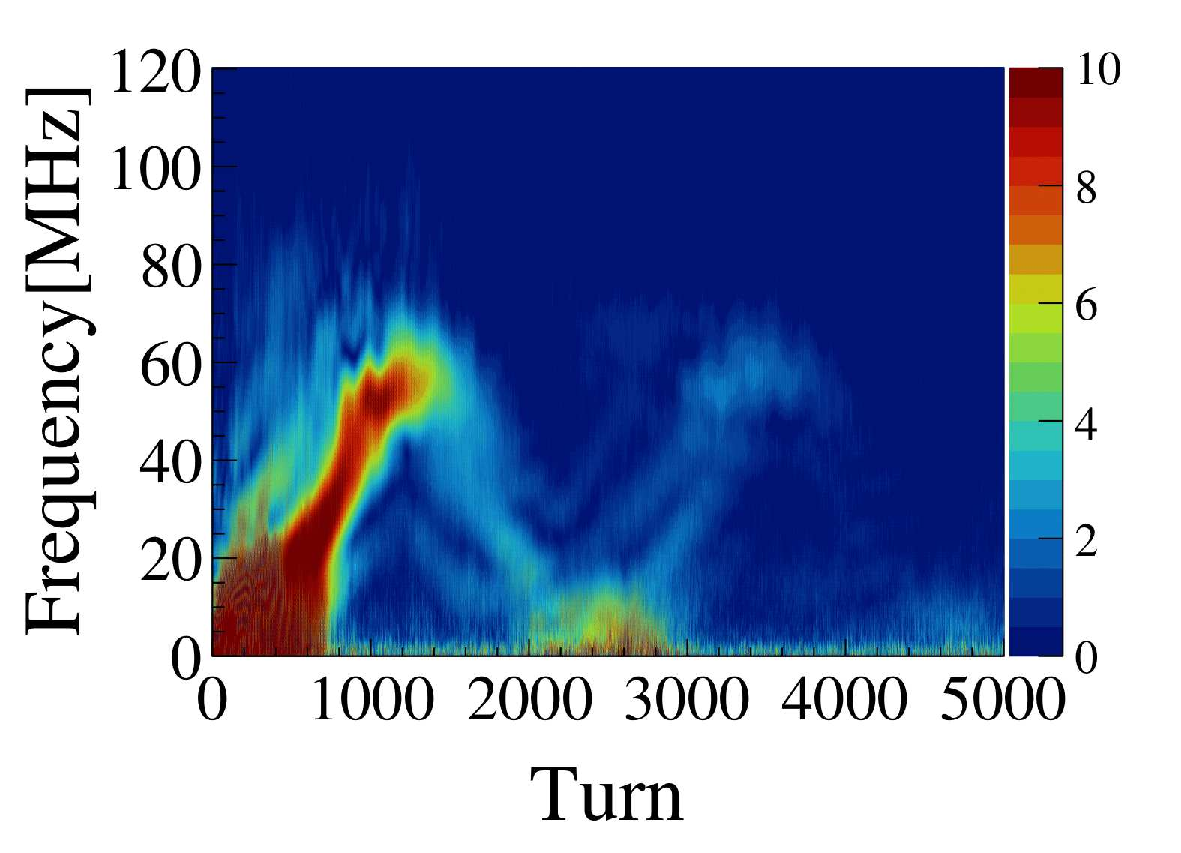}
      \subcaption{}
    \end{minipage} &
    \begin{minipage}[t]{0.20\hsize}
      \centering
      \includegraphics[width=1.5in]{FFT_230228_185.eps}
      \subcaption{}
    \end{minipage} \\

    \begin{minipage}[t]{0.20\hsize}
      \centering
      \includegraphics[width=1.5in]{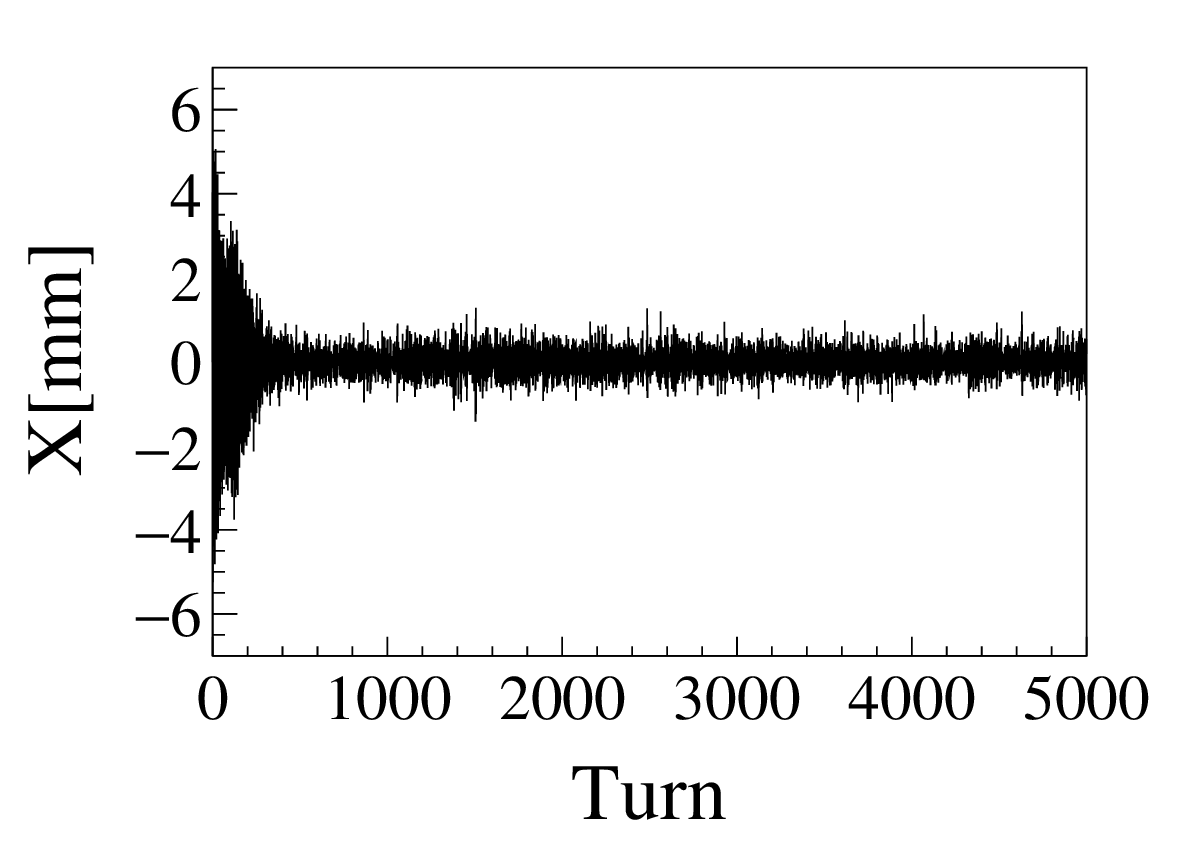}
      \subcaption{}
    \end{minipage} &
    \begin{minipage}[t]{0.20\hsize}
      \centering
      \includegraphics[width=1.5in]{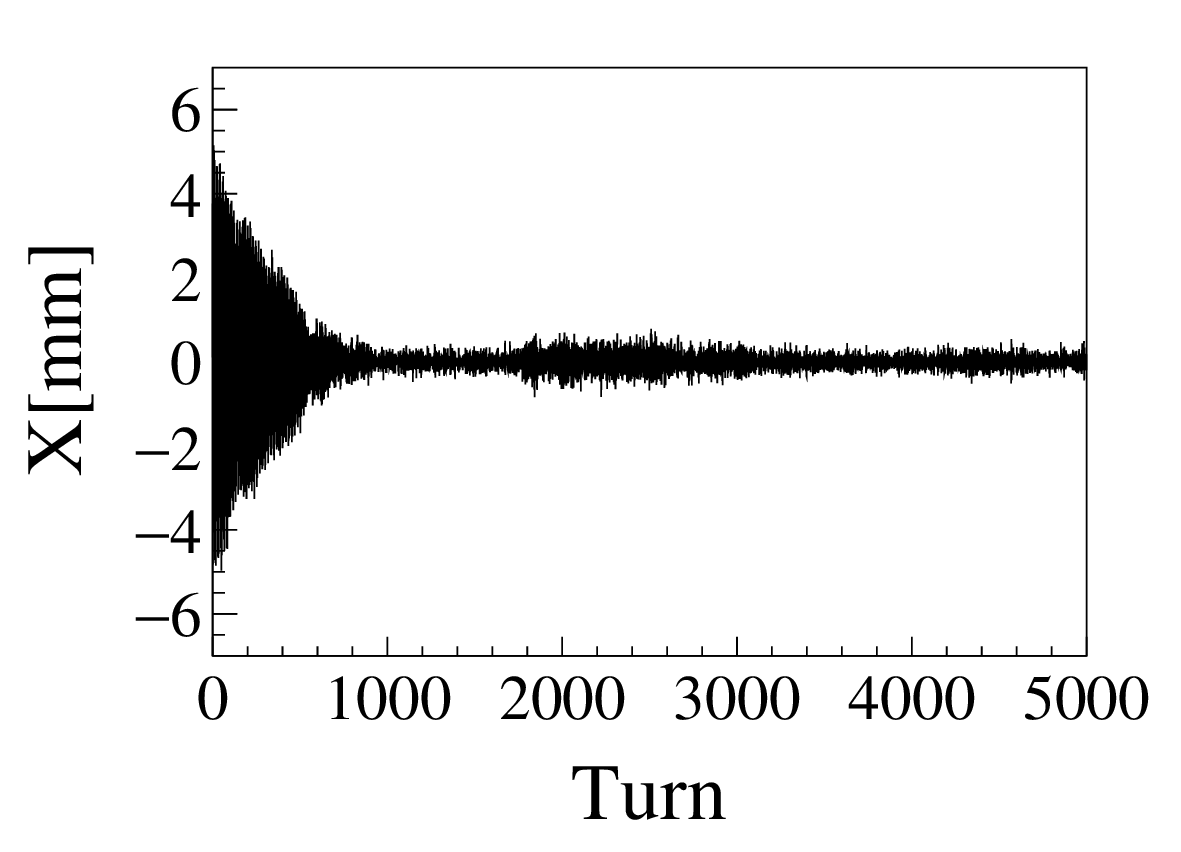}
      \subcaption{}
    \end{minipage} &
    \begin{minipage}[t]{0.20\hsize}
      \centering
      \includegraphics[width=1.5in]{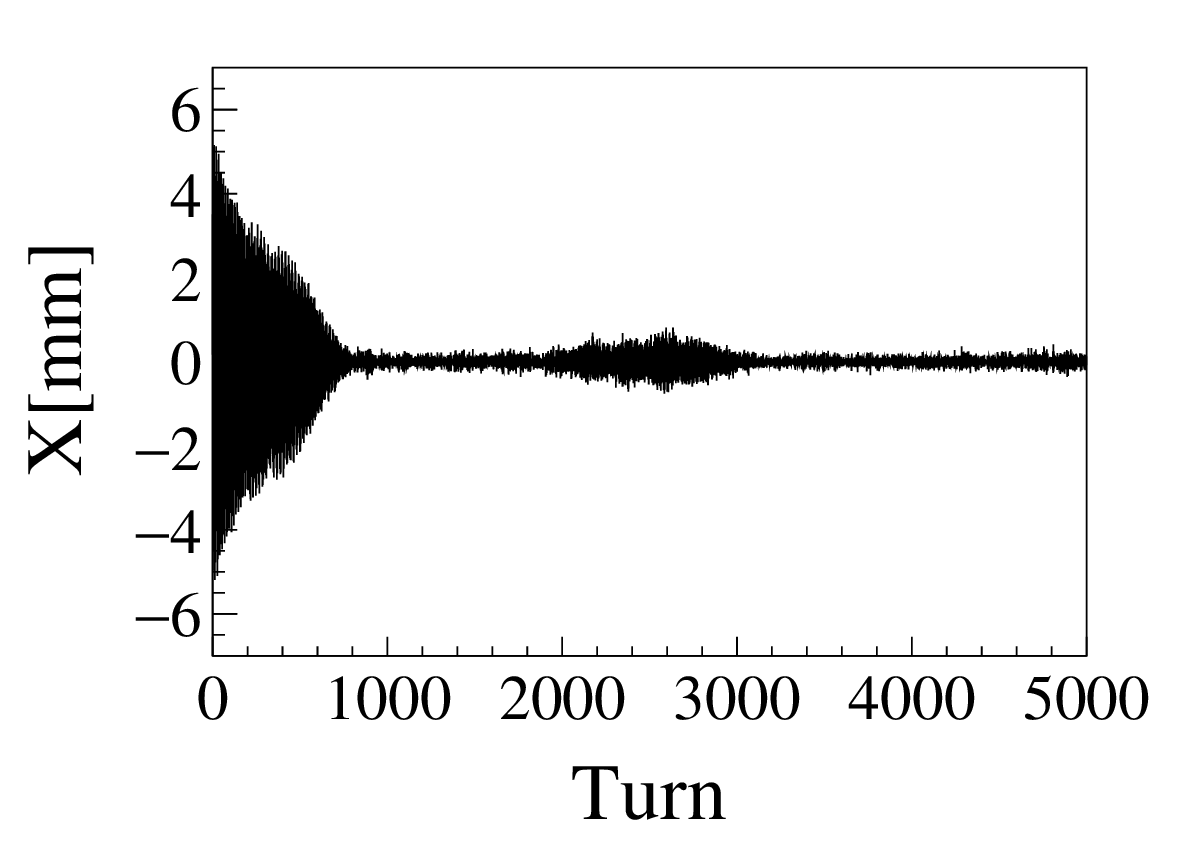}
      \subcaption{}
    \end{minipage} &
    \begin{minipage}[t]{0.20\hsize}
      \centering
      \includegraphics[width=1.5in]{position_230228_185.eps}
      \subcaption{}
    \end{minipage} \\

    \begin{minipage}[t]{0.20\hsize}
      \centering
      \includegraphics[width=1.5in]{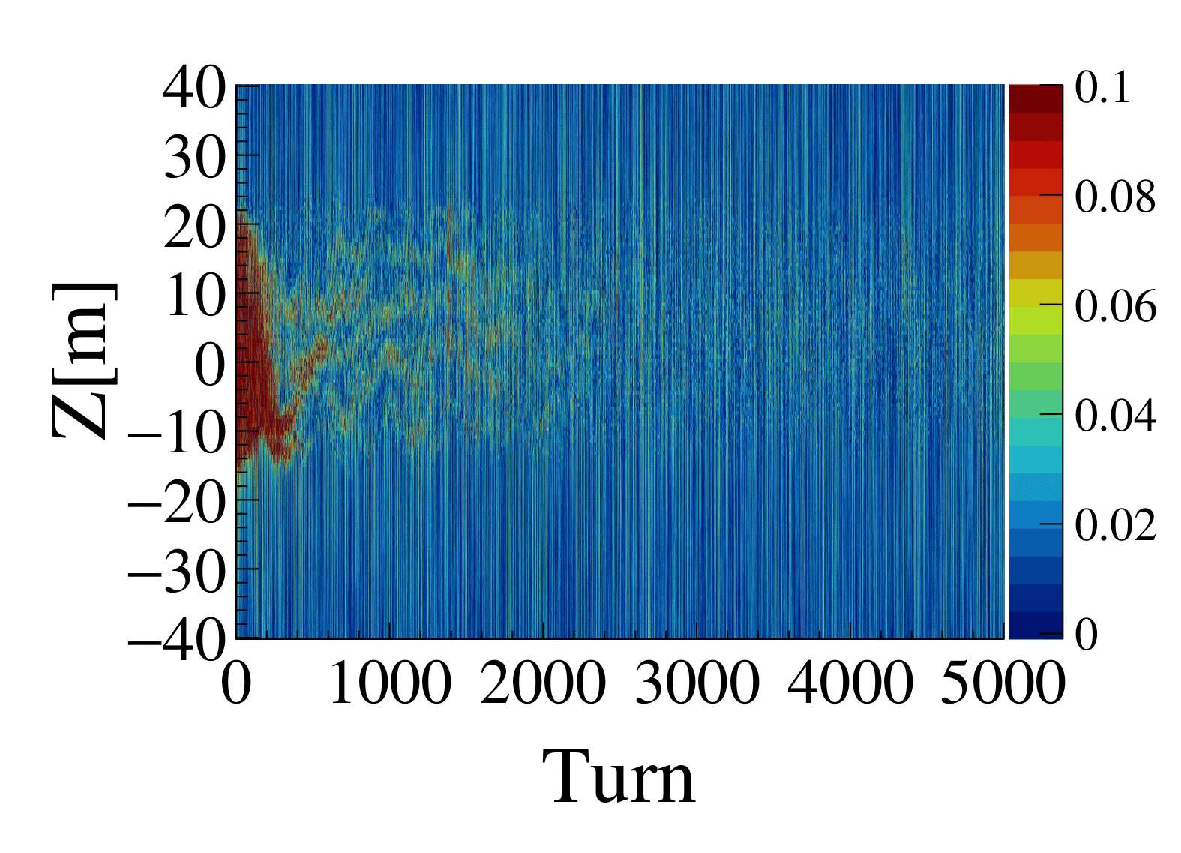}
      \subcaption{}
    \end{minipage} &
    \begin{minipage}[t]{0.20\hsize}
      \centering
      \includegraphics[width=1.5in]{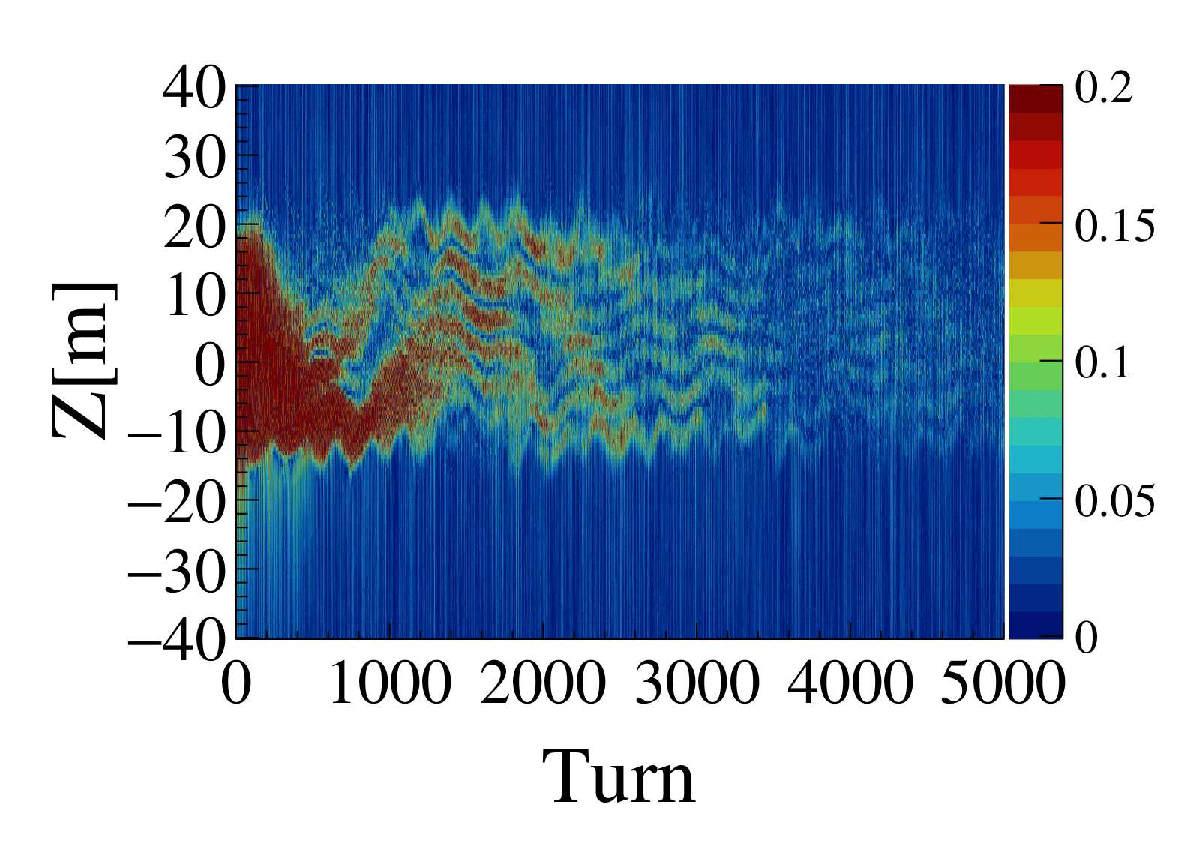}
      \subcaption{}
    \end{minipage} &
    \begin{minipage}[t]{0.20\hsize}
      \centering
      \includegraphics[width=1.5in]{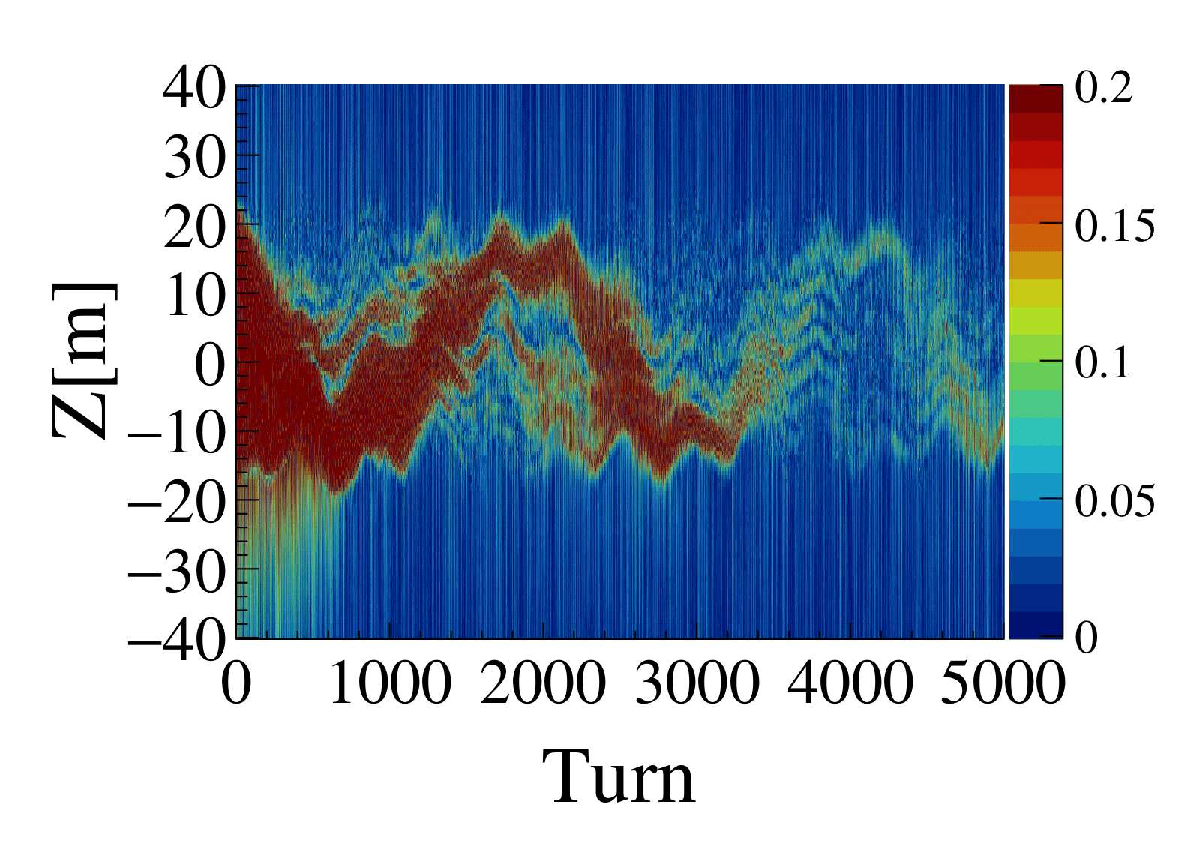}
      \subcaption{}
    \end{minipage}&
    \begin{minipage}[t]{0.20\hsize}
      \centering
      \includegraphics[width=1.5in]{dipoleintra_230228_185.eps}
      \subcaption{}
    \end{minipage}
  \end{tabular}
  \caption{Measurements of the transverse bunch motion in the time domain for the beam intensity: $N_B=1.0\times10^{12}, 2.4\times10^{12}, 4.2\times10^{12}$, and $8.6\times10^{12}$~ppb with the chromaticity $\xi_x=-7.9$ and the rf voltage $\VRF=263$~kV. (a) Frequency component of the dipole moment for $N_B=1.0\times10^{12}$~ppb, $\fmax=50.5$~MHz, $\NAM=1300$, $\NAM /N_s=3.2$, (b) frequency component of the dipole moment for $N_B=2.4\times10^{12}$~ppb, $\fmax=52.2$~MHz, $\NAM=2200$, $\NAM /N_s=5.4$, (c) frequency component of the dipole moment for $N_B=4.2\times10^{12}$~ppb, $\fmax=51.9$~MHz, $\NAM=2100$, $\NAM /N_s=5.1$, (d) frequency component of the dipole moment for $N_B=8.6\times10^{12}$~ppb, $\fmax=53.4$~MHz, $\NAM=1700$, $\NAM /N_s=4.1$, (e) average of the beam position $\overline{x_n}$ for $N_B=1.0\times10^{12}$~ppb, (f) average of the beam position $\overline{x_n}$ for $N_B=2.4\times10^{12}$~ppb, (g) average of the beam position $\overline{x_n}$ for $N_B=4.2\times10^{12}$~ppb, (h) average of the beam position $\overline{x_n}$ for $N_B=8.6\times10^{12}$~ppb, (i) dipole moment $|\Delta(z)|$ in the bunch for $N_B=1.0\times10^{12}$~ppb, (j) dipole moment $|\Delta(z)|$ in the bunch for $N_B=2.4\times10^{12}$~ppb, (k) dipole moment $|\Delta(z)|$ in the bunch for $N_B=4.2\times10^{12}$~ppb, and (l) dipole moment $|\Delta(z)|$ in the bunch for $N_B=8.6\times10^{12}$~ppb}
  \label{fig:22}
\end{figure}

Since, we have experimentally confirmed that the synchrotron period does not have a strong effect on the maximum intrabunch frequency and the normalized recoherence period $\NAM/N_s$ in Sec. \ref{10.3.2}, 
we handle the measured data with different synchrotron periods (rf voltages) simultaneously, from now on.
Accordingly, we will analyze the beam intensity and the chromaticity dependences after combining the results based on Tables \ref{tab:1}, \ref{tab:2}, and \ref{tab:3}.
Figure \ref{fig:22} shows the measurements of the transverse bunch motion in the time domain for the beam intensity: $N_B=1.0\times10^{12},2.4\times10^{12},4.2\times10^{12}$, and $8.6\times10^{12}$~ppb with fixed chromaticity $\xi_x=-7.9$, displaying the low to high-intensity beam results in left to right columns.
Just as in the simulation predictions, the recoherence signal becomes more visible when the beam intensity increases.
The measured values for $(\fmax, \NAM)$ were (50.5~MHz, 1300~turns) for $N_B=1.0\times10^{12}$~ppb, (52.2~MHz, 2200~turns) for $N_B=2.4\times10^{12}$~ppb, (51.9~MHz, 2100~turns) for $N_B=4.2\times10^{12}$~ppb, and (53.4~MHz, 1700~turns) for $N_B=8.6\times10^{12}$~ppb.

\subsubsection{Comparison and discussion}\label{10.4.2}
\begin{figure}[!h]
  \begin{tabular}{cc} 
    \begin{minipage}[t]{0.45\hsize}
      \centering
      \includegraphics[width=3.0in]{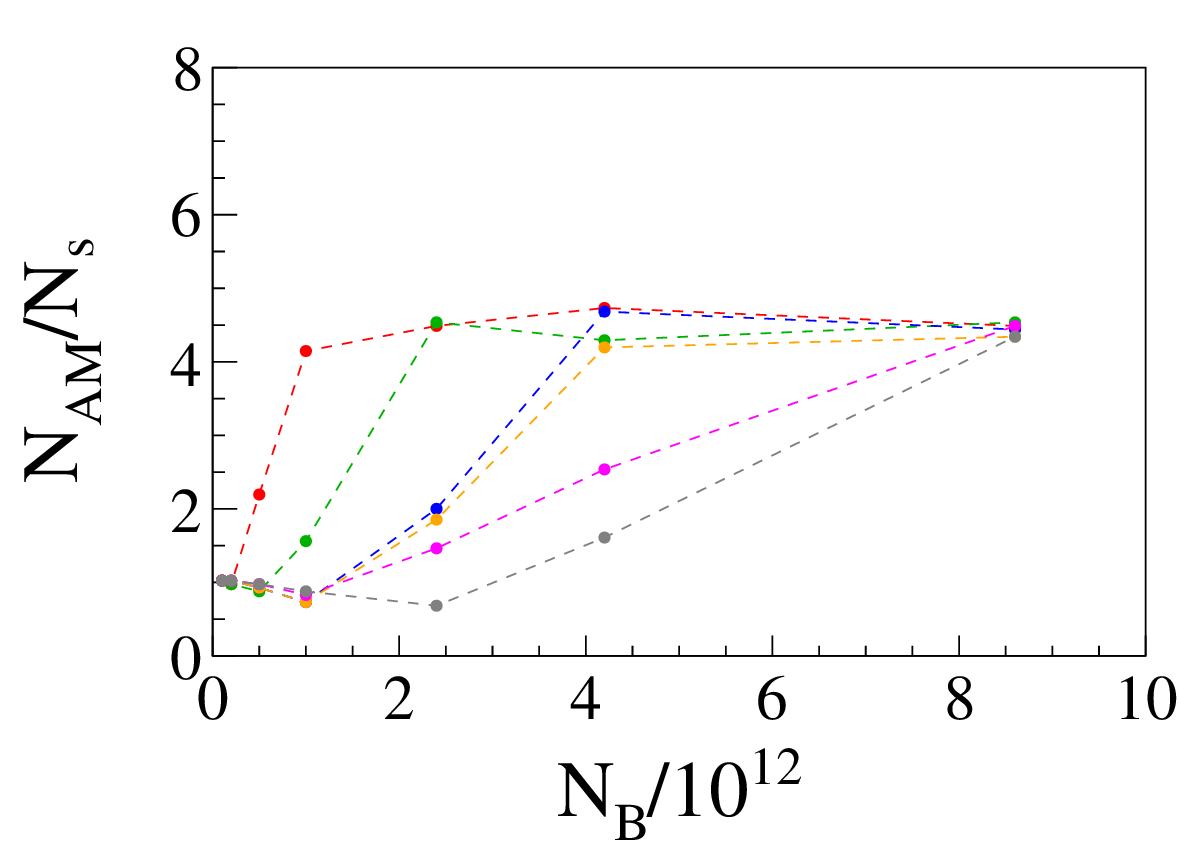}
      \subcaption{}
      \label{fig:23a}
    \end{minipage} &
    \begin{minipage}[t]{0.45\hsize}
      \centering
      \includegraphics[width=3.0in]{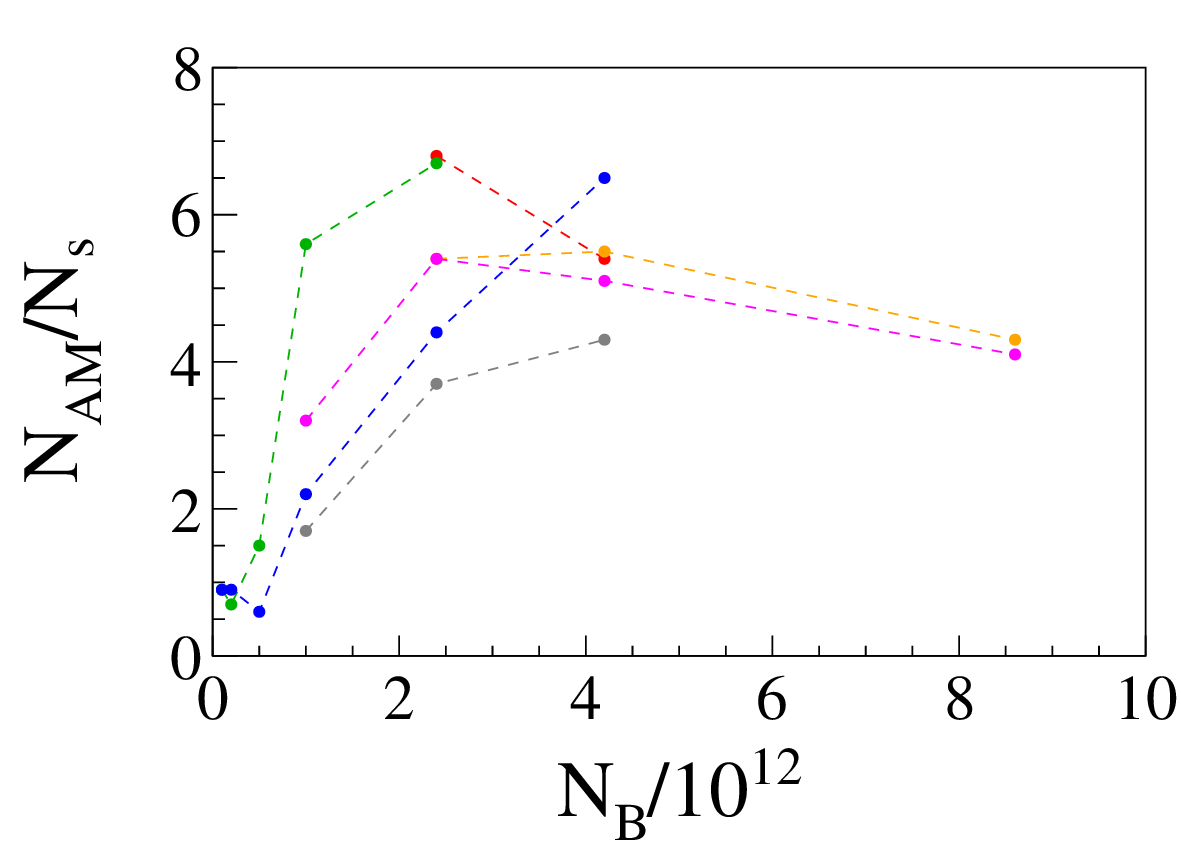}
      \subcaption{}
      \label{fig:23b}
    \end{minipage}
  \end{tabular}
  \caption{Normalized recoherence period $\NAM/N_s$ with several fixed chromaticities for different beam intensities. Here, the red, green, blue, orange, magenta, and gray curves in (a) denote the cases of $\xi_x=-1.5,-3.5,-5.5,-6.5,-8.0$, and $-12.5$, respectively. Here, the red, green, blue, orange, magenta, and gray curves in (b) denote the cases of $\xi_x=-1.5,-3.3,-5.7,-6.5,-7.9$, and $-12.5$, respectively. (a) Tracking simulation and (b) measurement.}
\end{figure}

Figure \ref{fig:23a} shows the simulations of the normalized recoherence period $\NAM /N_s$ for the beam intensity, which saturates for the high intensity extreme, regardless of the chromaticity.
This feature can be interpreted based on the tune spectrum results shown in Fig. \ref{fig:2}, similar to the discussion in the previous subsection.
When the separation of the excited tune shifts across different modes remains nearly identical within the given range of beam intensities, the normalized recoherence period becomes independent of the beam intensity.

For lower chromaticity, the lower head-tail modes are excited, and their degeneration levels rapidly saturate at high beam intensities.
In contrast, for higher chromaticity, the higher head-tail modes are excited, and their degeneration levels do not saturate easily, even at high beam intensities (refer to Fig. \ref{fig:2}).

As a result, achieving saturation of the normalized recoherence period for these beam intensity parameters requires a higher beam intensity in the higher chromaticity condition.
From the simulation point of view, Fig. \ref{fig:23a} reveals that the recoherence period is within $4N_s<\NAM<5N_s$ for the high-intensity beams.
In the J-PARC MR, the recoherence period is expected to lengthen to $1600<\NAM<2000$ due to the space charge effects for a beam characterized by the equal intervals of the head-tail modes $\Delta\nu$, within the range of $0.0005<\Delta\nu<0.0006$, based on the simulation results of Fig. \ref{fig:6d}.

Figure \ref{fig:23b} shows the measurements of the normalized recoherence period for the beam intensity, corresponding to the simulations in Fig. \ref{fig:23a}.
From a measurement perspective, it is challenging to identify $N_f$, where the highest frequency: $\fmax$ appears in the bunch with lower intensity under conditions of lower chromaticity like $\xi_x=-1.5$.
From a theoretical viewpoint, this difficulty arises because multiple lower head-tail modes with different mode differences are simultaneously excited in the low-intensity beam at the lower chromaticity setting as explained in Sec. \ref{7}.
This makes it difficult to define the recoherence period, $\NAM$.
As a result, we exclude the measurement point or the case of $N_B=1.0\times10^{12}$~ppb with $\xi_x=-1.5$ in Fig. \ref{fig:23b}, deeming it unreliable.

The recoherence period appears to approach about $\NAM=N_s$ for the weak beam intensity $N_B=0.13\times10^{12}$~ppb, close to the linear decoherence scenario.
As the beam intensity increases, the normalized recoherence period begins to rise, saturates at around $N_B=2.4\times10^{12}$~ppb, and then tends to decrease for higher-intensity beams in the measurements.

More specifically, Fig. \ref{fig:23b} shows that the saturation number of protons per bunch, at which the normalized recoherence period saturates, depends on the chromaticity.
Overall, the saturation number of protons per bunch increases as the chromaticity becomes higher, consistent with the simulation results in Fig. \ref{fig:23a}.

The decrease in the measured normalized recoherence period in the higher-intensity beams can be attributed to the same reason that explains the drop in normalized recoherence period for the case of 
$N_B=8.6\times10^{12}$~ppb, as shown in Fig. \ref{fig:19b}.
The longitudinal emittances of the injection beam into the J-PARC MR depend on the beam intensity, which enhances the bunching factor for 
$N_B=8.6\times10^{12}$~ppb compared to $N_B=4.2\times10^{12}$~ppb.
This effect effectively weakens the space charge effect on the bunch, resulting in a shortening of the recoherence period compared to the recoherence period of intermediate-strength beams.
As a result, the recoherence period is within the wider range of $4N_s<\NAM<7N_s$ for the high-intensity and high-chromaticity beams in the measurements, unlike the simulations.

\subsection{Chromaticity dependence}\label{10.5}
Finally, the dependence of intrabunch motion on chromaticity is investigated.

\subsubsection{Simulation predictions and measurement results}\label{10.5.1}
\begin{figure}[!h]
  \begin{tabular}{ccc}
    \begin{minipage}[t]{0.3\hsize}
      \centering
      \includegraphics[width=2.0in]{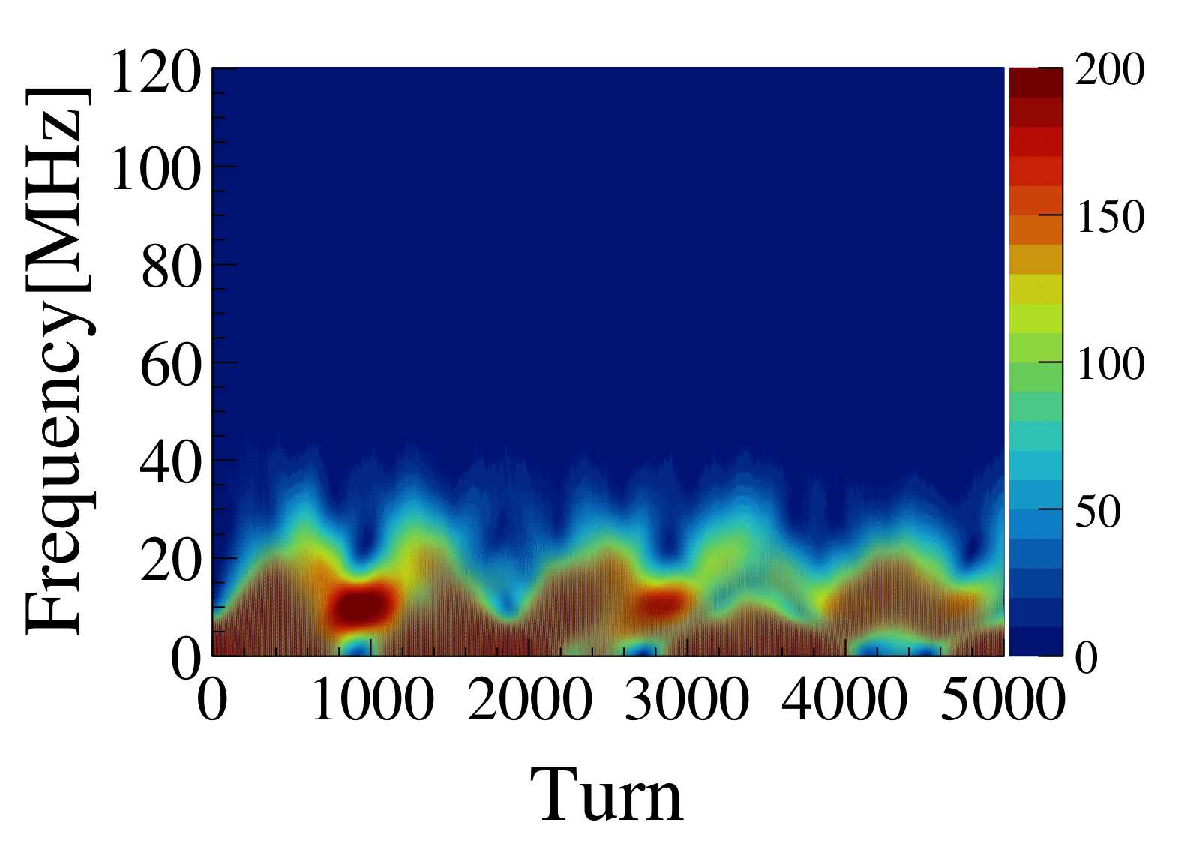}
      \subcaption{}
    \end{minipage} &
    \begin{minipage}[t]{0.3\hsize}
      \centering
      \includegraphics[width=2.0in]{FFT_bunch_scic_263kV_xi80_NB42E11.eps}
      \subcaption{}
    \end{minipage} &
    \begin{minipage}[t]{0.3\hsize}
      \centering
      \includegraphics[width=2.0in]{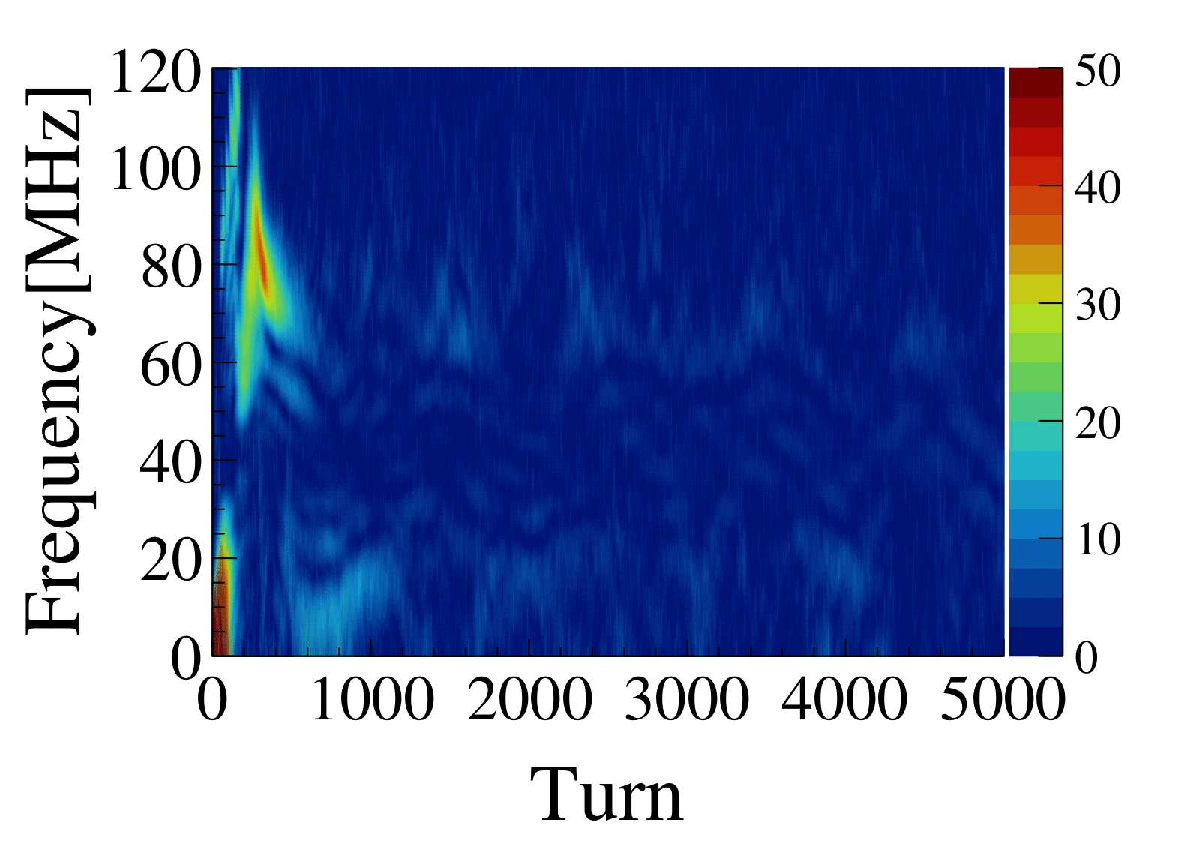}
      \subcaption{}
    \end{minipage} \\

    \begin{minipage}[t]{0.3\hsize}
      \centering
      \includegraphics[width=2.0in]{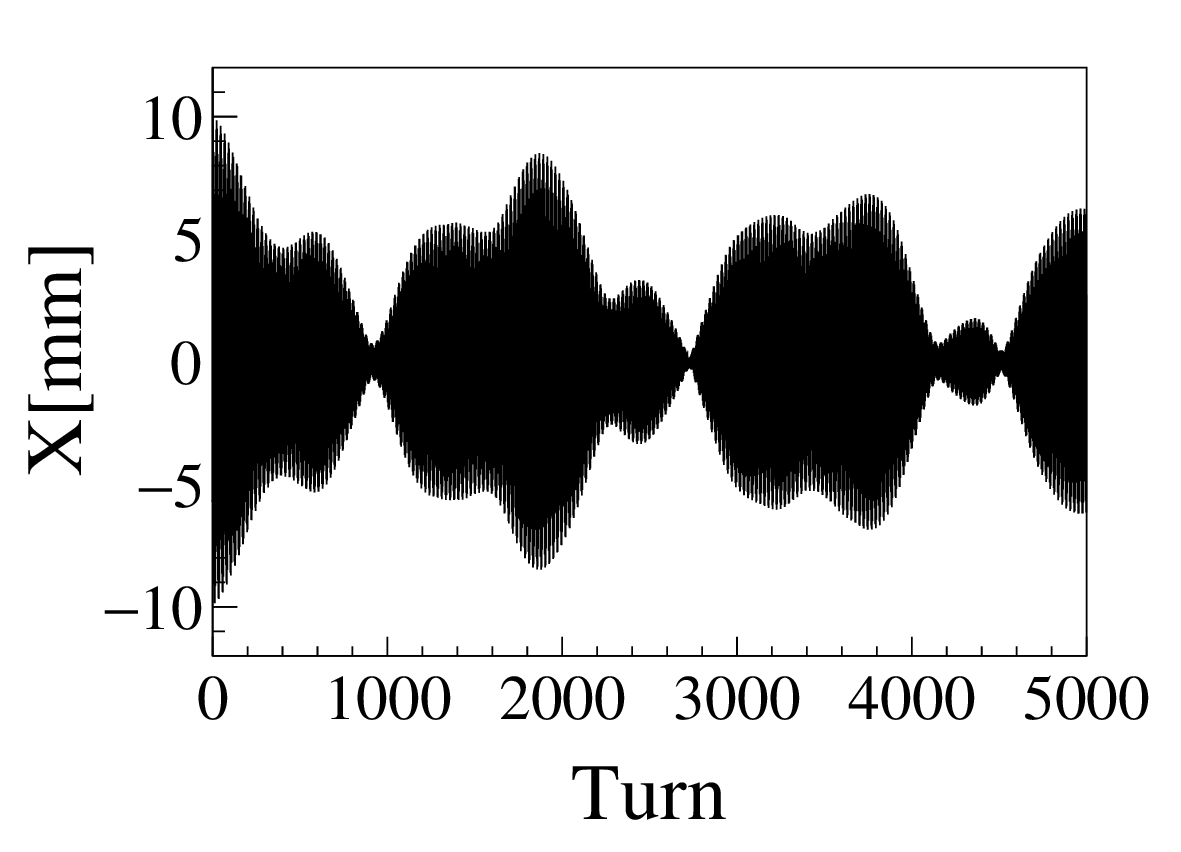}
      \subcaption{}
    \end{minipage} &
    \begin{minipage}[t]{0.3\hsize}
      \centering
      \includegraphics[width=2.0in]{offset_bunch_scic_263kV_xi80_NB42E11.eps}
      \subcaption{}
    \end{minipage} &
    \begin{minipage}[t]{0.3\hsize}
      \centering
      \includegraphics[width=2.0in]{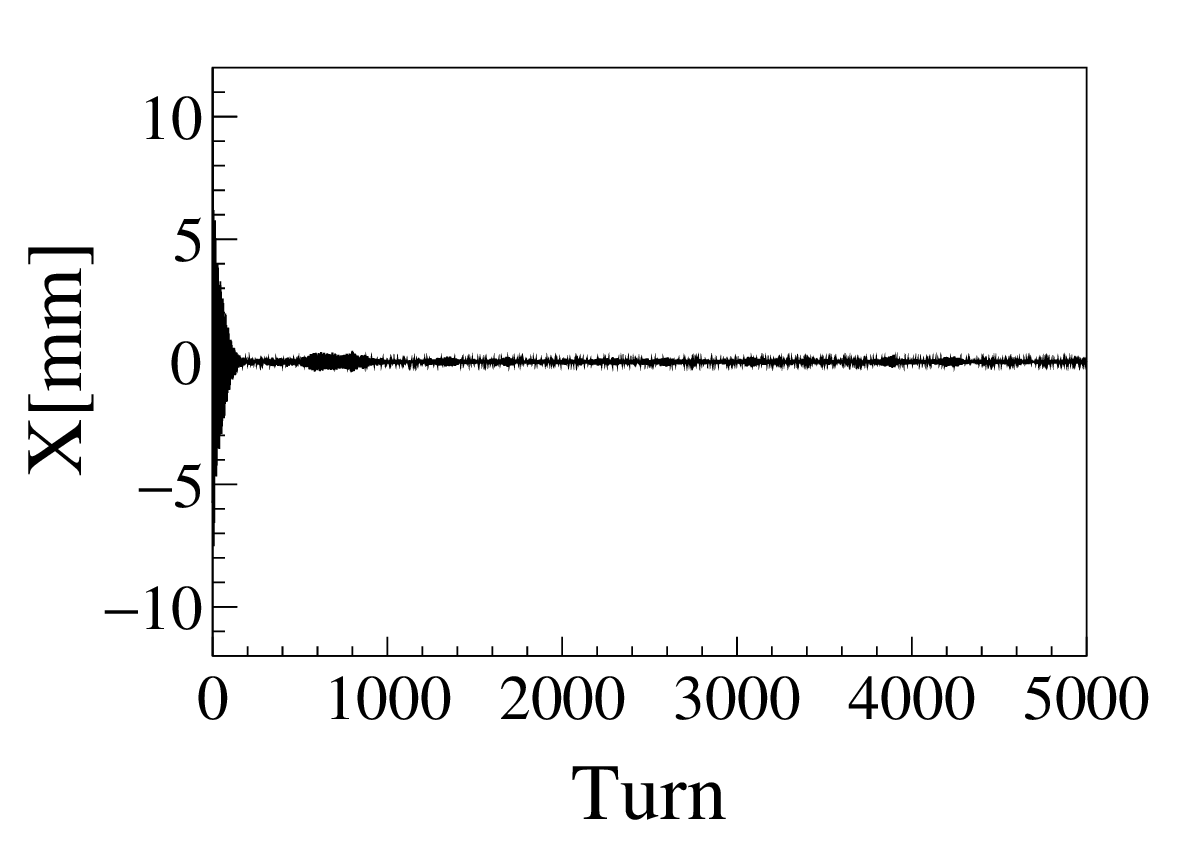}
      \subcaption{}
    \end{minipage} \\

    \begin{minipage}[t]{0.3\hsize}
      \centering
      \includegraphics[width=2.0in]{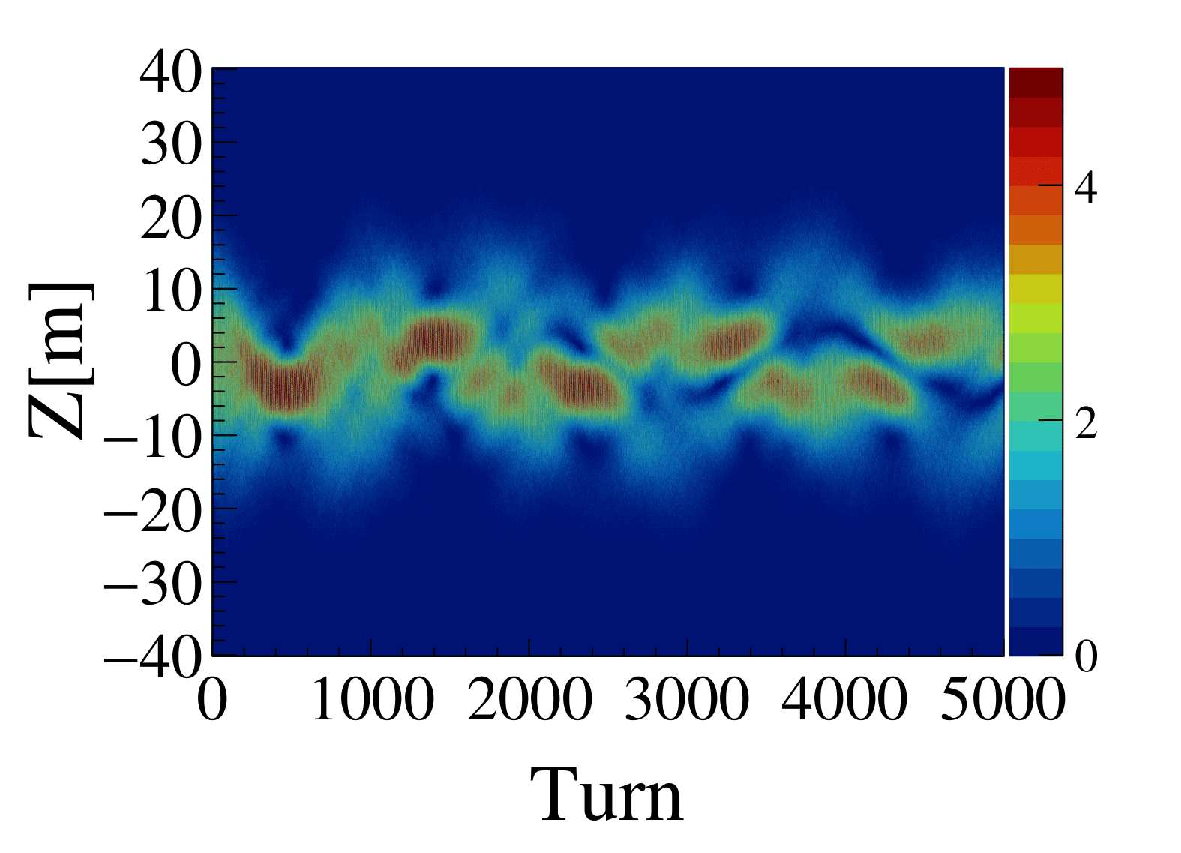}
      \subcaption{}
    \end{minipage} &
    \begin{minipage}[t]{0.3\hsize}
      \centering
      \includegraphics[width=2.0in]{dipole_bunch_scic_263kV_xi80_NB42E11.eps}
      \subcaption{}
    \end{minipage} &
    \begin{minipage}[t]{0.3\hsize}
      \centering
      \includegraphics[width=2.0in]{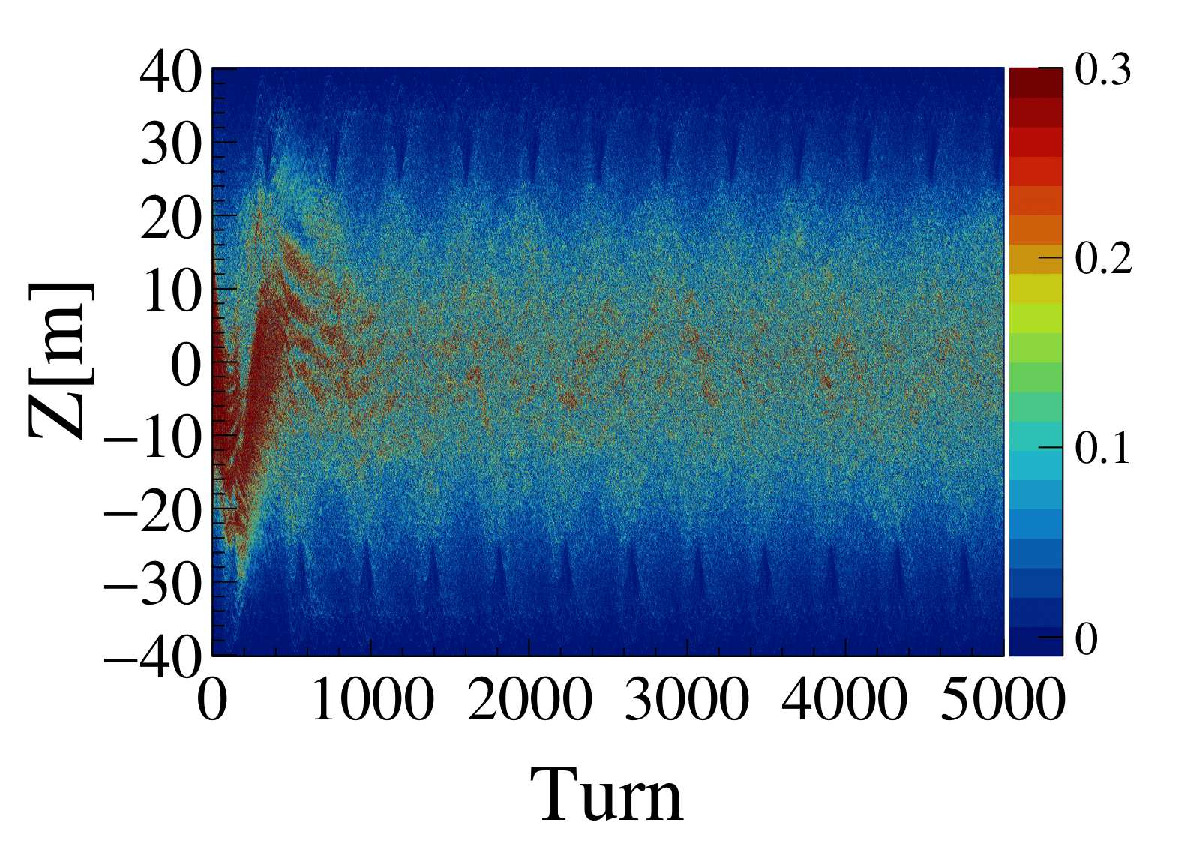}
      \subcaption{}
    \end{minipage}
  \end{tabular}
  \caption{Simulation of the transverse bunch motion in the time domain for the chromaticity: $\xi_x=-1.5,-8.0$, and $-12.5$ with the beam intensity $N_B=4.2\times10^{12}$~ppb and the rf voltage $\VRF=263$~kV. (a) Frequency component of the dipole moment for $\xi_x=-1.5$, $\fmax=9.5$~MHz, $\NAM=1940$, $\NAM /N_s=4.7$, (b) frequency component of the dipole moment for $\xi_x=-8.0$, $\fmax=51.3$~MHz, $\NAM=1040$, $\NAM /N_s=2.5$, (c) frequency component of the dipole moment for $\xi_x=-12.5$, $\fmax=76.5$~MHz, $\NAM=660$, $\NAM /N_s=1.6$, (d) average of the beam position $\overline{x_n}$ for $\xi_x=-1.5$, (e) average of the beam position $\overline{x_n}$ for $\xi_x=-8.0$, (f) average of the beam position $\overline{x_n}$ for $\xi_x=-12.5$, (g) dipole moment $|\Delta^{(k)}\sigma_\Delta/q|$ in the bunch for $\xi_x=-1.5$, (h) dipole moment $|\Delta^{(k)}\sigma_\Delta/q|$ in the bunch for $\xi_x=-8.0$, and (i) dipole moment $|\Delta^{(k)}\sigma_\Delta/q|$ in the bunch for $\xi_x=-12.5$.}
  \label{fig:24}
\end{figure}

We simulate the case where the chromaticity is scanned for the parameter set of rf voltage and particles per bunch described in the previous subsection.
Figure \ref{fig:24} shows the transverse bunch motion in the time domain
for the chromaticity: $\xi_x=-1.5,-8.0$, and $-12.5$ with the fixed beam intensity $N_B=4.2\times10^{12}$~ppb.
The extracted values for $(\fmax, \NAM)$ were (53.3~MHz, 340~turns) for $N_B=1.0\times10^{12}$~ppb, (9.5~MHz, 1940~turns) for $\xi_x=-1.5$, (51.3~MHz, 1040~turns) for $\xi_x=-8.0$, and (76.5~MHz, 660~turns) for $\xi_x=-12.5$.

\begin{figure}[!h]
  \begin{tabular}{ccc}
    \begin{minipage}[t]{0.3\hsize}
      \centering
      \includegraphics[width=2.0in]{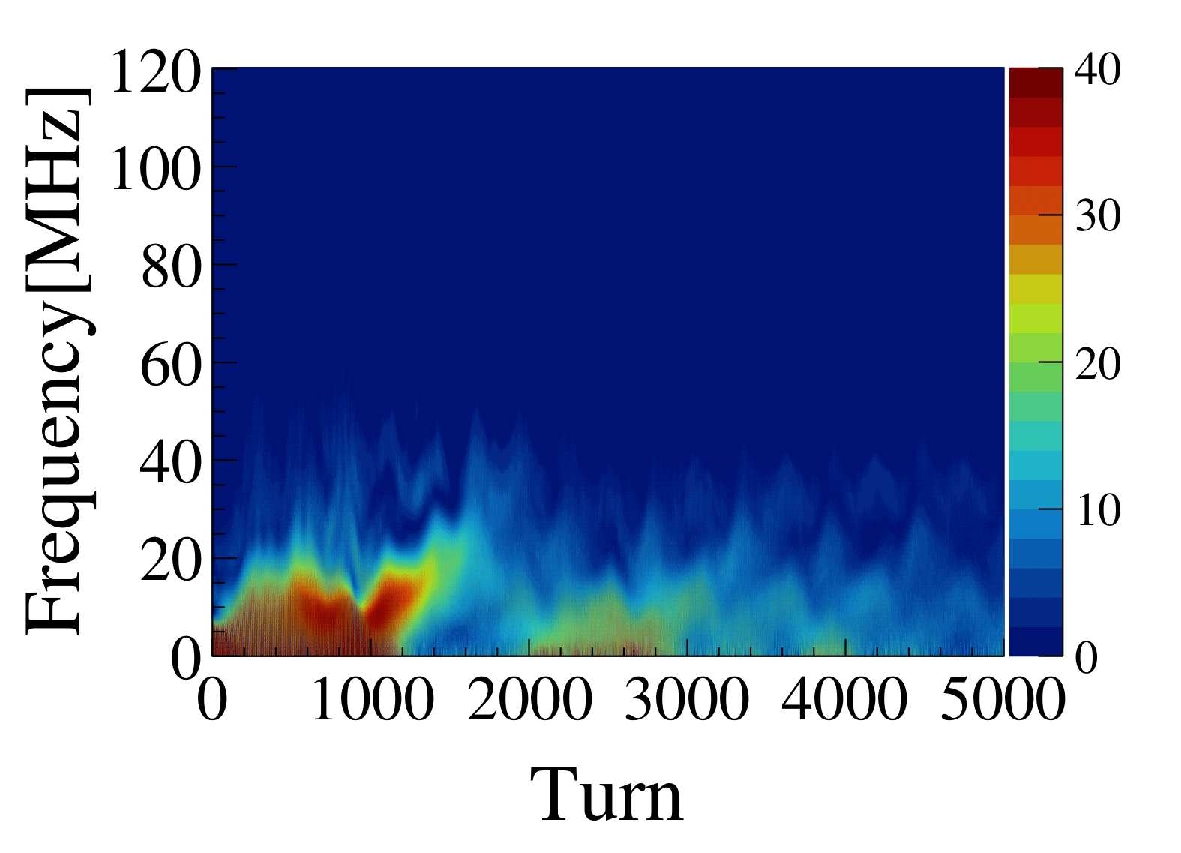}
      \subcaption{}
    \end{minipage} &
    \begin{minipage}[t]{0.3\hsize}
      \centering
      \includegraphics[width=2.0in]{FFT_230228_94.eps}
      \subcaption{}
    \end{minipage} &
    \begin{minipage}[t]{0.3\hsize}
      \centering
      \includegraphics[width=2.0in]{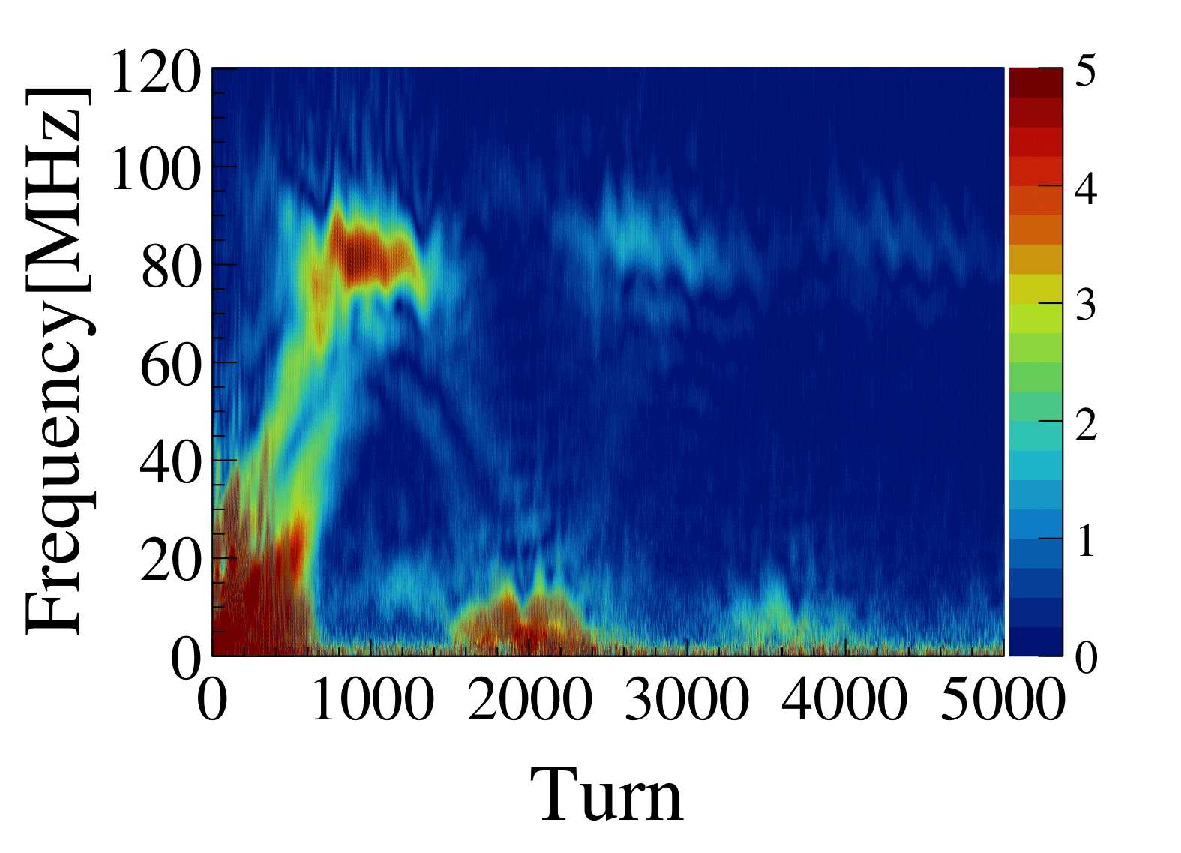}
      \subcaption{}
    \end{minipage} \\

    \begin{minipage}[t]{0.3\hsize}
      \centering
      \includegraphics[width=2.0in]{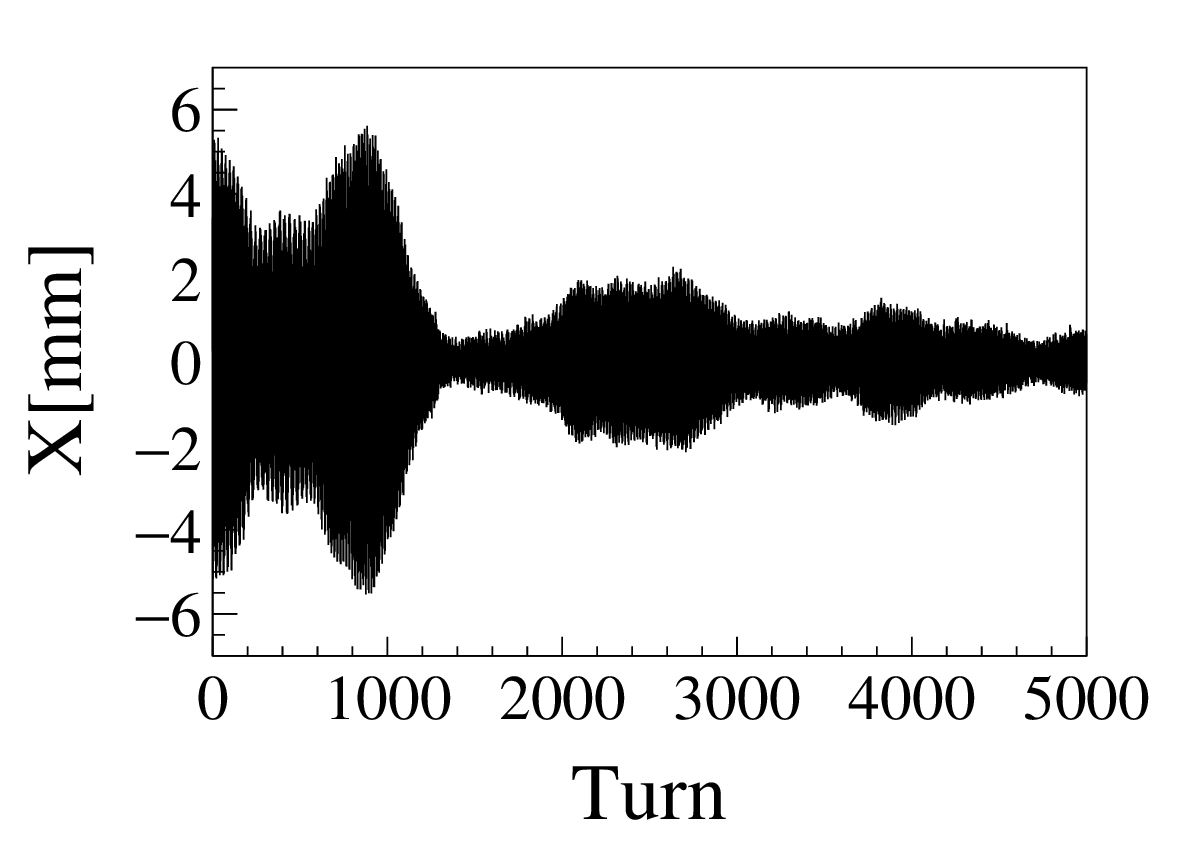}
      \subcaption{}
    \end{minipage} &
    \begin{minipage}[t]{0.3\hsize}
      \centering
      \includegraphics[width=2.0in]{position_230228_94.eps}
      \subcaption{}
    \end{minipage} &
    \begin{minipage}[t]{0.3\hsize}
      \centering
      \includegraphics[width=2.0in]{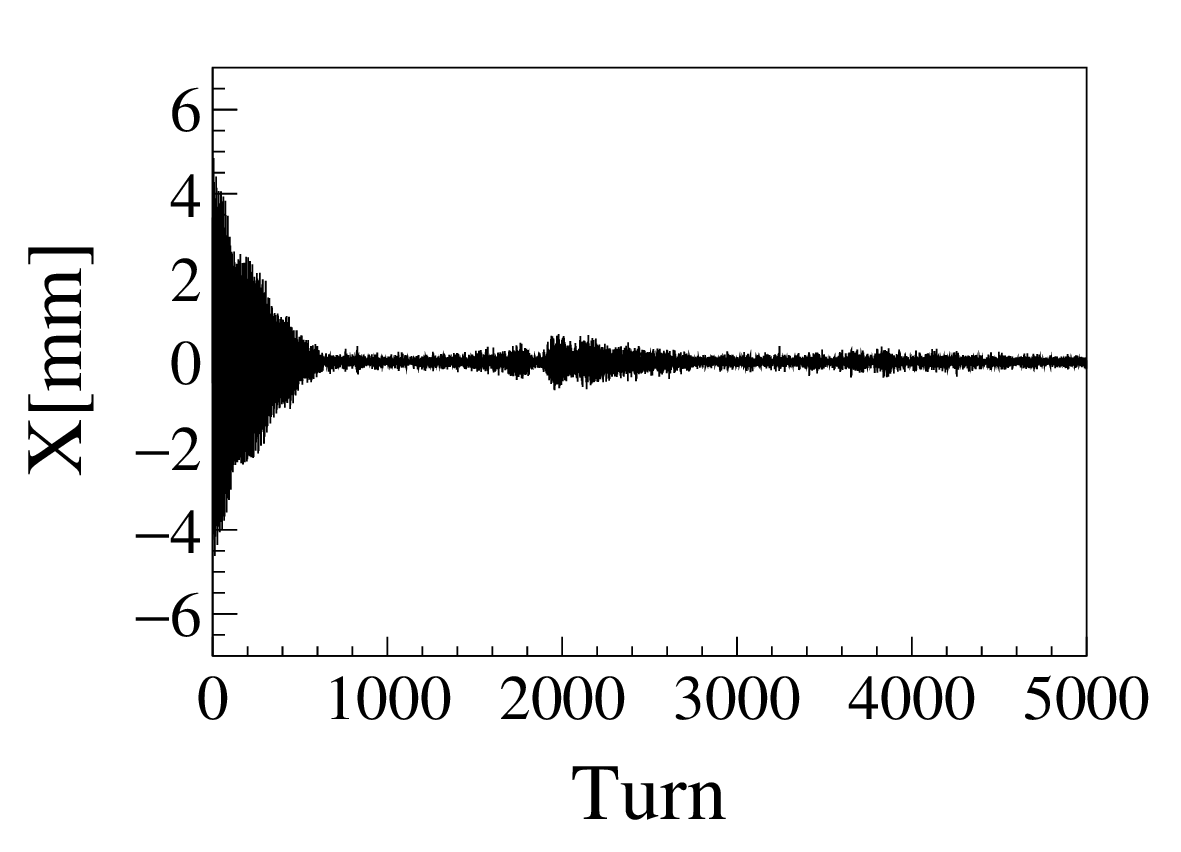}
      \subcaption{}
    \end{minipage} \\

    \begin{minipage}[t]{0.3\hsize}
      \centering
      \includegraphics[width=2.0in]{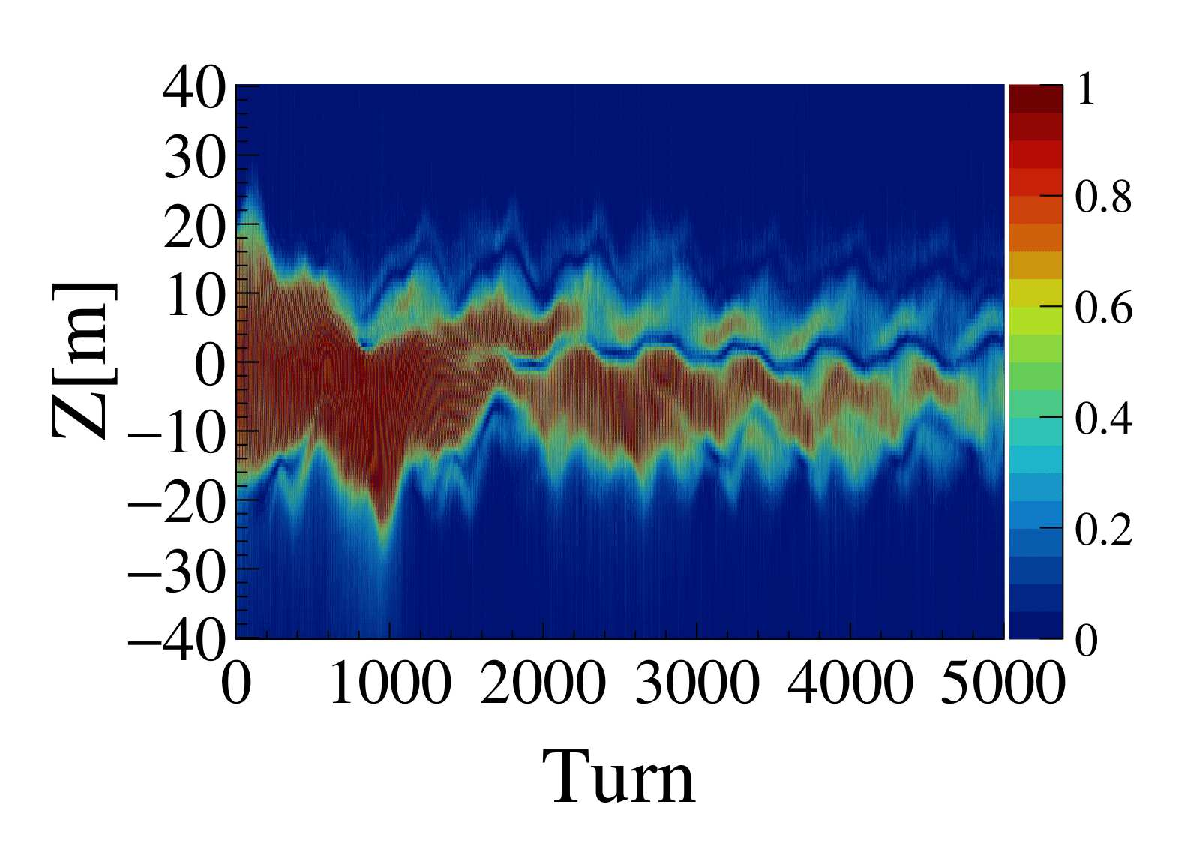}
      \subcaption{}
    \end{minipage} &
    \begin{minipage}[t]{0.3\hsize}
      \centering
      \includegraphics[width=2.0in]{dipoleintra_230228_94.eps}
      \subcaption{}
    \end{minipage} &
    \begin{minipage}[t]{0.3\hsize}
      \centering
      \includegraphics[width=2.0in]{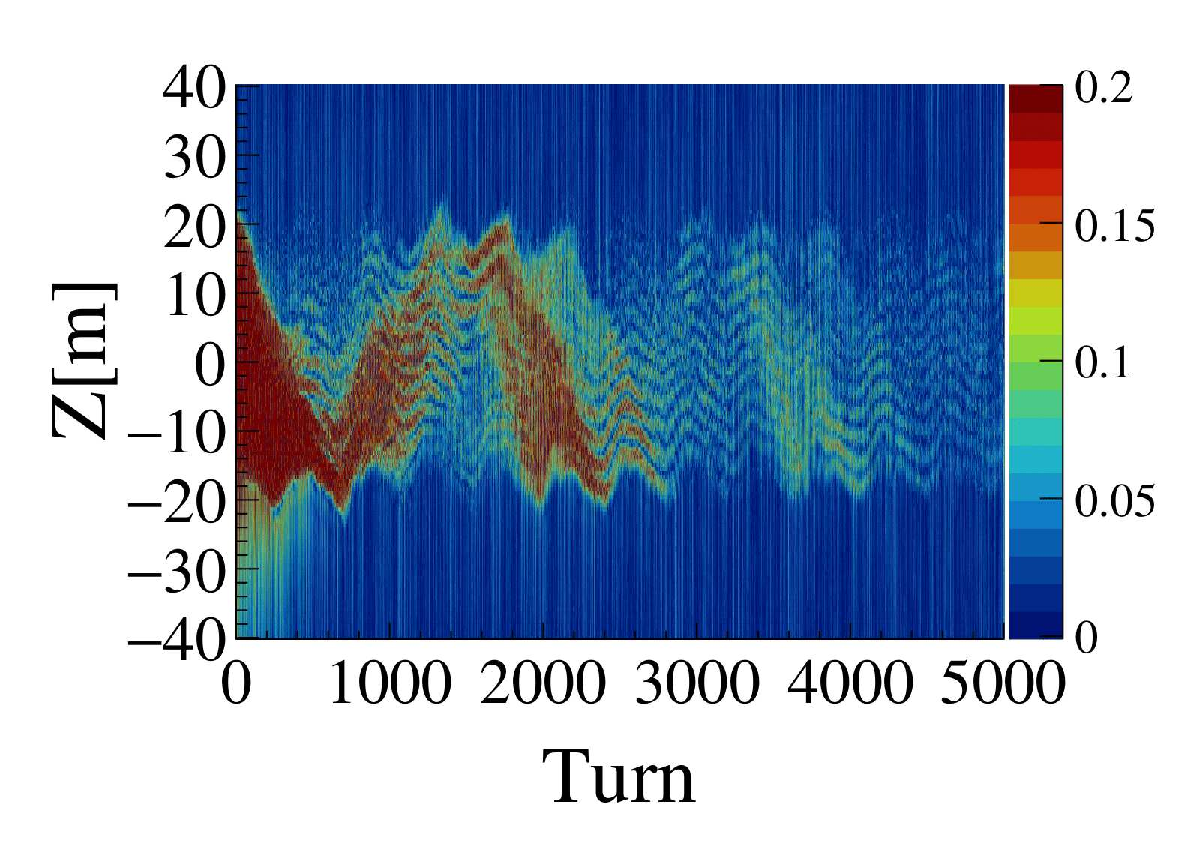}
      \subcaption{}
    \end{minipage}
  \end{tabular}
  \caption{Measurements of the transverse bunch motion in the time domain for the chromaticity: $\xi_x=-1.5,-7.9$, and $-12.5$ with the beam intensity $N_B=4.2\times10^{12}$~ppb and the rf voltage $\VRF=263$~kV. (a) Frequency component of the dipole moment for $\xi_x=-1.5$, $\fmax=10$~MHz, $\NAM=2200$, $\NAM /N_s=5.4$, (b) frequency component of the dipole moment for $\xi_x=-7.9$, $\fmax=51.9$~MHz, $\NAM=2100$, $\NAM /N_s=5.1$, (c) frequency component of the dipole moment for $\xi_x=-12.5$, $\fmax=81.3$~MHz, $\NAM=1800$, $\NAM /N_s=4.2$, (d) average of the beam position $\overline{x_n}$ for $\xi_x=-1.5$, (e) average of the beam position $\overline{x_n}$ for $\xi_x=-7.9$, (f) average of the beam position $\overline{x_n}$ for $\xi_x=-12.5$, (g) dipole moment $|\Delta(z)|$ in the bunch for $\xi_x=-1.5$, (h) dipole moment $|\Delta(z)|$ in the bunch for $\xi_x=-7.9$, and (i) dipole moment $|\Delta(z)|$ in the bunch for $\xi_x=-12.5$.}
  \label{fig:25}
\end{figure}

Figure \ref{fig:25} shows the measured transverse bunch motion for various chromaticities: $\xi_x=-1.5,-7.9$, and $-12.5$ with fixed beam intensity $N_B=4.2\times10^{12}$~ppb in the time domain.
The measured values for $(\fmax, \NAM)$ were (10~MHz, 2200~turns) for $\xi_x=-1.5$, (51.9~MHz, 2100~turns) for $\xi_x=-7.9$, and (81.3~MHz, 1800~turns) for $\xi_x=-12.5$.

\subsubsection{Comparison and discussion}\label{10.5.2}
\begin{figure}[!h]
  \begin{tabular}{cc}
    \begin{minipage}[t]{0.45\hsize}
      \centering
      \includegraphics[width=3.0in]{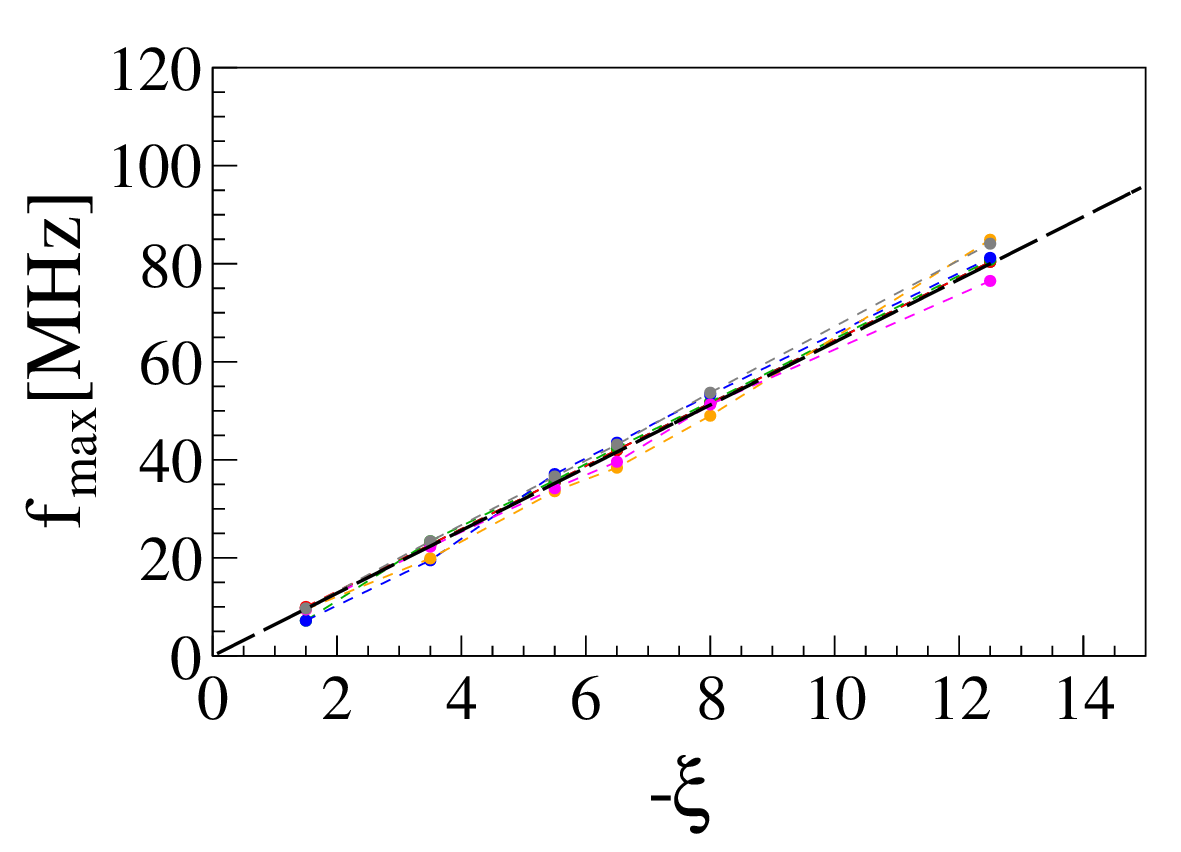}
      \subcaption{}
      \label{fig:26a}
    \end{minipage} &
    \begin{minipage}[t]{0.45\hsize}
      \centering
      \includegraphics[width=3.0in]{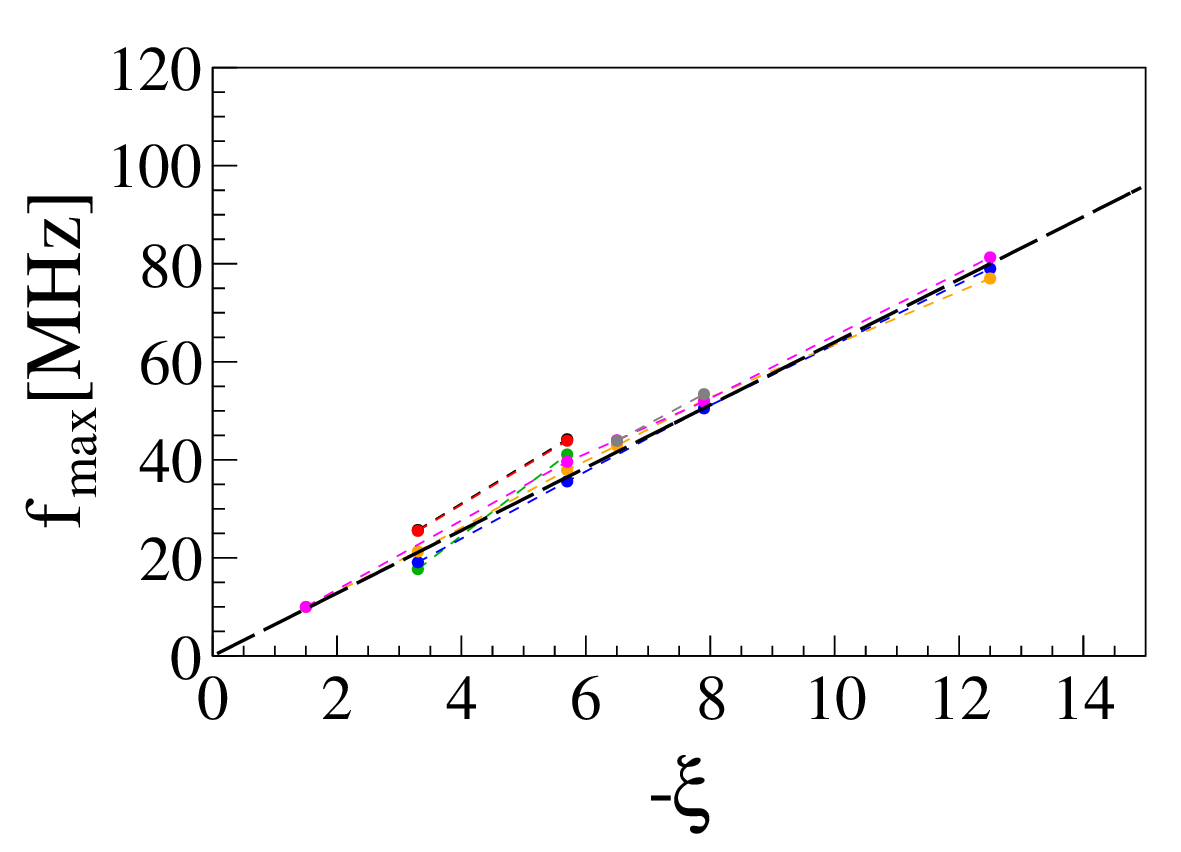}
      \subcaption{}
      \label{fig:26b}
    \end{minipage}
  \end{tabular}
  \caption{Maximum intrabunch frequency $\fmax$ with fixed beam intensity for different chromaticities. In (a), the black, red, green, blue, orange, magenta, and gray lines denote the cases of $N_B=1.0\times10^{11}, 2.0\times10^{11}, 5.0\times10^{11}, 1.0\times10^{12}, 2.4\times10^{12}, 4.2\times10^{12}$, and $8.6\times10^{12}$~ppb, respectively. In (b), the black, red, green, blue, orange, magenta, and gray lines denote the cases of $N_B=1.3\times10^{11}, 2.6\times10^{11}, 5.2\times10^{11}, 1.0\times10^{12}, 2.4\times10^{12}, 4.2\times10^{12}$, and $8.6\times10^{12}$~ppb, respectively. The black dashed lines in both figures are the result of Eq. \eqref{eq:140}. (a) Tracking simulation and (b) measurement.}
\end{figure}

Figure \ref{fig:26a} shows the maximum intrabunch frequency $\fmax$ dependence for the chromaticity in the tracking simulation for various beam intensities.
Regardless of the space charge effect, it is clear that the maximum intrabunch frequency follows $\fmax=6.4|\xi_x|$~MHz (refer to Eq. \eqref{eq:140} denoted by the black dashed line, identical to Eq. \eqref{eq:22} derived under the assumption of weak space charge effects.) and is proportional to the chromaticity because all the data points collapse onto a single line.

Figure \ref{fig:26b} shows the maximum intrabunch frequency $\fmax$ dependence for the chromaticity with the measurement data.
The black dashed line represents the analytical result given by Eq. \eqref{eq:140}.
The simulations in Fig. \ref{fig:26a} are in excellent agreement with the measurement in Fig. \ref{fig:26b}.
We observe the maximum intrabunch frequency follows the analytical result: $\fmax=6.4|\xi_x|$~MHz.
In the end, we find that the measurements confirm that space charge effects do not influence the relationship between the maximum intrabunch frequency and chromaticity, as predicted by the simulations.

\section{CONCLUSION}\label{11}
In this paper, we investigate the direct and indirect space charge effects on the intrabunch motion across wider chromatic phase regions, by examining the dipole excitation in detail, 
using theoretical analysis, simulations, and experimental measurements.
During the development of the simulation code, we clarify the limitations of applying the linear approximation to the space charge effects, namely that the beam size must be significantly smaller than the chamber radius.
We find that neither the longitudinal \cite{yoshimura2025space} nor the transverse impedance significantly influences the motions in the J-PARC MR unless beam instability occurs.

The consistency of our code to the previous studies is verified by examining the excitation patterns of radial and head-tail modes.
When space charge effects are neglected, the simulation reproduces beam instabilities consistent with the results of a well-established Vlasov solver.
Furthermore, we demonstrate that the indirect space charge effect suppresses beam instability, whereas the direct space charge effect enhances the growth rate.
As a result, our simulations, developed through careful examination of the fundamental aspects of the space charge effects, successfully reconcile two differing predictions regarding beam instability in the presence of the space charge.

Additionally, we reestablish the ABS theory to describe the space charge effects on the intrabunch motion from a fundamental perspective and extend it to partially account for a more realistic Gaussian beam within a harmonic potential model.
This theoretical framework allows us to better understand simulation features, including the extended recoherence time and the degeneration of head-tail modes toward the high-intensity beams.
The simulation code and theory are compared in the weak space charge region, showing excellent agreement that supports the extension of the simulations to stronger space charge regions.

In the ABS model, the head-tail modes depend on the chromaticity setting.
Higher head-tail modes survive in regions of higher chromaticity, which aligns with experimental observations.
However, in the simulations using a Gaussian beam with a harmonic potential, all head-tail modes are excited when the space charge effects are neglected.
This discrepancy with the observation is resolved when the strong space charge effects are considered in the harmonic potential model, especially in higher chromaticity regions.
Overall, the ABS model effectively describes simulations for Gaussian beams within a harmonic potential.
As a result, we can understand the simulations based on the theory.

For the high-intensity beams, the recoherence time saturates, because the indirect space charge effect mitigates the degeneration of head-tail modes.
In other words, the recoherence time becomes independent of beam intensity in higher chromaticity regions because the higher head-tail modes are excited and hard to degenerate.

Strictly speaking, it is challenging to precisely predict experimental outcomes because the longitudinal and transverse emittances of injection beams from upstream facilities vary with beam intensity.
For instance, if the emittances are broadened from the outset, the space charge effects on intrabunch motion are effectively mitigated during the storage mode in the ring, leading to a shortened recoherence time compared to the expected saturated value by the simulation.
Nevertheless, the relationship between the maximum frequency in the bunch and the chromaticity remains unaffected by space charge effects.

In high-intensity proton rings such as the J-PARC MR, the space charge parameter is large and the chromatic phase has a broad range to suppress beam instability, so it is necessary to understand the basics of intrabunch motion for stable operation.
We find a phenomenon in which the high-frequency component rises gradually as the beam intensity becomes higher.
If this characteristic is kept to the case when the beam instability occurs, it is quite beneficial to efficiently activate the IBFB.
Therefore, the next goal is to use this simulation to investigate the mechanism of beam instability at high beam intensities to determine the optimal conditions for the IBFB when the beam instability is excited at the J-PARC MR.

\section*{ACKNOWLEDGMENT}
We would like to express special gratitude to Professor T. Nakaya, who recommended the first author for research in accelerator science at the J-PARC as supervisor, for reading this paper critically and constructively.
We thank Dr. A. Kobayashi and Mr. M. Okada from the monitor group for the adjustments to the BPM and data acquisition system and for many helpful discussions.
The J-PARC MR commissioning group, with Professor Y. Sato and Dr. T. Yasui, is also acknowledged for coordinating beam conditions and the test schedule.
We also thank Dr. F. Tamura, Dr. Y. Sugiyama, and Dr. H. Okita from the rf group for their help with rf matching conditions, 
and Professor K. Ohmi for his involvement in setting the direction of this research at the early stage.
Finally, we would like to thank all members of the J-PARC Center for their hospitality and help.

This work was supported by JSPS KAKENHI Grants No. JP23H05434 and No. JP18H05537, 
JST SPRING, Grant No. JPMJSP2110,
and JST---the establishment of university fellowships toward the creation of science technology innovation, Grant No. JPMJFS2123.

\section*{DATA AVAILABILITY}
The data that support the findings of this article are openly available \cite{yoshimura2025zenodo}.

\bibliographystyle{unsrt}
\bibliography{yoshimura_250822.bib}

\begin{thebibliography}{10}

\bibitem{benedikt2004lhc}
M.~Benedikt, P.~Collier, V.~Mertens, J.~Poole, and K.~Schindl.
\newblock {LHC Design Report, volume III: The LHC injector chain}.
\newblock Technical report, 2004.

\bibitem{none1994fermilab}
{The Fermilab Main Injector Technical Design Handbook}.
\newblock Technical Report FERMILAB-DESIGN-1994-01, Fermilab, Batavia, IL, USA, 1994.

\bibitem{none1966bnl}
{Construction Completion Report Alternating Gradient Synchrotron Project}.
\newblock Technical report, Brookhaven National Laboratory, Upton, NY, USA, 1966.

\bibitem{kornilov2012transverse}
V.~Kornilov and O.~Boine-Frankenheim.
\newblock {Transverse decoherence and coherent spectra in long bunches with space charge}.
\newblock {\em Physical Review Special Topics-Accelerators and Beams}, 15(11):114201, 2012.

\bibitem{singh2013interpretation}
R.~Singh, O.~Boine-Frankenheim, O.~Chorniy, P.~Forck, R.~Haseitl, W.~Kaufmann, P.~Kowina, K.~Lang, and T.~Weiland.
\newblock {Interpretation of transverse tune spectra in a heavy-ion synchrotron at high intensities}.
\newblock {\em Physical Review Special Topics-Accelerators and Beams}, 16(3):034201, 2013.

\bibitem{karpov2016early}
I.~Karpov, V.~Kornilov, and O.~Boine-Frankenheim.
\newblock {Early transverse decoherence of bunches with space charge}.
\newblock 19(12):124201, 2016.

\bibitem{kornilov2010head}
V.~Kornilov and O.~Boine-Frankenheim.
\newblock {Head-tail instability and Landau damping in bunches with space charge}.
\newblock {\em Physical Review Special Topics-Accelerators and Beams}, 13(11):114201, 2010.

\bibitem{kornilov2010simulation}
V.~Kornilov and O.~Boine-Frankenheim.
\newblock {Simulation studies \& code validation for the head-tail instability with space charge}.
\newblock In {\em Proceedings of the 10th International Computational Accelerator Physics Conference}, volume San Francisco, CA, USA, pages 58--63, 2010.

\bibitem{boine2009transverse}
O.~Boine-Frankenheim and V.~Kornilov.
\newblock {Transverse Schottky noise spectrum for bunches with space charge}.
\newblock {\em Physical Review Special Topics-Accelerators and Beams}, 12(11):114201, 2009.

\bibitem{boine2006implementation}
O.~Boine-Frankenheim and V.~Kornilov.
\newblock {Implementation and validation of space charge and impedance kicks in the code PATRIC for studies of transverse coherent instabilities in FAIR rings}.
\newblock In {\em Proceedings of the 9th International Computational Accelerator Physics Conference}, pages 267--270, Chamonix, France, 2006.

\bibitem{igarashi2021accelerator}
S.~Igarashi et~al.
\newblock {Accelerator design for 1.3 MW beam power operation of the J-PARC Main Ring}.
\newblock {\em Progress of Theoretical and Experimental Physics}, 2021(3):033G01, 2021.

\bibitem{koseki2012beam}
T.~Koseki et~al.
\newblock {Beam commissioning and operation of the J-PARC main ring synchrotron}.
\newblock {\em Progress of Theoretical and Experimental Physics}, 2012(1):02B004, 2012.

\bibitem{igarashi2018high}
S.~Igarashi et~al.
\newblock {High-power beam operation at J-PARC}.
\newblock In {\em Proceedings of the 61st ICFA Advanced Beam Dynamics Workshop on High-Intensity and High-Brightness Hadron Beams}, pages 147--152, Daejeon, Korea, 2018.

\bibitem{koseki2018upgrade}
T.~Koseki et~al.
\newblock {Upgrade plan of J-PARC MR-toward 1.3 MW beam power}.
\newblock In {\em Proceedings of the 9th International Particle Accelerator Conference}, pages 966--969, Vancouver, BC, Canada, 2018.

\bibitem{Yasui:2023bfx}
T.~Yasui.
\newblock {J-PARC MR operation with the high repetition rate upgrade}.
\newblock In {\em Proceedings of the 14th International Particle Accelerator Conference}, pages 1294--1298, Venice, Italy, 2023.

\bibitem{yasui2023first}
T.~Yasui, S.~Igarashi, Y.~Sato, and H.~Hotchi.
\newblock {First Result of the High Repetition Operation in J-PARC MR}.
\newblock In {\em Proceedings of the 23rd International Workshop on Neutrinos from Accelerators}, volume~8, page~23, Salt Lake City, UT, USA, 2023.

\bibitem{abe2011indication}
K.~Abe et~al.
\newblock {Indication of electron neutrino appearance from an accelerator-produced off-axis muon neutrino beam}.
\newblock {\em Physical Review Letters}, 107(4):041801, 2011.

\bibitem{abe2014observation}
K.~Abe et~al.
\newblock {Observation of electron neutrino appearance in a muon neutrino beam}.
\newblock {\em Physical Review Letters}, 112(6):061802, 2014.

\bibitem{t2k2020constraint}
K.~Abe et~al.
\newblock {Constraint on the matter-antimatter symmetry-violating phase in neutrino oscillations}.
\newblock {\em Nature}, 580(7803):339--344, 2020.

\bibitem{abe2023updated}
K.~Abe et~al.
\newblock Updated {T2K} measurements of muon neutrino and antineutrino disappearance using $3.6 \times 10^{21}$ protons on target.
\newblock {\em Physical Review D}, 108(7):072011, 2023.

\bibitem{abe2018hyper}
K.~Abe et~al.
\newblock {Hyper-Kamiokande design report}.
\newblock 2018.

\bibitem{Chao:1993zn}
A.~W. Chao.
\newblock {\em {Physics of Collective Beam Instabilities in High Energy Accelerators}}.
\newblock John Wiley \& Sons, Inc., New York, 1993.

\bibitem{Ng:2002gr}
K.~Y. Ng.
\newblock {\em {Physics of Intensity Dependent Beam Instabilities}}.
\newblock World Scientific, Singapore, 2002.

\bibitem{chin2013analysis}
Y.~H. Chin.
\newblock {Analysis of transverse instabilities observed at J-PARC MR and their suppression using feedback systems}.
\newblock In {\em Proceedings of the 25th North American Particle Accelerator Conference}, pages 27--31, Pasadena, CA, USA, 2013.

\bibitem{Konstantinova:2013uba}
O.~Konstantinova, Y.~H. Chin, Y.~Kurimoto, T.~Obina, M.~Okada, K.~Takata, M.~Tobiyama, T.~Toyama, and Y.~Shobuda.
\newblock {Bunch by bunch intra-bunch feedback system for curing transverse beam instabilities at the J-PARC MR}.
\newblock In {\em Proceedings of the 4th International Particle Accelerator Conference}, pages 1739--1741, Shanghai, China, 2013.

\bibitem{nakamura2014intra}
K.~Nakamura, T.~Toyama, M.~Okada, M.~Tobiyama, Y.~H. Chin, T.~Obina, T.~Koseki, and Y.~Shobuda.
\newblock {Intra-bunch feedback system for the J-PARC main ring}.
\newblock In {\em Proceedings of the 5th International Particle Accelerator Conference}, pages 2786--2788, Dresden, Germany, 2014.

\bibitem{nakamura2014performance}
K.~Nakamura, Makoto T., T.~Toyama, M.~Okada, Y.~H. Chin, T.~Obina, T.~Koseki, H.~Kuboki, and Y.~Shobuda.
\newblock {Performance evaluation of the intra-bunch feedback system at J-PARC main ring}.
\newblock In {\em Proceedings of the 3rd International Beam Instrumentation Conference}, pages 727--730, Monterey, CA, USA, 2014.

\bibitem{Kurimoto2011THEBB}
Y.~Kurimoto, M.~Tobiyama, Y.~H. Chin, T.~Obina, and T.~Toyama.
\newblock {The bunch by bunch feedback system in J-PARC Main Ring}.
\newblock In {\em Proceedings of the 10th European Workshop on Beam Diagnostics and Instrumentation for Particle Accelerators}, pages 482--484, Hamburg, Germany, 2011.

\bibitem{toyamaanalysis}
T.~Toyama, M.~Okada, and A.~Kobayashi.
\newblock {Analysis and upgrade plan of the transverse intra-bunch feedback system in the J-PARC MR}.
\newblock In {\em Proceedings of the 16th Annual Meeting of the Particle Accelerator Society of Japan}, pages 1130--1133, 2019, (Japanese).

\bibitem{Toyama:2022gli}
T.~Toyama, A.~Kobayashi, T.~Nakamura, M.~Okada, Y.~Shobuda, and M.~Tobiyama.
\newblock {Beam instability issue and transverse feedback system in the MR of J-PARC}.
\newblock In {\em Proceedings of the 64th ICFA Advanced Beam Dynamics Workshop on High-Intensity and High-Brightness Hadron Beams}, pages 208--212, Batavia, IL, USA, 2022.

\bibitem{Yoshimura:2022idd}
N.~Yoshimura, T.~Toyama, A.~Kobayashi, T.~Nakamura, M.~Okada, Y.~Shobuda, and T.~Nakaya.
\newblock {Evaluation for updating the intra-bunch feedback at J-PARC Main Ring}.
\newblock In {\em Proceedings of the 19th Annual Meeting of the Particle Accelerator Society of Japan}, pages 936--941, 2022, (Japanese).

\bibitem{Nakamura2024intra}
T.~Nakamura, K.~Satou, T.~Toyama, M.~Okada, A.~Kobayashi, I.~Yamada, and Y.~Shobuda.
\newblock {Development of new RFSoC based intra-bunch transverse feedback system at J-PARC MR : design and initial evaluation}.
\newblock In {\em Proceedings of the 21th Annual Meeting of the Particle Accelerator Society of Japan}, pages 881--886, 2024, (Japanese).

\bibitem{Nakamura:823297}
T.~Nakamura, S.~Dat^^c3^^a9, K.~Kobayashi, and T.~Ohshima.
\newblock {Transverse bunch-by-bunch feedback system for the SPring-8 storage ring}.
\newblock In {\em Proceedings of the 9th European Particle Accelerator Conference}, pages 2649--2651, Lucerne, Switzerland, 2004.

\bibitem{Nakamura:2018orr}
T.~Nakamura.
\newblock {Transverse and Longitudinal Bunch-by-Bunch Feedback for Storage Rings}.
\newblock In {\em Proceedings of the 9th International Particle Accelerator Conference}, pages 1198--1203, Vancouver, BC, Canada, 2018.

\bibitem{blaskiewicz1998fast}
M.~Blaskiewicz.
\newblock {Fast head-tail instability with space charge}.
\newblock {\em Physical Review Special Topics-Accelerators and Beams}, 1(4):044201, 1998.

\bibitem{shobuda2017theoretical}
Y.~Shobuda, Y.~H. Chin, P.~K. Saha, H.~Hotchi, H.~Harada, Y.~Irie, F.~Tamura, N.~Tani, T.~Toyama, Y.~Watanabe, et~al.
\newblock {Theoretical elucidation of space charge effects on the coupled-bunch instability at the 3 GeV rapid cycling synchrotron at the Japan Proton Accelerator Research Complex}.
\newblock {\em Progress of Theoretical and Experimental Physics}, 2017(1):013G01, 2017.

\bibitem{saha2018simulation}
P.~K. Saha, Y.~Shobuda, H.~Hotchi, H.~Harada, N.~Hayashi, M.~Kinsho, F.~Tamura, N.~Tani, M.~Yamamoto, Y.~Watanabe, et~al.
\newblock {Simulation, measurement, and mitigation of beam instability caused by the kicker impedance in the 3-GeV rapid cycling synchrotron at the Japan Proton Accelerator Research Complex}.
\newblock {\em Physical Review Accelerators and Beams}, 21(2):024203, 2018.

\bibitem{burov2019convective}
A.~Burov.
\newblock {Convective instabilities of bunched beams with space charge}.
\newblock {\em Physical Review Accelerators and Beams}, 22(3):034202, 2019.

\bibitem{sabbi1995simulation}
G.~L. Sabbi.
\newblock {Simulation of single-bunch collective effects in LEP by linear expansion of the distribution moments}.
\newblock Technical Report CERN SL/95-25(AP), 1995.

\bibitem{sabbi1994trisim}
G.~L. Sabbi.
\newblock {TRISIM user's guide}.
\newblock Technical Report CERN SL/94-73(AP), 1994.

\bibitem{Meller:1987ug}
R.~E. Meller, A.~W. Chao, J.~M. Peterson, S.~G. Peggs, and M.~Furman.
\newblock {Decoherence of kicked beams}.
\newblock Technical Report SSC-N-360, 1987.

\bibitem{yoshimura2025space}
N.~Yoshimura, T.~Toyama, and Y.~Shobuda.
\newblock {The space charge effects on the intra-bunch motion under large chromaticity at the Main Ring in the Japan Proton Accelerator Research Complex}.
\newblock arXiv preprint arXiv:2501.16654, 2025.

\bibitem{kapchinskij1959limitations}
I.~M. Kapchinskij and V.~V. Vladimirskij.
\newblock {Limitations of proton beam current in a strong focusing linear accelerator associated with the beam space charge}.
\newblock In {\em Proceedings of the International Conference on High Energy Accelerators and Instrumentation}, page 274, 1959.

\bibitem{zotter1975tune}
B.~Zotter.
\newblock {Tune shifts of excentric beams in elliptic vacuum chambers}.
\newblock {\em IEEE Transactions on Nuclear Science}, 22(3):1451--1455, 1975.

\bibitem{abramowitz1968handbook}
M.~Abramowitz and I.~A. Stegun.
\newblock {\em {Handbook of Mathematical Functions With Formulas, Graphs, and Mathematical Tables}}.
\newblock Dover, New York, 1974.

\bibitem{laslett1963intensity}
L.~J. Laslett.
\newblock {On intensity limitations imposed by transverse space-charge effects in circular particle accelerators}.
\newblock In {\em Proceedings of the 1963 Summer Study on Storage Rings, Accelerators and Experimentation at Super-High Energies, eConf C630610}, pages 324--367, 1963.

\bibitem{Yoshimura:2023idd}
N.~Yoshimura, T.~Toyama, Y.~Shobuda, T.~Nakamura, K.~Omi, A.~Kobayashi, M.~Okada, Y.~Sato, and T.~Nakaya.
\newblock {Study on simulation code for transverse instabilities for the J-PARC MR}.
\newblock In {\em Proceedings of the 20th Annual Meeting of the Particle Accelerator Society of Japan}, pages 260--264, 2023, (Japanese).

\bibitem{shobuda2002resistive}
Y.~Shobuda and K.~Yokoya.
\newblock {Resistive wall impedance and tune shift for a chamber with a finite thickness}.
\newblock {\em Physical Review E}, 66(5):056501, 2002.

\bibitem{yokoya1993resistive}
K.~Yokoya.
\newblock {Resistive wall impedance of beam pipes of general cross section}.
\newblock {\em Particle Accelerators}, 41:221--248, 1993.

\bibitem{gluckstern1993coupling}
R.~L. Gluckstern, J.~van Zeijts, and B.~Zotter.
\newblock {Coupling impedance of beam pipes of general cross section}.
\newblock {\em Physical Review E}, 47(1):656, 1993.

\bibitem{kobayashiinvestigation}
A.~Kobayashi, T.~Toyama, Y.~Sato, S.~Igarashi, M.~Yoshii, and Y.~Sugiyama.
\newblock {The investigation on the time structure of the wake field at the J-PARC MR}.
\newblock In {\em Proceedings of the 16th Annual Meeting of the Particle Accelerator Society of Japan}, pages 223--227, 2019, (Japanese).

\bibitem{mounet2014delphi}
N.~Mounet.
\newblock {DELPHI: An analytic Vlasov Solver for impedance-driven modes}.
\newblock Technical Report CERN Technical Report No. CERN-ACC-SLIDES-2014-0066, 2014.

\bibitem{mounet2020direct}
N.~Mounet.
\newblock {Direct Vlasov solvers}.
\newblock arXiv preprint arXiv:2006.09080, 2020.

\bibitem{shobuda2016triangle}
Y.~Shobuda, Y.~H. Chin, K.~Takata, T.~Toyama, and K.~Nakamura.
\newblock {Triangle and concave pentagon electrodes for an improved broadband frequency response of stripline beam position monitors}.
\newblock {\em Physical Review Accelerators and Beams}, 19(2):021003, 2016.

\bibitem{yoshimura2025zenodo}
N.~Yoshimura.
\newblock {Data Release for ``The space charge effects on the intra-bunch motion under large chromaticity at the Main Ring in the Japan Proton Accelerator Research Complex''}.
\newblock Zenodo (v2), 10.5281/zenodo.17152919, 2025.

\end{thebibliography}

\end{document}